\RequirePackage[l2tabu]{nag}		
\documentclass[a4paper,11pt,openbib,oldfontcommands]{memoir} 
\usepackage{datetime}
\usepackage{ifpdf}
\ifpdf
\fi
\ifdraftdoc 
	\usepackage{draftwatermark}				
	\SetWatermarkScale{0.3}
	\SetWatermarkText{\bf Draft: \today}
\fi
\newsubfloat{figure}
\newsubfloat{table}
\settrimmedsize{297mm}{210mm}{*}
\settypeblocksize{634pt}{448.13pt}{*} 
\setulmargins{4cm}{*}{*} 
\setlrmargins{*}{*}{1.5} 
\setmarginnotes{17pt}{51pt}{\onelineskip} 
\setheadfoot{\onelineskip}{2\onelineskip} 
\setheaderspaces{*}{2\onelineskip}{*} 
\checkandfixthelayout
\usepackage{fouriernc}
\usepackage[T1]{fontenc}
\OnehalfSpacing 
\setsecnumdepth{subsection} 
\maxsecnumdepth{subsubsection}
\usepackage{calc,soul,fourier}
\makeatletter 
\newlength\dlf@normtxtw 
\newsavebox{\feline@chapter} 
\newcommand\feline@chapter@marker[1][4cm]{%
	\sbox\feline@chapter{%
		\resizebox{!}{#1}{\fboxsep=1pt%
			\colorbox{gray}{\color{white}\thechapter}%
		}}%
		\rotatebox{90}{%
			\resizebox{%
				\heightof{\usebox{\feline@chapter}}+\depthof{\usebox{\feline@chapter}}}%
			{!}{\scshape\so\@chapapp}}\quad%
		\raisebox{\depthof{\usebox{\feline@chapter}}}{\usebox{\feline@chapter}}%
} 
\newcommand\feline@chm[1][4cm]{%
	\sbox\feline@chapter{\feline@chapter@marker[#1]}%
	\makebox[0pt][c]{
		\makebox[1cm][r]{\usebox\feline@chapter}%
	}}
\makechapterstyle{daleifmodif}{

	\renewcommand\printchapternum{\null\hfill\feline@chm[2.5cm]\par}

} 
\makeatother 
\chapterstyle{daleifmodif}
\makepagestyle{myvf} 
\makeoddfoot{myvf}{}{\thepage}{} 
\makeevenfoot{myvf}{}{\thepage}{} 
\makeheadrule{myvf}{\textwidth}{\normalrulethickness} 
\makeevenhead{myvf}{\small\textsc{\leftmark}}{}{} 
\makeoddhead{myvf}{}{}{\small\textsc{\rightmark}}
\newcommand{\clearemptydoublepage}{\newpage{\thispagestyle{empty}\cleardoublepage}}
\makeindex
\usepackage{import}

\usepackage{lipsum}					
\usepackage{amsfonts} 					
\usepackage[centertags]{amsmath}			
\usepackage{stmaryrd}					
\usepackage{amssymb}					
\usepackage{amsthm}					
\usepackage{newlfont}					
\usepackage{layouts}					
\usepackage{graphicx}					
\usepackage{longtable,rotating}			
\usepackage[utf8]{inputenc}			
\usepackage{colortbl}					
\usepackage{wasysym}					
\usepackage{mathrsfs}					
\usepackage{float}						
\usepackage{verbatim}					
\usepackage{upgreek }					
\usepackage{latexsym}					
\usepackage[square,numbers,
		     sort&compress]{natbib}		
\usepackage{url}						
\usepackage[spanish,english]{babel}		
\usepackage{color}                    				
\usepackage[colorlinks=true,
		     allcolors=blue]{hyperref}              
\usepackage{memhfixc}					
\usepackage{enumerate}					
\usepackage{footnote}					
\usepackage{microtype}					
\usepackage{rotfloat}					
\usepackage{alltt}						
\usepackage[version=0.96]{pgf}			
\usepackage{tikz}						
\usetikzlibrary{arrows,shapes,decorations.pathmorphing,
		       automata,backgrounds,
		       petri,topaths}				
\usepackage{pdfpages}

\newcommand{\kk}{\ensuremath{\mathbf{k}}}
\newcommand{\xx}{\ensuremath{\mathbf{x}}}

\newcommand{\Msun}{\ensuremath{M_{\odot}}}

\newcommand{\Mpch}{\ensuremath{h^{-1}{\mathrm{Mpc}}}}
\newcommand{\hMpc}{\ensuremath{h\,{\mathrm{Mpc}}^{-1}}}
\newcommand{\kpch}{\ensuremath{h^{-1}{\mathrm{kpc}}}}
\newcommand{\kms}{\ensuremath{{\mathrm {km\,s}}^{-1}}}

\newcommand{\Om}{\ensuremath{\Omega_{\mathrm m}}}
\newcommand{\Ob}{\ensuremath{\Omega_{\mathrm b}}}

\newcommand{\ns}{\ensuremath{n_{\mathrm s}}}
\newcommand{\As}{\ensuremath{A_{\mathrm s}}}

\newcommand{\rsp}{\ensuremath{r_{\mathrm{sp}}}}
\newcommand{\Vpeak}{\ensuremath{V_{\mathrm{peak}}}}
\newcommand{\Nsm}{\ensuremath{N_{\mathrm{sm}}}}
\newcommand{\lfil}{\ensuremath{L_{\mathrm{fil}}}}
\newcommand{\rfil}{\ensuremath{r_{\mathrm{fil}}}}
\newcommand{\dcut}{\ensuremath{d_{\mathrm{cut}}}}
\newcommand{\avg}[1]{\ensuremath{\left\langle \,#1\, \right\rangle}}
\newcommand{\e}[1]{\ensuremath{{\mathrm e}^{#1}}}

\newcommand{\erf}[1]{\ensuremath{{\mathrm erf}\left(#1\right)}}

\newcommand{\eqn}[1]{equation~\eqref{#1}}
\newcommand{\eqns}[1]{equations~\eqref{#1}}
\newcommand{\be}{\begin{equation}}
\newcommand{\ee}{\end{equation}}

\newcommand{\Sinhagad}{\texttt{Sinhagad}}
\newcommand{\Sahyadri}{\texttt{Sahyadri}}

\newcommand{\filgen}{\texttt{FilGen}}
\newcommand{\filapt}{\texttt{FilAPT}}
\newcommand{\filtools}{\texttt{FilTools}}
\newcommand{\skeletor}{\texttt{Skeletor}}
\newcommand{\disp}{DisPerSE}

\newcommand{\myheading}[1]{%
    \par\addvspace{4mm}%
    \noindent\textbf{\large #1}\par
    \addvspace{2mm}%
}

\newcommand{\mysubheading}[1]{%
    \par\addvspace{0.5\baselineskip}%
    \noindent\textbf{#1}\par
    \addvspace{0.2\baselineskip}%
}
\newcommand{\pgftextcircled}[1]{                                                                    
    \setbox0=\hbox{#1}%
    \dimen0\wd0%
    \divide\dimen0 by 2%
    \begin{tikzpicture}[baseline=(a.base)]%
        \useasboundingbox (-\the\dimen0,0pt) rectangle (\the\dimen0,1pt);
        \node[circle,draw,outer sep=0pt,inner sep=0.1ex] (a) {#1};
    \end{tikzpicture}
}
\newcommand{\blackged}{\hfill$\blacksquare$}
\newcommand{\whiteged}{\hfill$\square$}
\newcounter{proofcount}

\let\oldsqrt\sqrt
\def\sqrt{\mathpalette\DHLhksqrt}
\def\DHLhksqrt#1#2{%
\setbox0=\hbox{$#1\oldsqrt{#2\,}$}\dimen0=\ht0
\advance\dimen0-0.2\ht0
\setbox2=\hbox{\vrule height\ht0 depth -\dimen0}%
{\box0\lower0.4pt\box2}}
\newcommand{\mycaption}[2][\@empty]{
	\captionnamefont{\scshape} 
	\changecaptionwidth
	\captionwidth{0.9\linewidth}
	\captiondelim{.\:} 
	\indentcaption{0.75cm}
	\captionstyle[\centering]{}
	\setlength{\belowcaptionskip}{10pt}
	\ifx \@empty#1 \caption{#2}\else \caption[#1]{#2}
}
\newcommand{\mysubcaption}[2][\@empty]{
	\subcaptionsize{\small}
	\hangsubcaption
	\subcaptionlabelfont{\rmfamily}
	\sidecapstyle{\raggedright}
	\setlength{\belowcaptionskip}{10pt}
	\ifx \@empty#1 \subcaption{#2}\else \subcaption[#1]{#2}
}
\usepackage{lettrine}
\newcommand{\initial}[1]{%
	\lettrine[lines=3,lhang=0.33,nindent=0em]{
		\color{gray}
     		{\textsc{#1}}}{}}
\theoremstyle{plain}

\theoremstyle{plain}

\theoremstyle{plain}
\theoremstyle{definition}

\theoremstyle{plain}

\theoremstyle{plain}

\theoremstyle{plain}

\newif\ifonechapter
\onechaptertrue   
\begin{document}
%
%
%
%
%
\frontmatter
\pagenumbering{roman}

%
%
%
%
%
%
\begin{titlingpage}
\begin{SingleSpace}
\calccentering{\unitlength} 
\begin{adjustwidth*}{\unitlength}{-\unitlength}
\vspace*{8mm}
\begin{center}
\rule[0.5ex]{\linewidth}{2pt}\vspace*{-\baselineskip}\vspace*{3.2pt}
\rule[0.5ex]{\linewidth}{1pt}\\[\baselineskip]
{\HUGE Cosmic Velocity Flows: }\\[4mm]
{\LARGE {from Theory to Observations}}\\
\rule[0.5ex]{\linewidth}{1pt}\vspace*{-\baselineskip}\vspace{3.2pt}
\rule[0.5ex]{\linewidth}{2pt}\\
\vspace{6.5mm}
{\large By}\\
\vspace{6.5mm}
{\large\textsc{Dhawalikar Saee Mahesh}}\\
\vspace{3mm}
{\large Supervisor: \textsc{Prof. Aseem Paranjape}}\\
\vspace{11mm}
\includegraphics[scale=0.06]{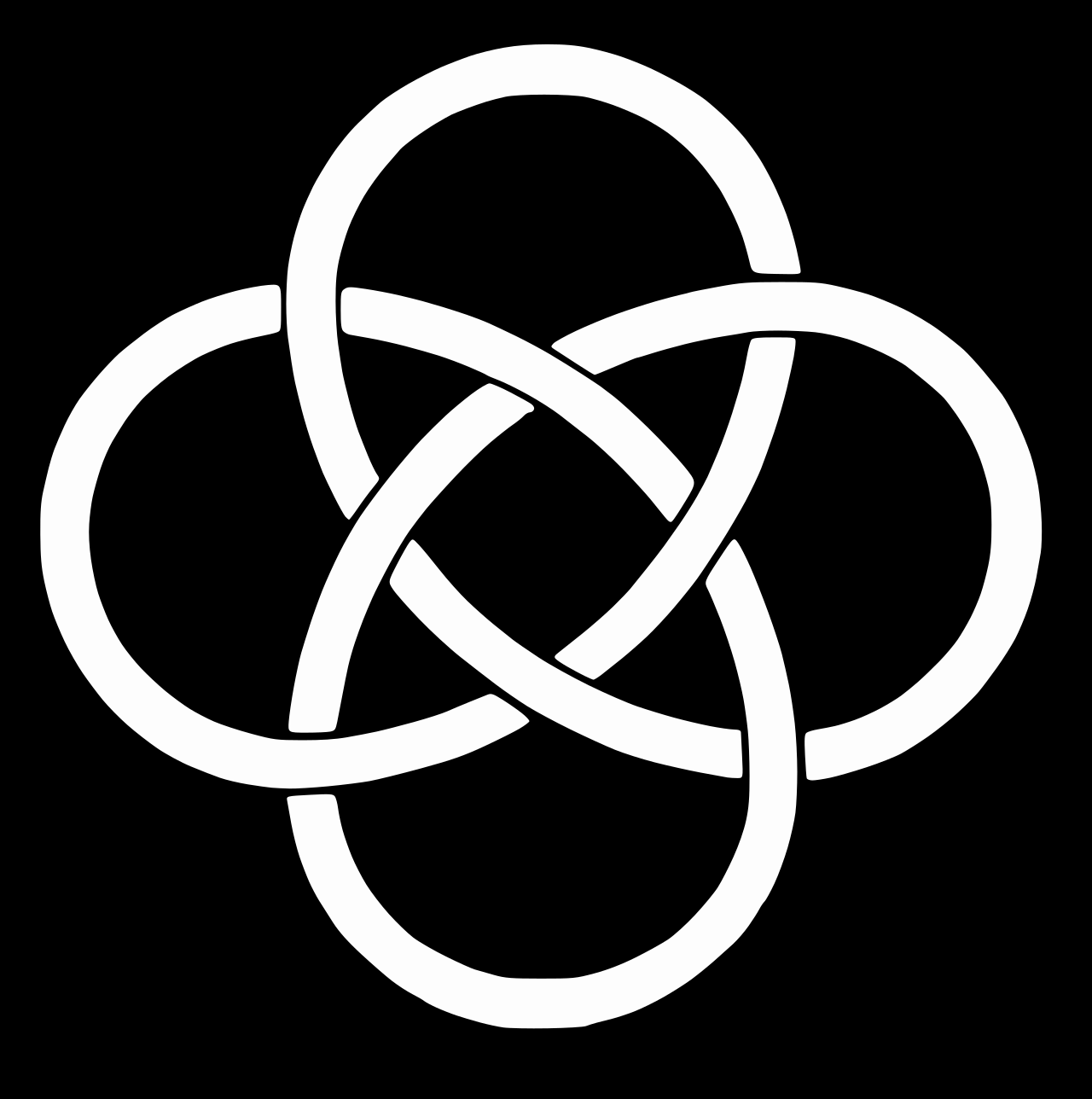}\\
\vspace{6mm}
{\large Inter-University Centre for Astronomy and Astrophysics (IUCAA)} \\
\vspace{11mm}
\renewcommand{\baselinestretch}{1.5}\selectfont
A thesis submitted to Jawaharlal Nehru University\\
in fulfillment of the requirements for the degree of\\
DOCTOR OF PHILOSOPHY (Physics)\\
\vspace{5mm}
\includegraphics[scale=0.15]{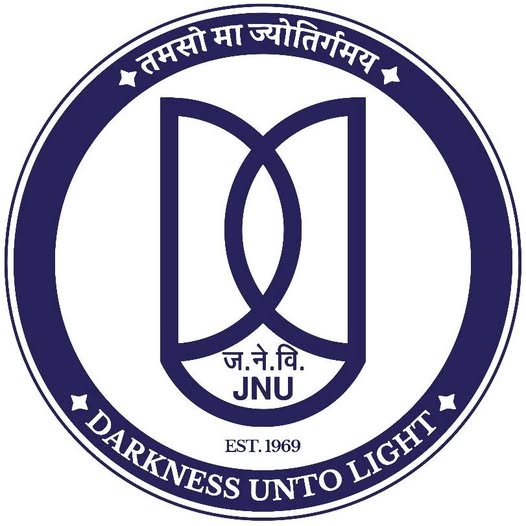}\\
\vspace{9mm}
{\large\textsc{July 2026}}
\vspace{12mm}
\end{center}
\begin{flushright}
\end{flushright}
\end{adjustwidth*}
\end{SingleSpace}
\end{titlingpage}
\clearemptydoublepage
\includepdf[pages=1,scale=0.9]{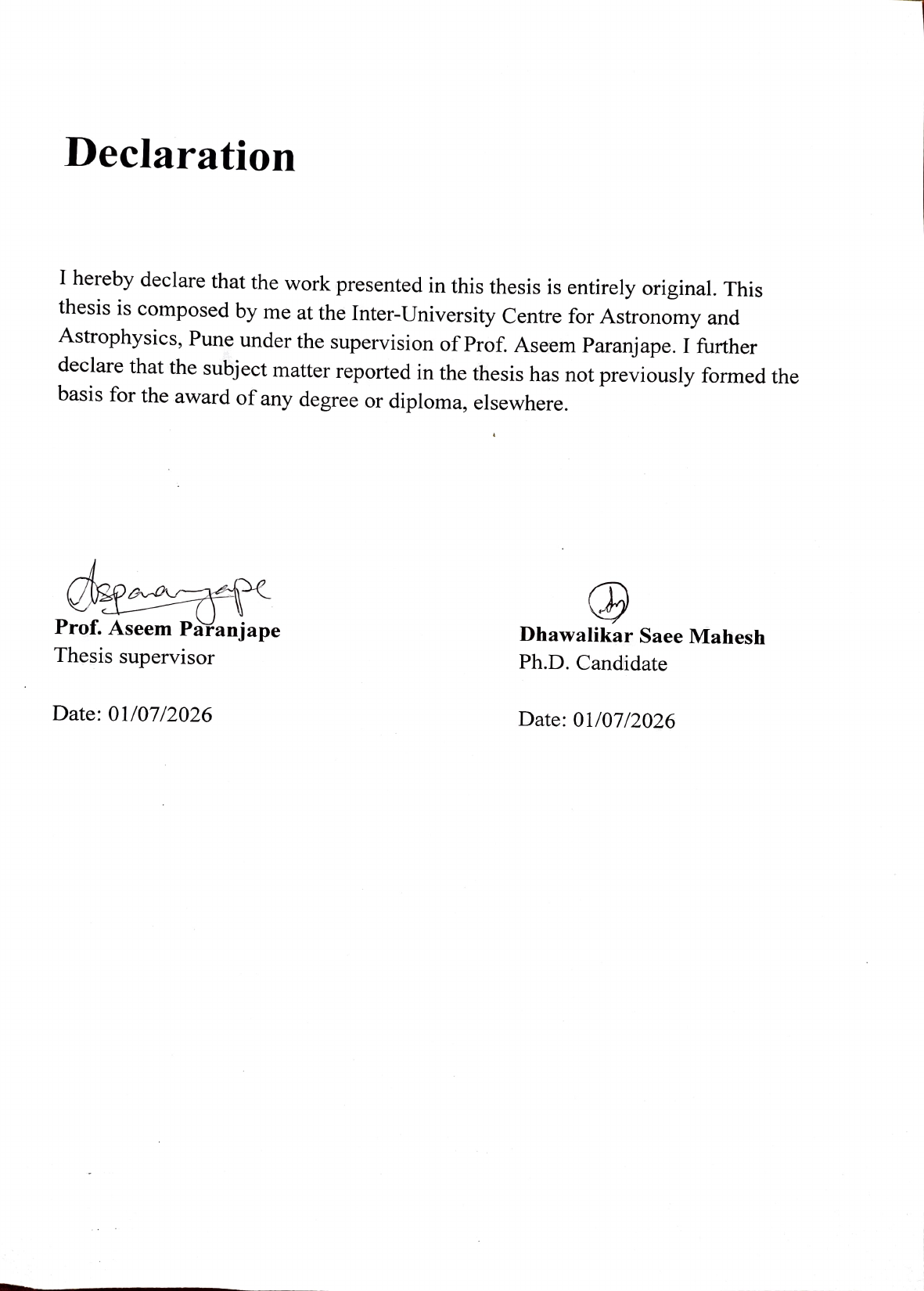}
\includepdf[pages=1,scale=0.9]{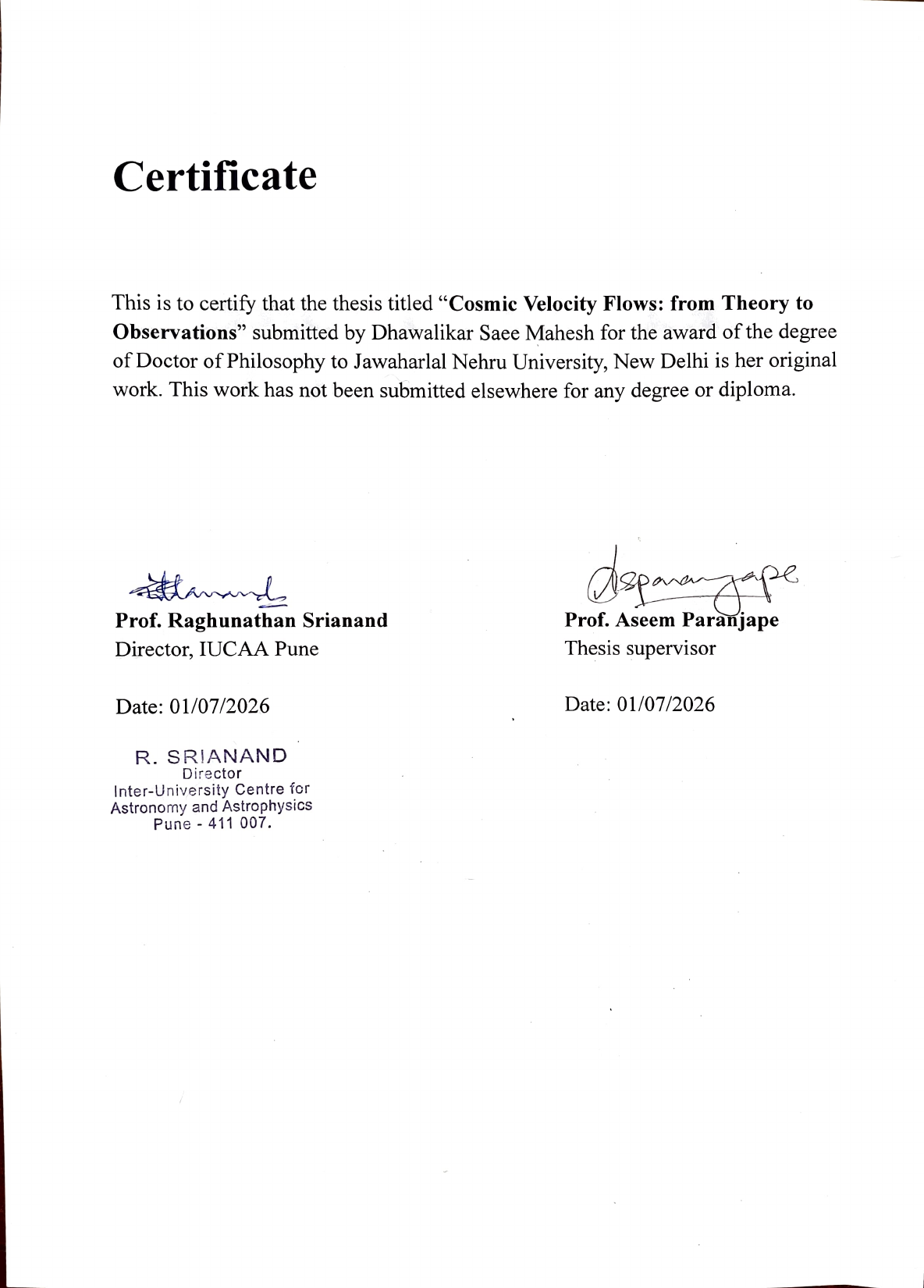}

\clearemptydoublepage

\chapter*{Acknowledgements}

Over the last few years I have been fortunate to learn from, work with, and be supported by many people. They have influenced this thesis in different ways: through discussions and collaborations, through advice and encouragement, and by making my time at IUCAA richer and more enjoyable.  It is a pleasure to acknowledge them here.

I would first like to thank my advisor, Prof. Aseem Paranjape, for guiding me throughout my Ph.D. Thank you for giving me the freedom to explore my own ideas, no matter how weird they sometimes seemed, and for always encouraging independent thinking. I have learned a great deal from our many discussions over the years, and I am grateful for your constant support and encouragement.

I am grateful to my collaborators, Prof. Shadab Alam and Prof. Arka Banerjee. I owe a special thanks to Shadab for his support throughout this work. I have learned a tremendous amount from him, particularly about observations. I have always appreciated how generously he shared his time, and how easily he could explain even complicated ideas without ever making them seem intimidating. I also thank my RAC members, Shadab and Prof.\ Surhud More, for their comments, suggestions, and encouragement over the course of my Ph.D. I had the opportunity to discuss parts of this work with Prof. Ravi Sheth, and those discussions provided valuable insights into several aspects of this thesis. Finally, I thank everyone who participated in the PMCAP meetings, where the idea of the Sahyadri simulation suite first took shape. The meetings provided a wonderful environment for discussions on a wide range of topics, many of which influenced different aspects of this thesis.

I thank the Director of IUCAA for providing a wonderful research environment. I would also like to acknowledge the Pegasus computing facility, without which a large part of the work presented in this thesis would not have been possible.

IUCAA has been a wonderful place to spend the last few years, thanks to my colleagues and friends. I especially thank my cosmology seniors, Prem, Divya and Bhaskar, for the many discussions we had, for patiently answering questions, and for always being willing to help, particularly during the early days of my PhD. I also thank Prathamesh, Sourabh, Prakash, Snehil, and all my batchmates for the many maths and physics discussions that rarely reached a conclusion, but certainly made the days more interesting.

Priyanka and Balpreet, thank you for all the fun, the food, and the many conversations that always gave me a different perspective on things. Eshita, I really do not know what I would have done without you. Thank you for the endless coffee breaks, the random conversations, for patiently listening to all my ``unga bunga'' thoughts, and for making IUCAA feel like home. Shreeya, thank you for being the one constant friend over the past decade. No matter the distance or the time-zone difference, you have always been there to listen, and I am incredibly grateful for that.

Prathamesh, these last few years would have been very different without you. From scientific debates to completely non-scientific ones, from good food to constant encouragement, I could always count on you. I couldn't have asked for a better partner. Mukai, you've always had a way of putting things back into perspective. Thank you for being the elder sister I could always rely on, and for patiently dealing with my tendency to overthink absolutely everything. Atharva, you're the person I know I can go to when everything else stops making sense. Somehow you always manage to make me laugh, and you've taught me, without ever trying to, not to take work and life quite so seriously.

Finally, Aai and Baba. Thank you for everything you have done for me, not just during these past few years but throughout my life. Thank you for encouraging me to follow my interests, for believing in me even when I doubted myself, and for always being there.

\clearemptydoublepage

\chapter*{List of Publications}
\begin{SingleSpace}
\vspace{1.5cm}

\noindent{\large\bfseries Publications contributing to this thesis}
\begin{enumerate}
    \item \textbf{``Towards unbiased recovery of cosmic filament properties: the role of spine curvature and optimized smoothing ''}\\
    \href{https://iopscience.iop.org/article/10.1088/1475-7516/2024/09/041}{Journal of Cosmology and Astroparticle Physics (JCAP), Volume 2024, Issue 09, id.041, 38 pp.}\\
    \href{https://arxiv.org/abs/2402.18669}{[arXiv:2402.18669]}\\
    \textbf{Saee Dhawalikar}, Aseem Paranjape

    \vspace{0.4cm}

    \item \textbf{``Sahyadri: A simulation suite for the cosmology dependence of the Cosmic Web''}\\
    Accepted by Journal of Cosmology and Astroparticle Physics (JCAP)\\
    \href{https://arxiv.org/abs/2601.07924}{[arXiv:2601.07924]}\\
    \textbf{Saee Dhawalikar}, Shadab Alam, Aseem Paranjape, Arka Banerjee
\end{enumerate}

\vspace{1cm}

\noindent{\large\bfseries Other publications}

\vspace{0.5cm}

\begin{enumerate}
    \item \textbf{`` Stabilizing simulation-based cosmological Fisher forecasts: a case study using the Voronoi volume function''}\\
    \href{https://iopscience.iop.org/article/10.1088/1475-7516/2026/03/040}{Journal of Cosmology and Astroparticle Physics, Volume 2026, Issue 03, id.040, 23 pp.} \\
    \href{https://arxiv.org/abs/2506.16408}{[arXiv:2506.16408]}\\
    \textbf{Saee Dhawalikar}, Aseem Paranjape, Shadab Alam
\end{enumerate}

\end{SingleSpace}
\clearpage
\clearemptydoublepage

\chapter*{Synopsis}
\begin{SingleSpace}
\initial{T}he Large-Scale Structure of the Universe is organised into a complex, interconnected network of nodes, filaments, sheets, and voids collectively known as the Cosmic Web. This structure was first revealed through early redshift surveys such as the CfA survey, which uncovered striking filamentary arrangements of galaxies interwoven with large underdense voids. These observations established that the Universe is not a simple distribution of isolated galaxies, but a highly structured and anisotropic network shaped by gravitational evolution across cosmic time.

The origin of this structure lies in tiny primordial density fluctuations generated in the early Universe. Through gravitational instability, these initially small perturbations evolve into the highly non-linear cosmic web observed today. Thus, the cosmic web encodes information about fundamental physics, including the nature of dark matter, dark energy, and gravity as well as primordial physics.

Modern galaxy surveys now probe this structure with increasing precision, mapping both the spatial distribution of galaxies and their redshifts, the latter being influenced by peculiar velocities. This represents a shift in cosmological inference: from relying primarily on spatial clustering statistics toward a more complete dynamical description of structure formation. While two-point statistics have been remarkably successful in establishing the $\Lambda$CDM model, they are approaching the cosmic variance limits on large scales and fail to fully describe the complexity of non-linear gravitational evolution on smaller scales. A significant fraction of cosmological information is therefore expected to reside in higher-order structure, geometry, and phase-space dynamics rather than in simple traditional clustering measures.

In the quasi-linear and non-linear regime, gravitational evolution produces a highly complex coupling between density and velocity fields, giving rise to filamentary structure and multi-streaming flows. While this regime is expected to contain a significant fraction of cosmological information, it is also difficult to model and interpret reliably. This is primarily due to the strongly non-linear nature of structure formation and the sensitivity of inferred structures to the methods used to reconstruct them from discrete and biased tracers.\\
\\
Addressing these challenges requires a combination of improved theoretical understanding, carefully designed cosmological simulations, and advanced statistical methods. The work presented in this thesis is centred on understanding the non-linear cosmic web through its filamentary structure and associated velocity flows. The primary goal is to develop robust methods for identifying and characterising filamentary structure, and to use these methods to study the phase-space behaviour of cosmic filaments using N-body simulations. Since filaments act as the primary channels through which matter flows across the cosmic web, particular emphasis is placed on understanding their velocity structure. These studies are complemented by the development of the \Sahyadri\ suite, which provides a high-resolution numerical framework for studying non-linear structure formation and its cosmological dependence. Together, these developments enable a detailed exploration of the geometry, hierarchy, and velocity structure of filaments within the non-linear cosmic web. \\
\\
This thesis is organised as follows. \textbf{Chapter~2} introduces and reviews cosmological simulations in general, summarising their role in studying non-linear structure formation and their importance for interpreting current and upcoming galaxy surveys. The chapter then presents the \Sahyadri\ suite of cosmological $N$-body simulations, developed in this work, designed to enable precision studies of the low-redshift Universe with next-generation spectroscopic surveys.

\Sahyadri\ includes systematic variations of six cosmological parameters around Planck 2018 constraints, with seed-matched initial conditions enabling cosmological parameter derivatives. Each simulation evolves $2048^3$ particles in a periodic box of side length $200 \Mpch$, yielding a particle mass of $m_p=8.1\times10^7 h^{-1}\Msun$ in the fiducial Planck 2018 cosmology. This resolution represents a factor of $\sim 25$ improvement over the AbacusSummit suite, and is over two orders of magnitude better than the Quijote and Aemulus suites in terms of minimum mass of resolved halos. The simulations are constructed with higher mass resolution than many existing parameter-varying suites while maintaining sufficiently large volumes to capture representative cosmic environments.

This combination allows simultaneous study of halo-scale structure and the surrounding cosmic web. Standard statistics such as the matter power spectrum and halo mass function are used for validation, while beyond 2-point statistics such as the Voronoi volume function and $k$th nearest neighbour statistics are explored, which are sensitive to non-Gaussian aspects of the matter distribution. These also show significant sensitivity to $\Om$ variations. Designed for precision cosmology, the \Sahyadri\ suite enables detailed studies of cosmological dependence of nonlinear structure formation and cosmic web statistics relevant for upcoming surveys.\\
\\
Building on this numerical foundation, \textbf{Chapter~3} addresses the problem of cosmic web reconstruction, with a particular focus on filament identification. A wide range of filament-finding approaches exists in the literature, including density-based, topological, graph-based, and phase-space methods. However, these approaches often yield significantly different filament networks when applied to the same underlying data, reflecting the absence of a unique physical definition of filaments. This introduces a fundamental “chicken-and-egg” problem: the true filamentary structure is not known a priori, making it difficult to quantify reconstruction biases in a controlled way.

To address this, we develop a calibration framework based on controlled filament realizations. The first component, the {\bf Fil}ament {\bf Gen}erator (\filgen), constructs mock filaments with known spine geometry, density profiles, and velocity structure, thereby providing a controlled “ground truth” for evaluating filament reconstruction methods. The second component, the {\bf Fil}ament {\bf A}nalysis and {\bf P}rocessing {\bf T}ool (\filapt), measures density and velocity profiles around reconstructed filament spines and enables systematic studies of reconstruction biases, including those arising from smoothing, sampling, and filament curvature.

Using this framework, we identify two important and previously unexplored sources of systematic uncertainty in filament studies. First, we demonstrate that filament curvature can significantly bias inferred density and velocity profiles through purely geometric effects, even when the filament spine is perfectly known. Second, we quantify the impact of reconstruction noise on filament profile measurements and introduce a novel Fourier-space smoothing framework to address it. Combined with an objective filament-by-filament optimization procedure, this approach substantially improves profile recovery while avoiding the ambiguities associated with conventional smoothing techniques. Together, these developments provide a robust framework for calibrating filament finders and for obtaining unbiased measurements of filament structure and dynamics.\\
\\
\textbf{Chapter~4} focuses on the development of a new filament-finding framework motivated by both physical and observational consideration. The cosmic web is intrinsically hierarchical, with filamentary structures spanning a continuum of scales and often embedded within larger filamentary environments. Any physically meaningful reconstruction must therefore account for this hierarchy while remaining robust to sparse and biased tracers. At the same time, observational applications demand methods that rely as little as possible on information from the underlying dark matter distribution. These considerations motivate the development of \skeletor, a novel Voronoi-based hierarchical filament finder that uses only tracer positions and masses as its primary inputs, incorporates hierarchical structure directly into its construction, and is complemented by a dedicated filament substructure classifier for identifying nested filamentary systems. The method is motivated by the local anisotropic structure of the cosmic web and identifies filamentary regions directly from discrete tracer distributions without requiring interpolation onto a grid. Local anisotropy is quantified using the Voronoi tessellation of the tracer field, enabling a tensor-based characterization of directional matter distribution that isolates filament-like environments. Connected filamentary skeletons are then reconstructed from these anisotropy-selected regions, providing a continuous representation of cosmic filament spines.

\skeletor\ also incorporates dark matter information, when available, in two complementary ways. First, it is used to refine the reconstructed filament spine by improving alignment with the underlying matter distribution. Second, it enables a dynamical estimate of filament extent through radial infall velocities, where the filament radius is defined via the location of maximum coherent infall. Illustrative results demonstrate both accurate reconstruction of filament spines and recovery of their radial profiles, consistent with previous studies.\\
\\
\textbf{Chapter~5} applies \skeletor\ to $N$-body simulations to study sub-filamentary structure within the cosmic web. Using the substructure classification framework introduced earlier, filamentary substructures embedded within larger parent filaments are systematically identified and characterised. These sub-filaments exhibit distinct statistical and geometric properties, highlighting how sub-filaments are distinctly different from their parent counterparts.

The analysis demonstrates how hierarchical decomposition can be used to resolve physically meaningful substructure within filaments, providing a more detailed view of filamentary environments in simulations. More generally, this chapter illustrates how hierarchical reconstruction approaches can be used to refine our description of filamentary environments and to support more detailed studies of structure formation in complex, non-linear regimes.\\
\\
The dynamics of filaments is then studied in \textbf{Chapter~6} using $N$-body simulations analysed with \skeletor. This chapter focuses on the phase-space structure of filaments, examining both density and velocity fields in detail. Radial and longitudinal density profiles are measured alongside radial and tangential velocity profiles and distributions. The results reveal coherent anisotropic inflows toward filament spines, consistent with ongoing non-linear gravitational collapse, as well as rich multistreaming behaviour within filament interiors. 

The combination of the analysis framework developed in Chapter~3 and the robust filament reconstruction provided by \skeletor\ enables a particularly clean view of filament phase space. Several features expected from theoretical models of anisotropic gravitational collapse, including coherent infall patterns, velocity transitions, and caustic-like structures, are clearly identified in the measured profiles. Velocity-based diagnostics, including maximum radial infall and transitions in velocity dispersion, are further used to define physically motivated filament boundaries, linking geometric reconstruction to underlying dynamical structure.

An important motivation for this work is the possibility that cosmic filaments may exhibit simple and potentially universal density and velocity structures, analogous to the NFW profile in dark matter halos. Identifying such universality would greatly simplify the description of filamentary environments and provide new insight into the physical processes governing the growth of the cosmic web.\\
\\
Finally, \textbf{Chapter~7} summarizes the principal results of the thesis and discusses their implications for future studies of the non-linear cosmic web. The thesis introduces the \Sahyadri\ suite for precision studies of non-linear structure formation, develops a novel framework for calibrating filament reconstruction methods and improving the robustness of filament analyses, and presents \skeletor, a filament-finding framework that explicitly accounts for the hierarchical nature of the cosmic web and enables the identification of embedded sub-filamentary systems. These developments are then brought together to study the density and velocity structure of cosmic filaments and to reveal dynamical features associated with anisotropic gravitational collapse and matter transport through the cosmic web.

The chapter concludes by discussing future directions, including the search for universal filament profiles, semi-analytical descriptions of filament dynamics, and the incorporation of filamentary environments into models of redshift-space distortions. As upcoming galaxy surveys increasingly probe the quasi-linear and non-linear regimes, robust methods for reconstructing and characterising cosmic web environments will become increasingly important. The tools and methodologies developed in this thesis provide a foundation for such studies and for future attempts to connect the geometry and dynamics of the cosmic web to fundamental cosmology.

\end{SingleSpace}
\clearpage
\clearemptydoublepage

%
\renewcommand{\contentsname}{Table of Contents}
\maxtocdepth{subsection}
\tableofcontents*
\addtocontents{toc}{\par\nobreak \mbox{}\hfill{\bf Page}\par\nobreak}
\clearemptydoublepage
\listoftables
\addtocontents{lot}{\par\nobreak\textbf{{\scshape Table} \hfill Page}\par\nobreak}
\clearemptydoublepage
\listoffigures
\addtocontents{lof}{\par\nobreak\textbf{{\scshape Figure} \hfill Page}\par\nobreak}
\clearemptydoublepage
%
%
\mainmatter
\ifonechapter
    %
%
\let\textcircled=\pgftextcircled
\chapter{Introduction}
\label{chap:introduction}

\initial{O}n sufficiently large scales, the Universe appears remarkably uniform. Yet observations over the past century have revealed that matter is distributed in an intricate network of nodes, filaments, sheets and voids collectively known as the Cosmic Web. Early redshift surveys such as the CfA survey revealed striking filamentary patterns and large underdense voids in the galaxy distribution, including structures such as the ``Great Wall'' extending over hundreds of Megaparsecs \citep{Lapparent+1986}. These observations established that matter in the Universe is organized into a complex interconnected network rather than a random distribution of galaxies. With increasingly deep redshift surveys, these patterns emerged in striking detail, revealing structures extending across hundreds of Megaparsecs and establishing the Large-Scale Structure of the Universe as one of the central probes of modern cosmology.

In the standard cosmological picture, the Universe evolved from an initially hot, dense and nearly homogeneous state. Tiny primordial density fluctuations, seeded in the early Universe, subsequently grew through gravitational instability to form the highly nonlinear web observed today. Because the growth of structure is governed by the properties of dark matter, dark energy, gravity and the initial conditions of the Universe, the cosmic web becomes a probe of fundamental physics. Modern galaxy surveys now measure the distribution and motions of galaxies with unprecedented precision, opening the possibility of probing structure formation deep into the quasi-linear and nonlinear regimes.

A significant fraction of the cosmological information accessible to these surveys lies precisely in these nonlinear regimes, where theoretical modelling becomes increasingly difficult. On the largest scales, measurements of the power spectrum are already approaching the cosmic variance limit, motivating growing interest in extracting information from smaller scales and from alternative probes of structure formation. Peculiar velocity flows, anisotropic gravitational collapse, galaxy bias, gas dynamics and baryonic feedback processes all contribute to the complexity of the observed galaxy distribution. Understanding these effects requires a combination of analytical modelling, numerical simulations and improved statistical descriptions of structure in both density and velocity fields. In particular, filamentary structure and the dynamics of the cosmic web provide information about nonlinear gravitational evolution and environmental dependence that is only partially captured by conventional clustering statistics. This thesis is concerned with these aspects of structure formation, with particular emphasis on filamentary structure, cosmic velocity flows and simulation-based studies of the Large-Scale Structure.

This chapter sets the stage for these discussions by outlining the broad theoretical and observational context. In particular, it highlights the role of peculiar velocities and redshift-space distortions as cosmological probes, the usefulness of filamentary structure as a tracer of the cosmic web, and the challenges involved in modelling the quasi-linear and nonlinear regimes. These challenges motivate the simulation-driven analyses and methodological developments, including the filament-finding and analysis framework developed in this work and the \Sahyadri\ simulation suite, presented in the chapters that follow.
\section{Large-Scale Structure in the Era of Precision Cosmology}
\label{sec:LSS_precision_cosmology}
The Large-Scale Structure (LSS) of the Universe encompasses the distribution of matter over scales ranging from individual galaxies and dark matter halos to the largest connected structures observed in the Universe. It provides a powerful observational probe of both cosmology and galaxy formation. Measurements of the LSS constrain the composition and expansion history of the Universe, the growth of structure under gravity, and the connection between galaxies and the underlying matter distribution. At the same time, the geometry and dynamics of the cosmic web offer insights into the assembly histories of dark matter halos and the processes governing galaxy evolution.

The large-scale dynamics of the Universe are described within the Friedmann–Lemaître–Robertson–Walker (FLRW) framework, based on the assumption that the Universe is statistically homogeneous and isotropic on sufficiently large scales \citep{Dodelson&Schmidt}. The corresponding spacetime interval ($ds$) is given by
\begin{equation}
ds^2 = -c^2dt^2 + a^2(t)\left[\frac{dr^2}{1-kr^2} + r^2 d\Omega^2 \right],
\end{equation}
where $c$ is the speed of light, $t$ is cosmic time, $a(t)$ is the scale factor describing the expansion of the Universe, and $(r,\theta,\phi)$ are comoving spatial coordinates, with $d\Omega^2 = d\theta^2 + \sin^2\theta\, d\phi^2$ representing the differential solid angle. The parameter $k$ determines the spatial curvature of the Universe, corresponding to open ($k<0$), flat ($k=0$) and closed ($k>0$) geometries. The expansion history is governed by the Friedmann equations:
\begin{equation}
    H^2(a) \equiv \left(\frac{\dot a}{a}\right)^2 = \frac{8\pi G}{3}\rho -
    \frac{kc^2}{a^2},
\end{equation}
and
\begin{equation}
    \frac{\ddot a}{a} = -\frac{4\pi G}{3}
    \left(\rho + \frac{3P}{c^2}\right),
\end{equation}
where $\dot{} \equiv d/dt$, $G$ is Newton's gravitational constant, $\rho$ is the total energy density, and $P$ is the corresponding pressure associated with its matter and energy components. Together, these equations describe the evolution of the scale factor under the influence of the matter and energy content of the Universe. Within the standard $\Lambda$CDM model, the energy density today is dominated by cold dark matter and dark energy, while structure formation proceeds through the gravitational amplification of initially small density perturbations. In the linear regime, the evolution of these perturbations can be described analytically using cosmological perturbation theory, allowing accurate predictions for the growth of structure on sufficiently large scales.

Over the last few decades, observations of the LSS, together with measurements of the Cosmic Microwave Background(CMB), weak lensing and Type-Ia supernovae, have established a cosmological model consistent with nearly scale-invariant primordial fluctuations evolving through gravitational instability in a Universe containing cold dark matter and a dominant late-time dark energy component \citep[e.g.][]{Planck18-VI-cosmoparam, SDSS_2017}. Large cosmological simulations within this framework have successfully reproduced many observed statistical properties of the matter distribution, including the abundance and clustering of galaxies, the emergence of the cosmic web and the baryon acoustic oscillation (BAO) feature in galaxy clustering \citep[e.g.][]{Springel2005,Angulo+2008}. At the same time, several fundamental questions remain unanswered, including the nature of dark matter and dark energy, the physical origin of the inflationary era and primordial perturbations, the neutrino mass hierarchy and the validity of general relativity on cosmological scales. Addressing these questions requires extracting cosmological information across a wide range of spatial scales. On large scales corresponding roughly to $k \lesssim 0.05\,\Mpch$, density fluctuations remain sufficiently small for linear perturbation theory to provide an accurate description of structure growth. Also, at these scales, the matter distribution is well described statistically by two-point measures such as the power spectrum and the two-point correlation function which have become central tools of cosmology through LSS.  On smaller scales, however, nonlinear gravitational evolution couples Fourier modes and progressively transfers information into higher-order correlations that are not fully captured by conventional two-point statistics. These effects motivate the study of the cosmic web as a complementary probe of large-scale structure.

\begin{figure}
    \centering
    \includegraphics[width=0.5\linewidth]{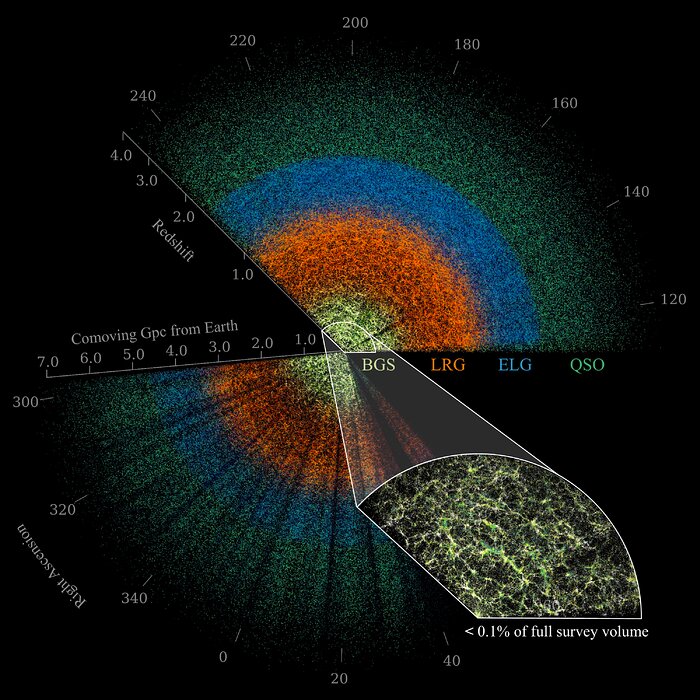}
    \caption{Visualization of the large-scale galaxy and quasar distribution mapped by the DESI survey. Colours correspond to different classes of tracers employed in the survey. The inset zooms into a small section of the survey volume, highlighting the filamentary patterns that make up the cosmic web. Credit: DESI Collaboration/DOE/KPNO/NOIRLab/NSF/AURA/C. Lamman.}
    \label{fig:DESI_slice}
\end{figure}

Large-scale structure observations are also closely connected to problems in galaxy formation and evolution. Halo and galaxy properties are known to correlate not only with mass, but also with environment and assembly history, giving rise to phenomena such as assembly bias and environmentally dependent galaxy evolution \citep[e.g.][]{Sheth&Tormen2004,Wechsler&Tinker2018, phs18}. These connections have motivated growing interest in understanding the role of the cosmic web in shaping the formation and evolution of galaxies and dark matter halos.

The observational study of LSS has undergone a major transformation with successive generations of increasingly large and precise galaxy surveys. The Sloan Digital Sky Survey (SDSS) established precision measurements of galaxy clustering and the baryon acoustic oscillation (BAO) feature using millions of galaxies distributed across cosmological volumes spanning several cubic gigaparsecs \citep{SDSS2000, SDSS_BAO_2005}. Subsequent surveys such as BOSS extended these measurements to redshifts $z\sim0.7$, while DESI is mapping the three-dimensional distribution of tens of millions of galaxies and quasars over roughly 14,000 square degrees, spanning redshifts from the nearby Universe to $z\gtrsim3$ using multiple tracer populations \citep{SDSS_2017, DESIDR1_cosmo}. Future surveys including Rubin-LSST \citep{Ivezic2019}, Euclid \citep{EuclidCollaboration2022} and the Square Kilometre Array (SKA) \citep{SKA2009} are expected to probe unprecedented cosmological volumes using imaging, spectroscopy and 21-cm observations, together mapping billions of galaxies across much of the observable Universe. These surveys now probe clustering over scales ranging from nonlinear halo environments on sub-megaparsec scales to baryon acoustic oscillation features at scales of roughly 150Mpc, enabling precision tests of cosmological models across an enormous dynamical range. The resulting galaxy maps provide a direct observational view of the cosmic web across a vast range of scales and redshifts, as illustrated in Figure~\ref{fig:DESI_slice}.

The statistical precision of modern surveys has now reached the level where theoretical uncertainties associated with nonlinear structure formation, galaxy bias and baryonic effects have become comparable to or larger than observational uncertainties on several scales. Moreover, measurements on the largest linear scales are approaching the cosmic variance limit, motivating growing interest in extracting additional information from smaller nonlinear scales and from statistics beyond the conventional two-point framework.

The increasing statistical power of these surveys has transformed the LSS into a precision cosmological probe. Measurements of the baryon acoustic oscillation (BAO) feature, imprinted by sound waves in the primordial baryon-photon plasma prior to recombination, provide a standard ruler for constraining the expansion history of the Universe, while weak gravitational lensing and redshift-space distortions probe the growth of structure and the dynamics of matter flows. Together, these observables provide stringent tests of the standard $\Lambda$CDM framework as well as possible extensions involving modified gravity, evolving dark energy or massive neutrinos \citep[e.g.][]{Weinberg+2013}. While much of the cosmological information extracted so far has relied on large-scale clustering statistics, modelling smaller nonlinear scales remains one of the major challenges in precision cosmology.

Extracting cosmological information from the nonlinear regime presents significant theoretical and numerical challenges. Nonlinear gravitational evolution generates strong mode coupling and non-Gaussian structure in the density and velocity fields, while galaxy bias, baryonic feedback and peculiar velocity flows complicate the connection between observable tracers and the underlying matter distribution. In particular, accurately modelling redshift-space distortions, environmental effects and baryonic modifications to matter clustering remains difficult in several regimes despite substantial theoretical and numerical progress \citep[e.g.][]{Joachimi+2015,Percival+2011,Daalen+2011}. 

A broad range of analytical approaches has been developed to model the weakly nonlinear evolution of large-scale structure. In Eulerian perturbation theory (EPT), the density and velocity fields are expanded perturbatively about the homogeneous background in Eulerian space, yielding systematic corrections to the linear matter power spectrum and higher-order clustering statistics \citep[e.g.][]{Bernardeauetal02}. Such approaches accurately describe matter clustering only on comparatively large scales, typically $k\leq 0.1\Mpch$ at low redshift, where density fluctuations remain small. Lagrangian perturbation theory (LPT), in contrast, follows the trajectories of fluid elements and naturally captures coherent bulk flows, anisotropic collapse and the early stages of filament and sheet formation \citep[e.g.][]{Matsubara2008}.

To extend predictions deeper into the quasi-linear regime, several improved perturbative frameworks have been developed, including Renormalized Perturbation Theory (RPT) \citep{Crocce&Scoccimarro2006a, Crocce&Scoccimarro2006b, Crocce&Scoccimarro2008} and Effective Field Theory descriptions of Large-Scale Structure (EFTofLSS) \citep[e.g.][]{Carlson+2009,Baumann+2012,Carrasco+2012}. These approaches account for nonlinear mode coupling and the influence of unresolved small-scale dynamics on large-scale observables. In practice, they substantially improve the modelling of matter clustering, baryon acoustic oscillations and redshift-space distortions, extending useful predictions to scales of roughly $k\sim 0.3\Mpch$, depending on redshift and observable. Such perturbative methods now form an important part of modern precision cosmology analyses and are routinely combined with simulations and emulators in the interpretation of survey data.
Their validity nevertheless degrades once shell crossing, virialisation and fully nonlinear gravitational evolution become dominant, where effects such as complex halo dynamics, nonlinear galaxy bias and baryonic physics are no longer accurately captured within a perturbative framework.

These challenges have motivated increasing interest in statistical descriptions of the LSS that extend beyond the traditional clustering statistics, including filamentary structure, void statistics and velocity-based characterisations of the cosmic web \citep[e.g.][]{Sheth&Weygaert, NEXUS2013, Libeskind_et_al2018}. These developments have highlighted the need to understand the LSS not only statistically, but also as an interconnected network whose geometry and environment influence the formation of galaxies and dark matter halos which reside in it. 

\section{The Cosmic Web and halo environments}
\label{sec: Cosmic web}
One of the most distinctive outcomes of cosmic structure formation is the emergence of the cosmic web: an intricate network of nodes, filaments, sheets and voids extending across a wide range of scales, as illustrated in Figure~\ref{fig:Millennium}. Rather than collapsing isotropically, matter flows preferentially along directions determined by the tidal gravitational field, producing a connected hierarchy of anisotropic structures. This picture is well established as a consequence of gravitational evolution from primordial density fluctuations \citep{Bond+1996}.

In the standard picture of structure formation, initially small density perturbations grow through gravitational instability in an expanding Universe. In the linear regime, different Fourier modes evolve independently, and the growth of structure is well described by linear perturbation theory. As perturbations grow, however, mode coupling and multistreaming become increasingly important, leading to the highly non-linear cosmic web observed today.
The Zel’dovich approximation \citep{Zeldovich1970} offers an intuitive description of this nonlinear evolution, in which matter follows trajectories dictated by the primordial displacement field. The collapse begins along the principal axis of compression, producing sheet-like structures or ``pancakes'', followed by collapse along a second axis that forms filaments, and finally along the third axis leading to the formation of virialized halos. This picture captures the progressive, anisotropic nature of gravitational collapse from initially small density fluctuations.

In cold dark matter cosmologies, structure formation proceeds in a hierarchical, bottom-up manner, with small-scale perturbations becoming nonlinear first and collapsing into low-mass halos. These halos subsequently grow through mergers and accretion, assembling into progressively larger structures such as groups, clusters, and the cosmic web. Numerical simulations reproduce the anisotropic network of structures observed in galaxy surveys, demonstrating that filamentary structure arises naturally through hierarchical gravitational collapse \citep[e.g.][]{Springel2005, Aragon-calvo+2010a}. The emergence of structure across a wide hierarchy of scales, from the cosmic web to virialized halos, is illustrated in Figure~\ref{fig:Millennium}.

Dark matter halos are virialized structures that have undergone gravitational collapse along all three spatial directions. These provide the deep potential wells within which galaxies form and evolve. In the simplest picture, halo properties are expected to depend primarily on halo mass. This assumption underlies many successful theoretical frameworks including the halo model and halo occupation distribution (HOD) approaches \citep{Cooray&Sheth2002,Zheng+2005}. However, increasingly precise numerical simulations and observations have demonstrated that halo and galaxy properties also depend on formation history and environment. Halo clustering, for example, is known to correlate not only with halo mass but also with quantities such as formation time, environment, concentration and spin, a phenomenon referred to as ``assembly bias''\citep[e.g.][]{Sheth&Tormen2004, Gao&White2007, Wechsler&Tinker2018}.

\begin{figure}
    \centering
    \includegraphics[width=0.5\linewidth]{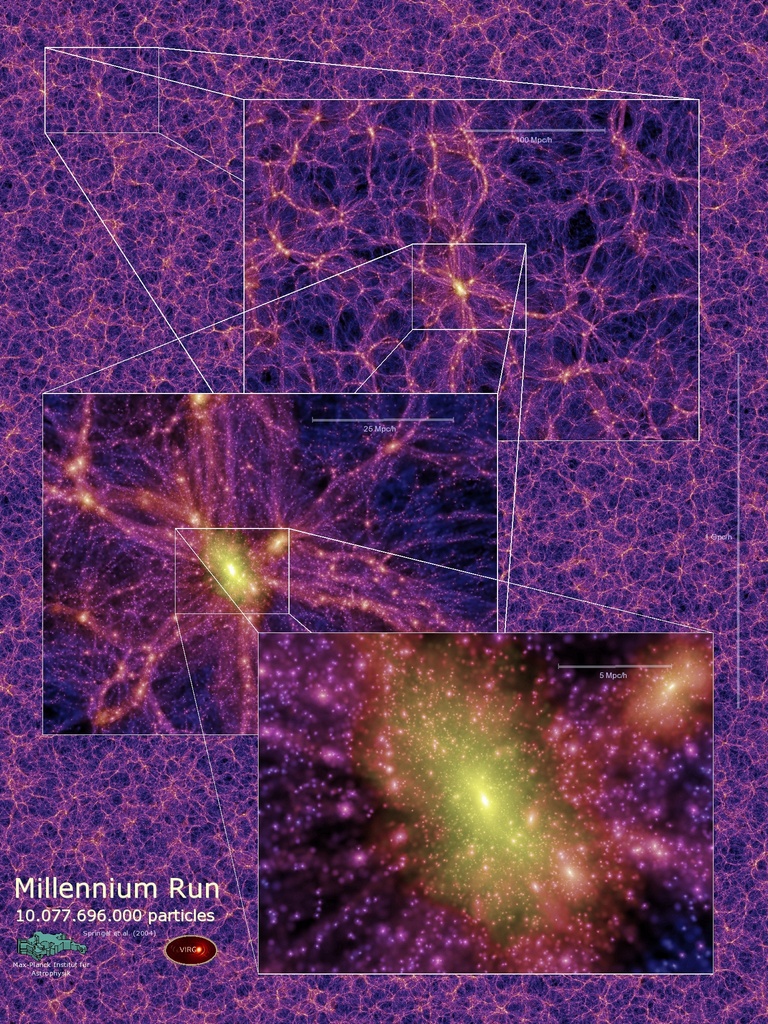}
    \caption{Dark matter density fields illustrating the cosmic web at multiple scales, adapted from the Millennium simulation \citep{Springel2005}.}
    \label{fig:Millennium}
\end{figure}
Observations also reveal strong environmental dependence in galaxy populations. High-density regions are dominated by red, quenched galaxies, while star-forming, blue galaxies are more common in underdense environments \citep[e.g.][]{Dressler1980,Blanton&Moustakas2009}. While much of this behaviour can be understood through the dependence of galaxy properties on halo mass and the distinction between central and satellite galaxies \citep[e.g.][]{Zehavi+2005,Zehavi+2011}, simulations show that the cosmic web provides the dynamical pathways through which matter is transported between different environments. In particular, filamentary accretion is expected to influence halo assembly, angular momentum acquisition, and satellite infall \citep[e.g.][]{Codis+2012,Laigle+2015}. Understanding this dynamical role of the cosmic web remains an active area of both theoretical and observational research.

Motivated by these connections, considerable effort has been devoted to quantifying the morphology and topology of the cosmic web. A variety of filament-finding and environment-classification techniques have been developed, based on density fields, tidal tensors, velocity shear, topological persistence and graph-based approaches \citep[e.g.][]{Libeskind_et_al2018}. Despite substantial progress, there remains no unique or universally accepted definition of a cosmic filament. Different approaches identify structures using distinct geometric, dynamical or topological criteria, often leading to systematically different filament populations and inferred properties \citep[e.g.][]{Libeskind_et_al2018}. Understanding these methodological differences and their physical implications has therefore become an important problem in cosmic web analysis. While the power spectrum and correlation function provide a nearly complete statistical description of a Gaussian random field, nonlinear structure formation transfers information into higher-order correlations and geometric features that are not fully captured by two-point statistics alone. Such statistics may capture aspects of anisotropic gravitational evolution and environmental dependence that are only partially encoded in conventional clustering measures. As observational uncertainties continue to decrease, a substantial fraction of cosmological information is expected to reside in these non-Gaussian aspects of the density and velocity fields. As a result, the characterization of the cosmic web and its environmental structure has become an important part of modern large-scale structure analysis.
These aspects are particularly relevant in the quasi-linear and nonlinear regimes considered in this thesis, and are discussed in more detail in the following chapters.

\section{Peculiar Velocities and Redshift Space}
In an expanding Universe, the large-scale motion of galaxies is dominated by the Hubble flow, so that their recession velocity increases approximately in proportion to distance \citep{Lemaitre1927, Hubble1929}.
\begin{equation}
\vec{v}_{H} = H(t)\vec{r},
\end{equation}
where $\vec{r}$ is the physical separation vector with respect to the observer, and $H(t)$ is the Hubble parameter. Superimposed on this smooth expansion are additional motions induced by local gravitational inhomogeneities. These departures from pure Hubble expansion are referred to as peculiar velocities. They arise from gravitational acceleration towards overdense regions, thereby directly tracing the growth of structure and the underlying matter distribution. In linear theory, the velocity divergence field is directly related to the growth rate of matter perturbations through the continuity equation, making peculiar velocities a sensitive probe of the LSS growth history \citep[e.g.][]{Peebles_cosmology, Peacock_cosmology}.

Peculiar velocities play a central role in LSS observations because distances along the line of sight are typically inferred from measured redshifts. In the absence of peculiar motions, the observed redshift would be determined completely by cosmic expansion. However, the line-of-sight component of the peculiar velocity contributes an additional Doppler shift, leading to a mismatch between the inferred redshift-space position and the true real-space position of a galaxy. Under the distant-observer approximation, the mapping between real-space and redshift-space coordinates can be written as
\begin{equation}
\vec{s} = \vec{r} + \frac{v_{\parallel}}{aH}\hat{n},
\end{equation}
where $\vec{r}$ and $\vec{s}$ denote the real- and redshift-space positions respectively, $v_{\parallel}$ is the line-of-sight peculiar velocity component, and $\hat{n}$ is the line-of-sight direction. As a result, the observed galaxy distribution in redshift surveys appears anisotropically distorted along the line-of sight.\
These redshift-space distortions (RSD) encode valuable cosmological information because peculiar velocities are sourced by the gravitational growth of density perturbations. In linear theory, coherent inflows toward overdense regions enhance clustering along the line of sight, producing the large-scale ``Kaiser effect'' \citep{Kaiser87}. The amplitude of this anisotropy is directly related to the logarithmic growth rate of structure, $f\equiv d \ln D/d \ln a$, where $D(a)$ is the linear growth factor. Measurements of RSD therefore provide direct constraints on the growth history of cosmic structure and constitute an important test of gravity on cosmological scales \citep[e.g.][]{Hamilton1998,Guzzo+2008}.

On smaller nonlinear scales, random virial motions within collapsed structures generate elongated features along the line of sight commonly known as the ``Finger-of-God'' effect. In the quasi-linear and non-linear regimes, the mapping between density and velocity fields becomes substantially more complicated due to nonlinear gravitational evolution, scale-dependent galaxy bias and mode coupling.
Considerable effort has therefore been devoted to constructing increasingly accurate models of redshift-space clustering, including perturbative approaches, streaming models, effective field theory descriptions and simulation-calibrated emulators capable of extending predictions deeper into the quasi-linear regime \citep[e.g.][]{Scoccimarro2004,TNS}.

Beyond their role in RSD, peculiar velocity fields themselves contain important information about the dynamics of structure formation. Since velocities remain more coherent and less nonlinear than the density field over a wider range of scales, velocity statistics can retain information about large-scale modes and growth history even in regimes where the density field has become strongly nonlinear.
In addition, peculiar velocities trace coherent matter flows within the cosmic web, including anisotropic accretion onto filaments and clusters as well as the outflows from under-dense voids. This makes the velocity field a direct probe of the dynamical assembly of structure across the cosmic web. This motivates many of the simulation-based investigations of velocity flows and filamentary environments presented later in this thesis.

\section{Precision Cosmology in the Nonlinear Regime}
As discussed earlier, a significant fraction of the cosmological information accessible to current and upcoming galaxy surveys resides in the quasi-linear and nonlinear regime. While large-scale clustering is accurately described by perturbation theory, much of the statistical power of modern surveys originates from smaller scales where nonlinear gravitational evolution becomes important. Extracting this information reliably therefore requires theoretical frameworks beyond linear theory together with numerical simulations capable of accurately modelling nonlinear structure formation.

Although approaches such as effective field theory and simulation-calibrated perturbative methods extend the predictive reach of perturbation theory into the quasi-linear regime up to roughly $k\sim0.3,\Mpch$, many of the scales probed by present and upcoming surveys lie beyond the regime where perturbative descriptions remain reliable. This has made cosmological N-body simulations an indispensable tool in modern LSS studies. Such simulations follow the nonlinear gravitational evolution of collisionless dark matter by integrating the equations of motion for billions of particles in an expanding Universe. Starting from initial conditions generated using perturbation theory at high redshift, they trace the growth of structure from an almost homogeneous early Universe to the highly nonlinear cosmic web observed today. Modern simulations routinely model cosmological volumes spanning several cubic gigaparsecs while simultaneously resolving halo populations across many orders of magnitude in mass.

While dark matter-only simulations successfully reproduce many observed properties of the large-scale matter distribution, increasing observational precision has also highlighted important modelling challenges. On small scales, baryonic processes such as gas cooling, star formation and feedback from supernovae and active galactic nuclei can significantly modify matter clustering relative to dark matter-only predictions \citep[e.g.][]{Daalen+2011,Chisari+2019}. In addition, accurately modelling galaxy bias, intrinsic alignments and redshift-space distortions in the nonlinear regime remains difficult despite substantial theoretical and numerical progress. These effects introduce systematic uncertainties that are becoming more and more important as observational errors continue to decrease.

The statistical complexity of nonlinear structure formation has also motivated growing interest in probes beyond the traditional two-point description. Linear density fields are well approximated by Gaussian random fields and are therefore largely characterised by the power spectrum or correlation function. Nonlinear evolution, however, transfers information into higher-order correlations, anisotropic structures and environmental dependence. Consequently, a wide variety of complementary statistics have been explored, including higher-order correlation functions, void statistics, Minkowski functionals, counts-in-cells measures, nearest-neighbour statistics such as k-nearest-neighbour (KNN) statistics, Voronoi volume function (VVF) statistics, filamentary structure and other topological or graph-based descriptors of the cosmic web \citep[e.g.][]{Sheth&Weygaert,NEXUS2013,Libeskind_et_al2018}. Many of these probes are designed to capture information associated with nonlinear gravitational evolution and environmental effects that are only partially encoded in conventional clustering measures.

The increasing complexity of both observations and theoretical modelling has further motivated a broader shift toward simulation-based inference methods in cosmology. Large ensembles of simulations are now routinely used to construct mock galaxy catalogues, calibrate emulators, estimate covariance matrices and model nonlinear observables in survey analyses \citep[e.g.][]{Manera+2013, DES_mocks2018, Heitmann+2014, Blot+2019, Cranmer+2020}. In many situations, simulation-based approaches provide the only practical route toward incorporating nonlinear gravitational evolution, survey geometry and observational systematics into precision cosmology pipelines \citep[e.g.][]{Angulo&Hahn2022, Grove+2022, Smith+2024}. As modern surveys continue to improve statistical precision and approach the cosmic variance limit on large scales, efficiently extracting information from nonlinear structure is expected to remain one of the central challenges of future LSS analyses.

Much of the work presented in this thesis is motivated by these developments, with particular emphasis on the cosmic web as a nonlinear probe of structure formation. The primary focus is the identification and characterization of filamentary structure using simulation-based methods. Supporting this analysis is the development of the \Sahyadri\ simulation suite, designed to study nonlinear structure formation and its cosmological dependence across a wide range of scales.

\section{Scope and organization of the Thesis}
This thesis focuses on the nonlinear large-scale structure of the Universe, with particular emphasis on the filamentary structure of the cosmic web. The central aim is to develop and apply robust methods for identifying and characterizing filaments in cosmological simulations, and to study the physical information encoded within these structures. Filaments are of particular interest because they trace anisotropic gravitational collapse, coherent matter transport and the hierarchical growth of halos and galaxies across a wide range of scales. As modern galaxy surveys increasingly probe the quasi-linear and nonlinear regime, these structures are expected to contain information that is not fully captured by conventional two-point clustering statistics alone.\
The analysis presented in this thesis is based primarily on numerical simulations, which provide controlled realizations of nonlinear structure formation across cosmological volumes. A secondary component of this work is the development of the \Sahyadri\ simulation suite, designed to study nonlinear structure formation and cosmic web and its cosmology dependence.

The thesis is organised as follows.\\
Chapter~2 provides a brief overview of the numerical techniques underlying modern cosmological N-body simulations. The chapter then motivates the need for large, high-resolution simulation suites with cosmology variations for studying nonlinear LSS, and concludes with a summary of the \Sahyadri\ simulation suite, including its numerical implementation, design choices and main results.

Chapter~3 addresses the problem of cosmic web reconstruction, with particular emphasis on filament identification and calibration. After reviewing existing filament-finding approaches and their associated challenges, the chapter introduces a calibration framework based on controlled filament realizations. The framework consists of the filament generator \filgen\ and the filament analysis and processing tool \filapt, which together enable systematic study of filament profiles, their biases and systematics. Particular attention is given to the effects of filament curvature, spine reconstruction errors, and smoothing procedures on inferred filament density and velocity profiles, leading to the development of a novel Fourier-space smoothing framework and optimized profile estimation techniques.

Chapter~4 presents \skeletor, the filament-finding framework developed in this thesis. Motivated by the hierarchical nature of the cosmic web and the need for methods applicable to sparse tracer distributions, \skeletor\ uses Voronoi tessellations to identify filamentary environments directly from discrete tracers. The chapter describes the underlying methodology, the incorporation of hierarchy through halo-based ordering, the identification of nested filamentary substructure, and the use of dark matter information for spine refinement and dynamical estimation of filament extent. 

Chapter~5 applies \skeletor\ to cosmological simulations to investigate sub-filamentary structure within the cosmic web. Using the hierarchical reconstruction framework developed in the previous chapter, filamentary substructures embedded within larger parent filaments are identified and characterized. The chapter explores their statistical and geometric properties and discusses the implications of filament hierarchy for understanding the organization of matter within the cosmic web.

Chapter~6 studies the phase-space structure of cosmic filaments using the filament reconstruction and analysis tools developed in this thesis. Radial and longitudinal density profiles, velocity profiles, and phase-space distributions are studied in detail. The chapter investigates coherent inflows, multistreaming behaviour, velocity transitions, and caustic-like features associated with anisotropic gravitational collapse. By revealing these dynamical signatures in simulations, the chapter provides new insight into the role of filaments as channels of matter transport and as fundamental components of nonlinear structure formation.

Finally, Chapter~7 summarises the principal results of the thesis and discusses future directions. Particular emphasis is placed on the use of filamentary structure and cosmic velocity flows as probes of nonlinear structure formation, the search for universality in the organization of filamentary phase space, and the incorporation of filamentary environments into models relevant for upcoming large-scale galaxy surveys.

Taken together, the work presented in this thesis spans the development of numerical simulations, the calibration and construction of filament-finding techniques, and the dynamical study of filamentary structure within the cosmic web. By combining high-resolution simulations, robust filament reconstruction methods, and detailed analyses of filament hierarchy and velocity flows, this work contributes towards a deeper, physically motivated description of the nonlinear Universe. More broadly, it provides a framework for connecting the geometry, hierarchy, and dynamics of the cosmic web, and lays the foundation for future studies of nonlinear structure formation in the era of precision cosmology.

    %
%
\let\textcircled=\pgftextcircled
\chapter{Simulating the non-linear cosmic web}
\label{chap:sahyadri}

\initial{C}osmological simulations have become an essential ingredient in modern studies of the Large-Scale Structure, particularly on nonlinear scales where analytical approaches become inadequate. Their role has become even more important in the era of precision cosmology, with upcoming surveys demanding accurate modelling of structure formation across a wide range of scales and environments. This chapter briefly reviews the basic principles of cosmological simulations, and discusses the motivation for developing new simulation suites aimed at precision studies of the low-redshift Universe.

The chapter then describes the \Sahyadri\ simulation suite developed as part of this thesis. Designed to probe the cosmological dependence of nonlinear structure formation, the suite combines high-resolution simulations with controlled variations of cosmological parameters around a fiducial model. The chapter outlines the numerical methodology underlying the simulations and discusses their use in studying halo populations, cosmic web environments, and statistical probes beyond conventional clustering measures.

\section{Introduction}

Precision studies of the low-redshift Universe rely heavily on accurate modelling of nonlinear structure formation. Many observables probed by current and upcoming surveys are sensitive to scales where perturbative techniques cease to be reliable, and where complex gravitational dynamics must be followed numerically. Numerical simulations therefore form a crucial bridge between cosmological theory and observations.

Over the last few decades, cosmological simulations within the standard $\Lambda$CDM framework have achieved remarkable success in reproducing a wide range of observed phenomena across vastly different scales \citep{Angulo&Hahn2022}. Large-volume dark matter simulations have successfully reproduced many observed statistical properties of large-scale structure, including the hierarchical structure of the cosmic web, the abundance and clustering of dark matter halos, and baryon acoustic oscillations, while also providing calibration and validation for theoretical descriptions of observables such as weak lensing and redshift-space distortions \citep[e.g.][]{Springel2005, Angulo+2008, Heitmann+2014, TNS}. High-resolution simulations have also played a central role in establishing the hierarchical picture of structure formation, studying the internal structure and assembly histories of dark matter halos, and calibrating theoretical frameworks such as the halo model and halo occupation distributions  \citep[e.g.][]{NFW97, Cooray&Sheth2002,Zheng+2005}.

Simulations have now become deeply integrated into the analysis pipelines of modern surveys. They are routinely used to generate mock catalogues, estimate covariance matrices, test inference pipelines and forecast cosmological constraints for surveys such as the Dark Energy Spectroscopic Instrument (DESI) \citep{DESICollaboration2016a, DESICollaboration2016b}, Euclid \citep{EuclidCollaboration2022} and the Rubin Observatory Legacy Survey of Space and Time (LSST)\citep{Ivezic2019}. In many cases, the precision demanded by these surveys exceeds the accuracy achievable with existing semi-analytical models, making simulations indispensable for precision cosmology.

At the same time, the increasing statistical power of observational data has exposed several limitations of current theoretical and numerical approaches. Modelling the galaxy-halo connection, assembly bias, environmental dependence, intrinsic alignments and redshift-space distortions remains challenging in several regimes \citep[e.g.][]{Wechsler&Tinker2018, Joachimi+2015, Percival+2011}. On smaller scales, baryonic feedback processes such as star formation, active galactic nuclei feedback and gas cooling can significantly alter the matter distribution, introducing substantial systematics and uncertainties in weak lensing and clustering observables \citep[e.g.][]{Daalen+2011, Semboloni+2011}. In parallel, there has been growing interest in extracting information beyond traditional two-point statistics through void statistics, topological descriptors, nearest-neighbour statistics and filament-based measures, many of which require simulations with both high mass resolution and large dynamic range.

Broadly, cosmological simulations can be classified into two categories: \emph{N-body simulations}, which follow the evolution of collisionless dark matter under gravity in an expanding background, and \emph{hydrodynamical simulations}, which additionally incorporate baryonic processes such as star formation, gas dynamics and feedback processes via sub-grid prescriptions. Both approaches offer distinct advantages, and serve complementary purposes.
\newline
N-body simulations are computationally less expensive, allowing efficient exploration of cosmological parameter space and the generation of large ensembles of realizations. Also, since they primarily model gravitational dynamics, they contain very few free parameters. Hydrodynamical simulations, on the other hand, attempt to directly model observable galaxies and gas, but at the cost of introducing numerous phenomenological sub-grid parameters. These parameters are typically calibrated against selected observations, and consequently different simulation suites, calibrated using different observables and methodologies, do not always agree with one another.

This chapter begins with a brief overview of the basic principles underlying cosmological simulations. We then discuss the growing need for high-resolution simulation suites tailored for precision studies of the low-redshift Universe in the era of next-generation spectroscopic surveys, motivating the development of \Sahyadri, a suite of cosmological $N$-body simulations designed to combine high mass resolution with controlled variations in cosmological parameters.

The remainder of the chapter describes the specifications of \Sahyadri, and explores its scientific capabilities using both standard and beyond-two-point summary statistics. These include the matter and halo power spectra, the halo mass function, the Voronoi volume function (VVF), and $k^{\mathrm{th}}$ nearest-neighbour ($k$NN) statistics. We also study correlations between halo environment and internal halo properties at $z=0$ and $z=1$. The chapter concludes with a summary of the main results and a discussion of future applications of the suite to precision large-scale structure analyses.

\section{Basics of Cosmological simulations}
\subsection{Cosmological N-body simulations}

Within the $\Lambda$CDM framework, structure formation on sub-horizon scales is driven predominantly by the gravitational evolution of cold dark matter in an expanding Universe. Dark energy influences this process mainly through its effect on the background expansion rate, while the clustering dynamics are well described by Newtonian gravity. Although the underlying spacetime is relativistic, the evolution of matter is commonly followed using Newtonian $N$-body simulations. This approximation has been shown to be consistent with General Relativity over the range of scales relevant for modern cosmological simulations, provided the particle trajectories are interpreted in an appropriate relativistic gauge and relativistic corrections are treated consistently when constructing observables \citep[e.g.][]{Challinor+2011, Fidler+2015, Yoo+2014}. Consequently, the dynamics may be described by the evolution of a large ensemble of collisionless particles interacting through gravity in an expanding Friedmann-Lemaître-Robertson-Walker background

The initial conditions for cosmological simulations are generated at a high redshift, when density perturbations remain small and perturbation theory is valid. Particles are initially placed on a regular lattice or glass-like distribution and displaced according to the Zel’dovich approximation or higher-order Lagrangian perturbation theory using an input linear matter power spectrum \citep{Crocce_2LPTIC, GADGET2005, Hahn&Abel_MUSIC}. A lattice configuration corresponds to particles placed on a uniform Cartesian grid, which is homogeneous but introduces preferred directions associated with the grid geometry. A glass-like configuration, on the other hand, is generated by evolving particles under repulsive gravity until the particles are relaxed to a nearly force-free state \citep{White1994}. This produces an isotropic distribution with suppressed small-scale clustering and reduced grid-alignment artefacts, although they are computationally more expensive to generate and can have residual small-scale correlations from the glass-making procedure. The initial conditions encode the cosmological parameters and primordial fluctuations that seed the formation of the large-scale structure.
The subsequent evolution is followed in comoving coordinates. Writing the physical position as
\begin{equation}
    \vec{r}(t) = a(t)\vec{x}(t)
\end{equation}
where $a(t)$ is the scale factor and $\vec{x}$ is the comoving coordinate, the equations of motion become
\begin{equation}
    \ddot{\vec{x}}+2H\dot{\vec{x}}= -\frac{1}{a^2}\vec{\nabla} \Phi
\end{equation}
together with the Poisson equation
\begin{equation}
    \nabla^2 \Phi = 4\pi G a^2 \bar{\rho}\delta
\end{equation}
where $H\equiv \dot a/a$ is the Hubble parameter, $\Phi$ is the peculiar gravitational potential, $\bar{\rho}$ is the mean matter density and $\delta$ is the matter overdensity. The cosmological expansion appears only through the scale factor and the Hubble friction term -- which is also typically eliminated by a suitable rescaling of the time coordinate -- while the gravitational dynamics themselves retain a Newtonian form. In practice, an $N$-body simulation therefore amounts to solving the coupled equations of motion for a very large number of particles interacting gravitationally in an expanding background.

A direct computation of forces between all particle pairs scales as $\mathcal{O}(N^2)$ and rapidly becomes computationally infeasible for realistic simulations containing billions of particles. Modern cosmological simulations therefore employ approximate force solvers that balance accuracy and computational efficiency.

One of the simplest and most widely used approaches is the Particle-Mesh (PM) method \citep{Barnes&Hut1986}. Here, particle masses are interpolated onto a regular grid to construct the density field, after which the Poisson equation is solved on the mesh, usually using Fast Fourier Transforms (FFTs). Gravitational forces are then interpolated back to particle positions. PM methods are computationally efficient and naturally compatible with periodic cosmological volumes, making them effective for computing long-range gravitational forces. Their spatial resolution, however, is limited by the grid spacing.\\
To improve force accuracy on smaller scales, many simulations employ tree algorithms. Here, particles are organized into a hierarchical structure, such that nearby particles are treated individually while distant groups are approximated collectively using multipole expansions. Tree methods scale approximately as $\mathcal{O}(N\log N)$ and provide substantially better small-scale force resolution than pure PM schemes. Most modern cosmological simulations therefore use hybrid TreePM methods, where long-range forces are computed using a PM solver while short-range interactions are calculated using a tree algorithm, combining computational efficiency with high force accuracy across a wide range of scales \citep{Bagla2002}.

Several highly successful cosmological simulation codes have been developed over the past few decades. Among the most widely used is \textsc{GADGET} \citep{GADGET2005,Gadget4}, a TreePM-based code that has been extensively employed in large cosmological simulation projects for studying structure formation across a wide range of scales. Another major development is \textsc{AREPO} \citep{AREPO2010}, which combines TreePM with a moving Voronoi mesh, providing a quasi-Lagrangian framework for cosmological simulations while maintaining accurate gravitational evolution.
\textsc{RAMSES} \citep{RAMSES2002} adopts an adaptive mesh refinement (AMR) approach, providing high spatial resolution in dense regions at controlled computational cost. In contrast, \textsc{PKDGRAV} \citep{PKDGRAV2001} is a parallel tree code designed for high-precision simulations of nonlinear gravitational evolution. \textsc{ABACUS} \citep{ABACUS_code2021} is optimised for fast, high-accuracy cosmological N-body simulations, particularly in the context of large ensembles for survey applications. \textsc{CUBEP$^3$M} \citep{CUBE3PM} combines a particle-mesh scheme with direct short-range force calculations to efficiently generate large-volume, moderate-resolution realizations of structure formation. These codes represent different methodological approaches to the same underlying problem of simulating gravitational structure formation.

The increasing scale of modern cosmological simulations has also driven the development of codes optimised for massively parallel and GPU-accelerated systems. Examples include \textsc{HACC} \citep{HACC2016}, designed for large-scale cosmological calculations on high-performance computing systems, \textsc{PKDGRAV3} \citep{Potter+2017}, which incorporates GPU acceleration for large cosmological simulations, and \textsc{SWIFT} \citep{SWIFT2024}, which focuses on efficient parallelisation for gravity and hydrodynamical simulations across modern computing architectures.

Over the past two decades, these methods have been used in a number of influential cosmological simulation projects. Examples include the Millennium Simulation \citep{Springel2005, Boylan-Kolchin2009}, Bolshoi \citep{Klypin2011}, IllustrisTNG \citep{Pillepich2018, Nelson2019}, EAGLE \citep{Schaye2015}, Quijote \citep{Villaescusa-Navarro2020}, Aemulus \citep{DeRose2019,McClintock2019}, and AbacusSummit \citep{AbacusSummit}. Together, these simulations cover a wide range of volumes, resolutions, and physical models, reflecting the diverse requirements of modern cosmological studies. Several of these simulation suites are discussed in greater detail later in this chapter.
\newline
Cosmological simulations are typically performed in a finite cubic volume with periodic boundary conditions (PBCs), in which opposite faces of the domain are identified. This eliminates boundary artifacts and preserves translational invariance. Periodicity is also naturally compatible with Fourier-based particle–mesh (PM) methods, where the density field is decomposed into discrete Fourier modes.

The primary outputs of an $N$-body simulation are snapshots of particle positions and velocities at different redshifts. From these particle distributions, gravitationally bound dark matter halos are identified using halo-finding algorithms. Since galaxies are expected to form within these halos, halo catalogues constitute one of the most important data products of cosmological simulations.

Halo-finding algorithms are traditionally motivated by either Friends-of-Friends (FoF) linking criteria or Spherical Overdensity (SO) definitions. In practice, however, most modern halo finders incorporate additional information such as gravitational boundedness, phase-space structure, or temporal consistency, leading to hybrid methodologies.

The simplest and historically most widely used approach is the Friends-of-Friends (FoF) algorithm \citep{Davis+1985}, which links together particles separated by less than a chosen linking length to define overdense regions. FoF halos are straightforward to construct and naturally accommodate non-spherical morphologies, although they can sometimes artificially connect neighbouring structures via thin particle bridges.\\
An alternative approach is provided by Spherical Overdensity (SO) methods, which locate density peaks and grow spherical regions around them until a specified overdensity threshold is reached with respect to either the background or critical density. SO-based definitions are often more closely aligned with analytical halo models and observational mass definitions.

Modern halo finders further resolve substructure within halos. Widely used examples include \textsc{SUBFIND} \citep{Springel+2001} and \textsc{ROCKSTAR} \citep{Behroozi2019}, among several other phase-space and bound-structure algorithms. \textsc{ROCKSTAR} uses phase-space information to identify self-bound subhalos embedded within larger parent structures.

A complementary component of cosmological simulations is the construction of merger trees, which track the progenitors and descendants of halos across simulation snapshots. These trees encode the assembly history of structure over cosmic time and are central to studies of hierarchical structure formation, halo growth, and merger histories.

Together, these methods allow cosmological $N$-body simulations to follow the nonlinear evolution of structure from nearly Gaussian initial conditions to the cosmic web at late times. They form a key component of large-scale structure analyses and are widely used in precision cosmology studies.

\subsection{Cosmological hydrodynamical simulations}
Here we briefly summarise cosmological hydrodynamical simulations for completeness. While they are not the primary focus of this thesis, they form an important component of modern large-scale structure modelling, particularly in regimes where baryonic physics significantly modifies the matter distribution on nonlinear scales. For comprehensive reviews of recent developments in cosmological hydrodynamical simulations and galaxy formation modelling, see \citep[e.g.][]{Crain+2023, Vogelsberger+2020}.
Although dark matter dominates the matter density and drives gravitational clustering, most observables are traced by baryonic components such as galaxies, the intergalactic medium, and the circumgalactic medium. Accurately modelling these requires simulations that extend beyond collisionless dynamics and incorporate hydrodynamics coupled with astrophysical processes. Cosmological hydrodynamical simulations therefore combine N-body evolution for dark matter with numerical solvers for gas dynamics, along with sub-grid prescriptions for unresolved baryonic physics.

These sub-grid models include radiative cooling, star formation and evolution, metal enrichment, supernova and active galactic nucleus (AGN) feedback, and black hole growth. Since these processes occur well below the resolution scale of the cosmological simulations, they are not derived from first principles but are instead calibrated against a limited set of observables, including but not limited to: stellar mass functions, galaxy size distribution, star formation histories.\\
\newline
Three broad numerical approaches are commonly used to solve the hydrodynamical equations. Smoothed Particle Hydrodynamics (SPH) represents gas using particles with smoothing kernels and estimates fluid quantities via kernel-weighted interpolation, though classical formulations have known limitations in capturing fluid mixing and instabilities. Grid-based methods solve the Euler equations on fixed or adaptively refined meshes (AMR), providing improved treatment of shocks and discontinuities. A third class of moving-mesh and mesh-free finite-volume methods combines aspects of both approaches, offering improved Galilean invariance and accuracy in complex flows. Modern large-scale simulation suites such as IllustrisTNG \citep{Nelson2019}, EAGLE \citep{Schaye2015}, and SIMBA \citep{Dave2019} demonstrate that, with calibrated sub-grid models, a broad range of galaxy population statistics can be reproduced within observational uncertainties, although differences remain in detailed predictions across simulations.

In particular, these simulations reproduce key observables such as the stellar mass function at low and intermediate redshift, the cosmic star formation rate density, and large-scale clustering and halo statistics, as well as several integrated properties of gas in and around halos. At the same time, tensions remain in detailed predictions of the thermodynamic state of the circumgalactic medium, and in simultaneously matching galaxy clustering with baryonic observables across different mass scales and gas phases.

A key limitation of hydrodynamical simulations is their sensitivity to sub-grid modelling choices. Different implementations of feedback and star formation can yield comparable agreement with selected observables while producing noticeable differences in others, particularly in the nonlinear regime where baryonic effects suppress the matter power spectrum by $\sim 10–20\%$ at $k \geq 1 h/\mathrm{Mpc}$ (e.g \citep{Daalen+2011, Chisari+2019}). This introduces a significant astrophysical systematic for precision cosmology, especially for weak lensing and small-scale clustering analyses.

As a result, while hydrodynamical simulations are indispensable for understanding galaxy formation and baryonic effects in the nonlinear regime, their computational cost and model complexity make them less suitable for extensive exploration of cosmological parameter space. This makes high-resolution N-body simulations the standard tool for cosmological inference studies focused on the dark matter–dominated large-scale structure.

\section{Motivation for a new suite of N-body simulation}

Ongoing and upcoming large-scale structure (LSS) surveys such as the Dark Energy Spectroscopic Instrument \citep[DESI;][]{DESICollaboration2016a}, the 4-metre Multi-Object Spectroscopic Telescope \citep[4MOST;][]{4MOST2019}, the Subaru Prime Focus Spectrograph \citep[PFS;][]{Takada2014}, \textit{Euclid} \citep{EuclidCollaboration2022}, and the Vera C. Rubin Observatory's Legacy Survey of Space and Time \citep[LSST;][]{Ivezic2019} will map the Universe with unprecedented precision. These surveys will probe scales ranging from the quasi-linear to the deeply non-linear regime across a wide range of redshifts, providing access to a wealth of cosmological information. Extracting this information requires accurate theoretical predictions, robust summary statistics, and comprehensive modeling frameworks that can account for complex non-linear physics and baryonic effects.

Over the past two decades, numerous simulation suites have been developed with diverse objectives and specifications—spanning different volumes, mass resolutions, cosmological parameter coverage, and baryonic prescriptions—tailored to specific science goals. High-resolution $N$-body and hydrodynamical simulations such as Uchuu \citep{Ishiyama2021}, SIMBA \citep{Dave2019}, EAGLE \citep{Schaye2015}, Illustris \citep{Vogelsberger2014}, IllustrisTNG \citep{Nelson2019,Pillepich2018}, the Millennium suite \citep{Springel2005,Boylan-Kolchin2009}, Bolshoi \citep{Klypin2011}, and MultiDark \citep{Klypin2016} have revolutionized our understanding of galaxy formation and evolution. However, these simulations typically explore a fixed or limited set of cosmological parameters, restricting their utility for constraining cosmology through LSS observations. The CAMELS suite \cite{camels-2021} bridges these categories by varying both cosmological and astrophysical parameters at a mass resolution comparable to that considered in the present work, but with volumes limited to $(50\,h^{-1}\,{\mathrm Mpc})^3$ which is 64 times smaller than what we use below, thus restricting large-scale structure and sample variance studies.

Conversely, simulation suites designed explicitly for cosmological inference—including Quijote \citep{Villaescusa-Navarro2020}, Aemulus \citep{DeRose2019,McClintock2019}, Dark Quest \citep{Nishimichi2019}, Abacus Cosmos \citep{Abacus_cosmo2018}, and AbacusSummit \citep{Maksimova2021}—systematically vary cosmological parameters to enable derivative calculations and emulator construction. Table~\ref{tab:suite_comparison} summarizes the specifications of these suites. 
While these efforts have substantially advanced cosmological modeling, they are constrained by practical trade-offs between volume, resolution, and parameter coverage. For instance, with a particle mass of $m_{\mathrm p} = 2.0 \times 10^{9}\,h^{-1}\,M_{\odot}$, AbacusSummit—currently the highest-resolution suite with comprehensive parameter coverage—resolves halos down to $M_{\mathrm min} \approx 8.0 \times 10^{10}\,h^{-1}\,M_{\odot}$ using a conservative 40-particle threshold. Similarly, Quijote ($m_{\mathrm p} = 6.6 \times 10^{11}\,h^{-1}\,M_{\odot}$), Aemulus ($m_{\mathrm p} = 3.5 \times 10^{10}\,h^{-1}\,M_{\odot}$), Dark Quest ($m_{\mathrm p} = 1.0 \times 10^{10}\,h^{-1}\,M_{\odot}$), and Abacus Cosmos ($m_{\mathrm p} = 4.0 \times 10^{10}\,h^{-1}\,M_{\odot}$ and $1.0 \times 10^{10}\,h^{-1}\,M_{\odot}$) resolve halos down to approximately $M_{\mathrm min} \approx 2.6 \times 10^{13}$, $1.4 \times 10^{12}$, $4.0 \times 10^{11}$, and $1.6 \times 10^{12}$/$4.0 \times 10^{11}\,h^{-1}\,M_{\odot}$, respectively. In addition, most AbacusSummit simulations terminate at $z = 0.1$, whereas \Sahyadri\ continues to $z = 0$, capturing the full low-redshift range relevant for galaxy surveys such as DESI BGS. As we show below, these mass thresholds remain insufficient for comprehensively modeling the faint galaxy populations targeted by state-of-the-art low-redshift spectroscopic surveys \cite{Alam_2021}. 

\begin{table*}
\centering
\begin{tabular}{lccccc}
\hline\hline
Suite & $L_{\mathrm box}$ & $N_{\mathrm part}$ & $m_{\mathrm p}$ & $M_{\mathrm min}$ (40 particles) \\
 & $[h^{-1}\,{\mathrm Mpc}]$ &  & $[h^{-1}\,M_{\odot}]$ & $[h^{-1}\,M_{\odot}]$  \\
\hline
Quijote & 1000 & $512^3$ & $6.6 \times 10^{11}$ & $2.6 \times 10^{13}$  \\
Aemulus & 1050 & $1400^3$ & $3.5 \times 10^{10}$ & $1.4 \times 10^{12}$  \\
Dark Quest & 1000 & $2048^3$ & $1.0 \times 10^{10}$ & $4.0 \times 10^{11}$  \\
Abacus Cosmos & 1100/720 & varies & $4.0 \times 10^{10}/1.0\times10^{10}$ & $1.6 \times 10^{12}/4.0\times10^{11}$  \\
AbacusSummit & 2000 & $6912^3$ & $2.0 \times 10^{9}$ & $8.0 \times 10^{10}$ \\
\hline
\textbf{Sahyadri} & \textbf{200} & $\mathbf{2048^3}$ & $\mathbf{8.1 \times 10^{7}}$ & $\mathbf{3.2 \times 10^{9}}$ \\
\hline
\end{tabular}
\caption{Comparison of cosmological simulation suites with parameter variations. $M_{\mathrm min}$ denotes the minimum halo mass resolved with at least 40 particles. \Sahyadri\ achieves a factor of $\sim$25 improvement in mass resolution compared to AbacusSummit, the previous highest-resolution suite with comprehensive cosmological parameter coverage.} 
\label{tab:suite_comparison}
\end{table*}

Here, we present \Sahyadri, an $N$-body simulation suite specifically designed to address the resolution requirements of current and next-generation low-redshift spectroscopic surveys while maintaining the capability to compute cosmological derivatives.  \Sahyadri\ builds upon the \Sinhagad\ pilot suite introduced in \cite{Fisher_stabilization2026}, increasing the particle count by a factor of 512 (from $256^3$ to $2048^3$ in the same $200\,h^{-1}\,{\mathrm Mpc}$ volume) to achieve the mass resolution necessary for comprehensive modeling of low-redshift galaxy samples. With a particle mass of $m_{\mathrm p} = 8.1 \times 10^{7}\,h^{-1}\,M_{\odot}$ in its default cosmology, \Sahyadri\ resolves dark matter halos down to $M_{\mathrm min} = 3.2 \times 10^{9}\,h^{-1}\,M_{\odot}$ (40 particles), representing a factor of $\sim$25 improvement in mass resolution compared to the default AbacusSummit simulation, and over 2 orders of magnitude better than Quijote and Aemulus. This enhanced resolution is particularly critical for modeling the DESI Bright Galaxy Survey \citep[BGS;][]{Hahn2023}, which targets galaxies at $z \lesssim 0.5$ with peak number density at $z < 0.2$, and similar low-redshift samples from 4MOST \citep{Yildiz2020}. Table~\ref{tab:suite_comparison} compares \Sahyadri\ with the existing cosmological variation simulation suites.

Figure~8 of \cite{Smith+2024} shows that achieving $\sim99\%$ completeness for the DESI BGS sample with an $r$-band magnitude limit of $M_r=-22$ requires resolving halos down to masses of $\sim5\times10^{10},\Msun/h$. While this halo mass is below the resolution limit of AbacusSummit, it is well within the mass resolution of \Sahyadri. This indicates that \Sahyadri\ is able to resolve essentially the full halo population hosting DESI BGS galaxies at this luminosity threshold.

A quantitative estimate of this improvement is presented in \cite{Sahyadri2026}, which finds that approximately $40\%$ of BGS galaxies at $z<0.15$ reside in halos below the AbacusSummit resolution limit but within the reach of \Sahyadri. The detailed analysis also shows that the BGS spectroscopic sample is within \Sahyadri's optimal resolution regime while extending significantly beyond AbacusSummit's capabilities.

\begin{figure}
    \centering
    \includegraphics[width=\textwidth,]{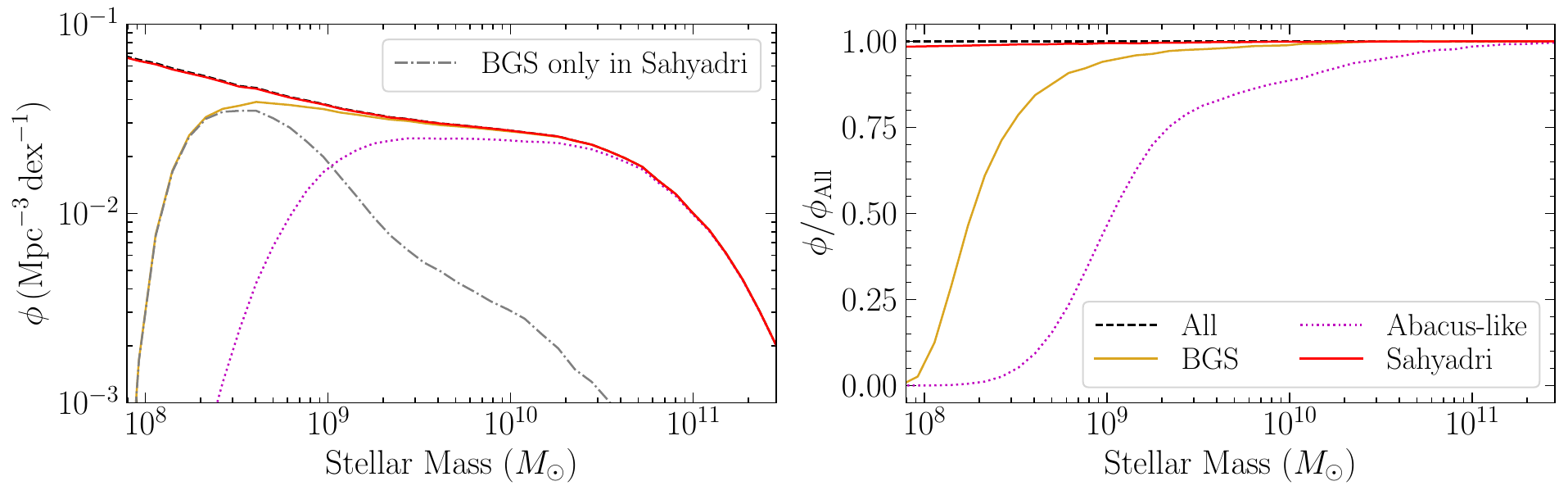}
    \caption{Stellar mass function using a stellar-to-halo mass relation and a conditional $r$-band magnitude distribution at redshift $z=0.15$ based on Alam et al. (in prep). We show that the default \Sahyadri\ simulation allows us to  model much lower stellar masses as compared to AbacusSummit which misses $\sim 40\%$ of galaxies in BGS at these low redshifts.}
    \label{fig:smf}
\end{figure}
This mass resolution opens up new avenues for precision studies of the low-mass galaxy population, including the faint end of the stellar mass function \citep{Baldry2012}, environmental quenching in low-mass satellites \citep{Wetzel2013}, and AGN activity in dwarf galaxies \citep{Reines2013}. Furthermore, the low-redshift Universe ($z < 0.4$) represents a regime of considerable contemporary interest, not only because surveys achieve their highest number densities there, but also because several outstanding cosmological tensions may have roots in low-redshift physics. Both the Hubble tension, the discrepancy between local distance ladder measurements \citep{Riess2022} and CMB-inferred values \citep{Planck18-VI-cosmoparam}, as well as hints of evolving dark energy from type Ia supernovae and BAO measurements \citep{DESIDR1_cosmo, DESIDR2_cosmo} suggest potential new physics operating at $z \lesssim 0.4$. Current analyses of large-scale structure at these redshifts have approached cosmic variance limits in two-point statistics given available volumes \citep{Wadekar2020,DESIDR1_cosmo}. Extracting additional cosmological information necessitates probing non-linear scales and higher-order statistics, both of which require the combination of high mass resolution and cosmological parameter variations that \Sahyadri\ provides.

The simulation volume of $(200\,h^{-1}\,{\mathrm Mpc})^3$ represents a carefully chosen compromise. While smaller in volume than suites optimized for Baryon Acoustic Oscillation (BAO) studies, this volume is sufficient for convergence in large-scale clustering statistics such as the linear halo bias $b_1$ (see Appendix~\ref{App: sahyadri b1-voldep}), while enabling the enhanced mass resolution critical for low-redshift galaxy surveys
\section{Sahyadri:A simulation suite for the cosmology dependence of the Cosmic Web}
\subsection{Suite Specifications}
\label{sec:specs}

\begin{table}[t]
    \centering
    \begin{tabular}{lccc}
    \hline\hline
        Parameter & Fiducial value ($\theta_{\mathrm{f}}$) & Variation magnitude ($\Delta$) & Status \\
        \hline
        $\Omega_{\mathrm m}$ &  $0.3138$ & $0.05\,\Omega_{\mathrm{m,f}}$ & Complete \\
        $n_{\mathrm s}$ &  $0.9649$ & $0.05\,n_{\mathrm{s,f}}$ & Complete \\
        $h$ &  $0.6736$ & $0.05\,h_{\mathrm{f}}$ & Complete\\
        $A_{\mathrm s}$ &  $2.0989\times 10^{-9}$ & $0.1\,A_{\mathrm{s,f}}$ & Complete \\
        $w_{\parallel }$ &  $0$ & $0.125$ & In progress\\
        $\Omega_{\mathrm k}$ &  $0.0$ & $0.05$ & Planned\\
        \hline
    \end{tabular}
    \caption{Fiducial cosmological parameters and variation magnitudes for the \Sahyadri\ simulation suite. The subscript ``f'' denotes fiducial values. Each parameter is varied independently while maintaining all other parameters at their fiducial values. }
    \label{tab:cos_params}
\end{table}

The \Sahyadri\ simulation suite adopts fiducial cosmological parameters consistent with the Planck 2018 analysis \citep{Planck18-VI-cosmoparam}. Table~\ref{tab:cos_params} lists the fiducial parameter values ($\theta_{\mathrm f}$) and their respective variation magnitudes ($\Delta$). This configuration enables Fisher matrix analyses, wherein each of the six cosmological parameters is varied individually while holding the remaining parameters fixed at their fiducial values. 

\begin{figure}
    \centering
    \includegraphics[width=\linewidth]{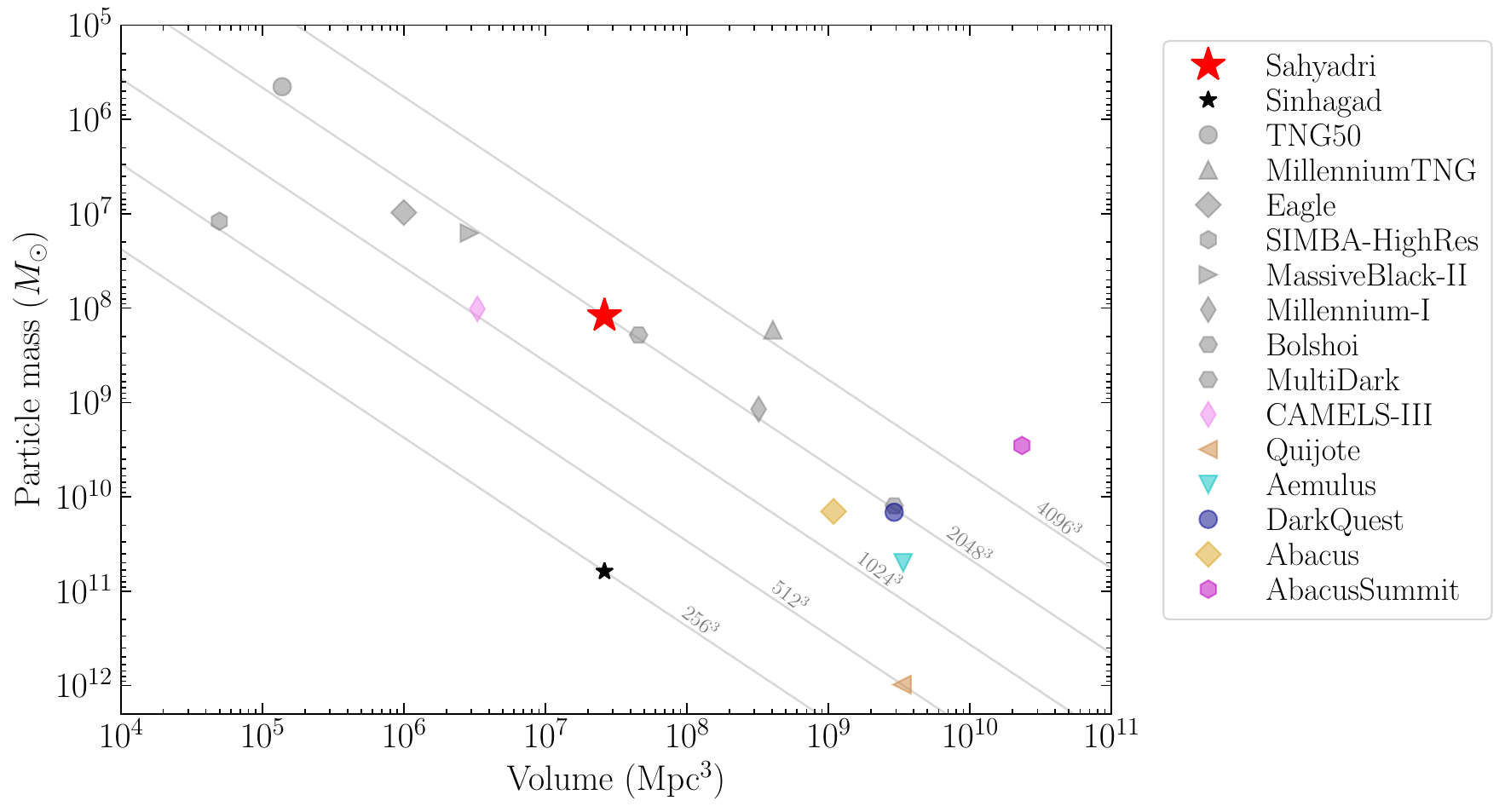}
    \caption{Comparison of the \Sahyadri\ simulation suite with other large-scale structure simulation efforts. Grey lines indicate simulations with constant particle number assuming Planck 2018 cosmology. Coloured points highlight suites with cosmology variations. The red (black) star marks the \Sahyadri\ (\Sinhagad) simulation configuration, highlighting the state-of-the-art mass resolution achieved by our suite. }
    \label{fig:intro}
\end{figure}

Figure~\ref{fig:intro} places the \Sahyadri\ suite in context with other contemporary simulation efforts, illustrating the trade-off between box size and mass resolution. The grey lines represent simulations with constant particle numbers assuming Planck 2018 cosmology, while the black (red) star denotes the \Sahyadri\ (\Sinhagad) configuration. 

The matter density parameter \Om\ is precisely constrained in $\Lambda$CDM, with Planck 2018 yielding $\Omega_{\mathrm m} = 0.315 \pm 0.007$ \citep{Planck18-VI-cosmoparam}. However, recent analyses reveal tensions in $\Omega_{\mathrm m}$ across different probes.  Interesting discrepancies have emerged between weak lensing (CMB and galaxy) and primary CMB measurements along the \Om\ axis (see Figure 12 and associated discussion in \citep{Hang_2020}), while \citep{Ishiyama2024} demonstrated that the BAO feature locations are nearly identical between Planck 2018 $\Lambda$CDM and Planck 2018 with DESI-preferred values of the dark energy equation-of-state parameters $(w_0, w_a)$. This indicates that apparent dark energy evolution cannot be attributed to dynamics alone but reflects differences in \Om\ and the sound horizon $r_{\mathrm d}$. These findings highlight the critical role of precise non-linear structure modeling at $z \lesssim 1$, where \Om\ variations produce interesting signatures.

Therefore, the analyses in this chapter focus on \Om\ variations, demonstrating a methodology that is applicable to the complete suite. At the time of writing this thesis, we have completed the fiducial cosmology and full variations for \Om, $h$, \ns\, and \As, with the remaining parameters ($w_\parallel$ and $\Omega_k$) in progress. In total, the \Sahyadri\ suite will comprise 13 simulations: one fiducial and two variations for each of the six cosmological parameters. For the dark energy equation of state, we define $w_\parallel$ along the CMB degeneracy $w_a = A(w_0 + 1)$ (where $A = -3.68$) via $w_0 = -1 + w_\parallel/\sqrt{1 + A^2}$ and $w_a = A w_\parallel/\sqrt{1 + A^2}$ (setting $w_\perp = 0$), allowing efficient sampling of $D_M(z_\star)$-consistent models (see Lodha et. al. in prep. for more details).

\myheading{Simulations}
\label{subsec:simulations}

All simulations in the \Sahyadri\ suite were executed in a periodic comoving box of side length $L_{\mathrm box} = 200\,h^{-1}\,{\mathrm Mpc}$ containing $2048^3$ dark matter particles. This corresponds to a particle mass of $m_{\mathrm p} = 8.1 \times 10^{7}\,h^{-1}\,M_{\odot}$ in the fiducial cosmology. At this resolution, as mentioned earlier, halos with masses $M > 3.2\times10^{9}\,h^{-1}\,M_{\odot}$ are resolved with more than 40 particles, satisfying standard convergence criteria for halo identification and statistical analysis. We note that massive neutrinos are not included in these simulations.

\myheading{Halo and environmental catalogs}
\label{subsec:halos_vahc}
Dark matter halos in the simulations were identified using the six-dimensional phase-space Friends-of-Friends algorithm implemented in \textsc{rockstar} \citep{Rockstar2013},\footnote{\url{https://bitbucket.org/gfcstanford/rockstar/}} modified by us to allow for $\Omega_{\mathrm k}$ variations (which are not discussed in this work). Halo merger trees were constructed with the \textsc{consistent-trees} code \citep{Consistent_trees2013}.\footnote{\url{https://bitbucket.org/pbehroozi/consistent-trees/}} The high density of stored snapshots ensures reliable tracking of merger and accretion relations between halos across cosmic time.

Alongside the primary halo catalogs produced by \textsc{rockstar} and \textsc{consistent-trees}, we calculate a number of environmental quantities for each halo. These are recorded separately in a `value added' halo catalog for each snapshot. The details are described in Appendix~\ref{App: sahyadri vahc}. The halo and value added catalogs were compressed by storing them in FITS format, achieving a factor $\sim4$ gain relative to the native ASCII.

Finally, our default post-processing pipeline (written in Python) also allows for estimates of the matter auto, halo-matter cross and halo auto power spectra and mass functions for halos resolved with $\geq40$ particles, as well as beyond 2-point statistics such as the Voronoi volume function (VVF) and $k^{\mathrm th}$ nearest neighbour ($k$NN) statistics (see below). In the interest of reproduceability, our entire pipeline for performing the simulations, generating halo and environment catalogs and other post-processing products, along with the data compression described in \citep{Sahyadri2026}, is publicly available at \url{https://github.com/a-paranjape/sahyadri-sandbox}. The Pythonic post-processing code, in particular, is modular and easily extendable to include other statistics.

The final compressed data products occupy approximately 9 TB of storage for each cosmology variation, about $96\%$ of which corresponds to the snapshot storage. The complete suite thus occupies approximately 117 TB of data. The average run time of each variation, including halo finding and post-processing, was approximately 0.44 million CPU hours, with the `$+$' (`$-$') variations typically being slower (faster) than the default box. The simulations and analysis were performed on the Pegasus cluster at IUCAA, Pune.\footnote{\url{http://hpc.iucaa.in}}

\myheading{Halo samples for present analysis}
\label{subsec:halo_samples}
For all analyses in this work,  the halo catalogs are cleaned according to the degree of relaxation of each halo, quantified by $\eta \equiv 2T/|U|$, where $T$ and $U$ are the total kinetic and potential energies of the halo. Following \citep{Bett+2007}, we retain only halos with $0.5 \leq \eta \leq 1.5$, a criterion that we refer to as the QE cut. Halo mass is defined as $M_{200\mathrm b}$, the gravitationally bound mass enclosed within a radius $R_{200\mathrm b}$ corresponding to 200 times the mean matter density of the Universe.

We work with two distinct halo samples. The first, designed for calculating the VVF and $k$NN statistics, is observationally motivated. Halos are selected based on a threshold in their maximum circular velocity along the main progenitor branch of the merger tree, denoted \Vpeak\ \citep{Reddick+2013}, which correlates closely with stellar mass in subhalo abundance matching \cite{SHAM_2015, SHAM2016, SHAM_RSD2022}. Only halos resolved with at least 40 particles and satisfying the QE cut are included. The \Vpeak\  thresholds are then chosen to ensure fixed tracer number densities across simulations, with three samples at $2\times10^{-4}$, $2\times10^{-3}$, and $2\times10^{-2}\, (\mathrm{Mpc})^{-3}$, respectively. In the default cosmology, for $z=0$, this corresponds to \Vpeak\ thresholds of $422.5,196.6,74.6\,\kms$, respectively. The highest number density is set such that the fraction of unresolved halos (i.e., halos having $<40$ particles enclosed within $M_{\mathrm{200b}}$) at the corresponding \Vpeak\ threshold is $\lesssim 0.5\%$. We refer to this as the \Vpeak-selected sample below. 

The second sample is more theoretically motivated and intended for comparison with analytic fitting functions. Here, in addition to the QE cut and the 40 particle mass cut, we exclude all subhalos and retain only parent halos in order to remove 
the effects of substructure. We refer to this as the mass-selected sample below.

\subsection{Cosmic web studies with \Sahyadri: Highlights}
\label{sec:highlights}
As mentioned in the Introduction, \Sahyadri\ opens up multiple avenues for studies of the cosmology dependence of the small-scale distribution of faint objects. In this section, we showcase some of these possibilities by presenting visualizations of various aspects of the cosmic web defined by our observationally motivated halo samples and explicitly displaying the \Om-dependence of beyond 2-point observables such as the VVF and $k$NN statistics.

\myheading{Overview and visualizations}
\label{subsec:visualizations}

We begin by showing the redshift evolution of the dark matter density field for the fiducial cosmology, before moving on to comparisons with existing simulations and tracer-dependent visualizations of the cosmic web.
These Figures are intended to provide qualitative insight and to motivate the statistical analyses presented later in this section.

\begin{figure}
\centering
\includegraphics[width=\linewidth]{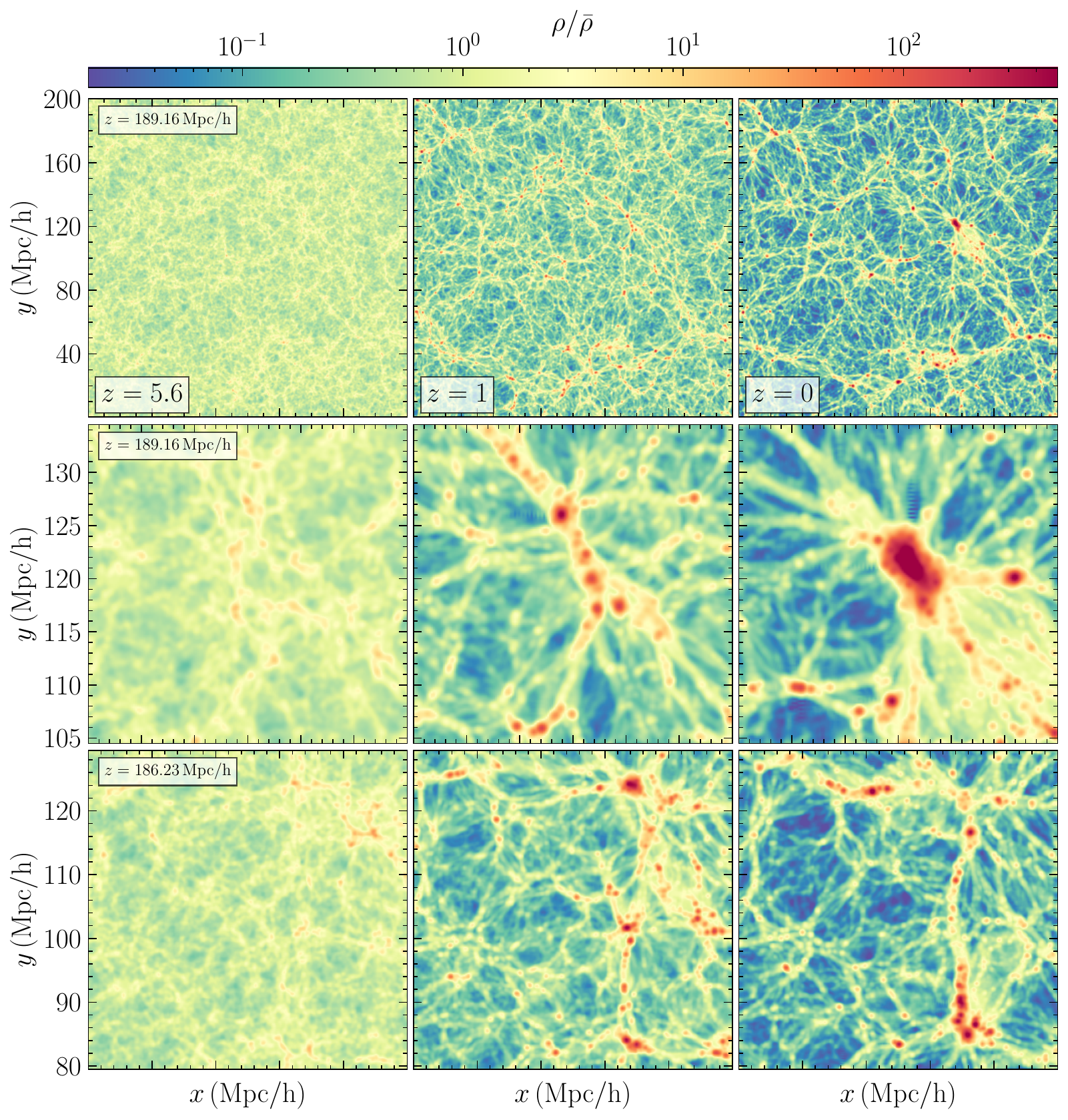}
\caption{Visualization of the evolution of the dark matter density field as a function of redshift for the fiducial cosmology. The density field is evaluated on a $1024^3$ grid and smoothed with a Gaussian kernel of radius $185\,\kpch$. The \emph{left, middle and right columns} correspond to the same single-cell spatial slices at redshift $z=5.6, 1, 0$ respectively. The \emph{top panel} shows a slice through the entire simulation box containing the most massive halo at $z=0$. The \emph{middle panel} provides a zoom-in on this halo, while the \emph{bottom panel} shows the evolution of a region that later develops into a prominent filament. The centers of the slices are indicated in each column. Each panel’s $x$- and $y$-axes span the same length, ensuring uniform tick mark size in the $x$ and $y$ directions. }
\label{fig:z_comparison}
\end{figure}

Figure \ref{fig:z_comparison} shows the evolution of the dark matter density field from high redshift to the present epoch for the fiducial cosmology. The \emph{top row} displays a slice through the full simulation volume containing the most massive halo at 
$z=0$, while the \emph{middle} and \emph{bottom rows} focus on regions that evolve into the most massive cluster in the box, and a prominent filament respectively. The dark matter density fields are estimated using CIC interpolation onto a $1024^3$ grid. The resulting field is further smoothed with a Gaussian kernel of radius $0.95$ times the grid size, corresponding to $\approx 185\,\kpch$.  Throughout, we display single slices of this density grid. The Figure highlights the well-known emergence of the cosmic web from initially diffuse density fluctuations, the growth of filaments and nodes, and the increasing density contrast due to gravitational evolution towards low redshifts.

\begin{figure}
    \centering
    \includegraphics[width=\linewidth]{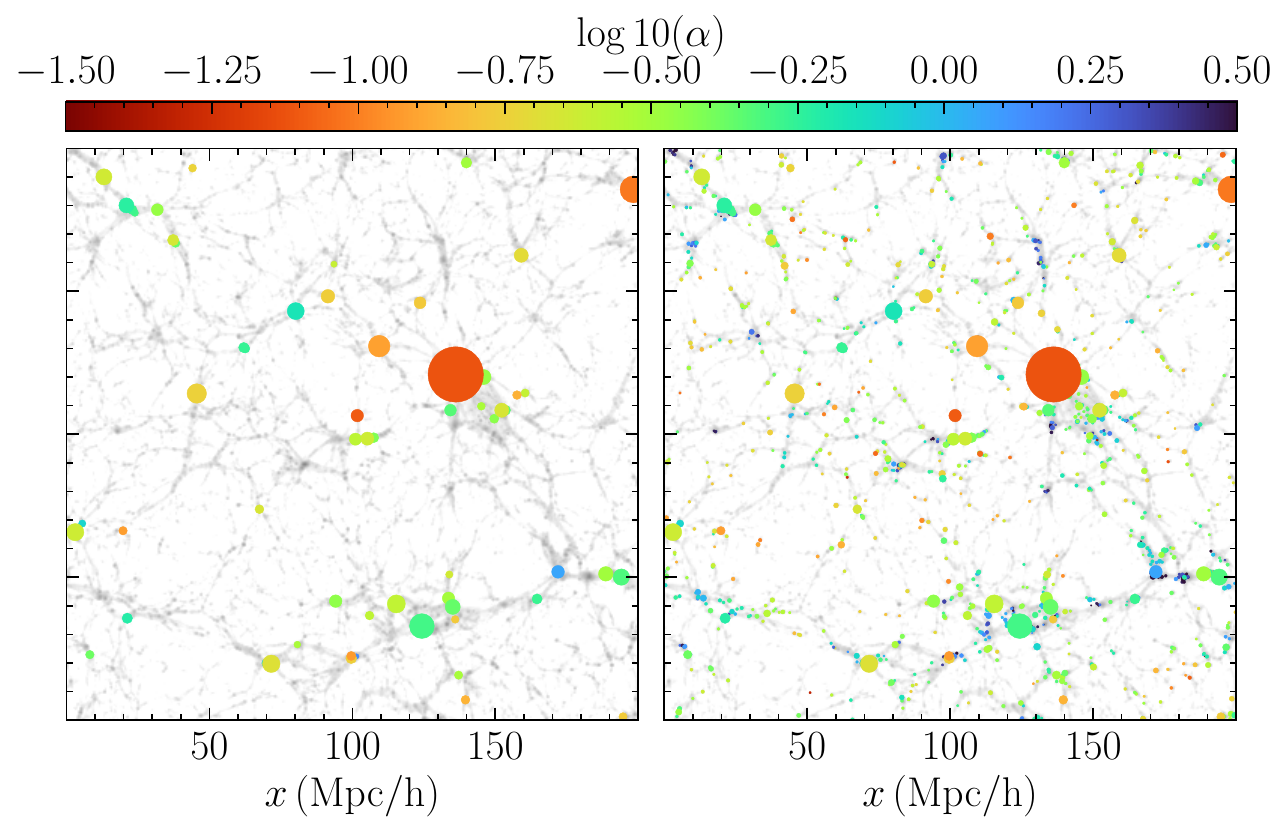}
    \caption{Comparison between AbacusSummit-like and \Sahyadri\, halos. \textit{Right panel} shows the highest number density halo sample used in this work, with $n=2\times 10^{-2} \mathrm{Mpc}^{-3}$ (see section~\ref{subsec:halo_samples} for the selection criterion), from the default \Sahyadri\, simulation at $z=0$. \textit{Left panel} shows the sample of halos from the same simulation, but after degrading the particle mass to that of AbacusSummit and applying the same selection criterion, leading to $n=6.5\times 10^{-4} \mathrm{Mpc}^{-3}$.
    Both panels display the same spatial slice as the top right panel of Figure~\ref{fig:z_comparison}, including halos in this slice and the two neighboring slices, overplotted on the underlying density field. Circles mark halo positions, with radii equal to $4R_{\mathrm{200b}}$, and are colored by $\log_{10}(\alpha)$, where $\alpha$ is the tidal anisotropy defined in \eqn{eq:alpha-def}. The  $x$- and $y$-axes span the same length, ensuring uniform tick mark size in each direction. The higher resolution of \Sahyadri\ allows significantly more low-mass halos to be resolved, improving the sampling of all cosmic environments.}
    \label{fig:alpha_comparison}
\end{figure}

\begin{figure}
    \centering
    \includegraphics[width=\linewidth]{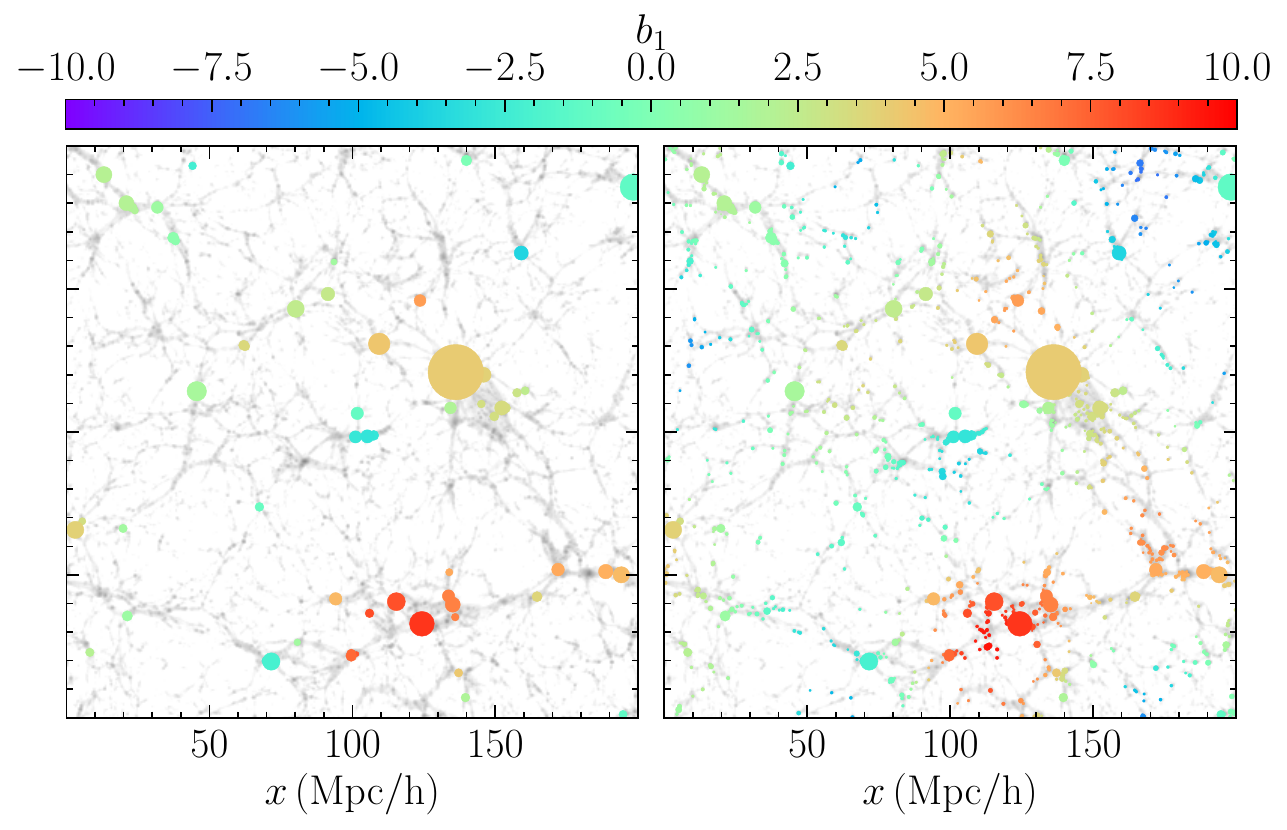}
    \caption{Same as  Figure~\ref{fig:alpha_comparison}, but with halos are coloured by the halo-by-halo bias $b_1$. We see that the most massive halo does not necessarily correspond to the largest bias, which instead traces the larger-scale density field, as established in the literature. See text for a discussion.} 
    \label{fig:b1_comparison}
\end{figure}

To visualize the impact of mass resolution on resolved halo populations, Figures \ref{fig:alpha_comparison} and \ref{fig:b1_comparison} compare a halo sample drawn from the same simulation at $z=0$, analyzed at the native \Sahyadri\ resolution, with an AbacusSummit-like sample.
The \emph{right panels} show the highest number-density \Vpeak-selected halo sample used in this work, while the respective \emph{left panels} show the corresponding sample obtained after degrading the particle mass to that of AbacusSummit and applying the same selection criterion. The increase in the abundance of low-mass halos in \Sahyadri\ is clearly visible across all environments of the cosmic web. In particular, the additional halos populate filaments, cluster outskirts, and underdense regions that are sparsely sampled in lower-resolution simulations.

Following the picture established in \citep{phs18}, Figure~\ref{fig:alpha_comparison} shows that filamentary environments are predominantly populated by halos with large tidal anisotropy $\alpha$ (equation~\ref{eq:alpha-def}).  On the other hand, large halos have very low $\alpha$, and smaller halos in their outskirts, which experience strong tidal influence, have correspondingly larger $\alpha$. Figure~\ref{fig:b1_comparison} highlights the strong environmental dependence of large-scale bias, consistent with the established view that the halo bias is not simply a monotonic function of halo mass, but is better thought of as a specific smoothing of the large-scale density environment of the halo, as discussed in \citep{phs18} and references therein.

All of these qualitative trends have been reported in earlier studies. The new aspect here is that they can be followed to substantially lower halo masses. The increased mass resolution of \Sahyadri\ results in a substantially larger population of low-mass halos across the entire web, which is directly visible in the Figures. In particular, the dense cluster region around $(x,y)=(120,30) \Mpch$ is populated by numerous high-bias halos, while extended underdense regions such as $(170,170) \Mpch$ and $(10,130) \Mpch$ exhibit a large number of strongly negative-bias halos that are largely absent at lower resolution. This leads to a smoother and more continuous representation of the cosmic web and extends the validity of these trends well beyond the mass scales previously accessible.

The role of tracer density in resolving the cosmic web is illustrated in Figure \ref{fig:voronoi_field}, which shows Voronoi-based density fields constructed from the \Vpeak-selected tracer populations with increasing number density that were described in section~\ref{subsec:halo_samples}. For the Voronoi density estimation, we adopt the Monte Carlo algorithm described in \citep{VVF2020} to compute cell densities. These densities are then mapped onto a grid using nearest-grid-point (NGP) interpolation.
As the tracer density increases, the structure of the cosmic web become progressively better resolved (cf., the top right panel of Figure~\ref{fig:z_comparison}).

\begin{figure}
\centering
\includegraphics[width=\linewidth]{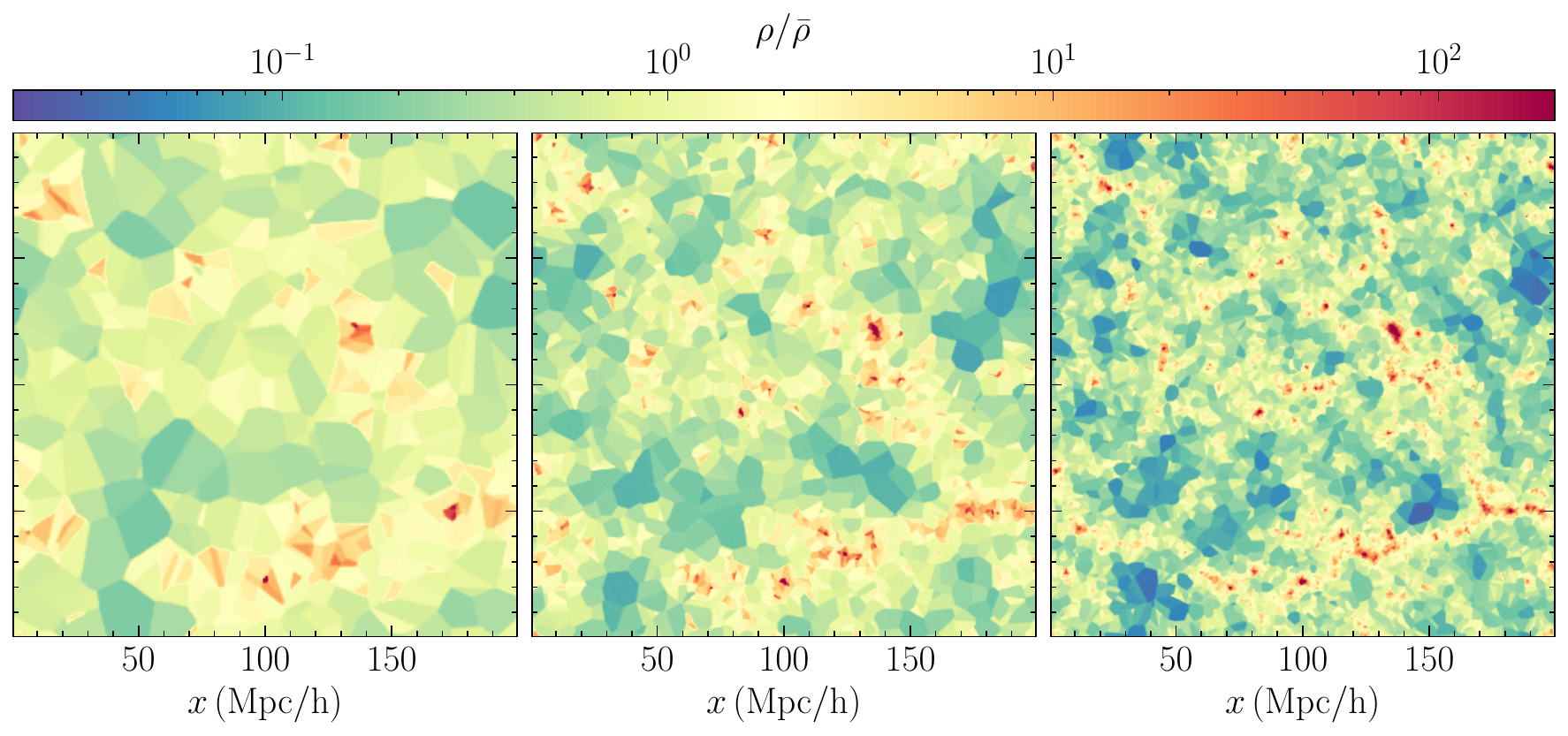}
\caption{Visualizations of single-cell slices of the tessellated density field at $z=0$. The panels from \emph{left} to \emph{right} show density estimates obtained from a Voronoi tessellation of tracers with number density $2\times 10^{-4}, 2 \times 10^{-3}, 2\times 10^{-2}\, (\mathrm{Mpc})^{-3}$ respectively, evaluated on a $256^3$ grid for the lowest number density and $1024^3$ grid for the two higher number densities. All the slices are centered at the same position in the simulation box, the same as the top right panel in Figure~\ref{fig:z_comparison}, and the $y$-axis spans the full box length from 0 to 200 Mpc/h. See text for details.  
}
\label{fig:voronoi_field}
\end{figure}

Overall, these visualizations demonstrate a key capability of the \Sahyadri\ suite: the resolution of low-mass halos across a wide range of cosmic environments. This enables cosmic web analyses based on dense,  well-sampled tracer populations. In the following section, we quantify the cosmological information content of a set of clustering and environment-sensitive statistics.

\myheading{Cosmological dependence of statistics}
\label{subsec:cosmo_dependence}
We now turn to a quantitative analysis of the cosmological dependence of several dark-matter and halo-based statistics. These include both conventional measures of clustering and abundance as well as statistics that probe the geometric and environmental properties of the cosmic web.

We use the mass-selected sample for analysing the matter auto power spectrum and halo mass function, and the \Vpeak-selected samples for the halo auto power spectrum, VVF and $k$NN statistics (see section~\ref{subsec:halo_samples} for the sample definitions).
Results are presented at $z=0$ and $z=1$.

\begin{figure}
    \centering
    \includegraphics[width=\linewidth]{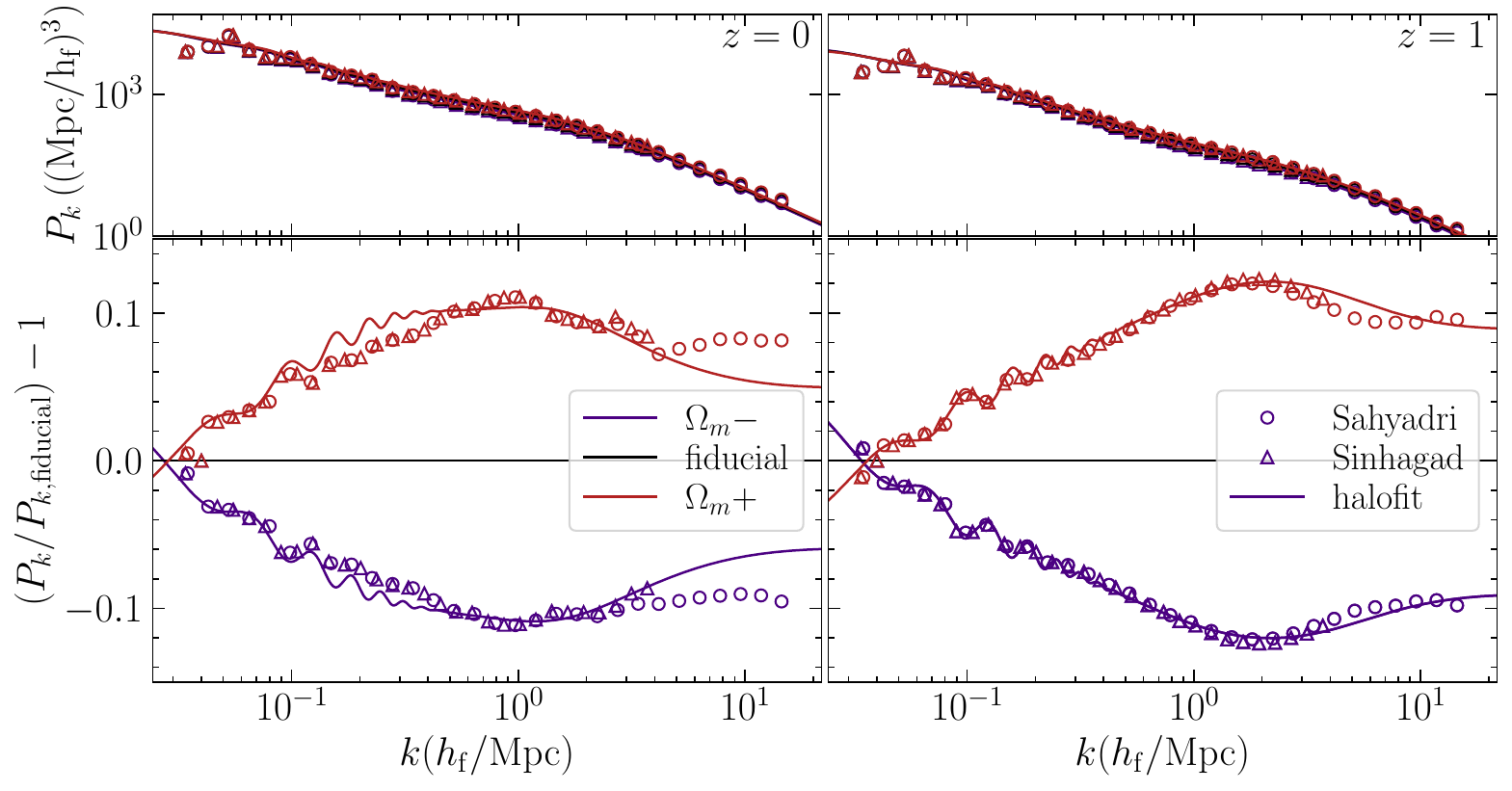}
    \caption{Variation of the matter power spectrum as a function of \Om. Bottom panels show the ratio of the matter power spectrum for each cosmology to that of the fiducial cosmology, and top panels show the full matter power spectrum. Red (blue) corresponds to the $\Om+$ ($\Om-$) variations, while black represents the fiducial cosmology. The left (right) panel shows the results at $z=0$ ($z=1$). Solid lines indicate the Halofit expectations, and circles (triangles) denote results from \Sahyadri\ (\Sinhagad). The simulations show good agreement with the model, particularly at $z=1$.}
    \label{fig:Pk}
\end{figure}

\begin{figure}
    \centering
    \includegraphics[width=\linewidth]{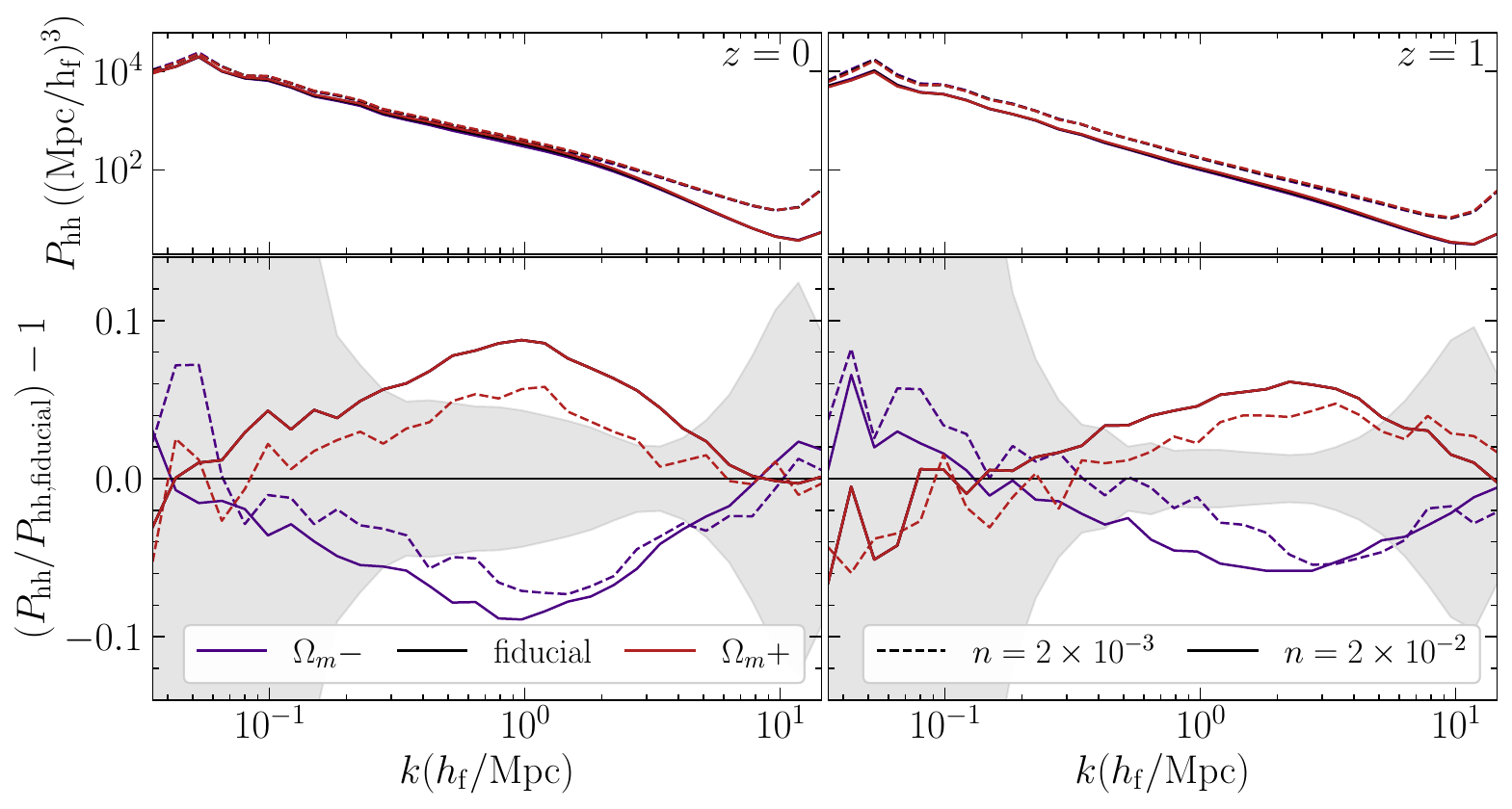}
    \caption{
    Comparison of the halo power spectrum ($P_{\mathrm{hh}}$) for different tracers, \Om values, and redshifts. The top panels show $P_{\mathrm{hh}}$ for two tracer populations with different number densities, thresholded on \Vpeak (see text for details). Solid lines correspond to the fiducial cosmology, while dashed (dotted) lines indicate the $\Om+$ ($\Om-$) variations. The bottom panels show the ratio of each $P_{\mathrm{hh}}$ to its fiducial counterpart, following the same conventions as Fig.~\ref{fig:Pk}. Grey bands show the jackknife errors on the highest number density sample halo power spectrum
    }
    \label{fig:Pk_hh}
\end{figure}

Figure~\ref{fig:Pk} shows the matter auto power spectrum and its response to variations in \Om. The measured power spectra from \Sahyadri\, are in good agreement with expectations from \textsc{halofit} \citep{Takahashi2012} across the range of scales shown, and the expected cosmological dependence is clearly recovered. This demonstrates that the simulations robustly capture both the amplitude and scale dependence of clustering in the quasi-linear and mildly non-linear regime. At $z=0$, deviations from \textsc{halofit} begin to appear for $k\gtrsim  4\,\Mpch$. However, this behaviour is expected, given that \textsc{halofit} was calibrated using lower-resolution simulations and relies on interpolation across cosmological parameter space.  Appendix~\ref{App:sahyadri fitfuncs} shows a detailed comparison of the \Sahyadri\ matter auto power spectra with \textsc{halofit} expectations. 

Figure \ref{fig:Pk_hh} shows the halo auto power spectra for the 
\Vpeak-selected tracer samples. We focus on the two higher number density samples, as the lowest density sample is dominated by noise. Among these, the highest density sample exhibits the clearest and most robust cosmological trends across cosmic epoch.
For comparison, we estimate the uncertainties using the delete-one jackknife procedure in which the simulation volume is divided into $10\times10$ cuboidal regions in the $x-y$ plane, each spanning the full extent of the box along the $z$-direction. The halo power spectrum is calculated by leaving one region out at a time, and accounting for the reduced volume effect while calculating the density contrast. The resulting jackknife errors from these 100 realizations for the highest-density sample are shown as a shaded grey band in the figure. These errors are  for illustrative purpose only and are not intended for quantitative inference. We see from the Figure that, owing to its high resolution, \Sahyadri\ probes deeply non-linear, highly non-Gaussian scales up to $k\gtrsim 10\Mpch$.

\begin{figure}
    \centering
    \includegraphics[width=\linewidth]{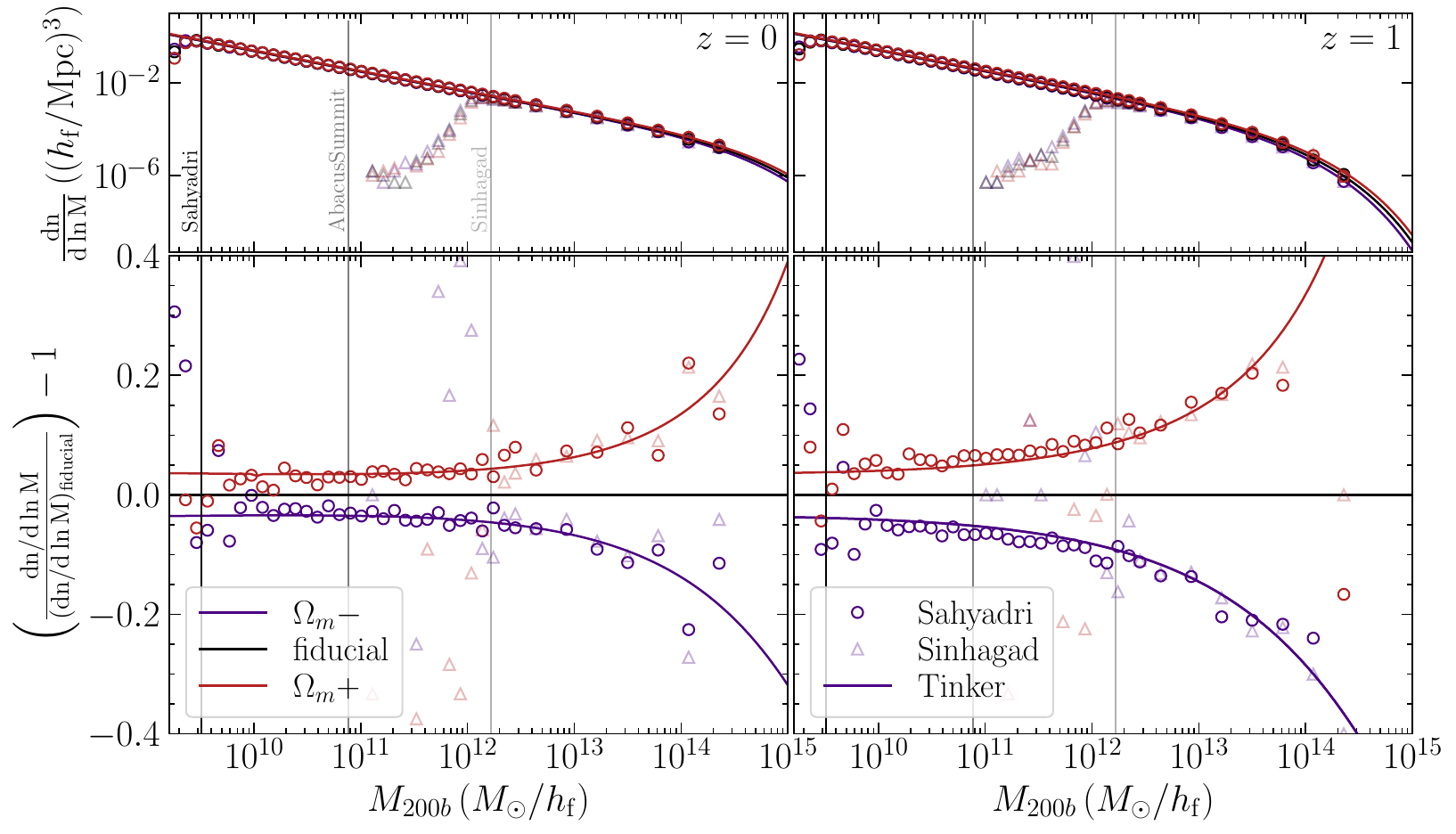}
    \caption{Variation of the mass function as a function of \Om. The arrangement of colors and panels follows Fig.~\ref{fig:Pk}. Solid lines show the Tinker mass function expectation, while the grey vertical lines mark the 40-particle limit for the fiducial \Sahyadri, AbacusSummit and \Sinhagad\, simulation. 
    }
    \label{fig:mf}
\end{figure}

Figure \ref{fig:mf} shows the variation of the halo mass function with \Om, for mass-selected samples. The measured mass functions are in good agreement with the Tinker \citep{Tinker+2008} prediction over the resolved mass range, and the impact of cosmological variations is clearly detected. The extended low-mass reach of \Sahyadri\ allows these trends to be explored well below the mass thresholds accessible to existing simulation suites. Appendix~\ref{App:sahyadri fitfuncs} shows a detailed comparison with Tinker expectation.

\begin{figure}
    \centering
    \includegraphics[width=\linewidth]{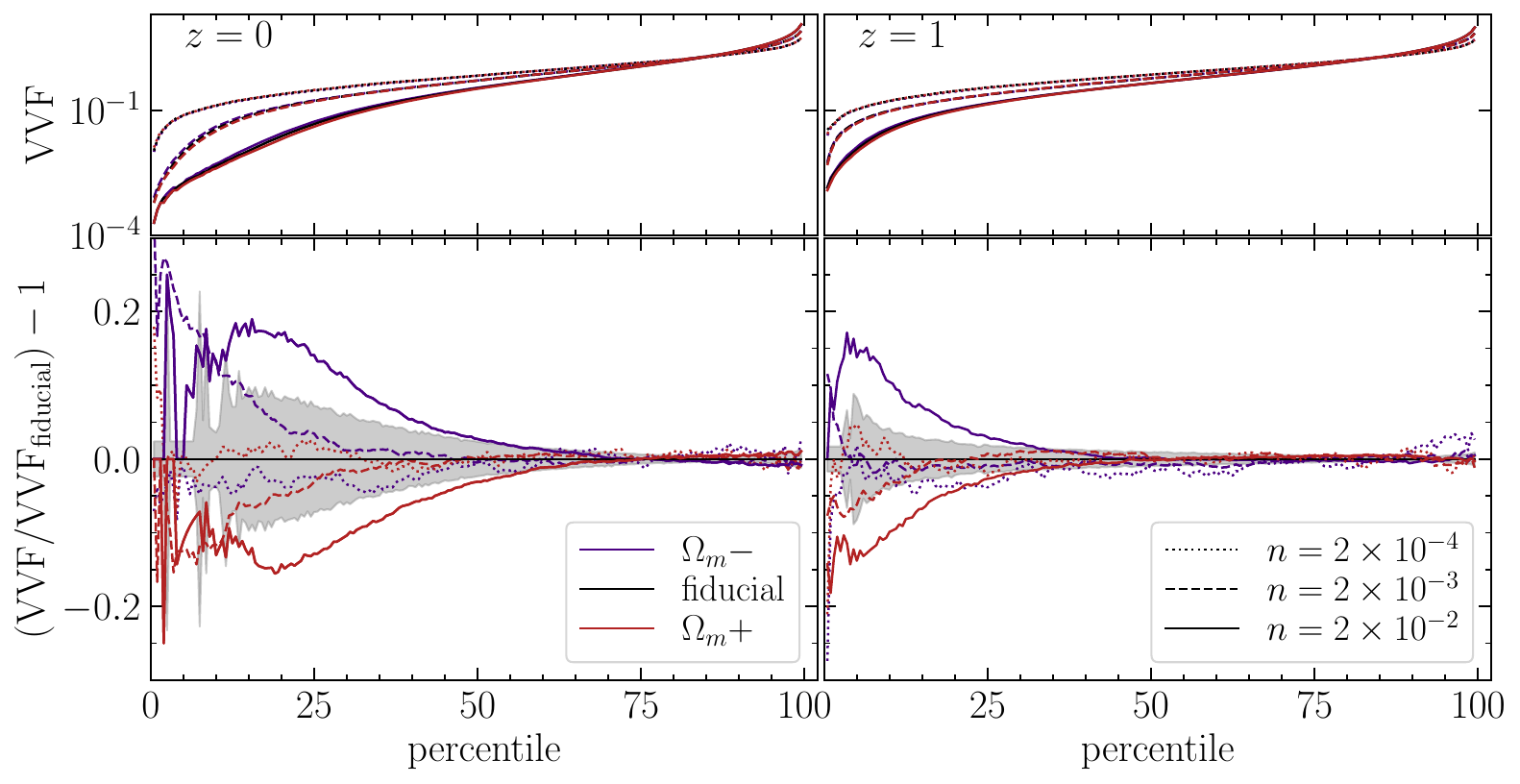}
    \caption{Same as Figure~\ref{fig:Pk_hh}, but for VVF. Here, we show all the three number density tracers considered in this work. Grey bands show the jackknife errors on the higher number density VVF. The results suggest that the VVF, particularly for the higher density tracers, has significant potential to constrain \Om.}
    \label{fig:vvf_100}
\end{figure}

We now turn to statistics that probe the geometry of the cosmic web beyond standard number counts or 2-point clustering measures. Figure \ref{fig:vvf_100} shows the cosmological dependence of the Voronoi volume function (VVF) for three \Vpeak-selected tracer samples. The VVF, introduced in \citep{VVF2020}, is defined as the inverted cumulative distribution function of Voronoi cell volumes \citep{Voronoi1908} and is sensitive to the full hierarchy of higher-order correlations encoded in the tracer distribution.

The improved mass resolution of \Sahyadri\ allows us to compute the VVF for tracer samples with substantially higher number densities than previously explored. The physical length scales corresponding to the VVF percentiles are shown in Appendix \ref{App: sahdyadri vvf-scale}. For the highest-density sample, the statistic probes deeply non-linear scales down to $\sim 0.1\mathrm{Mpc}$.
The delete-one jackknife errors for the highest density tracer sample are shown as the grey band in Figure \ref{fig:vvf_100}. The VVF exhibits a clear and systematic response to variations in \Om, with the strongest sensitivity observed for the highest density tracer sample, where the cosmological signal significantly exceeds the estimated uncertainties over a wide range of percentiles. As shown in \citep{Fisher_stabilization2026}, lower density samples suffer from increased noise, requiring stabilization techniques to identify subsets of percentiles that provide robust Fisher constraints. In contrast, the high-density samples accessible with \Sahyadri\ display smooth and coherent variations of the VVF with cosmology, enabling substantially cleaner measurements and improved constraining power.

\begin{figure}
    \centering
    \includegraphics[width=\linewidth]{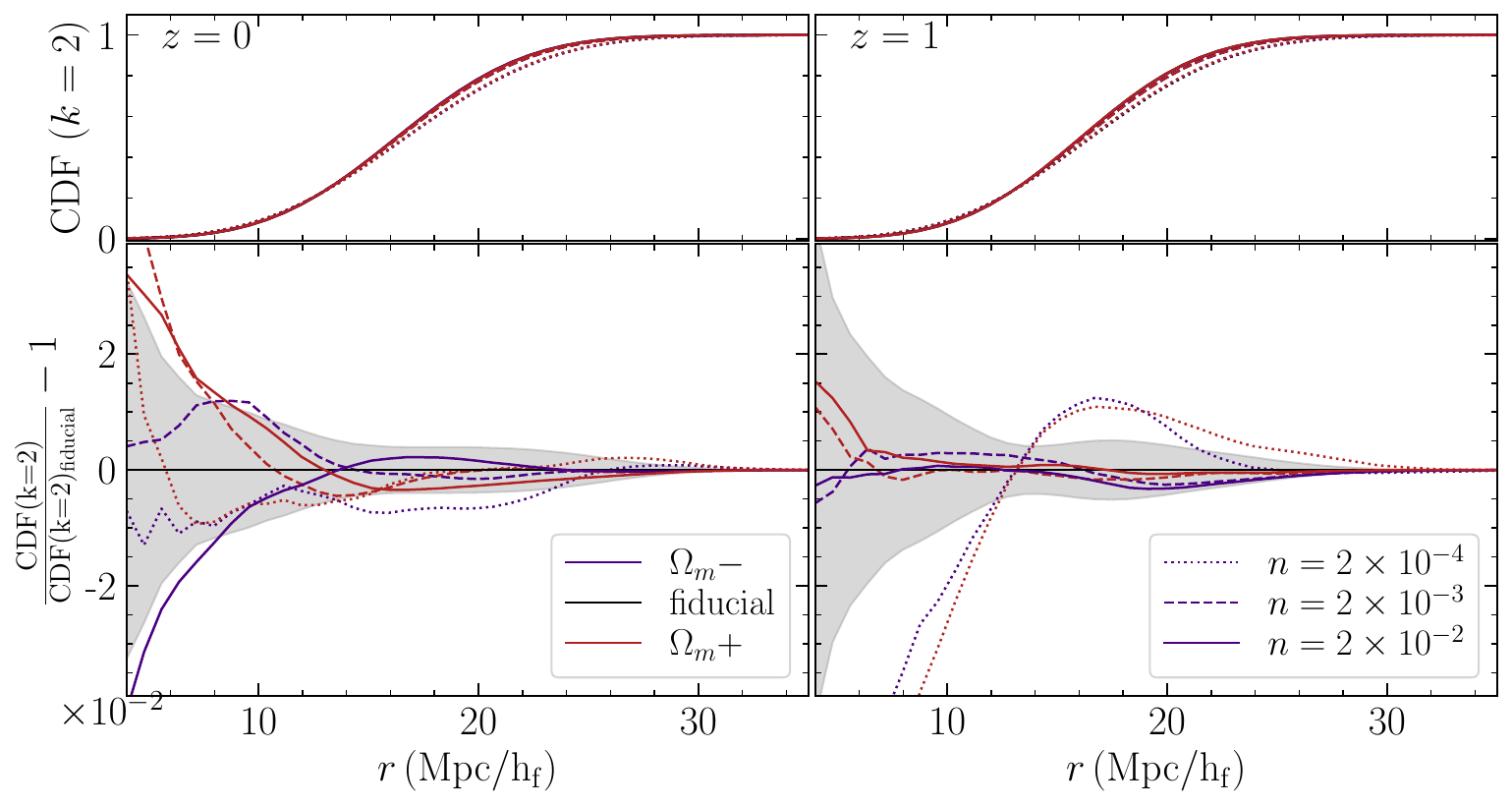}
    \caption{Same as Figure~\ref{fig:vvf_100}, but for the $k$NN CDF at $k=2$.}
    \label{fig:knn}
\end{figure}

A related set of summary statistics that can be used to quantify differences in higher-order halo clustering across cosmologies and samples are the $k$-Nearest Neighbor ($k$NN) distributions. Introduced in \cite{Banerjee:2020}, these statistics are also formally sensitive to all connected $N$-point functions of the underlying clustering \cite{Banerjee:2020, Banerjee:2021, Banerjee:2022} and can be directly related to the geometric features of the cosmic web~\cite{Gangopadhyay:2025}. The $k$-Nearest Neighbor Cumulative Distribution functions ($k$NN-CDFs) are computed by first populating the simulation box with a random, volume filling set of query points, and then creating a sorted list of distances from these query points to the $k$-th nearest neighbor data (halo) points. Using a $k$-d tree structure, these calculations can be sped up to $\mathcal O (N \log N)$ scaling. In Figure \ref{fig:knn}, we plot the results for $k=2$ at redshifts $z=0$ \emph{(left panels)} and $z=1$ \emph{(right panels)} over the radial range $[5,35]\,h^{-1}$Mpc. The $k$NN-CDFs are sensitive to the number density; therefore, to plot them on the same radial scale, we have split halos from each sample into sets with fixed number densities $1\times 10^{-4} (\hMpc)^3$ before performing the calculations. The bottom panels in each column plot the residuals with respect to measurements of the $k$NN-CDFs of that halo sample in the fiducial cosmology. At $z=0$, the sample with an inherent number density of $n=2\times 10^{-2}$ show a clear trend with $\Omega_m$ that is statistically significant. Interestingly, for the other samples, the tradeoff between the underlying clustering of the matter field (controlled by $\Omega_m$), and the change in the bias values of the tracers at a fixed inherent number density nearly cancels each other out over a certain range of scales in the plot. The $k$NN-CDFs are sensitive to both of these effects~\cite{Banerjee:2021cmi}, and  by varying the number density, can be used to break potential degeneracies.

The VVF and $k$NN results above highlight the potential of higher-order, cosmic web–based statistics to extract cosmological information from the quasi-linear and non-linear regimes enabled by the resolution of \Sahyadri.

Finally, following the approach of \citep{Ramakrishnan+2019}, Figures \ref{fig:correlations} and \ref{fig:correlations_z} examine the correlations between halo environment and internal halo properties in our mass-selected samples, and their dependence on cosmology and redshift. We consider three internal properties: the halo concentration $c_{\mathrm vir}$, the mass ellipsoid axis ratio $c/a$, which characterizes halo shape, and the dimensionless spin parameter $\lambda$, which quantifies halo angular momentum. To ensure robust measurements of these quantities, we restrict the analysis to halos resolved with at least 500 particles. For the $\Om +$ cosmology, which has the largest particle mass, this corresponds to a minimum halo mass of $4.2\times10^{10}\,\Msun/h$; this mass threshold is adopted uniformly throughout the analysis.

Halo environment is characterized using the tidal anisotropy $\alpha$ (equation~\ref{eq:alpha-def}) and the halo-by-halo large-scale bias $b_1$, both evaluated on a $1024^3$ grid (see section~\ref{subsec:halos_vahc} for details). Following Appendix A1 of \citep{phs18}, we further require that halos be resolved by at least eight grid cells within a sphere of radius $4R_{\mathrm 200b}$. For the $\Om +$ cosmology, this corresponds to a mass threshold of $1.7\times10^{10}\,\Msun /h$, which we again apply consistently across all cosmologies. Correlations are quantified using Spearman’s rank correlation coefficient.
Figures \ref{fig:correlations} and \ref{fig:correlations_z} present the resulting correlations \emph{(top panels)}, together with the conditional correlation coefficients with distributions conditioned on $\alpha$ \emph{(bottom panels)}. The qualitative trends are consistent with those reported by \citep{Ramakrishnan+2019}, and with related studies demonstrating that correlations between halo environment and internal halo properties encode much of the observed assembly bias, with tidal anisotropy $\alpha$ playing a key role, extending across mass, redshift, and cosmology \citep[e.g.][]{Ramakrishnan+2025, RamakrishnanVelmani2022,RPS2021}.
However, owing to the substantially improved mass resolution of \Sahyadri, we are able to extend this analysis to halos that are smaller by factors of $\sim45$ for the $b_1\leftrightarrow\alpha$ correlation and $\sim18$ for correlations involving internal halo properties. 
We see that $\alpha$ is positively correlated with $b_1$ and $\lambda$, with the correlation strength increasing with halo mass. The correlation between $\alpha$ and $c_{\mathrm{vir}}$ is positive at lower masses, and crosses over to becomes negative for $M_{\mathrm{200b}} \gtrsim 10^{13}\Msun /h$ \cite{phs18,Ramakrishnan+2019}. The correlation with $c/a$ remains positive and comparatively independent of mass. The \emph{top right panel} shows that the correlation of internal properties with $b_1$ is much weaker than that with the tidal anisotropy $\alpha$, while the \emph{bottom panel} highlights the central result of \citep{Ramakrishnan+2019}:  conditional correlation coefficients are significantly smaller relative to their unconditional counterparts, indicating that $\alpha$ is largely responsible for all of these assembly bias trends. No strong trends are seen with variations in \Om.

Overall, the enhanced resolution and tracer density of \Sahyadri\, increase the reach of standard clustering statistics and, at the same time, allow reliable measurements of higher-order, environment-sensitive statistics that probe the cosmic web in the quasi-linear and non-linear regimes.

\begin{figure}
    \centering
    \includegraphics[width=\linewidth]{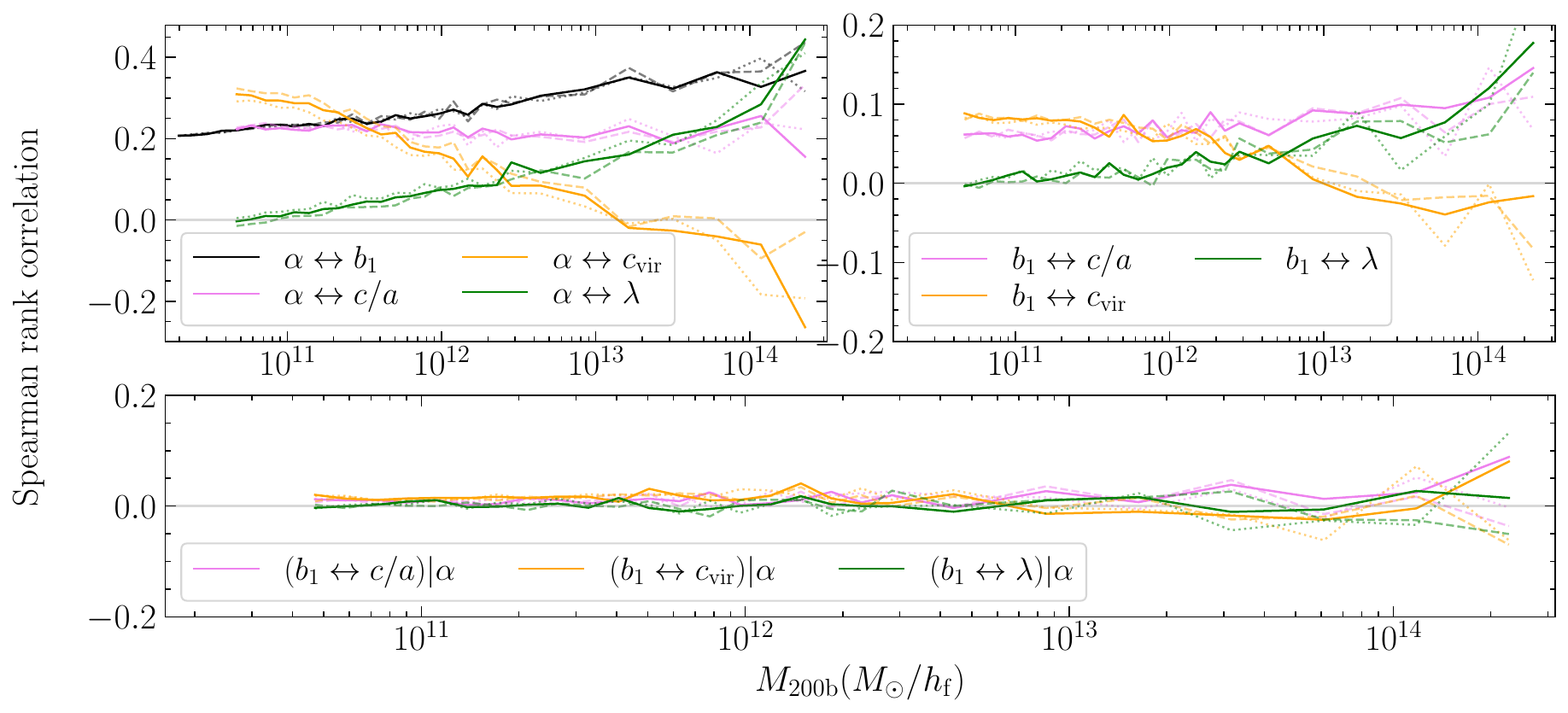}
    \caption{Spearman rank correlations between halo environment and internal properties, as a function of halo mass. Solid lines correspond to the fiducial cosmology, while dashed (dotted) lines indicate the $\Om+$ ($\Om-$) variations. With Sahyadri, these correlations can now be measured for halos a factor of $\sim18–45$ smaller than previously accessible (see text for details).
    }
    \label{fig:correlations}
\end{figure}

\begin{figure}
    \centering
    \includegraphics[width=\linewidth]{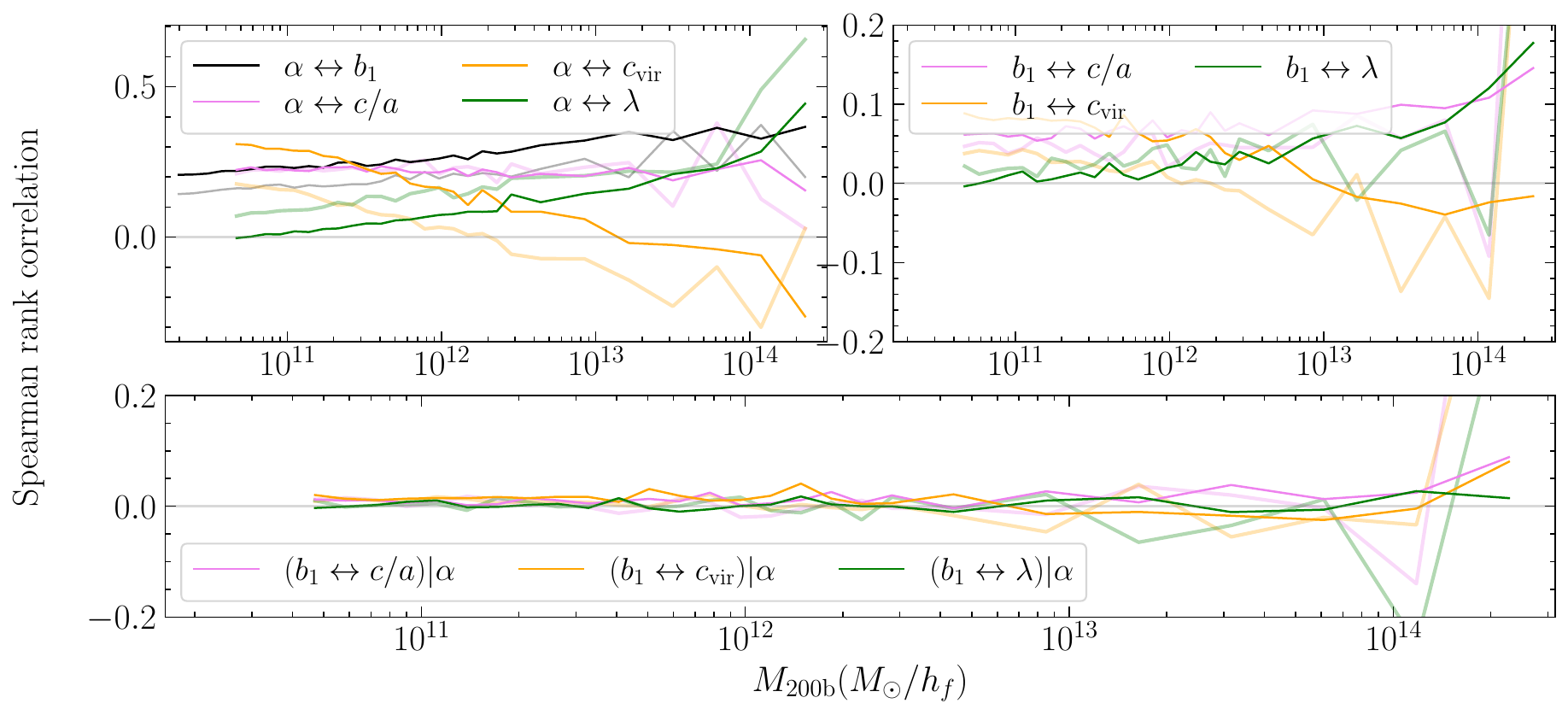}
    \caption{Redshift dependence of correlations for the fiducial cosmology. The panels are the same as in Fig~\ref{fig:correlations}, with dark (pale) curves representing $z=0 \, (1)$.}
    \label{fig:correlations_z}
\end{figure}

\section{Summary and Future Outlook}
\label{sec:summary}

In this chapter, we have motivated and summarized the role of cosmological simulations in modelling the nonlinear large-scale structure. We have presented \Sahyadri, a new suite of high-resolution $N$-body simulations specifically designed to bridge the gap between the mass resolution required for modeling low-redshift spectroscopic surveys and the cosmological parameter coverage needed for precision inference. The key specifications and results of this work can be summarized as follows:

\Sahyadri\ simulations evolve $2048^3$ dark matter particles in periodic boxes of side length $200\,h^{-1}\,{\mathrm Mpc}$, achieving a particle mass of $m_{\mathrm p} = 8.1 \times 10^{7}\,h^{-1}\,M_{\odot}$ and resolving halos down to $M_{\mathrm min} = 3.2 \times 10^{9}\,h^{-1}\,M_{\odot}$ (40 particles). This represents a factor of $\sim$25 improvement in mass resolution compared to AbacusSummit \citep{Maksimova2021} and enables comprehensive modeling of $\sim$40\% more DESI BGS galaxies at $z < 0.15$ than previously possible. The suite systematically varies six cosmological parameters ($\Omega_{\mathrm m}$, $h$, $n_{\mathrm s}$, $A_{\mathrm s}$,  $w_{\parallel}$, $\Omega_{\mathrm k}$) around Planck 2018 values using seed-matched initial conditions to enable derivative calculations for Fisher matrix analyses.

We employ \textsc{gadget-4} with 2LPT initial conditions at $z=49$, producing 101 snapshots uniformly spaced in scale factor between $z=12$ and $z=0$. Our custom particle data compression scheme reduces storage requirements by a factor of $\sim$3 while maintaining clustering accuracy at the sub-percent level, achieving a total data footprint of $\sim$117 TB for the complete suite (including FITS-compressed halo and value-added catalogs). The simulations utilize seed-matched initial conditions across cosmological variations to suppress sample variance and isolate cosmological effects.

We identify halos using the \textsc{Rockstar} phase-space halo finder \citep{Rockstar2013} and implement quality control criteria to ensure robust mass estimates and remove spurious objects.
 Our convergence tests demonstrate that finite volume effects on the linear halo bias become negligible ($\lesssim 1\%$) for halos with $M \gtrsim 10^{12}\,h^{-1}\,M_{\odot}$ (Appendix~\ref{App: sahyadri b1-voldep}). The measured matter power spectra and halo mass functions show excellent agreement with Halofit and Tinker predictions, respectively, validating our numerical methods and force resolution choices (Appendix~\ref{App:sahyadri fitfuncs}).

We have showcased \Sahyadri's capabilities through:
\begin{itemize}
\item Visualizations of cosmic web evolution from $z=5.6$ to $z=0$, highlighting the emergence of filaments, nodes, and the increasing density contrast.
\item Direct comparisons with AbacusSummit-resolution halos, demonstrating the enhanced sampling of filaments, cluster outskirts, and underdense regions enabled by \Sahyadri's improved mass resolution.
\item Voronoi-based density field reconstructions estimates showing progressively better resolution of cosmic web structure with increasing tracer number density.
\item Measurements of the cosmological dependence of standard clustering, abundance, and environment statistics — including the matter and halo auto power spectra, halo mass function, and correlations between halo environment (tidal anisotropy and halo bias) and internal halo properties (concentration, spin, and shape) as a function of mass at $z=0$ and $z=1$. While these trends are consistent with earlier work, \Sahyadri\ extends such analyses to significantly lower halo masses than previously accessible, with the matter power spectrum included mainly as a consistency check. 
\item First measurements of the Voronoi volume function (VVF) and $k^{\mathrm th}$-nearest neighbour ($k$NN) statistics  for high-density halo samples ($n \sim 10^{-2}\,{\mathrm Mpc}^{-3}$), revealing significant sensitivity to $\Omega_{\mathrm m}$ variations.
\end{itemize}

The enhanced resolution of \Sahyadri\ is particularly well-suited for addressing several pressing questions in contemporary cosmology. The low-redshift Universe ($z < 0.5$) is the regime where surveys achieve their highest number densities and where potential solutions to the Hubble tension and hints of evolving dark energy may reside. Traditional 2-point statistics in this regime have begun to approach cosmic variance limits given available volumes. \Sahyadri\ enables the exploration of higher-order statistics and smaller-scale clustering in a controlled framework with cosmological parameter variations, opening new avenues for extracting cosmological information from the non-linear regime.

The \Sahyadri\ simulations enable diverse applications including halo occupation distribution modeling for DESI BGS and 4MOST, studies of environmental quenching and AGN activity in low-mass halos, and validation of analytical models bridging quasi-linear and non-linear scales. The enhanced resolution facilitates Fisher matrix forecasts for beyond-two-point statistics (VVF, $k$NN, Minkowski functionals, etc.), improved estimates of fiber collision effects in spectroscopic surveys through better sampling of faint galaxies, and construction of training datasets for machine learning-based galaxy-halo connection studies in the non-linear regime.

At the time of writing this thesis, we have completed the fiducial cosmology simulation along with the full parameter variations for \Om, $h$, \ns, and \As. The remaining parameter variations ($w_{\parallel}$, $\Omega_{\mathrm k}$) are in progress. We will also release higher-level data products, including detailed merger trees, pre-computed summary statistics (power spectra, correlation functions, VVF, $k$NN statistics, etc.) across all cosmologies, mock galaxy catalogs using empirical galaxy-halo connection models calibrated to DESI, SDSS, and GAMA observations, and cut-sky catalogs for direct comparison with survey geometries.

All \Sahyadri\ data products—including particle snapshots, halo catalogs, compression tools, and analysis scripts will be made publicly available to the community through a dedicated web portal. We anticipate that \Sahyadri\ will become a valuable resource for the broader community working on low-redshift large-scale structure, galaxy evolution, and non-linear cosmology, complementing existing simulation suites and enabling new science in the transition regime between quasi-linear theory and full hydrodynamical modeling.

    
%
\let\textcircled=\pgftextcircled
\chapter{Cosmic web reconstruction: Methods and challenges}
\label{chapt:filtools}

\initial{C}osmic filaments are the most visually striking feature of the LSS, forming an interconnected network that channels matter, gas and galaxies across the cosmic web. Their properties are closely linked to anisotropic gravitational collapse and coherent matter flows, making them important probes of nonlinear structure formation. Despite their prominence in both observations and simulations, however, filaments remain considerably more difficult to define and characterize than virialized dark matter halos. Different filament-finding methods often identify substantially different structures even when applied to the same input data, leading to significant ambiguities in filament reconstruction and interpretation.

This chapter discusses the problem of cosmic web reconstruction, with particular emphasis on filament identification and calibration. A broad overview of commonly used filament-finding approaches is presented, followed by a discussion of the conceptual and numerical challenges associated with reconstructing filamentary structure. The chapter then introduces a framework developed to systematically calibrate filament finders and quantify as well as alleviate biases arising from filament spine reconstruction, smoothing procedures and geometric effects.

\section{Introduction}
\label{sec: Introduction}
Cosmic filaments are the most prominent features of the cosmic web, forming elongated bridges that connect massive halos and clusters across a wide range of scales. Over the past few decades, they have been studied from several different perspectives, including their geometry, topology, environmental dependence and dynamical evolution
\citep[e.g.][]{Colberg_et_al2005, Aragon-calvo+2010a, NEXUS2014, Martizzi_et_al2019, Malavasi_et_al2020}. Many properties of galaxies and dark matter halos, including their accretion history, angular momentum and velocity environment, are known to correlate with their location within the cosmic web \citep[e.g.][]{Hahn_et_al2007b, Dubois_et_al2014, Kraljic_et_al2018}
. In particular, the impact of filaments on the formation and evolution of low-mass halos is now well established \citep[e.g.][]{hahn+09, zomg-I, phs18, musso+18}. Filaments are also believed to act as cosmic highways that channel matter and gas onto halos, thereby influencing galaxy growth and star formation activity \citep[e.g.][]{Keres_et_al2005, Raychaudhury&Porter2005, Kirk_et_al2013,Konyves_et_al2015, Seth&Raychaudhury2020}. Since they trace coherent anisotropic matter flows across the Universe, filaments naturally occupy a central place in studies of nonlinear structure formation and cosmic velocity flows.

Despite decades of work, however, fundamental properties regarding the evolution and demographics (e.g., length and thickness distributions as a function of node halo properties) are not widely agreed upon \cite{Aragon-calvo+2010a, 2pop2020, Zhu_et_al2021, wang+24}. Considering the well-known universality of density structure of dark matter halos \cite{NFW97, Einasto1965, merritt+06} and similar properties recently emerging for cosmic voids \cite{Pan_et_al2012, Hamaus_et_al2014, Nadathur_et_al2015}, it seems natural to expect the structure of cosmic filaments to also show a similar universality \cite{Yang_et_al2022, Xu+2026}.  Such universality, if present, could potentially provide new probes of cosmology and dark matter physics. Addressing this question, however, is complicated by the multiscale nature of the filamentary network and by the substantial differences between existing filament-finding approaches \citep{Libeskind_et_al2018}.

Unlike dark matter halos, which admit relatively clear physical definitions associated with overdensity and virialization; filaments are diffuse, anisotropic and intrinsically multiscale structures. They are neither linear objects that can be treated perturbatively, nor fully virialized systems with a simple dynamical definition. Their morphology and internal structure are shaped by nonlinear gravitational collapse occurring simultaneously across a wide range of scales, making them considerably more difficult to characterize theoretically and observationally. As a result, there is currently no universally accepted definition of what constitutes a filament. Existing filament-finding algorithms are based on a variety of fundamentally different ideas, including Hessian-based classifications of density or tidal fields, topological skeleton reconstruction, graph-based approaches and tessellation-based methods \citep{Libeskind_et_al2018}. Some methods identify filaments as extended three-dimensional regions, whereas others directly reconstruct one-dimensional filament spines. Importantly, different filament finders often identify rather different filaments even when applied to the same input data. Quantities such as thickness and phase-space profiles can therefore depend sensitively on the choice of filament finder, smoothing scale, tracer population and numerical resolution.

This poses a significant challenge for attempts to systematically study filament demographics and internal structure. In particular, uncertainties in filament spine reconstruction can directly propagate into measurements of density and velocity profiles, thereby biasing attempts to characterize filament phase-space structure and matter flows along the cosmic web. 

Since the true filament spines are not known \emph{a priori}, it becomes difficult to assess which features are physical and which arise from reconstruction systematics. Understanding and controlling these effects is therefore essential before filament statistics can be reliably used for precision studies of the nonlinear large-scale structure.

In this chapter, we first present a broad overview of commonly used filament-finding techniques and discuss the conceptual difficulties associated with cosmic web reconstruction. We then introduce a framework designed to systematically calibrate filament finders and quantify biases arising from filament spine reconstruction, smoothing procedures and geometric effects. Particular emphasis is placed on understanding how reconstruction uncertainties affect inferred filament density and velocity structure, and how the biases can be reduced.

\section{Cosmic web finders: methods and challenges}

\subsection{A broad overview}
Over the years, a large number of methods have been proposed to identify and classify structures in the cosmic web. These approaches differ not only in their implementation, but also in what they fundamentally regard as a  filament, wall or node. Some define filaments through the geometry of the density field, others through topology or connectivity, while certain methods attempt to follow the underlying dynamical evolution more directly. As a result, different filament finders often recover rather different filamentary networks even when applied to the same data \citep{Libeskind_et_al2018}. This diversity reflects the fact that filaments do not possess a unique physical definition analogous to virialized dark matter halos. In this section, we briefly review the broad classes of filament-finding approaches commonly used in studies of the cosmic web, along with their associated limitations and systematic uncertainties.

\myheading{Hessian based methods}
One of the most widely used classes of cosmic web finders is based on the local properties of the density field or related tensor quantities. In these approaches, the particle or galaxy distribution is first interpolated onto a grid to obtain a continuous field, which is subsequently smoothed on one or more scales. The Hessian matrix of the density field, gravitational potential or velocity shear tensor is then computed, and structures are classified according to the eigenvalues and eigenvectors of these tensors.

The basic idea behind these methods is that the local eigenstructure of the Hessian field encodes the preferred directions of gravitational collapse. Regions collapsing along three directions are identified as nodes, collapse along two directions corresponds to filaments, collapse along one direction gives sheets or walls, while regions with no collapsing directions are classified as voids. Representative examples include the tidal tensor method \citep{Hahn_et_al2007b}, the velocity shear-based V-web formalism \citep{Hoffman+2012}, and the multiscale NEXUS framework \citep{NEXUS2013,NEXUS2014}.

These methods are computationally efficient and naturally provide a classification of the entire simulation volume. They are also relatively straightforward to implement on gridded fields and are effective at recovering the large-scale morphology of the cosmic web. At the same time, the identified structures can depend strongly on the choice of smoothing scale and grid resolution, motivating the development of multi-scale approaches such as NEXUS, which combine information across multiple smoothing scales to identify more robust structures. In addition, although these methods successfully identify filamentary regions, they do not always directly produce well-defined filament spines, which are required for studying filament profiles and phase-space structure.

\myheading{Topological and skeleton-based methods}

A second major class of filament finders is based on topology and Morse theory. These approaches attempt to reconstruct the filamentary skeleton of the input field by identifying critical points and tracing topological connections between them. Among the most widely used examples is \disp\ \citep{Disperse_theory,Disperse_illustration}, which constructs the Morse-Smale complex of the density field and identifies filaments as integral lines connecting maxima and saddle points.

Topological methods are widely used because they naturally recover connected filamentary networks without requiring an explicit geometric description of filaments. They can also be applied directly to discrete tracer distributions, avoiding the need to first construct a regular grid-based density field. In practice, persistence thresholds are usually introduced to remove structures arising from shot noise and sparse sampling.

At the same time, these methods are not free from ambiguities. The recovered filamentary network can depend sensitively on the choice of persistence threshold, tracer density and smoothing procedure, especially in case of sparse samples. Moreover, the reconstructed filament spines are often noisy and may require additional smoothing or post-processing before reliable measurements of filament properties can be made.
\myheading{Graph-based approaches}
Another broad class of filament finders treats the cosmic web as a network connecting discrete tracers such as galaxies or dark matter halos. In these approaches, filaments are identified through connectivity patterns in the tracer distribution, often using graph-theoretic constructions such as minimal spanning trees or related network algorithms \citep[e.g.][]{Barrow+1985,Pereyra+2020}.

Such methods are particularly useful for studying the connectivity of the cosmic web. Some approaches explicitly model filaments as chains of connected halos or galaxies, while others reconstruct filamentary backbones through optimization or network-building procedures. Since these methods operate directly on discrete tracers, they do not require reconstruction of a continuous density field.

At the same time, the recovered filamentary network can depend significantly on the adopted linking criteria, tracer density and graph construction procedure. In addition, graph-based methods often prioritize connectivity and network structure over the detailed morphology of the underlying matter distribution.

\myheading{Phase-space and multistream methods}
More recently, several approaches have attempted to characterize the cosmic web using information from the full phase-space structure of dark matter rather than relying solely on the density field. Since cold dark matter initially occupies a thin three-dimensional sheet in six-dimensional phase space, nonlinear gravitational evolution leads to shell crossing and the formation of multistream regions that encode dynamical information about structure formation.
Methods based on multistream counting or phase-space tessellations identify structures according to the number of overlapping streams at a given location \citep{Shandarin+2012,Abel+2012}. Such approaches provide a more dynamical characterization of the cosmic web and are closely connected to the underlying gravitational evolution itself. In principle, they can distinguish genuinely nonlinear structures from transient density fluctuations more naturally than purely geometric methods.

Their main limitation is that they require access to the full dark matter phase-space information, making them difficult to apply directly to observational galaxy catalogues. In addition, the computational cost associated with phase-space reconstruction can become significant for large simulations.

\subsection{Common systematics and reconstruction challenges}
Although these methods are based on very different principles, they share several common systematic uncertainties. One of the most important is tracer bias. Observationally, filaments are identified using galaxies, which are biased tracers of the underlying dark matter distribution. The recovered filamentary network can therefore depend strongly on galaxy selection, number density and survey completeness.

Smoothing also plays a central role in most filament-finding approaches. Small smoothing scales preserve fine substructure but are more sensitive to noise and discreteness effects, whereas large smoothing scales produce cleaner filamentary networks at the cost of washing out physically relevant small-scale structure. Since the cosmic web is intrinsically multiscale, there is generally no uniquely preferred smoothing scale \citep{Aragon_calvo+2024}.

Additional complications arise in observational analyses due to redshift-space distortions. Peculiar velocities distort the inferred galaxy distribution along the line of sight, producing elongated structures such as Fingers-of-God and large-scale anisotropies associated with coherent infall. These distortions can significantly affect filament identification and inferred filament properties, particularly for methods based directly on spatial geometry or local density gradients.

Numerical resolution and discreteness effects also play an important role, especially in low-density environments. Finite particle number, grid resolution and shot noise can all influence the recovered filamentary network. Furthermore, different filament finders often respond differently to these effects, complicating comparisons across methods and datasets (c.f. Figure~2 in \citep{Libeskind_et_al2018}).

Another important difficulty is that most filament finders contain one or more free parameters, such as smoothing scales, persistence thresholds or linking lengths, which must be chosen before the filamentary network can be reconstructed. In practice, these parameters are often selected heuristically, based on whether the resulting network appears visually reasonable or qualitatively consistent with expectations. Since the true filamentary structure is not known \emph{a priori}, however, there is generally no objective way to determine the optimal parameter choices beforehand. This leads to a somewhat circular problem: one requires a filament finder to identify filaments, but one also requires some prior knowledge of filamentary structure in order to calibrate the filament finder itself.

More fundamentally, the lack of a unique physical definition of a filament implies that different filament finders may identify substantially different structures even for the same underlying field. This finder-to-finder variation remains one of the central challenges in cosmic web studies and complicates attempts to compare filament statistics across different analyses.
\subsection{Motivation for calibration and robustness studies}
The absence of a uniquely defined filament population makes it difficult to determine which reconstructed structures correspond to physically meaningful filaments and which arise from methodological choices or numerical systematics. In particular, uncertainties in filament spine reconstruction can directly propagate into measurements of filament density and velocity profiles, thereby biasing attempts to characterize phase-space structure and matter flows along the cosmic web.

This motivates the need for systematic calibration and robustness studies of filament-finding methodologies. Ideally, one would like to understand how reconstruction uncertainties depend on intrinsic filament properties, tracer sampling, smoothing procedures and observational effects. However, unlike dark matter halos, the true filament spines are not known \emph{a priori}, making direct calibration difficult.

Most filament finders also contain one or more free parameters, such as smoothing scales, persistence thresholds or linking criteria, which must be chosen before the filamentary network can be reconstructed. In practice, these parameters are often selected heuristically, based on whether the recovered structures appear visually reasonable or qualitatively consistent with expectations. Since the underlying filamentary structure is itself unknown, however, there is generally no objective way to determine the optimal parameter choices beforehand.

The remainder of this chapter discusses a framework developed to address this problem by constructing controlled filament realizations with known underlying spine geometries and phase-space structure. Such realizations provide a means to systematically quantify reconstruction biases, evaluate filament-finding performance and study the impact of smoothing and geometric effects on inferred filament properties.

\section{\filtools: Calibration of filaments and systematics}
\label{sec: filtools}
\subsection{Tools and techniques}
\label{subsec: tools and techniques}
Here, we present a framework designed to systematically study and calibrate filament finders using controlled filament realizations with known geometric and phase-space structure. The framework consists of two complementary tools. The first, the {\bf Fil}ament {\bf Gen}erator (\filgen), generates mock filament realizations with specified spine geometries, density profiles and velocity structure. The second, the {\bf Fil}ament {\bf A}nalysis and {\bf P}rocessing {\bf T}ool (\filapt), estimates filament density and velocity profiles from reconstructed filament spines and enables curvature-based analyses of filament populations.

In addition, we introduce a robust Fourier-space based smoothing approach along with optimized smoothing procedures for reconstructed filament spines. These methods are designed to mitigate biases introduced by reconstruction noise and to obtain more robust estimates of filament density and velocity structure.

\mysubheading{Filament Generator \filgen}
The basic philosophy underlying \filgen\ is to construct particle realizations around analytically specified filament spines, such that the underlying filament geometry and phase-space structure are known exactly. These realizations then serve as ''the truth'' against which filament finders and post-processing procedures can be evaluated. Since the true spine is known by construction, biases arising from smoothing, curvature, tracer sampling and reconstruction noise can be quantified directly.

The generated filaments can include specified radial density and velocity profiles, curved filament spines, halo-like nodes, their outskirts, and background environments. This flexibility allows systematic investigation of how inferred filament properties depend on intrinsic filament geometry and dynamical structure. Technical details of the filament generation procedure are presented in Appendix:~\ref{App: filtools filgen}. Figure~\ref{fig:filament_illustration} shows a mock filament generated using \filgen.
\begin{figure}
\centering
\includegraphics[width=\textwidth]{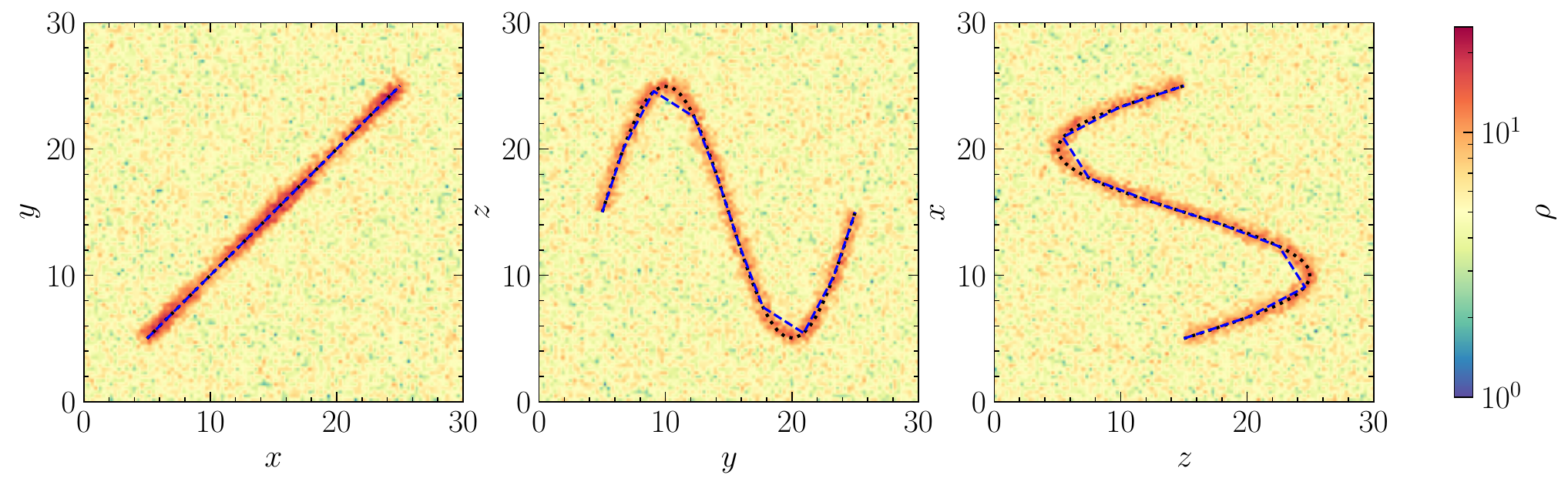}
\caption{Plots showing number density projections of the fiducial filament along the three Cartesian axes for one realization. All axes are in units of \Mpch, and the number density is given in units $(h/\mathrm{Mpc})^3$, with input background density being $5$. Black dotted curves show the actual spine of the filament, while blue dashed curves show the spine reconstructed using sparse tracers. The discrepancies become more significant in regions of high curvature, illustrating the impact of sampling and reconstruction noise on filament spine identification.}
\label{fig:filament_illustration}
\end{figure}

\mysubheading{Filament Analysis and Processing Tool \filapt}
The tool \filapt\ estimates density and velocity profiles around reconstructed filament spines and enables statistical analyses of filament populations. In particular, it allows filament properties to be studied separately in regions of different curvature, which becomes important because geometric effects alone can bias inferred filament profiles, as we show later in this chapter.

When the local radius of curvature becomes comparable to the filament thickness, different parts of the filament cross-section can overlap during profile estimation, distorting recovered density and velocity profiles even if the spine itself is perfectly known. Restricting analyses to relatively straight, low-curvature segments therefore becomes important for obtaining robust filament statistics.

Technical details of the profile estimation and curvature-based analysis are presented in Appendix~\ref{App: filtools filapt}.

\mysubheading{Spine processor /smoother}
We have stressed in the discussion so far that a robust estimation of the spine is necessary for the extraction of unbiased profiles. However, it is a widely acknowledged fact in the literature that even in the most ideal of circumstances, the filament spines recovered by any filament finder will have errors. For particle-based filament finders, these errors originate from both the imperfections in the method of filament finding, as well as the discreteness of the tracers used. For grid-based filament finders, they originate from the finite size of the grid and Poisson noise in the simulations used to estimate the fields on the grid itself. Note that the errors will persist even when the parameters of the filament finder are optimally calibrated. These errors on the spine will in turn lead to systematic errors in the estimation of various profiles of the filaments (since the recovered spine is not exactly aligned with the actual spine, the inferred radial and tangential directions will not be exact), as well as in estimates of the filament lengths. Thus, it is essential that the spines identified by filament finders are processed properly before being used to calculate any statistics.

One of the most widely used methods of smoothing the spines obtained from particle-like data using \disp\, is to average the position of every point on the spine by those of its immediate neighbours, keeping the end points fixed \cite{Flows_around_galaxiesI, ramsoy+21}. This is done iteratively \Nsm\ number of times, until the filament appears smooth enough, or matches with the visually apparent filaments.  Thus, the number of iterations is highly subjective. Also, if the same value of \Nsm\ is used for all the filaments (as is common), the filaments with less number of points on the spine will invariably be over-smoothed as compared to the other filaments. Another problem with this kind of smoothing is that the outlier points with large errors tend to pull the other points progressively away from the actual spine with each iteration, when \Nsm\ is small. However, this effect vanishes at high enough \Nsm. On the other hand, there is no convergence with increasing \Nsm\ (i.e., more smoothing), since the smoothed spine at large enough \Nsm\  always tends to a straight line joining the end points, regardless of the actual original shape. 

Here, we present a new approach that optimizes the smoothing parameters for the chosen smoothing method separately for each individual filament. The idea is to iteratively smooth the filament by varying the smoothing parameters and estimating the density profile after each iteration. Both under- and over-smoothing will tend to broaden the density profile, and thus the narrowest density profile should correspond to the optimum smoothing. This removes any ambiguity related to the choice of smoothing parameters, as well as optimally smooths every filament separately. We also present an alternative method to the neighbour smoothing: Fourier smoothing, where the spines are processed in Fourier space by applying a low pass filter. We explain both the Fourier smoothing and optimization in Appendix~\ref{App: filtools spine_smoothing}, and illustrate the improvement provided by optimization in section~\ref{sec:optimized_smoothing}. The smoothing of filament spines using optimized Fourier smoothing is implemented by the module \filapt\texttt{.SmoothSpine}. Although we illustrate our results using spines obtained via \disp\, and compare our Fourier smoothing with neighbour smoothing which is widely used by \disp\, users, the optimized smoothing introduced in this chapter can be used on discrete spines obtained by any filament finder. We also note that there exist other techniques to smooth the filament spines such as the one employed in T-ReX, the robustness of which we have not explored in this work.

\subsection{Results and Discussion}
\label{sec:results}
Here, we illustrate the results of \filgen\ and \filapt\ using some examples. We generate different filaments with densities comparable to those expected in a $600\, \Mpch$ N-body simulation with $1024^3$ particles. The velocity profiles are also similar to what one might expect in typical cosmological filaments (see section~\ref{sec:fiducial} for details). First, we illustrate a thin filament with a sinusoidal spine, in order to sample regions of varied curvature. The spine is finely sampled, with zero noise to study the working of the profile generator in ideal conditions. This is our fiducial model. Then, we generate a thick filament, where $r_f$ is equal to $R_\kappa$ of the spine at the point of its maximum curvature, and illustrate various interesting effects that appear in such situations. Next, we use the fiducial filament, but with a sparsely sampled spine and illustrate the advantages of curvature splitting. For all the aforementioned examples, the spine provided to the estimator was perfect, with no error and noise, which is far from what one expects in real situations. To account for realistic noise, we estimate the profiles of the fiducial filament by using a spine identified using \disp. We show that this leads to unsatisfactory estimations of the profiles, and there is a need to post-process the noisy spines. We do this using the two methods, and compare the results. 

\begin{figure}
    \centering
    \includegraphics[width=\textwidth]{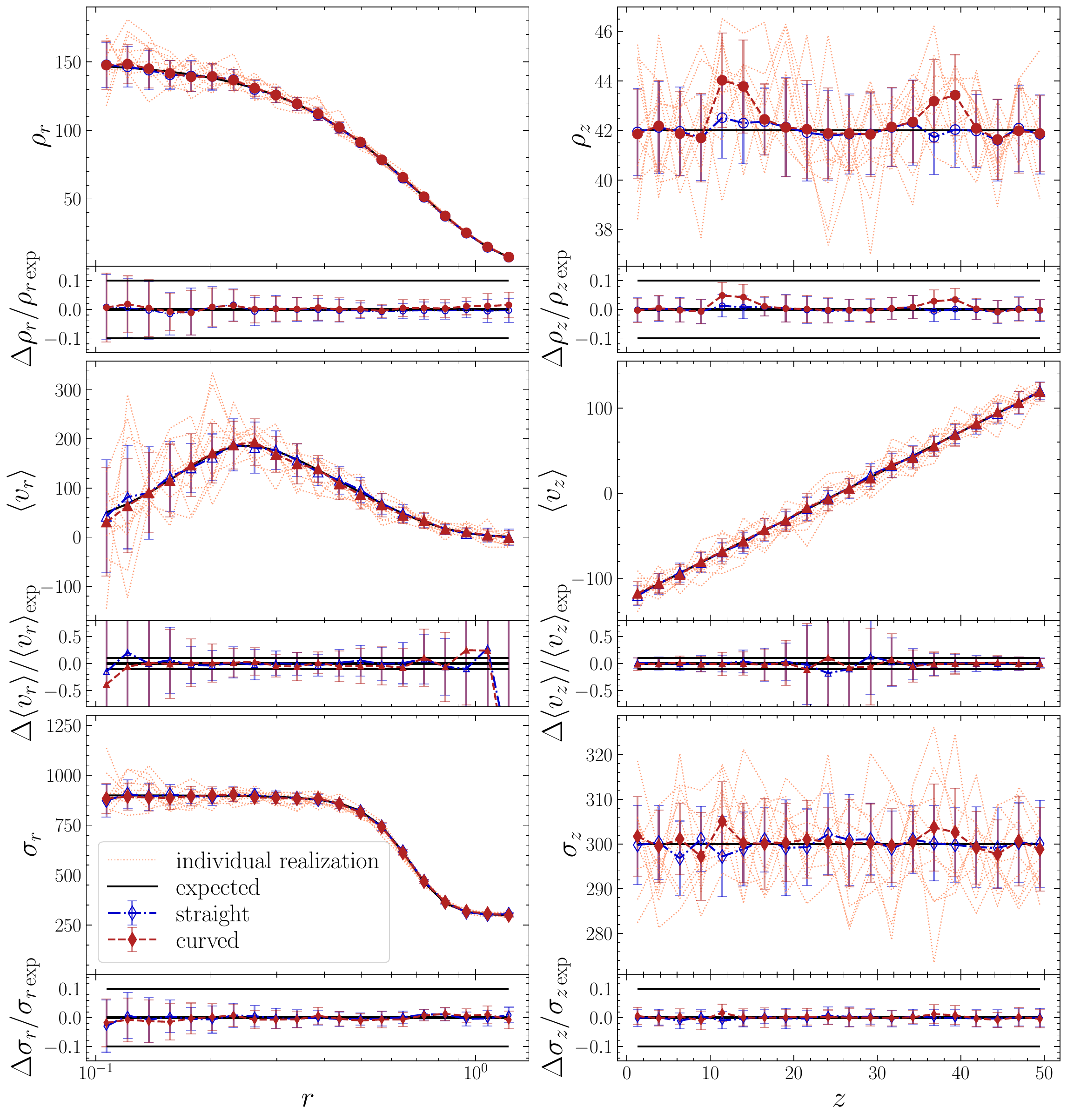}
    \caption{Various profiles for the fiducial filament model. The left (right) panels show longitudinally (radially) averaged radial (longitudinal) profiles. All horizontal axes are in units of \Mpch. See the text for a detailed explanation. The top, middle and bottom left (right) panels show the density, mean radial (longitudinal) velocity and radial (longitudinal) velocity dispersion profiles respectively. Densities have units $(\Mpch)^{-3}$, and velocities are given in \kms. The markers and error bars show the mean and standard deviations across $50$ realizations. Solid red markers represent the fiducial curved filament, whereas open blue markers represent the corresponding straight filament. 10 random realizations of the curved filament are shown with red dotted lines for reference. The solid black lines represent the expected analytical profiles. The narrow panels below each main panel show the relative errors with respect to the expected profiles to give better intuition about the error bars. Thin horizontal black lines in these panels show the $\pm 10 \%$ levels. We see that there is an excellent agreement between the generated and the analytical profiles.}
    \label{fig:fiducial}
\end{figure}

\subsubsection{The fiducial filament}
\label{sec:fiducial}
We generate an aperiodic box of side $30\,\Mpch$, with a curved filament having radius $r_f=0.5\,\Mpch$, concentration $c_f=30$ and radial profile described in Appendix~\ref{App: filtools filgen}. Since we are working in real space at this stage, there is no need to model the node halos and their outskirts, as they will
be cut off in any case while estimating the profiles. Therefore, for all the real space examples we only model the filament and the background. The region of maximum curvature of the spine has $R_\kappa = 2\, \Mpch$, so that $R_\kappa \gg r_f$. Hence, the effect due to the overlap of particles because of curving of the spine will be negligible, and we can safely compare the profiles of the curved filament with the corresponding inputs. Figure~\ref{fig:filament_illustration} shows the density projections of this fiducial filament along the three Cartesian axes. Note that the density variations seen along the spine in the left panel are purely projection effects.

Although considerable work has been done in literature to study the density profiles of cosmological filaments see, e.g., \cite{NEXUS2014, Colberg_et_al2005, 2pop2020, Yang_et_al2022, Galarraga_Espinosa_et_al2023a}), not much attention has been given to their velocity profiles (although, see, \cite{Rost_et_al2024, ramsoy+21}). However, it is expected that the matter flows radially onto the filament, and longitudinally away from its centre towards the nodes. Thus, we adopt the following ansatz for the velocity profiles: 
\begin{align}
    \langle v_z \rangle & = V_{0z} \times \frac{(z-l_f/2)}{l_f} \,, \label{eq:vz_fil}\\
    \langle v_r \rangle & = \frac{V_{0r}}{2}\bigg[1+ \erf{c\left(r/R_f-a\right)}\bigg]\,\bigg[1+\erf{b\left(a-r/R_f\right)}\bigg]\,, \label{eq:vr_fil}\\
    \sigma_r &= \sigma_0\big[2-\tanh(u(r/R_f-g))\big]\,.\label{eq:sigvr_fil}
\end{align}
Here, $l_f$ is the length of the filament, and $\langle v_i \rangle, \sigma_i$ are the mean and standard deviation profiles of the velocities, where $i \in \{ z, r\}$, and $\erf{x}$ is the error function. $\sigma_z$ is assumed to be constant along the spine, with a value $300 \,\kms$, same as that for the background 1D dispersion. Here, we assume $V_{0z}=V_{0r}=250 \, \kms, \sigma_0= 300\,\kms, a=0.125, \, b=2.5, \,c=15,\, u=8,\, g=0.5$, which approximately match the results of \cite{Rost_et_al2024} for the radial velocity profiles. The particular form of $\langle v_r \rangle$ is chosen to get a profile shape in which the mean infall peaks at some particular $r$ slightly away from the filament spine, and smoothly approaches $0$ at both the extremes of $r$, the decrease in $\langle v_r \rangle$ being shallower in the outer regions of the filament as compared to the inner ones.

We generate $50$ realizations of this filament, and estimate the density and velocity profiles for each one separately, taking a finely sampled spine with $50$ equi-distant points along its length. We have checked that our results for the fiducial filament are converged with respect to increasing these numbers. The mean and error profiles are calculated to be the mean and standard deviation across the realizations. The velocity dispersion statistics are calculated for $\sigma_i^2$, and error propagation is used to get the error on $\sigma_i$. The maximum radius probed is $R_f$, so that any longitudinal profile as a function of $z$ is averaged over a disc of radius $R_f$ perpendicular to the spine at the location $z$. The same procedure will be followed for all the other examples, and we shall only mention the quantity or parameter that is changed relative to the fiducial filament.

Figure~\ref{fig:fiducial} shows the 1D radial (left panels) and longitudinal profiles (right panels) for the fiducial filament, as well as the corresponding straight filament. The expected (input) profiles are shown as solid black curves. We have also shown $10$ randomly chosen individual realizations with dotted curves, to give an idea of the variation across realizations. We also show the relative errors on the profiles, for better visualization of the errors. As seen from the plots, the recovered mean profiles are in excellent agreement with the input profiles. However, note that the individual profiles can be noisy; stacking of comparable filaments would clearly be advantageous.  The only noticeable deviation of the mean profiles from the expected ones is the slight increase in the longitudinal density profiles at two points along the spine, which correspond to regions of highest curvature. We will return to this point below.

Even though the mean values are slightly higher, they are still in agreement with the input profile within the error bars. Another point worth mentioning is that the errors on both the density and velocity dispersion profiles are almost always well within $10 \%$. The large errors on $\langle v_z \rangle$ near the centre of the filament are not because of anything physical, but because the denominator approaches zero at this point.  Also, the large errors on $\langle v_r \rangle$ closer to the spine (small $r$) are because the radial velocity dispersion is high in this region. The increase in errors away from the spine is, again, because the denominator approaches zero. Thus, we have shown that both  \filgen\ and \filapt\ perform extremely well when $r_f$ is much smaller than the maximum $R_\kappa$ of the spine, and the filament spine is well estimated.

\subsubsection{Effect of filament radius}
\label{sec:effect of radius}
\begin{figure}
    \centering
    \includegraphics[width=\textwidth]{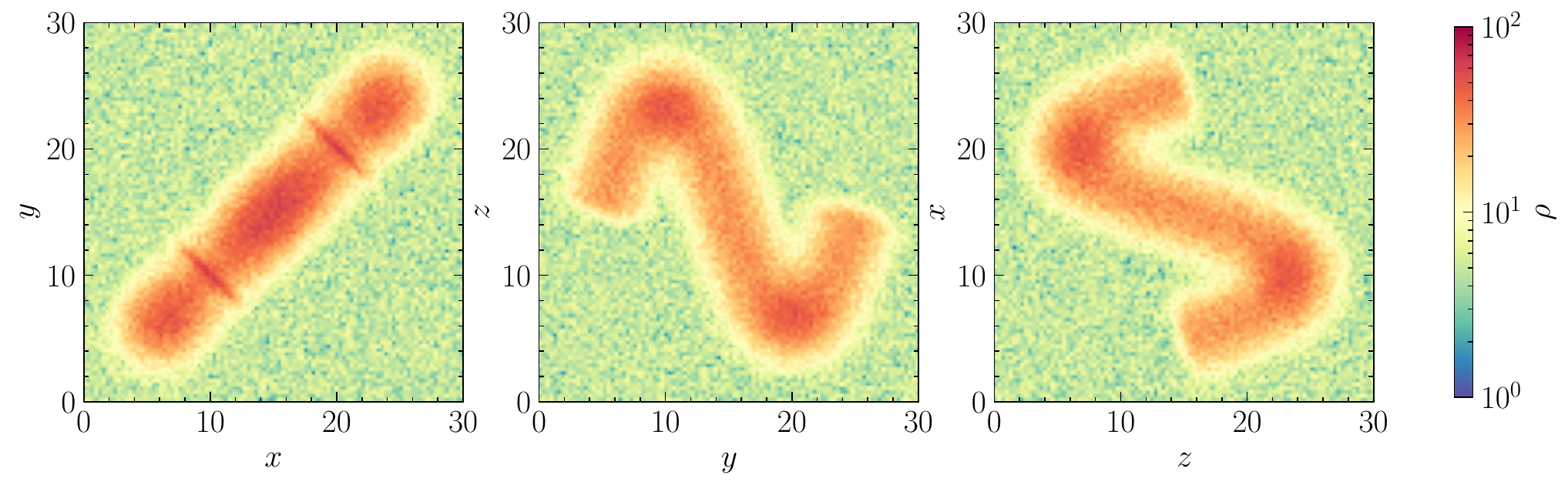}
    \caption{Projected number density for the thick filament. The panels are the same as Figure~\ref{fig:filament_illustration}. With larger $r_f$, the radial density variation in the filament is clearly visible. The density variation along the spine in the left panel is a projection effect, but that in the other two panels is real. See section~\ref{sec:effect of radius} for details.} 
    \label{fig:fat_fila_illustration}
\end{figure}

\begin{figure}
    \centering
    \includegraphics[width=\textwidth]{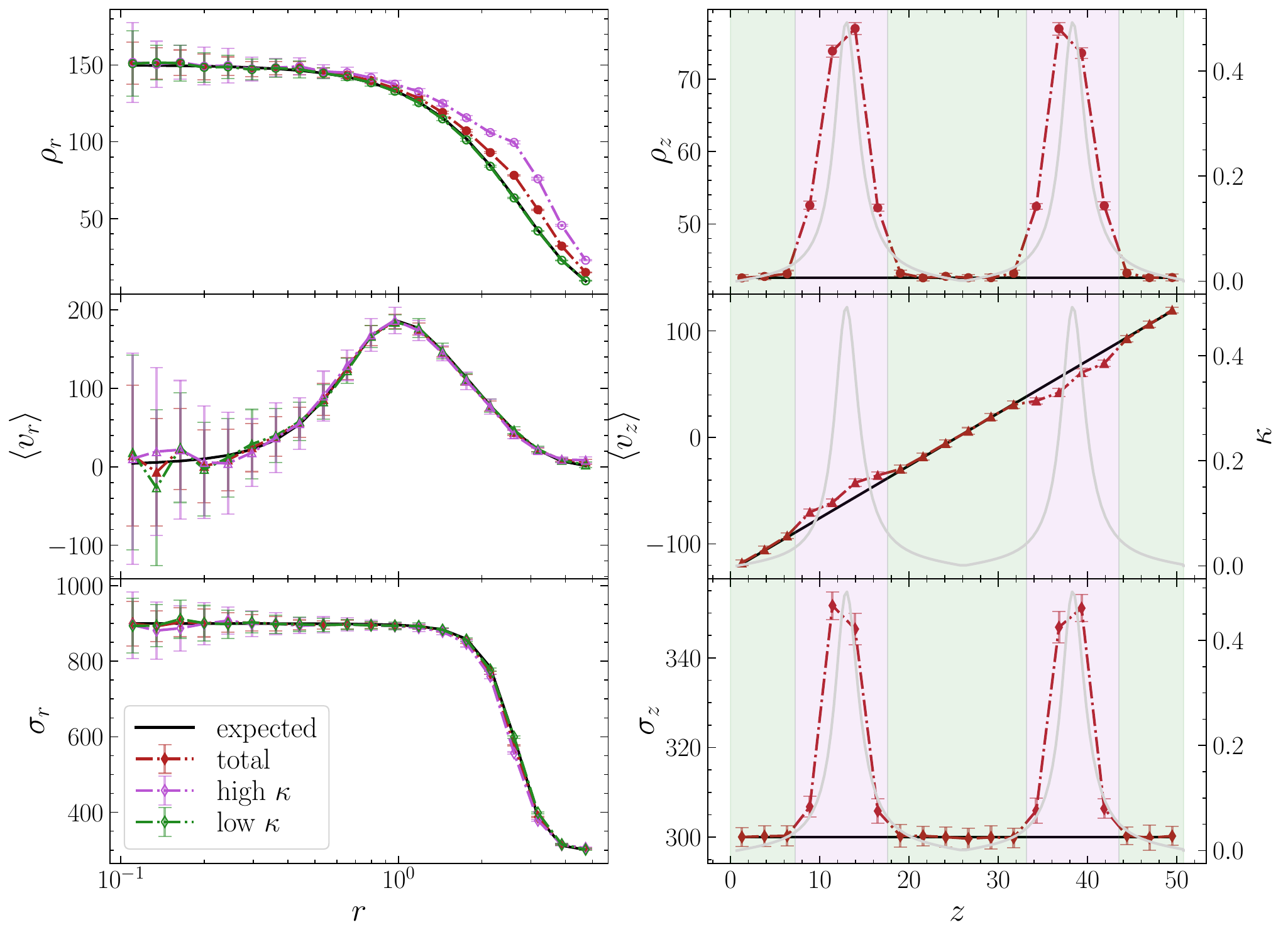}
    \caption{Radial and longitudinal density and velocity profiles for the thick filament. The panels are the same as in Figure~\ref{fig:fiducial}; except the error panels are skipped, as the error bars are qualitatively similar to the fiducial case. The curvature $\kappa$ of the filament is over-plotted in grey solid lines in the right panels. The regions of high (low) $\kappa$ are shaded with purple (green). Red colour represents the profiles for the whole filament. Purple (green) colours show the average radial profiles obtained for segments having high (low) $\kappa$. We see that the major contribution towards the biasing of the density profile arises from the high curvature regions.}
    \label{fig:curvature_split1}
\end{figure}

As we saw in the case of the fiducial filament, the regions of high curvature show some effects worth studying. Since we first create a straight filament with the given profiles and then curve it, there will be an overlap in the regions where $R_\kappa \sim r_f$, creating deviations from the expected profiles. Therefore, one has to be cautious while using \filgen. Here, we illustrate these effects using a filament identical to the fiducial one, except now $r_f=2 \,\Mpch$, which is equal to the minimum $R_\kappa$ of the spine. Figure~\ref{fig:fat_fila_illustration} shows the projected density for this thick filament. It is evident that overlap in regions of high curvature is giving rise to deviations from the input density profile. We now study the profiles quantitatively in detail. 

Figure~\ref{fig:curvature_split1} shows the density and velocity profiles of the thick filament. The panels are identical to those shown for the fiducial filament, except that we do not show the residuals since these are qualitatively similar to those obtained for the fiducial filament. The black curves show the expected profiles whereas the red markers show the profiles obtained from the realizations constructed using \filgen. In the right panels, the curvature $\kappa$ of the spine as a function of length along it is overlaid on the profile plots. It is seen that regions of high $\kappa$ (shaded purple) show substantial deviation from the expected profiles, whereas in regions of low $\kappa$ (shaded green), there is a good agreement between the expected and recovered profiles. Although the deviation is most prominent in the longitudinal profiles, one can also see corresponding deviations in the radial density profile. The effect is more prominent in the outer regions of the filament, where the overlap is more.

In order to recover unbiased profiles we split the spine into two sets comprising of segments of high ($\kappa \geq \kappa_{\mathrm th}$) and low curvature ($\kappa <\kappa_{\mathrm th}$), with the threshold set to $\kappa_{\mathrm th}=1/(10r_{f})$.
The high and low curvature regions are represented by purple and green shaded regions respectively in Figure~\ref{fig:curvature_split1}. The radial profiles are calculated separately for both the sets, and are plotted in the left panels. As expected, the high curvature regions show maximum deviation from the expected profiles, and the low curvature regions recover the expected profiles perfectly. Also, note that the decrease in signal-to-noise ratio is not substantial (here, the volume of the low curvature regions is $\approx 60 \%$ of the total volume, leading to the signal-to-noise ratio of $\approx 0.77$ times the original). \emph{Therefore, removing the high curvature regions can substantially increase the accuracy of the profile estimate without significantly affecting the precision.} In general, rather than discarding the high curvature region, it will be beneficial to study the different curvature regions separately. 

We would like to emphasize that the effects discussed here are a result of the way the filaments were constructed in the first place: starting with a straight filament with the expected profiles, and then curving it. This will not always be so in the case of real cosmological filaments. The spines may be inherently curved and the radial growth with accretion might occur on this curved filament. We do not presume that in such cases the high-$\kappa$ regions will necessarily lead to higher densities, but differences in accretion rates and possibly other physical processes might be expected in such regions. However, Figure~$8$ in \cite{Galarraga_Espinosa_et_al2023a} shows that filament spines can, in fact, curve over time. In such cases, one might expect effects similar to what we have illustrated here. In any case, it appears prudent to split the spines by curvature to account for curving after filament generation, or different accretion due to different geometry. This will increase the robustness of the inferred filament profiles. Also, we show below that the high-$\kappa$ regions are most affected due to sparse sampling of the spine as well as noise; in these cases as well, splitting by curvature gives more unbiased and robust results.

Another point to note is that when $r_f \gtrsim R_\kappa$, although the filaments generated using \filgen\ do not exactly recover the input profiles, they are still perfectly valid filaments. While using these, however, the correct profiles will no longer be the input analytical profiles; instead, they have to be determined by using a very finely sampled spine, and numerically calculating the profiles using \filapt.

We would like to point out that the curvature threshold is set to $\chi/r_f$, where the choice of the factor $\chi$ is not fixed based on any physical argument. We have checked for various choices of $\chi$, and find monotonic trends in the behaviour of the recovered filament profiles. The results become more robust with more stringent cuts and we find convergence at the chosen value of $\chi=1/10$ for our choice of filament properties.

For more realistic filament populations, such as those considered in later chapters, an estimate of individual filament radius may not always be available. In such cases, the curvature threshold must instead be defined using a characteristic scale of the sample, for example the typical filament radius or a suitable percentile of the curvature distribution of the selected filaments.

\begin{figure}[h]
    \centering
    \includegraphics[width=\textwidth]{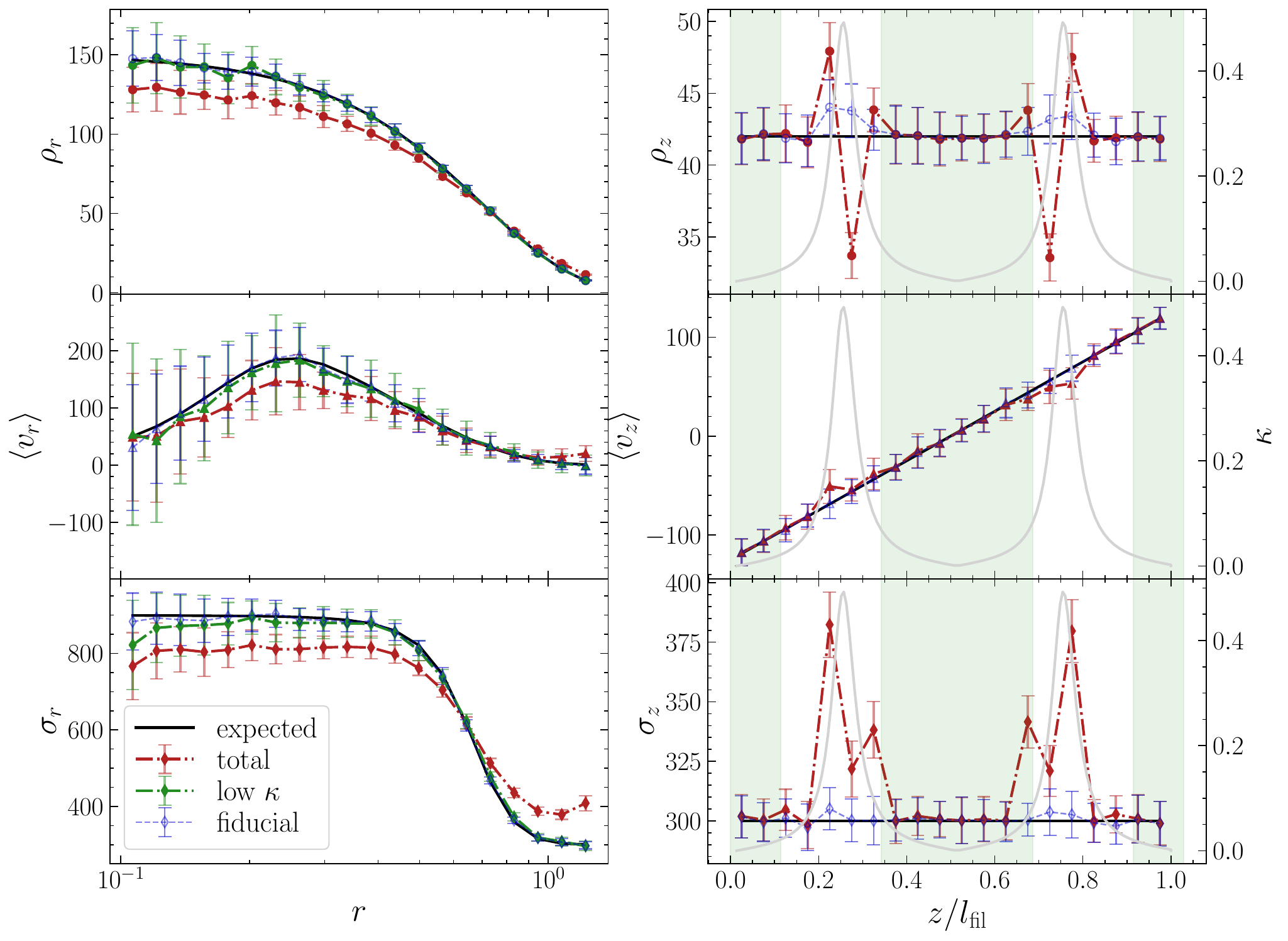}
    \caption{Effect of sampling of the spine on the density and velocity profiles. The panels are the same as in Figure~\ref{fig:curvature_split1}, except the length along the spine ($z$) is scaled by the measured length $l_{\mathrm{fil}}$ of the discrete spine (sparsely and finely sampled for the corresponding cases). Black solid curves show the expected profiles, whereas blue open markers show the profiles obtained for the fiducial model. Solid red markers show the profile obtained for the fiducial filament, but when the number of points tracing the spine is reduced by a factor $5$. Green shaded regions in the right panel are the regions of low curvature, and the green markers in the left panels show the radial profiles calculated for only these regions. It is seen that sparse sampling of the spine induces biases in the inferred profiles, which can be alleviated by using only regions of low curvature. }
    \label{fig:ds_spine}
\end{figure}

\subsubsection{Effect of spine sampling}
\label{sec:spine_sampling}
The filaments in observations as well as simulations will be traced by discrete and finite tracers, and thus the filament spine can never be infinitely finely sampled. Even if one assumes that there is no error on the recovered spine (which is far from reality, and we shall explore this effect in the next section), the discreteness of the spine sampling can induce biases in the recovered profiles. The effect of sampling on the filaments recovered using the Bisous filament finder has been studied by \cite{Muru_tempel2021}. The effect of under-sampling will be more prominent when the tracers are massive halos (as opposed to less massive halos or dark matter particles), which are highly biased but few in number. Being strongly clustered, the more massive halos will tend to reside closer to the filament spines, which are regions of higher densities. Thus, the errors on the recovered spine are expected to be relatively low, but the effect of poor sampling of the spine would be prominent. Here, we quantify this effect and show that recovering the information from only low curvature regions can alleviate the biases, and the gain in accuracy is much more dominant over the loss of signal-to-noise ratio due to discarding high curvature regions.  

Here, we use the same filament configuration as in the fiducial case, the only difference being that the spine sampling is reduced by a factor $5$, keeping the sampling points equi-distant along the spine length. This spine is shown using blue dashed curves in Figure~\ref{fig:filament_illustration}. Note that in the case of biased tracers, there might be a local clustering along the spine (especially in high curvature regions), which we have not accounted for here. The red markers in Figure~\ref{fig:ds_spine} show the profiles recovered for the sparsely sampled spine, as opposed to the blue ones for the fiducial sampling. It is seen that the density in the inner regions is underestimated, and so is the mean radial infall. The radial dispersion profile is smoothed out. Also, both the density and tangential velocity dispersion estimates deviate strongly from the expectation in regions of high curvature. This is because although all the points sampling the spine lie exactly on top of the correct spine, the segments joining them deviate from the spine. This will be more prominent when the separation between the points is not much smaller than $R_\kappa$ at that point. Thus, the inferred radial distance as well as the radial direction of a region from the filament spine will have errors. Therefore, the regions that are considered for calculating the density very close to the spine will actually have contributions from outer regions of the filament, reducing the inferred density. Also, the radial vectors that are to be added while calculating velocity statistics will in reality have tangential components as well, thus smoothing out the statistics. 

In order to remove these biases, we split the spine based on curvature, placing the curvature threshold at $2/3$ of the inverse mean segment length. Again, the factor $2/3$ is open to choice, and we see monotonic trends in the recovered low curvature profiles when changing this factor, with convergence seen at the chosen value. We obtain the profiles only for the low curvature regions. This is shown using green markers in Figure~\ref{fig:ds_spine}, and the regions chosen are shown using green shaded areas in the right panels. There is a clear reduction of bias, with excellent recovery of the density profile and the outer regions of both velocity profiles. Some residual bias remains in the inner velocity dispersion profiles, which is nevertheless within the error bars. Thus, we have shown that splitting by curvature can lead to a more robust and unbiased recovery of profiles of filaments when the spine is sparsely sampled.
\begin{figure}[h]
    \centering
    \includegraphics[width=\textwidth]{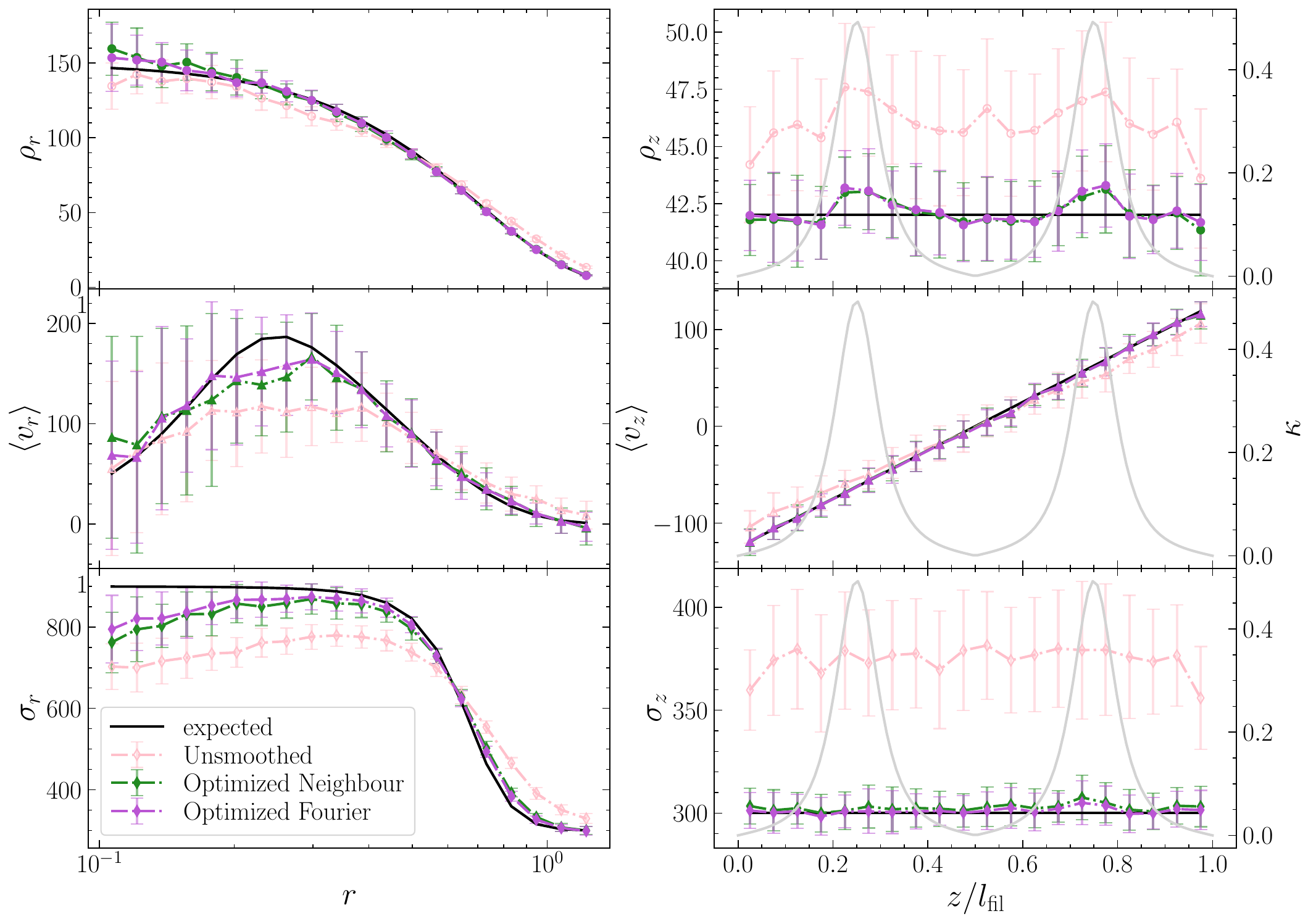}
    \caption{Effect of error in spine extraction on the estimated density and velocity profiles. The panels are the same as in Figure~\ref{fig:curvature_split1}, except the length along the spine ($z$) is scaled by the measured length of the spine used  $l_{\mathrm{fil}}$. Black solid curves show the expected profiles, whereas open pink markers show the profile obtained for the fiducial filament, with a noisy spine. It is seen that strong biases are induced in the profiles due to the errors on the spine. Solid green (purple) markers show the profiles obtained using optimized neighbour (Fourier) smoothing, and retaining only low curvature regions for calculation of the radial profiles.  It is evident that proper smoothing of spines is necessary before predicting any profiles. See section~\ref{sec:error on the spine} for details.}
    \label{fig:noisy_spine}
\end{figure}

\begin{figure}[h]
    \centering
    \includegraphics[width=\textwidth]{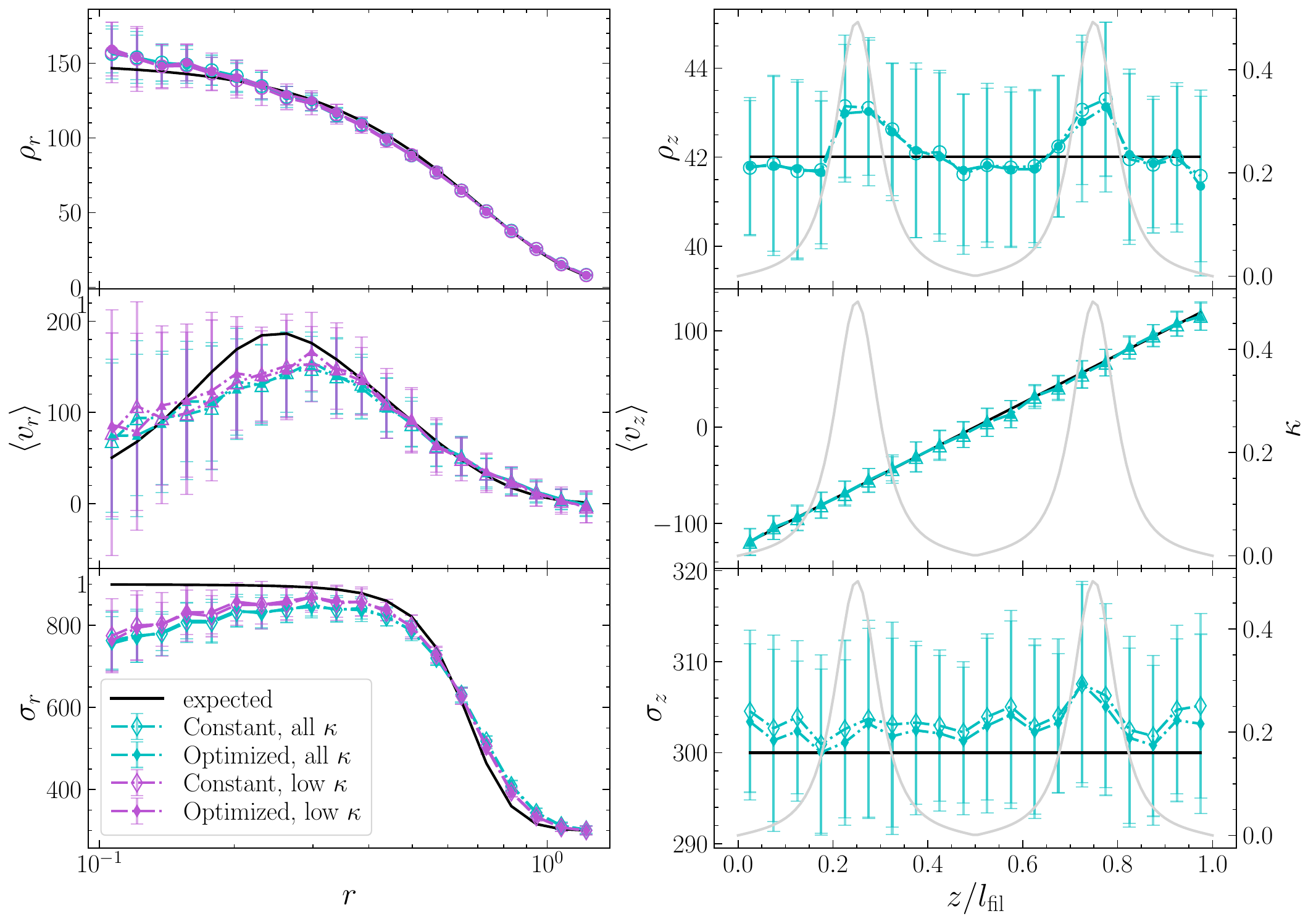}
    \caption{Comparison of optimized and constant smoothing and curvature segregation on the estimated profiles. The panels are the same as in Figure~\ref{fig:noisy_spine}. Black solid curves show the expected profiles, whereas coloured markers show the profiles obtained using neighbour smoothing. Solid (open) markers represent optimized (constant) smoothing. Cyan colour shows the radial profiles obtained for the whole filament, whereas purple colour shows those obtained for only the low curvature regions. It is seen that optimization does not provide much improvement when all the filaments belong to the same model.  See section~\ref{sec:error on the spine} for details.}
    \label{fig:NS_comparison}
\end{figure}

\begin{figure}
    \centering
    \includegraphics[width=0.7\textwidth]{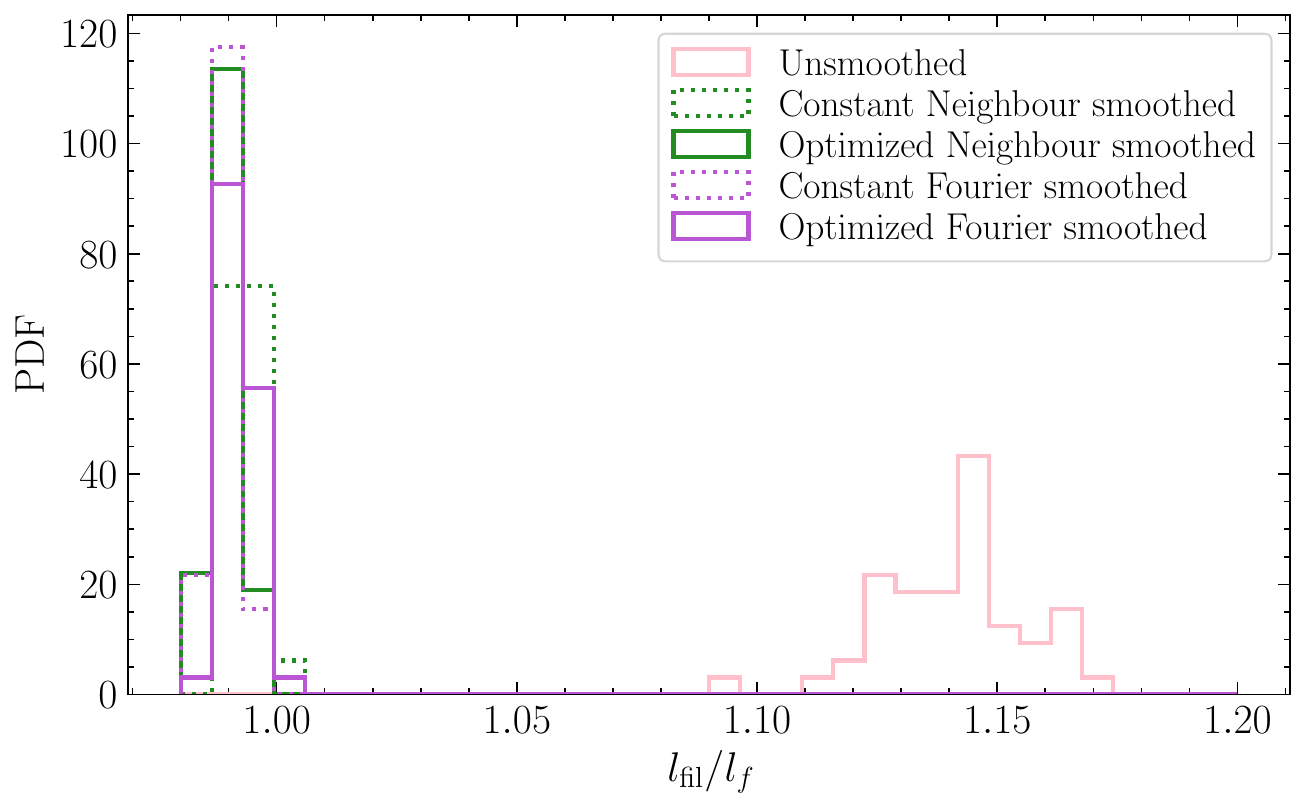}
    \caption{PDFs of lengths $l_{\mathrm{fil}}$ of the \disp\, inferred spines in units of the length $l_f$ of the actual spine. Pink colour represents the unsmoothed case, while green and purple colours represent the spines after applying a constant neighbour smoothing and optimized Fourier smoothing, respectively. Solid (dotted) lines show optimized (constant) smoothing. We see that unprocessed spines lead to substantial errors on the lengths of the spines.}
    \label{fig:length_distribution}
\end{figure}

\subsubsection{Effect of error on the spine}
\label{sec:error on the spine}
As explained in section~\ref{subsec: tools and techniques}, there will always be errors on the spines recovered using filament finders, which will propagate into errors and biases in the estimated profiles. Here, we quantify this effect, and illustrate the improvements due to using an optimally smoothed spine. 

We use the same filaments as discussed in section~\ref{sec:fiducial}, but with spines recovered using \disp\ (see Appendix~\ref{app: filtools filament spine identification} for details of spine recovery). The pink markers in Figure~\ref{fig:noisy_spine} show the profiles recovered using the noisy spines. Because of the noise, the inferred length of the spine is high, and one cannot compare the longitudinal profiles with the input profiles directly. Hence, the $x$-axis shows the length along the spine in units of the \emph{measured} length of the filament $l_{\mathrm{fil}}$  (which will always be longer than the true length $l_f$ in case of the noisy spine). We see that the deviation in the longitudinal profile is large not only in regions of high curvature (as seen before in the noise-free case), but now also in the low curvature regions. Thus, splitting by curvature alone will not help in this case. All the radial profiles also show strong biases. The density is over-estimated in the outer parts of the filaments, and is under-estimated in the inner parts. Both the mean and dispersion profiles of the radial velocity are smoothed out. These effects are sourced by mis-estimates, due to noise in the spine, of the distance from the true spine and, more importantly, the direction of the radial vector. 

As motivated in section~\ref{subsec: tools and techniques}, smoothing can alleviate these biases. We smooth the filament spines by using two kinds of smoothings: (a) the widely used neighbour smoothing, and (b) Fourier smoothing (see Appendix~\ref{App:filtools fourier_smo}, \ref{App: filtools smoothing_optimization} for details), and illustrate the results in Figure~\ref{fig:noisy_spine}. The smoothing parameters in both the cases have been optimized using our optimization technique, and only low curvature regions with $\kappa < 1/(10r_f)$ have been retained. The green markers show the profiles obtained by implementing the neighbour smoothing (see Appendix~\ref{App: filtools neighbour_smo} for details). It is seen that this provides a huge improvement over the unsmoothed case, especially in the longitudinal profiles and radial density profiles. Fourier smoothing provides an equivalent performance. It is seen that except for the radial velocity dispersion ($\sigma_r$), all other profiles are in agreement with the expected profiles within the error bars. For $\sigma_r$, the deviations occur mainly close to the filament spine (small $r$). 
Slight deviations are to be expected, since one can never perfectly recover the exact spine, but the improvement with smoothing is encouraging.
Note that since here we are considering multiple realizations of a single filament model, our filament-by-filament optimization does not offer much advantage over constant smoothing, where the smoothing parameter is fixed by visual optimization of a single filament. To illustrate this, we smooth the spines using optimized and constant neighbour smoothing, and show the results in figure~\ref{fig:NS_comparison}, with and without discarding the high curvature regions. For constant smoothing, we have fixed $\Nsm=30$, which works well visually for this particular filament model (i.e., visual optimization). It is seen that profiles from optimized and constant smoothings are visually almost indistinguishable, whereas splitting by curvature provides a slight improvement (see the $\sigma_r$ profiles at large radii). One should note that even in the case of constant smoothing, some kind of optimization is still being performed; and the chosen value of $\Nsm$ is not arbitrary. Our technique prescribes a systematic way of optimization, with quantifiable justification. The importance of this kind of filament-by-filament optimization is more evident when the sample contains filaments with varied properties, as we demonstrate in section~\ref{sec:optimized_smoothing}.

To study the errors on the inferred lengths of the filament spines, the probability density function (PDF) of the lengths of the spines is calculated for the unsmoothed spines as well as processed spines using the two kinds of smoothings discussed above, with and without optimization. These PDFs are shown in Figure~\ref{fig:length_distribution}, where the lengths are given in units of $l_f$, the actual spine length. If there was no error on the spine, one would expect a Dirac-$\delta$ centred on unity. For our noisy spines, we see that the filament length is overestimated by $\gtrsim 10 \%$, with a wide spread. This error reduces to $\lesssim 2\%$ after smoothing. The width of the PDFs also reduce substantially. Here, where all the filaments belong to the same model, there is no preferred method of smoothing, and all the methods perform equally well. Thus, smoothing of the spines is essential even when one is dealing with statistics related to the lengths of the filament spines.

\section{Applications of \filtools}
\label{sec:Applications}
Here, we discuss two applications of the methods and tools in this chapter. The first is an application of optimized smoothing to a set of different filaments. The second is the effect of redshift space distortions (RSD) \cite{Kaiser87} on two simple cases: straight filaments aligned parallel and perpendicular to the line of sight (los).

\subsection{Optimized smoothing}
\label{sec:optimized_smoothing}

\begin{figure}
    \centering
    \includegraphics[width=\textwidth]{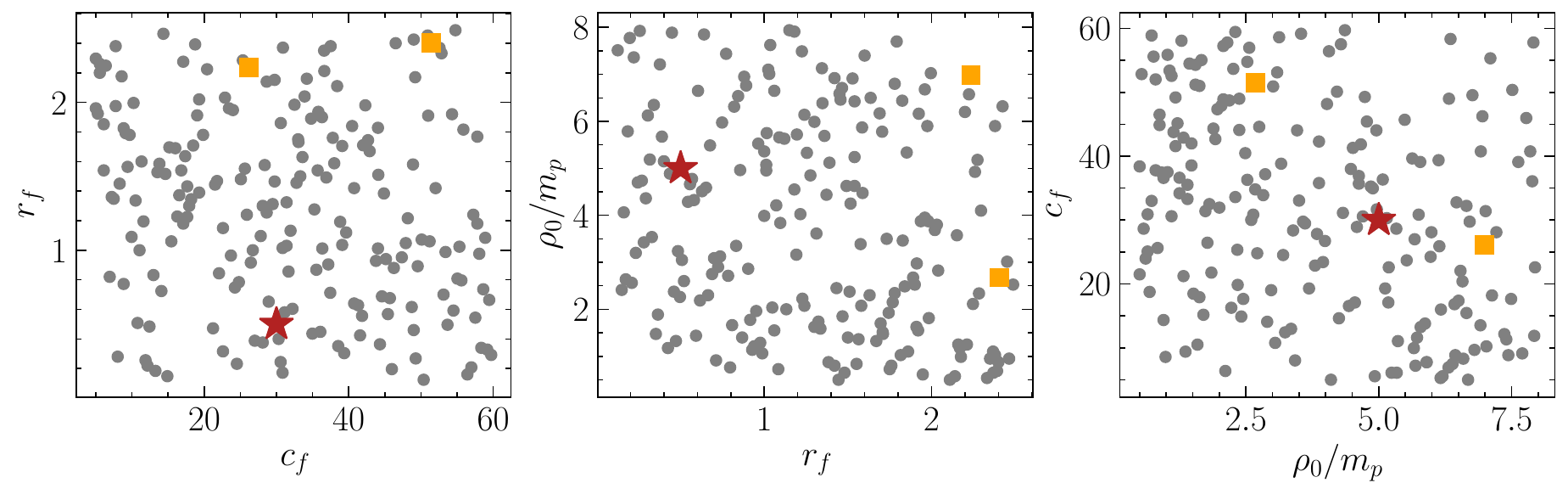}
    \caption{Distribution of the parameters of the filaments, $\{c_f, r_f, \rho_0/m_p \}$, used to generate a set of different filaments to study the application of optimized smoothing. The red star and orange squares represent the fiducial filament and the two filaments whose profiles are illustrated in Figure~\ref{fig:optimization_pdf} respectively (see section~\ref{sec:optimized_smoothing} text for details).}
    \label{fig:LHC_scatterplots}
\end{figure}

\begin{figure}
    \centering
    \includegraphics[width=\textwidth]{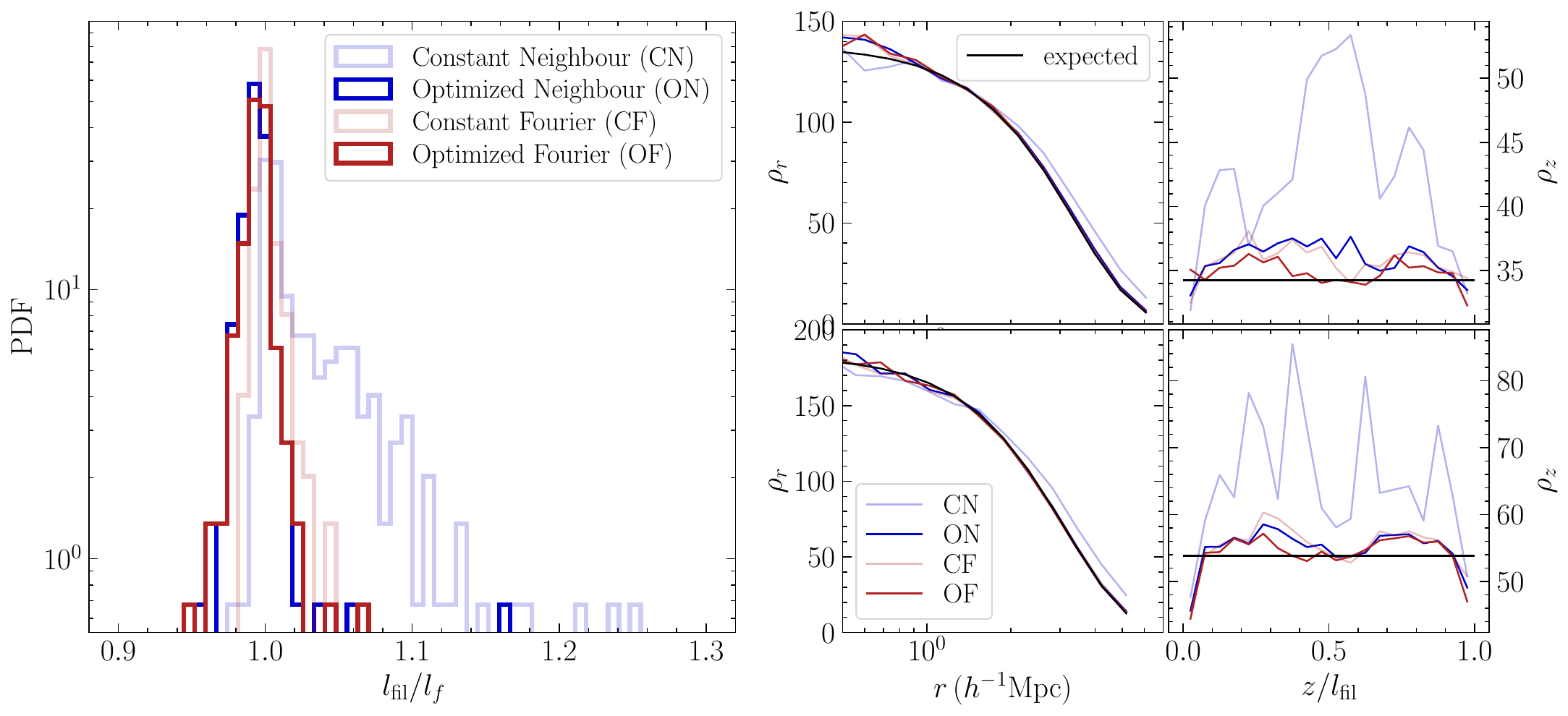}
    \caption{Illustration of need for optimized smoothing for a diverse filament set. The left panel shows the PDFs of filament spine lengths calculated over different filament models, after using various kinds of smoothings. The axes are the same as in Figure~\ref{fig:length_distribution}. Red (blue) colour represents Fourier (neighbour) smoothing, whereas dark (pale) lines show optimized (constant) smoothing. The four small panels on the right show the density profiles, each row corresponding to a different filament (see section~\ref{sec:optimized_smoothing} text for details). The colour scheme is the same as the one used in the left panel. We see that optimization gives excellent results for both neighbour and Fourier smoothing.}
    \label{fig:optimization_pdf}
\end{figure}

We have motivated the need for optimized smoothing in section~\ref{subsec: tools and techniques}, and provided a detailed explanation in Appendix~\ref{App: filtools smoothing_optimization}. As seen in section~\ref{sec:error on the spine}, optimization does not provide much improvement over constant smoothing when all the filaments belong to the same filament model. However, this will not be the case in cosmological simulations. Here, we consider different models of filaments with the same $l_f$, and study the effect of multiple kinds of smoothings on their length PDFs. We generate $200$ filaments with the same sinusoidally curved input spine (see Figure~\ref{fig:smoothing_comparison} for visualization), but different filament parameters $(c_f, r_f)$  and different background number densities $\rho_0/m_p$. The distribution of these parameters is shown in Figure~\ref{fig:LHC_scatterplots}. The filament spines are identified using \disp\ (see Appendix~\ref{app: filtools filament spine identification} for details).

The left panel in figure~\ref{fig:optimization_pdf} shows the length PDFs obtained for this set of $200$ filaments, smoothed using both optimized and constant, neighbour and Fourier smoothings. The constant smoothing parameters are fixed by visual optimization for the fiducial case, which correspond to $\Nsm=30, \, k_0=0.1 \,\hMpc.$ It is seen that the same constant neighbour smoothing does not work for all the filaments, giving rise to a wide spread in the inferred lengths $l_{\mathrm{fil}}${; whereas constant Fourier smoothing performs much better}. The spread substantially reduces as one shifts to optimized neighbour smoothing, with the peak of the PDF lying very close to unity, as is desirable. Optimized Fourier smoothing also gives excellent results.

To further study the effects of optimized smoothing, we focus on two particular filaments whose lengths are strongly overestimated in case of constant neighbour smoothing, with $l_{\mathrm{fil}}/l_f >1.2$. The parameters of these filaments are shown using orange squares in figure~\ref{fig:LHC_scatterplots}, whereas their radial and longitudinal density profiles are plotted in the four small panels in figure~\ref{fig:optimization_pdf}. The two rows represent the two filaments. In the case of radial profiles, we focus only on larger radii, where the error on the profiles is expected to be low. It is evident that constant neighbour smoothing leads to a strong bias in the recovered profiles, which reduces substantially with optimization. Comparatively, constant Fourier smoothing fares much better, improving further with optimization. Overall, both neighbour and Fourier smoothing with optimization lead to excellent recovery of the radial density profiles, and are visually indistinguishable from the expected profiles in the figure.

Note that constant Fourier smoothing produces much better length PDFs as well as density profiles as compared to constant neighbour smoothing, as the noise always contributes towards high $k$ modes, which are cut off in Fourier smoothing. For this reason, we prefer and recommend Fourier smoothing over neighbour smoothing. Although both the optimized and constant Fourier smoothings produce comparable length PDFs and very similar density profiles, we recommend optimized smoothing over its constant counterpart (see Appendix~\ref{App: filtools spine_smoothing} for details).

In the following chapters, however, we use a different smoothing procedure when identifying filaments from discrete tracers such as dark matter halos. In this case, particularly for sparsely sampled filaments, the spine traced by the halos can be systematically offset from the underlying matter filament. This cannot, in general, be corrected by Fourier or neighbour smoothing alone. Instead, we use a physically motivated smoothing algorithm that iteratively minimizes the mass dipole about the spine, thereby shifting it towards the centre of the filament. Thus, although optimized Fourier smoothing is the preferred choice among purely geometric smoothing schemes, the mass-dipole method is better suited to the filament catalogues analysed in the remainder of this thesis.

\begin{figure}[h]
    \centering
    \includegraphics[width=\textwidth]{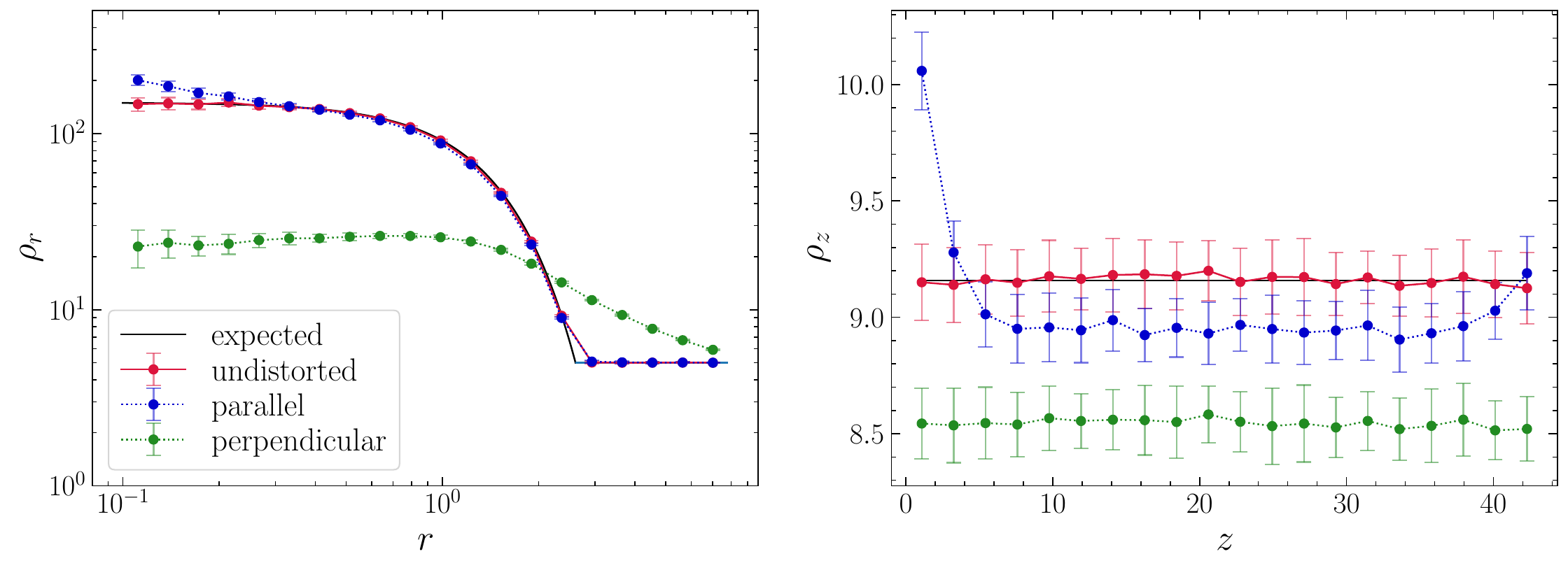}
    \caption{ Effect of redshift space distortions on the density profiles of straight filaments with nodes at both ends. The left (right) panel shows the radial (longitudinal) density profile. The black curves show the analytical expected profiles, whereas the red markers show the profiles obtained in real space.  The green and blue colours represent the redshift space distorted profiles when the line of sight is, respectively, perpendicular and parallel to the axis of the filament. The markers and error bars are obtained from the mean and standard deviation across $50$ realizations. See section~\ref{sec: RSD} for details.}
    \label{fig:rsd}
\end{figure}

\subsection{Redshift space distortions}
\label{sec: RSD}
Until now, we have focused on the recovery of profiles of filaments in real space. In real data, distances along the line of sight are inferred from observed redshifts which are affected by peculiar motions, leading to redshift space distortions (RSD). Here, we create simplistic but physically motivated velocity profiles for the filaments as well as the node halos and their environment, and study the effect of RSD on two simple cases: a straight filament aligned along and perpendicular to the line of sight, respectively. Since modelling of the velocity profiles is completely in our hands, we can switch off particular parts of the flows (e.g. dispersion in filaments, mean infall around halos, etc) and separately study the effect of various parameters on the RSD recovered filaments.

We consider an aperiodic box of side $80\, \Mpch$, and place a straight cylindrical filament of length $l_f=50 \, \Mpch$ and radius $r_f=1 \, \Mpch$. The filament concentration is $c_f=30$, and its velocity profiles are the same as those used for the fiducial filament. An NFW halo is placed at each end of the filament; the radii ($R_{200c}$) and concentrations of the two halos are $R_1=0.6 \, \Mpch, \, R_2=0. 4 \, \Mpch$ and $c_1=5, \, c_2=8$ respectively. The halo outskirts are chosen to have a power law density profile $\rho_{2h} \propto \rsp^{-3}$, and the normalization is chosen such that the densities of the NFW halo and the halo outskirts match at the halo boundary. The velocity dispersion in the halo outskirts is assumed to be isotropic, and matches that of the background, with 1D velocity dispersion of $300\, \kms$. The mean tangential velocity in the halo outskirts is assumed to vanish, and the mean radial infall is modelled using $\langle v_r \rangle =-D/ \chi$, with $D_1=500 \, \kms, \, D_2=400\, \kms$, and $\chi$ is the value of $\rsp$ in units of radius of the respective halo. Although in reality, the halo outskirts will not be radially symmetric and simple, here our main goal is just to illustrate qualitatively how \filgen\ and \filapt\ can be used to study RSD in filaments, and the choice of profiles is not very important. The filament profiles are estimated by taking a spine that starts and ends at the outer boundary of the halo outskirts, so that the halos and their outskirts would be excluded from the profiles in real space. 

The straight filament axis is taken to be along the $z$-direction. We consider two cases, one where the line of sight is along the $z$-axis and the other where it is along the $x$-axis. Let the Cartesian coordinate along the line of sight be represented by $\alpha$ ($\alpha=z, x$ in the two cases, respectively). The positions of the particles along the line of sight will get distorted, and are given by:
\begin{equation}
\label{eq:RSD}
    \alpha_{\mathrm{RSD}}= \alpha + \frac{v_\alpha}{aH} \,,
\end{equation}
where $a, H$ are the scale factor and Hubble parameter respectively.\footnote{For this demonstration, we work at the current epoch ($a=1$), so that $H=H_0=100h\,\kms{\mathrm Mpc}^{-1}$, the Hubble constant. Since our units of length are \Mpch, the value of the Hubble constant is irrelevant.} We shift the positions of all the particles according to this equation, and calculate the profiles, excluding regions up to $4R_{200m}$ on either side of the spine corresponding to the respective nodes and their outskirts (c.f., the discussion in section~\ref{subsec: tools and techniques}). The profiles are calculated by taking a finely sampled set of points on the input spine of the filament; we do not consider the noisy spines obtained by applying a filament finder. We only show the radial and longitudinal density profiles in Figure~\ref{fig:rsd}. Here, we have probed radii well beyond the boundary of the filament $R_f$, since the filament after being distorted can get spread out.

First, looking at the case where the line of sight is perpendicular to the filament axis (green markers), one sees that both the longitudinal and inner radial density is lowered, whereas the outer radial density has increased. Two effects will play opposing roles in this case. The radial infall will tend to compress the filament along the line of sight increasing the density, whereas the radial dispersion will tend to spread it out (similar to the usual Fingers-of-God (FoG) effect), decreasing the density. For our choice of parameters, the second (FoG smearing) effect is dominant. This leads to a decrease in longitudinal density, which is calculated by averaging over a disc of fixed radius $R_f$. FoG also smears the radial profile, increasing the density in filament outskirts, and decreasing it inside.

Second, considering the case where the line of sight is along the axis of the filament (blue markers), many effects contribute, leading to complicated changes in the filament profiles. The halo outskirts have a strong radial infall, hence these particles will predominantly be shifted towards the nodes, not contributing to any changes in the filament profiles. The strong velocity dispersion of the nodes, however, will spread them out (the usual FoG effect), substantially increasing the density in the filament, especially at its edges, 
which is seen in the longitudinal profile. The asymmetry in the increase is because one of the nodes is more massive than the other. This effect also leads to an increase in the density close to the axis of the filament, as seen from the radial profile. The radial infall on top of the filament is perpendicular to the line of sight, and thus will not contribute to any changes in redshift space. Finally, the mean flow of the particles away from the filament centre towards the nodes causes a decrease in density in the central part of the filament, as seen from the longitudinal plot. 

One has to note that the RSD effects will vary as one changes the density and velocity profiles of the filaments, as well as nodes and their outskirts. Here, we have only illustrated that \filgen\ and \filapt\ can in principle be used to study the filaments in redshift space. We leave a detailed discussion of redshift space applications for future work.

\section{Conclusion}
\label{sec:conclusion}
Cosmic filaments are visually the most striking features of the cosmic web. Unlike halos and voids, however, their potential in helping understand galaxy evolution or cosmology remains largely untapped. It is not clear, e.g., whether filaments exhibit universality in their density and/or velocity profiles similar to that seen for halos and voids. In part, this is due to a lack of consensus on what exactly constitutes a filament. For example, multiple filament finders exist in the literature, and each one identifies rather different filaments for the \emph{same} data set \cite{Libeskind_et_al2018}. In the absence of a notion of a `true' spine, it is then difficult to objectively choose a filament finder and robustly fix its parameters; the general practice is to rely on visual inspection. 

In this work, we have taken steps towards rectifying this lack of objectivity in the identification and analysis of cosmic filaments, by introducing two tools. The first -- the {\bf Fil}ament {\bf Gen}erator, or \filgen\ -- generates the complete phase space information of particles in realizations of mock filaments with a \emph{known} spine and some input density and velocity properties, all of which can be made arbitrarily realistic/complex (see section~\ref{subsec: tools and techniques}). Operationally, \filgen\ first samples the user-specified density and velocity profiles assuming a straight filament, and then self-consistently curves the filament according to a user-specified spine curve. Thus, \filgen\ synthesizes particle realizations of the `truth', solving the problem of arbitrariness of judgement when subsequently using a filament finder. For simplicity, in this work, we have focused on generating filaments whose radial density profiles can be described by a single Gaussian as a function of perpendicular distance $r$ from its spine.

Although knowing the `truth' regarding the filament underlying a given particle data set is important, it is equally important to identify robust statistics that maximise the information retained after applying a filament finder. E.g., it is unlikely that the `true' radial or longitudinal density and velocity profiles of a filament would be accurately recovered if the spine identified by the filament finder is very noisy. This issue is addressed by our second tool -- the {\bf Fil}ament {\bf A}nalysis and {\bf P}rocessing {\bf T}ool, or \filapt\ -- which provides three main post-processing modules for use on the data obtained from a filament finder (see section~\ref{subsec: tools and techniques}): \begin{itemize}
\item \filapt\texttt{.ExtractProfiles} extracts the profiles of density along with mean and dispersion statistics for different velocity components, as a function of perpendicular distance from the spine or length along the spine. This module inherently assumes that the spine being used is `perfect', thus allowing its output to be used in various quality tests in combination with \filgen.
\item \filapt\texttt{.SmoothSpine} \emph{optimally} smooths each filament spine identified by the filament finder. The optimization uses a novel method which minimizes the estimated filament radius, for reasons discussed in section~\ref{subsec: tools and techniques}. Not all filaments in the cosmic web are the same, leading to different error properties on the individual inferred spines, and we have demonstrated that optimization is therefore essential. E.g., filament lengths can be substantially overestimated in the absence of optimization (Figure~\ref{fig:optimization_pdf}). Based on the results of tests in section~\ref{sec:optimized_smoothing}, we also recommend smoothing in Fourier space, rather than the commonly used neighbour smoothing. For the halo-based filament catalogues analysed in later chapters, however, we employ a different optimization scheme that uses the surrounding mass distribution to refine the spine, as this is better suited to sparsely sampled filaments.
\item \filapt\texttt{.ComputeCurvature} estimates the local curvature $\kappa$ along the spine (see Appendix~\ref{App: filtools curvature}). \filapt\ further provides for each filament to be split by $\kappa$, thus revealing the curvature dependence of the extracted profiles. We have shown in section~\ref{sec:spine_sampling} that regions with the highest $\kappa$ along any filament are the major contributors of biases in the extracted radial profiles, and discarding these can substantially clean the profiles. The definition of ``high $\kappa$'' is not completely unambiguous, and the value has to be decided taking into consideration the radius of the filament as well as the mean length of the segments of the filament spine. In general, rather than discarding the high curvature profiles, one would like to study them separately. To our knowledge, this is the first recognition that selecting on spine curvature may be critical in obtaining unbiased filament demographics.
\end{itemize}

As a simple initial application of \filgen\ and \filapt, we have shown how redshift space distortions can potentially bias the inference of the underlying filament properties (section~\ref{sec: RSD}). Ultimately, we are interested in uncovering universality in the phase space profiles of filaments in the cosmic web. For this purpose, one would like to first work in real-space dark matter filaments in simulations. In this case, the biases will arise primarily from errors on the inferred spines, spine sampling and curvature. We have discussed all these biases and ways to alleviate them in this chapter. We also briefly mention a more detailed application using N-body simulations that we are currently pursuing, at the end of the section. 

There are, however, several potential improvements possible in \filapt\ and \filgen. We list these caveats here and will address them in future work.
\begin{itemize}
\item As mentioned at the end of section~\ref{sec:effect of radius}, the radial profile produced by \filgen\ in regions when the local radius of curvature $\kappa^{-1}$ is comparable to or smaller than the filament thickness $r_f$, does not match the input profile. If the goal is to generate a specific radial profile in regions of high curvature, an iterative procedure would be necessary, where the
azimuthal profile of the straight filament is changed iteratively so that after curving of the spine, the radial profile corresponds to the required one. 
\item \filgen\ currently does not model the multi-scale nature of filaments; the presence of substructure (halos in filaments, `sub-filaments') is not accounted for. Similarly, \filapt\ currently assumes each filament to be a monolithic structure.
\item Filaments traced by biased tracers such as halos or galaxies are not yet modelled by \filgen. In principle, it should be possible to include density fluctuations and use excursion-set inspired biased point processes such as density peaks \cite{ps12} as proxies for halos/galaxies.
\item The halo outskirts, relevant for modelling and understanding redshift space effects, are currently not modelled well by \filgen. This is primarily due to a lack of analytical understanding of the so-called 1-halo to 2-halo transition regime. It should, however, be possible to mitigate this issue using controlled calibrations of halo surroundings in N-body simulations.
\item \filapt\texttt{.SmoothSpine} currently optimizes the smoothing parameters (e.g., the cutoff wave number in case of Fourier smoothing) for each filament assuming that the radial density profile is well described by a \emph{single} Gaussian (as is currently modelled by \filgen). Filaments in simulations are typically more complex \cite{wang+24, 2pop2020}, and we are currently exploring the performance of Gaussian mixture models for the optimization step.
\end{itemize}

Taken together, \filgen\ and \filapt\ can be used to calibrate any given parametrised filament finder before applying it to simulation data. This would enable a systematic and simultaneous exploration of multiple filament properties including length, thickness, mean overdensity, concentration, velocity anisotropy, etc., in N-body simulations, with the ultimate aim of uncovering possible universality in filament demographics. An even more exciting prospect is to be able to robustly distinguish between filament properties for different dark matter models \cite[e.g.,][]{aragon-calvo24,bl24}.

    %
%
\let\textcircled=\pgftextcircled
\chapter{\skeletor: A Voronoi-based Hierarchical filament finder}
\label{chapt:skeletor}
\initial{T}he identification of filamentary structures in the cosmic web remains challenging due to their intrinsically hierarchical nature, the absence of well-defined boundaries, and the presence of embedded substructure across a wide range of scales. In this chapter, we present \skeletor, a hierarchical filament finder based on Voronoi tessellations and local anisotropy in the tracer distribution. The method exploits the fact that filamentary environments leave characteristic signatures in the geometry of Voronoi cells, allowing filaments to be identified directly from discrete tracer populations.

Using a tensor description of Voronoi cell geometry, \skeletor\ reconstructs filamentary skeleton through a hierarchical procedure that progressively incorporates lower-mass tracers. The recovered filaments are subsequently refined using dark matter information, and physically motivated filament radii are estimated from the surrounding velocity field. We further develop a substructure classifier to identify nested filamentary structures and separate large-scale filament spines from smaller embedded filaments. We conclude by presenting illustrative examples of the reconstructed filament population and examining some of their basic statistical properties.
\newpage
\section{Introduction and Motivation}
One of the central difficulties in identifying filamentary structures in the cosmic web arises from their intrinsically multiscale nature. Filaments do not possess sharply defined boundaries or a universal characteristic scale, but instead form a continuously connected network spanning a wide range of densities, lengths, and environments. As discussed in the previous chapter, different filament-finding methods often recover substantially different filament populations from the same underlying matter distribution. Much of this ambiguity stems from the absence of a unique physical definition of a filament and from the complex morphology of the nonlinear cosmic web itself.

The interpretation of filament properties is further complicated by the presence of structure on many scales. Large filaments frequently contain smaller embedded secondary filamentary structures, while dense environments often host overlapping structures that are difficult to disentangle. As a result, stacked filament statistics can mix together physically distinct environments, obscuring trends in their phase-space structure. Any attempt to identify robust or universal filament properties must therefore account for the hierarchical organization of the filament network.

Motivated by these considerations, this chapter presents \skeletor, a filament finder based on the local geometry of the tracer distribution. The method exploits the fact that filamentary environments are intrinsically anisotropic, with matter preferentially distributed along a dominant spine direction. To quantify this anisotropy, \skeletor\ uses Voronoi tessellations constructed directly from discrete tracers. The shapes and orientations of the resulting Voronoi cells provide adaptive local probes of the surrounding environment without requiring interpolation onto regular grids or the introduction of explicit smoothing scales. By characterizing the geometry of these cells through tensor-based estimators, the method identifies candidate filamentary structures directly from the local anisotropy of the tracer distribution.

A key feature of \skeletor\ is its explicit incorporation of cosmic web hierarchy into the reconstruction procedure. Rather than attempting to identify all structures simultaneously, the algorithm progressively reconstructs the filament network across multiple tracer mass scales. The largest tracers first define the dominant filamentary spines, while lower-mass tracers are subsequently introduced to recover progressively smaller structures. This hierarchical approach stabilizes the anisotropy-based reconstruction and naturally extends the filament network to smaller scales.

The basic filament finder operates using tracer information alone and is therefore directly adaptable to observational datasets. Additional dark matter and velocity information are incorporated only during subsequent refinement stages. This data is used to improve the positioning of the reconstructed spines and to identify physically motivated filament boundaries through the radial behaviour of the surrounding phase-space profiles. The resulting filament radii provide a dynamical characterization of filament extent that will be explored in detail in the following chapter.

The hierarchical reconstruction naturally produces a population of both large-scale filaments and smaller embedded structures. To characterize these separately, we introduce a filament substructure classifier that distinguishes between parent filaments and subfilaments. The motivation is analogous to the distinction between halos and subhalos in hierarchical structure formation, where treating the two populations separately has proven essential for understanding halo properties and obtaining cleaner statistical relations \citep[e.g.][]{Gao+2004}. We expect a similar situation for cosmic filaments. Parent filaments and subfilaments may occupy different dynamical environments and need not exhibit identical density or velocity structure. Identifying these populations separately therefore provides a cleaner framework for studying filament universality and interpreting stacked filament profiles.\\

The chapter is organized as follows. In Section~\ref{sec:local anisotropy}, we discuss the relationship between local anisotropy and Voronoi geometry, introduce the tensor formalism used to characterize Voronoi cell shapes, and motivate the hierarchical interpretation of the filament network. Section~\ref{sec: Skeletor algorithm} presents the \skeletor\ filament-finding algorithm, including the construction of the mass hierarchy, the identification of filament-like cells, the reconstruction of filamentary skeleton, and the refinement of filament spines using dark matter information. In Section~\ref{sec: substructure finder}, we describe the filament substructure finder used to classify subfilaments and parent filaments. Section~\ref{sec: results} presents some illustrative examples and statistical properties of the recovered filament population. The chapter concludes with a discussion of the main limitations of the method, its caveats, and several directions for future work. The sensitivity to the free parameters and choices in the algorithm are discussed in the Appendix.
\section{Local anisotropy and Voronoi geometry}
\label{sec:local anisotropy}
The filamentary structures of the cosmic web are intrinsically anisotropic environments. Matter within filaments is preferentially distributed along a dominant direction corresponding to the filament spine, while the transverse directions remain comparatively compressed, where the density falls quickly as one moves away from the spine. As a result, the local distribution of tracers around a filament spine is highly directional: neighbouring tracers are typically more aligned to the filament spine than to the transverse directions, reflecting the rapid decrease in density away from the spine. This directional asymmetry provides a natural geometric signature of filamentary environments and suggests that filament identification can be approached as a problem of detecting coherent local anisotropy in the tracer distribution.

The central idea underlying \skeletor\ is that this anisotropy should be reflected in the local geometry of Voronoi cells constructed around the tracers. A Voronoi tessellation uniquely partitions the space into a set of non-overlapping cells associated with a discrete set of tracers. Each Voronoi cell consists of the region of space closer to a given tracer than to any other tracer in the distribution. The geometry of the resulting cells therefore depends entirely on the local arrangement of neighbouring tracers, making Voronoi tessellations naturally sensitive to the morphology of the underlying matter distribution.

Voronoi tessellations possess several features that make them particularly useful for studies of the LSS. Since the tessellation is constructed directly from the tracer distribution, it automatically adapts to the local density and geometry without requiring interpolation onto a regular grid or the introduction of an explicit smoothing scale. This adaptive nature has been widely exploited in cosmological analyses, most notably in the Delaunay Tessellation Field Estimator (DTFE), where Voronoi and Delaunay tessellations are used to reconstruct continuous density and velocity fields from discrete tracers \citep{Schaap+2000,Schaap2007,Diaz+2007}. Tessellation-based methods are  effective at preserving anisotropic structures such as filaments and walls, which can be significantly degraded by fixed-scale smoothing procedures. Consequently, Voronoi geometry has found applications across a broad range of LSS problems, including density estimation, void finding, environmental classification, phase-space reconstruction, and cosmic web analysis \citep{ZOBOV2008,Platen+2007,Sutter+2015,vdWeygaert+2009,Disperse_theory}. More recently, the Voronoi Volume Function (VVF), defined as the distribution of Voronoi cell volumes, has been shown to contain substantial information beyond conventional two-point statistics and has emerged as a promising probe of higher-order clustering \citep{VVF2020}. These applications illustrate the sensitivity of Voronoi geometry not only to local density but also to the spatial arrangement of tracers, making it a natural framework for studying the multiscale morphology of the cosmic web.

In this work, we use the Monte-Carlo Voronoi tessellation procedure introduced by \citep{Alam+2019}. A large number of random points are generated uniformly throughout the simulation volume. Each random point is then assigned to the nearest tracer particle or halo. In the limit of sufficiently dense random sampling, the collection of random points associated with a tracer provides a Monte-Carlo realization of its Voronoi cell. This approach provides a discrete representation of the cell geometry from which shape statistics can be measured directly.

The geometry of Voronoi cells varies systematically across different cosmic web environments. In underdense and comparatively isotropic regions such as void interiors, Voronoi cells tend to be relatively symmetric and extended. In contrast, near filamentary structures the cells become anisotropic due to the directional distribution of neighbouring tracers. Cells near filament spines are typically compressed along the filament direction while remaining comparatively extended perpendicular to it. Near dense nodes and cluster environments, the cells may become highly compressed and irregular due to the large local tracer density and the intersection of multiple structures. The shapes and orientations of Voronoi cells therefore provide a direct geometric probe of the local anisotropy.

Figure~\ref{fig:voronoi_filament} illustrates an example Voronoi tessellation in a region containing a prominent filament in an N-body simulation. The Voronoi cells associated with tracers along the filament are visibly compressed along the filament direction, reflecting the highly anisotropic distribution of matter in these environments. As progressively lower-mass tracers are included, this signature becomes less pronounced around the dominant filament, while smaller filamentary structures begin to emerge. This behaviour, discussed further below, motivates the hierarchical reconstruction adopted in \skeletor\ and forms the geometric basis of the filament finder.
\begin{figure}
    \centering
    \includegraphics[width=\linewidth]{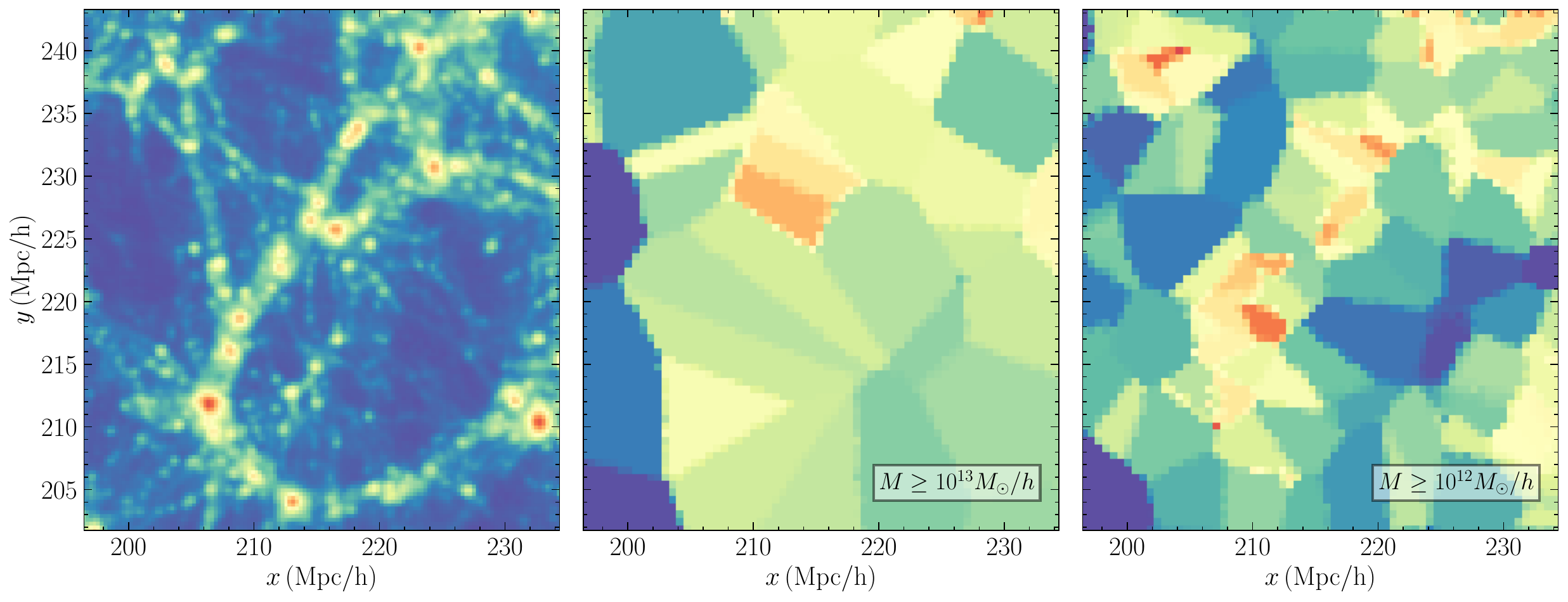}
    \caption{Illustration of the geometric motivation behind \skeletor. The left panel shows a dark matter density slice of thickness 7\Mpch\ from an N-body simulation, centred on a prominent visually chosen filament. The middle and right panels show Voronoi tessellations constructed using tracer populations with the mass thresholds indicated in the lower-right corners. Tracers lying along filament spines are associated with highly anisotropic Voronoi cells that are preferentially compressed along the filament direction. As the mass threshold is lowered, the anisotropy of cells near the dominant spine decreases because of denser local sampling, while progressively smaller filamentary structures become visible (e.g., see the filamentary structure appearing at $x \approx 225\Mpch$,  $y \approx 230 \Mpch$).}
    \label{fig:voronoi_filament}
\end{figure}

To quantify the geometry of a Voronoi cell, we construct a mass tensor using the random points associated with it. Let $\mathbf{r}(n)$ denote the position of the $n$th random point relative to the center of mass of the cell, where the center of mass is computed from all random points belonging to that cell. The mass tensor is defined as
\begin{equation}
    T_{ij} = \frac{5}{N}\sum_{n=1}^{N} r_i(n)\,r_j(n),
\end{equation}
where $N$ is the number of random points in the cell. The normalization is chosen such that, for a uniformly populated ellipsoid, the eigenvalues of $T_{ij}$ are equal to the squares of the semi-principal axes, while the corresponding eigenvectors define the principal directions of the cell. 

The set of three eigenvalues therefore provides a simple description of cell shape. Isotropic cells have comparable eigenvalues along all three directions, whereas anisotropic cells exhibit one or more preferred axes. In filamentary environments, the Voronoi cells are expected to be compressed along the filament axis while remaining relatively extended in the transverse directions. One therefore expects the eigenvalue corresponding to the filament direction to be significantly smaller than the other two, while the latter remain broadly comparable. To quantify the anisotropy, we define
\begin{align}
    q &\equiv \sqrt{\frac{1}{2}\left[(c-b)^2+(b-a)^2+(a-c)^2\right]},\\
    \alpha &\equiv \frac{q}{a+b+c},
\end{align}
where $a \geq b \geq c$ are the eigenvalues of the mass tensor. This definition is inspired by the tidal anisotropy parameter constructed from the tidal tensor in \citep{phs18}. Large values of $\alpha$ correspond to strongly anisotropic environments, and such cells can be selected by imposing a lower threshold on $\alpha$:
\begin{equation}
\label{eq:alpha_th}
    \alpha \geq \alpha_{\mathrm{th}};\hspace{1cm}\alpha_{\mathrm{th}}=0.476,
\end{equation}
The exact value of the threshold will be justified in section~\ref{subsec:filament_like_cells}.

Anisotropy alone, however, is insufficient to identify filamentary cells. Wall-like structures can also produce large values of $\alpha$, but their eigenvalue hierarchy differs from that expected for filaments. For a filamentary environment, the cylindrical symmetry implies $c \ll b \approx a$, whereas wall-like structures are characterized by $c \approx b < a$. We therefore impose an additional requirement,
\begin{equation}
\label{eq:pancake}
    \frac{c}{b} < \frac{b}{a},
\end{equation}
which selects cells that are preferentially squished along a single axis. Cells satisfying both the anisotropy and shape criteria are identified as filament-like cells and are used in the subsequent reconstruction of the filament network. Both quantities are independent of the overall normalization of the mass tensor, which is chosen only so that the eigenvalues have a direct geometric interpretation.

However, a purely local anisotropy-based filament finder encounters an important limitation in highly sampled regions. As progressively smaller tracers are included, the local sampling around dense filament spines becomes increasingly \emph{isotropic}. Although the filament itself remains physically anisotropic on larger scales, the immediate local neighbourhood around the spine becomes densely populated from all directions, causing the corresponding Voronoi cells to lose the strong anisotropic signatures expected in sparse sampling regimes, as seen in the rightmost panel in figure~\ref{fig:voronoi_filament}. Consequently, the simple identification of highly anisotropic cells is no longer sufficient to robustly recover filament spines across all tracer populations and scales.\\
This behaviour reflects the fundamentally hierarchical nature of the cosmic web. The dominant large-scale filamentary spine is typically traced by more massive halos, while progressively smaller tracers populate smaller-scale filaments in lower density regions and substructures embedded within the larger network. Motivated by this picture, \skeletor\ incorporates an explicit hierarchical reconstruction procedure. Rather than identifying filaments using a single tracer population, the algorithm progressively reconstructs the filament network across multiple mass scales. At each hierarchy level, the most massive halos act as nodes, whereas the filamentary structures connecting them are traced by lower-mass halos arranged along the spine like beads on a string. Progressively lowering the mass thresholds reveals finer filamentary structures and substructures.

The hierarchical framework therefore plays a central role in stabilizing anisotropy-based filament identification. By separating the reconstruction into multiple mass levels, the algorithm isolates the large-scale coherent filamentary structures before incorporating smaller-scale structure, allowing the underlying filament network to be identified more robustly across a wide range of environments and tracer densities.
\section{The \skeletor\ filament finder algorithm}
\label{sec: Skeletor algorithm}
Having established the geometric connection between filamentary environments and the anisotropic shapes of Voronoi cells, we now describe the filament-finding procedure implemented in \skeletor. The algorithm reconstructs connected filamentary skeleton directly from discrete tracer distributions while accounting for the hierarchical nature of the cosmic web.
\begin{figure}
    \centering
    \includegraphics[width=\linewidth]{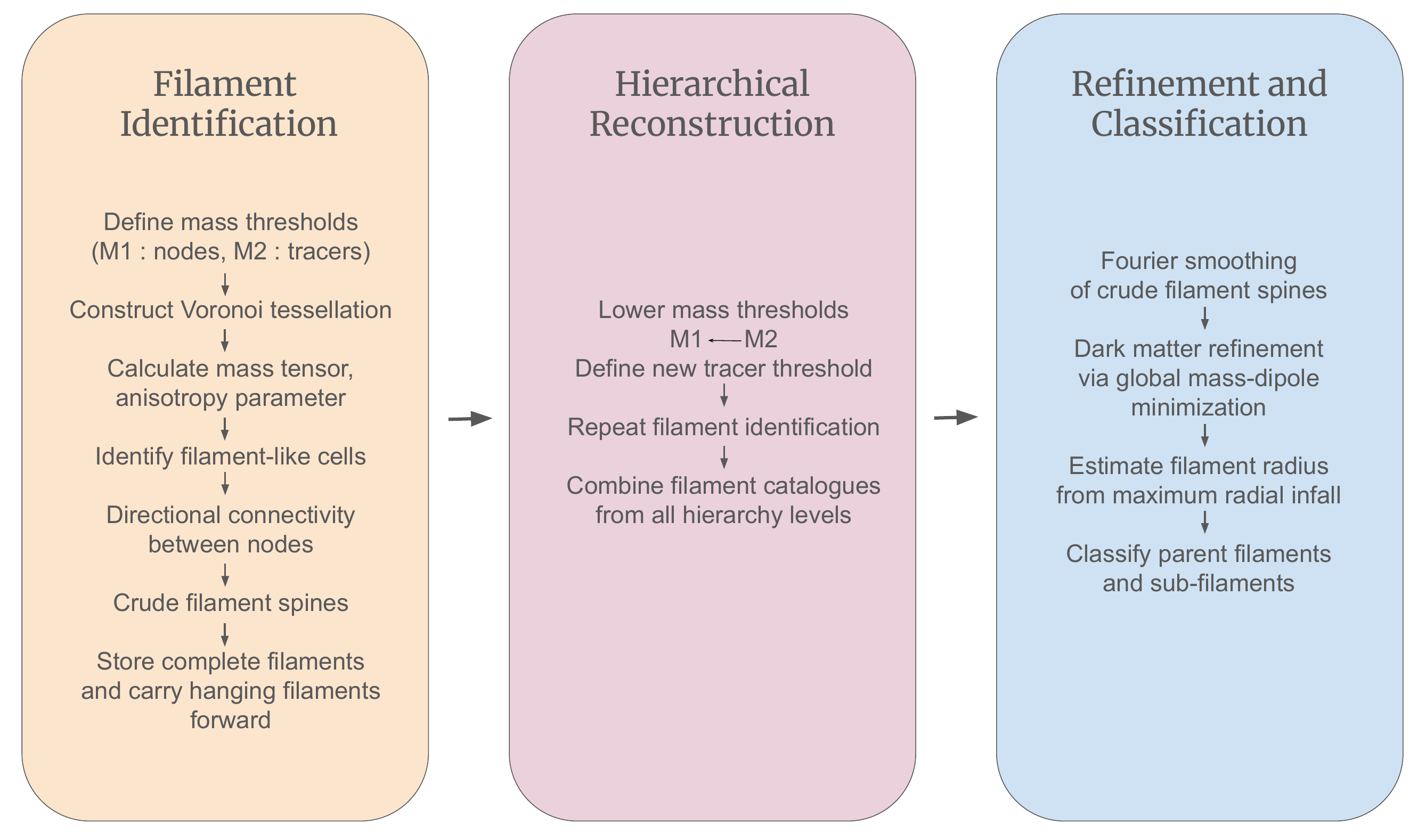}
    \caption{Schematic overview of the \skeletor\, algorithm.}
    \label{fig:skeletor_flowchart}
\end{figure}

The \skeletor\ algorithm consists of three main components. The first is the identification of filament-like Voronoi cells based on their local geometric anisotropy. These cells are then connected through a directional linking procedure to connect massive nodes. The procedure is subsequently repeated at progressively lower mass thresholds, allowing smaller-scale filamentary structures to be identified. A schematic overview of the algorithm is shown in Figure~\ref{fig:skeletor_flowchart}.

\subsection{Construction of the mass hierarchy}
When applied to an N-body simulation, \skeletor\ uses dark matter halos as tracers of the filament network. Prior to filament identification, the halo catalogue is cleaned using the virialization criterion described in Chapter~\ref{chap:sahyadri}, and subhalos are retained. Throughout this work, halo masses and radii are defined using  $M_{\mathrm{200b}}$ and $R_{\mathrm{200b}}$ respectively. However the `mass hierarchy' construction can in principle also be done with observed samples using some mass proxy like galaxy luminosity or stellar mass.

The reconstruction at a given hierarchy level is specified by two mass thresholds, $M_1>M_2$. Halos with masses exceeding $M_1$ are designated as potential nodes, while all halos with masses exceeding $M_2$ are used to construct the Voronoi tessellation and estimate the local anisotropy associated with each tracer. Candidate filament tracers are then selected from the population satisfying $M_2 \leq M < M_1$, applying the anisotropy and shape criteria discussed above (equations~\ref{eq:alpha_th}, \ref{eq:pancake}). These objects act as the building blocks of the filamentary skeleton connecting massive nodes. Once the filament network at a given hierarchy level has been reconstructed, the mass thresholds are lowered and the procedure is repeated.

The hierarchy is constructed by progressively lowering the mass thresholds. Once the filament network has been reconstructed at a given level, the tracer threshold $M_2$ is adopted as the node threshold for the next level, and a new lower tracer threshold is chosen. This procedure continues until the minimum halo mass permitted by the simulation resolution, or observational completeness limit, is reached. The highest node mass threshold is chosen such that a statistically significant population of primary filaments is obtained where the large-scale filament population is well-sampled, while the lowest threshold is fixed by the available halo catalogue. Intermediate thresholds are selected to balance two competing requirements. If the tracer population is too dense, the local environments near filament spines become increasingly isotropic and the anisotropy-based identification becomes less effective. Conversely, if too few tracers are used, the filamentary skeleton becomes poorly sampled and the connectivity reconstruction becomes unreliable.

For the applications presented in this work, we adopt mass thresholds of $10^{14}$, $10^{13}$, $10^{12.5}$ and $10^{12.3},h^{-1}M_\odot$, resulting in three levels of hierarchy. The first level identifies the dominant large-scale filamentary spine connecting the most massive nodes -- called the ``primary filaments'' -- while subsequent levels progressively recover smaller filamentary structures associated with lower-mass halos. Although these fixed mass thresholds are adopted throughout this thesis, Appendix~\ref{app: mass sensitivity} shows that the resulting filament population is largely insensitive to small variations in their values.

The motivation for this hierarchical reconstruction follows directly from the multiscale nature of the cosmic web. The largest and most coherent filaments are typically traced by more massive halos and connect the most massive halos and clusters, whereas lower-mass halos populate both these large-scale structures and the smaller filaments either present in low-density regions like voids, are embedded in bigger filaments. Reconstructing the network progressively from high to low masses therefore allows the dominant filamentary spine to be identified before incorporating increasingly fainter filaments and substructures.
\subsection{Identification of filament-like cells}
\label{subsec:filament_like_cells}
For each Voronoi cell associated with a tracer satisfying $M_2 \leq M < M_1$, we compute the mass tensor defined in the previous section and evaluate the corresponding anisotropy parameter $\alpha$. Large values of $\alpha$ indicate strongly anisotropic local environments and therefore represent potential filamentary regions.

To identify significantly anisotropic cells, we compare the measured $\alpha$ values to those obtained from a uniformly random distribution of particles. The anisotropy threshold $\alpha_{\mathrm{th}}$ in equation~\ref{eq:alpha_th} is defined as a fixed percentile of the $\alpha$ distribution for such a random catalogue. As expected, for sufficiently large particle numbers, the $\alpha$ distribution of a uniform random distribution is largely independent of number density. Throughout this work, we adopt $\alpha_{\mathrm{th}}$ corresponding to the $95^{\mathrm{th}}$ percentile of the random distribution. Here, it corresponds to the value $0.476$, which is very close to the value $0.5$ of the tidal tensor $\alpha$ motivated to represent filamentary environments in \citep{phs18}. Appendix~\ref{app:alpha_sensitivity} demonstrates that the resulting filament reconstruction is largely insensitive to the precise choice of this threshold.

\begin{figure}
    \centering
    \includegraphics[width=0.7\linewidth]{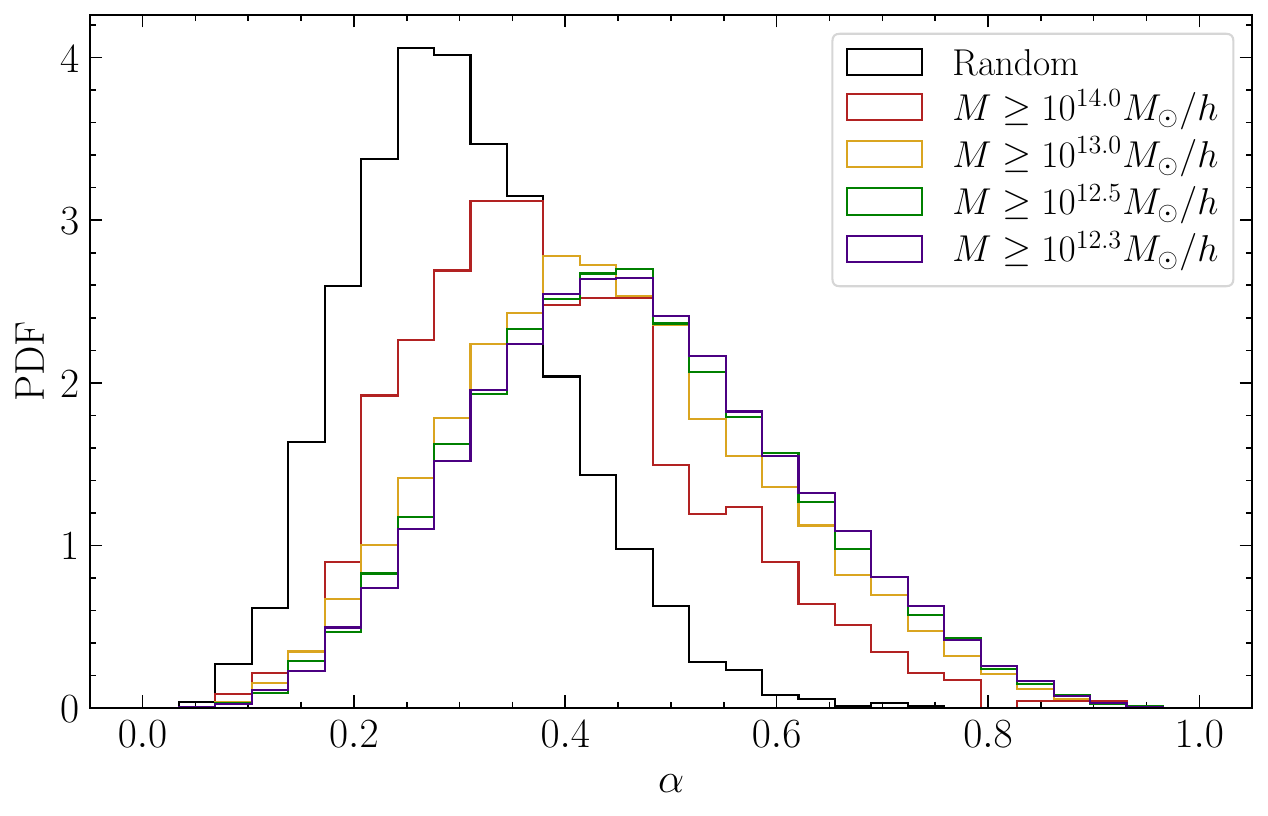}
    \caption{Distribution of the anisotropy parameter $\alpha$ for the mass-thresholded halo samples considered in this work, compared to a uniform random distribution (black curve). Halo populations exhibit systematically larger values of $\alpha$ and broader distributions than the random sample, reflecting the anisotropic environments generated by gravitational clustering.}
    \label{fig:alpha_distribution}
\end{figure}
Figure~\ref{fig:alpha_distribution} compares the distributions of $\alpha$ for the tracer population and the corresponding random catalogue. The clustered tracer population exhibits systematically larger values of $\alpha$ together with a broader distribution, reflecting the presence of anisotropic structures generated by gravitational clustering. The most massive halo sample, with $M>10^{14}\Msun/h$, shows comparatively lower values of $\alpha$, while the distributions for the lower mass-thresholded samples are remarkably similar. This suggests that the anisotropy properties of the tracer population are broadly similar across the lower mass thresholds considered here.

As discussed in the previous section, anisotropy alone is insufficient to distinguish filamentary environments from wall-like structures. We therefore impose the additional shape criterion given by equation~\ref{eq:pancake}. This requirement preferentially selects pancake-like cells that are strongly compressed along a single direction, as expected for filamentary environments, while rejecting wall-like configurations.
Cells satisfying both the anisotropy threshold and the shape criterion are identified as filament-like cells. These form the fundamental building blocks of the filament reconstruction procedure described below.
\subsection{Connectivity and filament reconstruction}

Once the filament-like cells have been identified, they are connected to reconstruct filamentary skeleton linking massive nodes. The nodes are first ordered by decreasing mass, and the reconstruction proceeds iteratively beginning with the most massive node.

As in the case of friends-of-friends halo finders, a characteristic linking scale is required to determine whether two objects can be connected. We therefore define a maximum connection distance, \dcut. Throughout this work, \dcut\ is chosen to be equal to the mean inter-tracer separation of the tracer population at the corresponding hierarchy level. Since filaments are typically denser than the mean tracer distribution, this choice provides a natural upper limit on physically plausible filament connections. Appendix~\ref{app:dcut_sensitivity} provides a study of sensitivity of the filament finder to this parameter.

Starting from a given node, the nearest available filament-like cell is identified. If this cell lies within \dcut, it is connected to the node and forms the first element of a candidate filament. This initial connection defines a local propagation direction,  pointing from the node to the cell. The reconstruction then proceeds by searching for the nearest filament-like cell lying in the forward hemisphere defined by the current direction of travel. If such a cell exists within \dcut, it is connected and the procedure is repeated.

The process continues until one of three outcomes is reached. First, if another node is encountered within \dcut, it is connected and the filament is considered complete. Second, if no valid forward connection can be found, the structure is labelled as a hanging filament. Third, if the growing filament encounters the endpoint of a previously identified hanging filament, the two structures are joined and the filament is again classified as complete. 

Once a complete or hanging filament has been identified, all filament-like cells belonging to that filament are removed from the pool of available cells and the search is restarted from the original node. This procedure is repeated until no further filament-like cells can be connected to the node. This node is then removed, and the algorithm then proceeds to the next most massive node and repeats the reconstruction. Iteration continues until all nodes at the current hierarchy level have been processed. This iterative feature makes this part of the code necessarily serial, which is perhaps a disadvantage.

The resulting set of structures consists of both complete filaments and hanging filaments. Complete filaments define the filament network at the current hierarchy level, while hanging filaments are carried forward to subsequent hierarchy levels where additional lower-mass tracers may allow them to connect to neighbouring structures.  This organization of filamentary structure is different than that in \disp, which we will study in detail in the future.

The reconstructed filaments obtained through this procedure constitute discrete chains of tracer positions. To obtain smooth filament spines suitable for further analysis, all complete filaments are subsequently smoothed using the Fourier-based procedure described in Chapter~\ref{chapt:filtools}.
\subsection{Hierarchical reconstruction of the filament network}
The hierarchical reconstruction is particularly important in dense regions where a large number of small tracers may otherwise obscure the dominant filamentary geometry. By first identifying the large-scale coherent structures and subsequently incorporating smaller-scale information, the algorithm avoids the isotropization problem discussed in the previous section and preserves the stability of the filament reconstruction across a wide range of tracer densities.

\subsection{Dark matter refinement and spine smoothing}

The filament skeleton reconstructed from the halo distribution provides only a discrete approximation to the underlying dark matter filament. Since the reconstructed spine is constrained to pass through a sparse set of halo positions, it need not coincide with the centre of the surrounding dark matter distribution. To obtain a more physically meaningful representation of the filament spine, we perform an additional refinement using the dark matter field.\\
The refinement begins with the crude filament spine obtained from the connectivity algorithm. This spine is first smoothed using the Fourier smoothing procedure described in Chapter~\ref{chapt:filtools}. In addition, the regions lying within the radii of the endpoint nodes are removed in order to minimize contamination from the halo interiors. The resulting curve acts as the initial estimate for the minimization procedure.

A characteristic filament radius, (\rfil), is then estimated from the enclosed dark matter density profile around the current spine. We compute the enclosed cylindrical density profile and define \rfil\ as the radius at which the mean enclosed density falls to 20 times the mean background density. This radius provides an approximate estimate for filament region and determines the scale over which the subsequent refinement is performed.

The minimization is carried out locally along the filament. For each point on the spine, the two neighbouring filament segments are considered simultaneously, forming a short filament section spanning three consecutive spine points. A cylinder passing through the endpoints of this section is constructed, and all dark matter particles lying within a distance \rfil\ from the cylinder axis are selected. If the current spine is displaced from the centre of the filament, the enclosed dark matter distribution will exhibit a non-zero transverse mass dipole. The center of mass (CM) of the selected particle distribution therefore provides an estimate of the displacement required to reduce this dipole. Repeating this procedure for every point on the spine point yields a set of local displacement vectors along the filament, each vector pointing from the point considered to the CM.

Directly shifting each spine point to its corresponding CM would make the reconstruction highly sensitive to shot noise and local fluctuations in the particle distribution. Instead, the displacement vectors are treated collectively and used to construct a smooth global correction to the filament spine. To achieve this, each cartesian component of the displacement field is fitted with a 5th order polynomial as a function of arc length along the filament. The resulting smooth deformation approximately minimizes the transverse mass dipole along the entire filament while suppressing spurious small-scale oscillations. This global minimization substantially improves the stability of the reconstruction.

The characteristic filament radius is recomputed after each iteration and the minimization is repeated until convergence. In practice, convergence is assumed when the maximum displacement of any spine point between successive iterations becomes smaller than $\rfil / 50$. A maximum of five iterations is imposed to avoid unnecessary computations. 
The final reconstructed spine traces the centre of the surrounding dark matter distribution while retaining the large-scale geometry identified by the halo-based filament finder. 

\begin{figure}
    \centering
    \includegraphics[width=0.65\linewidth]{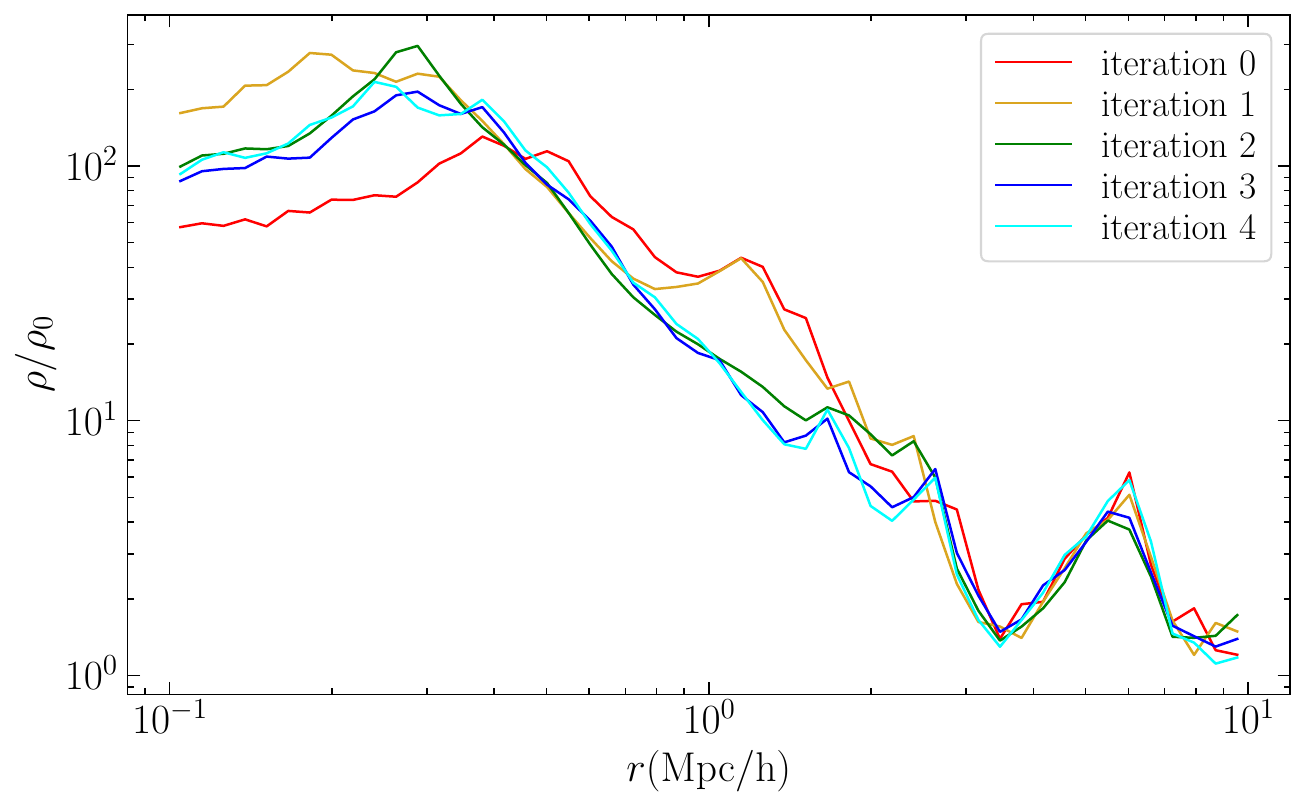}
    \caption{Example of the dipole-minimization smoothing procedure. The radial density profile of a representative filament is shown after successive smoothing iterations, with different colours corresponding to different iterations.}
    \label{fig:dipole_smoothing}
\end{figure}

Figure~\ref{fig:dipole_smoothing} illustrates the effect of the smoothing procedure on the radial density profile of a representative filament. Before smoothing, the spine is slightly offset from the centre of the underlying density field, causing the density profile to peak away from the spine. After only two iterations, the spine becomes well aligned with the density field, increasing the central density while reducing the density at intermediate radii. Beyond this point, additional iterations produce very little change in the profile, indicating that the procedure has effectively converged. As expected, the density at large radii remains largely unaffected by the smoothing.
\subsection{Estimation of filament radii}
Once the final filament spine has been obtained, we estimate the physical extent of the filament using its radial velocity profile. The radial infall velocity around filaments typically reaches a maximum at the transition between the filament interior and its surrounding environment. We therefore use the radius corresponding to the largest radial infall as the filament boundary.

For each filament, the radial velocity profile is measured relative to the refined spine. A second-order polynomial is then fitted to the velocity profile in the vicinity of the global minimum of the noisy radial velocity profile, and the location of the minimum of the fitted polynomial is identified. The corresponding radius is adopted as the final filament radius, $R_v$. In Figure~\ref{fig:vrfit}, the red curve shows the polynomial fit to the velocity dip, while the solid vertical line marks the resulting filament radius. Appendix~\ref{app:skeletor adaptmin} explains an adaptive procedure developed in this work to find this minimum in a noisy profile.

Not all filaments possess sufficiently well-sampled velocity profiles for a robust determination of the infall minimum. This is particularly common for faint filaments in higher level of hierarchy, where shot noise can significantly affect the measured velocity profile. In such cases, filaments with similar node masses are grouped together and stacked. The average radial velocity profile of the stack is then used to determine the characteristic location of the infall minimum. The resulting radius is subsequently assigned to the individual filaments belonging to the corresponding stack.
This approach allows physically motivated filament radii to be estimated even when the velocity information of individual filaments is insufficient for a reliable measurement.

\begin{figure}
    \centering
    \includegraphics[width=0.75\linewidth]{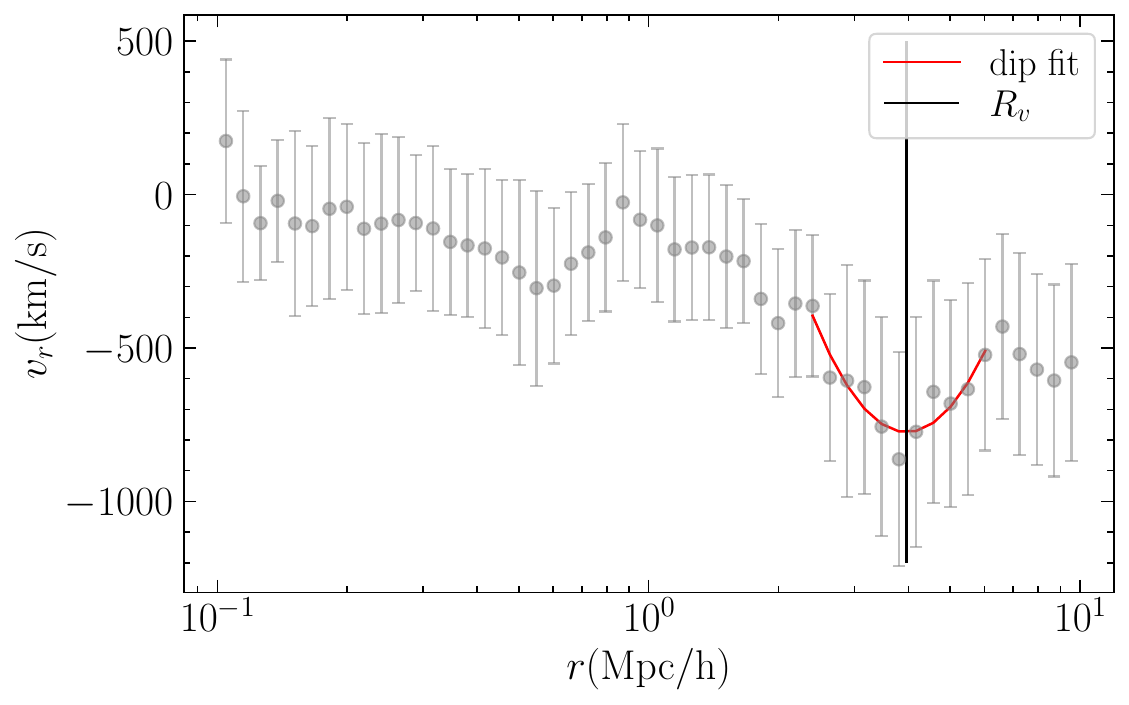}
    \caption{Example illustrating the estimation of the filament radius. The plot shows the radial velocity profile for an individual filament, with error bars estimated from the scatter across $z, \phi$ bins. The red curve shows the quadratic fit used to locate the velocity dip. The solid vertical line indicates the final filament radius $R_v$, defined by the location of largest radial infall.}
    \label{fig:vrfit}
\end{figure}

\section{Filament substructure finder}
\label{sec: substructure finder}

As the filament network is reconstructed across multiple hierarchy levels, many of the structures identified at higher levels lie partially or entirely within larger filaments identified at earlier stages of the reconstruction. To quantify these relationships, \skeletor\ includes a substructure classification procedure that identifies parent filaments and embedded subfilaments.

The classification is based on the filament radii estimated from the radial velocity profiles. Since these radii approximately define the dynamical extent of a filament, they provide a natural scale for assessing whether one filament is embedded within another.

For each filament at a given hierarchy level, we compare its spine to all filaments identified at lower (i.e., previous) hierarchy levels. For every point along the spine, the minimum distance to the lower-level filament spines is computed. A point is considered embedded if it lies within the radius of a lower-level filament. We then calculate the fraction of the filament length satisfying this condition. If this overlap fraction exceeds a specified threshold, taken to be $0.5$ in this work, the filament is classified as a subfilament of the corresponding parent structure. Otherwise, it is retained as an independent filament.

This criterion provides a simple geometric definition of filament substructure while preserving the full filament catalogue. To accelerate the search, spatial trees constructed from the lower-level filament spines are used when evaluating distances and overlap fractions.

The same framework can also be used to compare filament catalogues obtained from different filament finders or from different parameter choices. In this case, the overlap fraction is computed between two independent filament populations. Adopting a more stringent threshold, for example $90\%$, allows corresponding filaments to be matched across catalogues. A robust match is established when the overlap criterion is satisfied in both directions, providing a simple measure of agreement between filament networks.

The reconstruction and classification steps together produce a filament catalogue containing the geometry, hierarchy, and substructure of the network. This catalogue forms the basis for the analyses presented in the remainder of this chapter and in the following chapters.

\section{Computational cost}
\begin{table}[t]
\centering
\begin{tabular}{lccc}
\hline\hline
Code component & Number of cores & Peak memory (GB) & Wall time (min)\\
Voronoi Tessellation & & &\\
Primary level & 32 & 101 & 7.7 \\
Secondary level & 32 & 370 & 33 \\
Tertiary level & 32 & 336 & 44\\
Filament construction & 1& 4.3 & 0.37\\
Refinement of a single filament & 1 & 346 & 1-2\\
Subfilament classification & 1 & 0.2 & 0.05\\
\hline
\end{tabular}
\caption{Summary of computational cost of each component of the code.} 
\label{tab:computational_cost}
\end{table}

Table~\ref{tab:computational_cost} summarizes the computational cost of the different stages of the algorithm. The overall cost is dominated by the construction of the Monte Carlo Voronoi tessellation. For the fiducial configuration used in this chapter, the tessellation requires approximately 1.5 hours on 32 CPU cores and a peak memory of 336,GB. The primary and secondary levels were run with an average of $120,000$ random points per tracer, while this was reduced to $60,000$ for the tertiary level. For tracers with higher number densities, convergence of $\alpha$ and the eigenvalue ratios is achieved with fewer random points. For a fixed random-point fraction, the peak memory scales approximately linearly with the number of tracers $N$, while the wall time scales roughly as $N\log N$.

In comparison, the filament reconstruction itself is extremely fast. The shape determination and directional connectivity steps used to construct the crude filament skeleton typically require less than a minute on a single CPU core, while the subsequent substructure classification requires only a few seconds.

The subsequent filament refinement and profile estimation are more computationally demanding, requiring roughly 1-2 minutes per filament on a single CPU core. However, these calculations are independent for different filaments and are therefore trivially parallelizable. The reported peak memory is relatively high because the full dark matter particle positions and velocities are loaded into memory before the calculation begins.

Although the overall computational expense is currently dominated by the Voronoi tessellation and profile estimation, ongoing work on a machine-learning-based fast Voronoi shape estimator is expected to accelerate the code substantially. The basic filament finder itself remains extremely inexpensive in terms of both CPU time and memory.

\section{Illustrative results}
\label{sec: results}
We now present a few illustrative results obtained using \skeletor. The primary purpose of this section is to demonstrate that the algorithm identifies physically meaningful filamentary structures and produces a sensible hierarchical decomposition of the cosmic web. These results serve primarily as an illustration of the filament population recovered by \skeletor. The physical interpretation of filament profiles, boundaries, and hierarchical substructure is explored in detail in the next two chapters.

The analysis is performed using N-body simulations described in \citep{phs18}. The simulations evolves $1024^3$ particles in a periodic box of side length $L_{\mathrm{box}}=300, \Mpch$, corresponding to a particle mass resolution of $1.93 \times10^9,h^{-1}\Msun$, using the tree-PM code \textsc{GADGET-2} \citep{GADGET2005}. Halos are identified using the \textsc{ROCKSTAR} halo finder \citep{Rockstar2013}. We retain both host halos and subhalos, and the resulting catalogue is   cleaned using the virialization criteria described in Chapter~\ref{chap:sahyadri}, and $M_{\mathrm{200b}}$ is adopted as the halo mass definition throughout.
The filament hierarchy is constructed using the mass thresholds $10^{14}, 10^{13}, 10^{12.5}, 10^{12.3} h^{-1}\Msun$, producing three levels of filament hierarchy. At each level, the higher mass threshold defines the node population, while the lower threshold determines the tracer population used in the Voronoi tessellation and filament reconstruction.

For all profile measurements, the portions of the filament lying within $R_{\mathrm{200b}}$ of the endpoint nodes are removed. This is necessary because the halo interiors can significantly contaminate measurements intended to characterize the filament itself. Furthermore, after the node regions are removed, filaments whose remaining length is smaller than the sum of the radii of the two endpoint nodes are discarded. Such objects generally correspond to diffuse bridges between neighbouring halo outskirts rather than well-defined filamentary structures and are therefore excluded from the profile analysis.

\begin{figure}
    \centering
    \includegraphics[width=\linewidth]{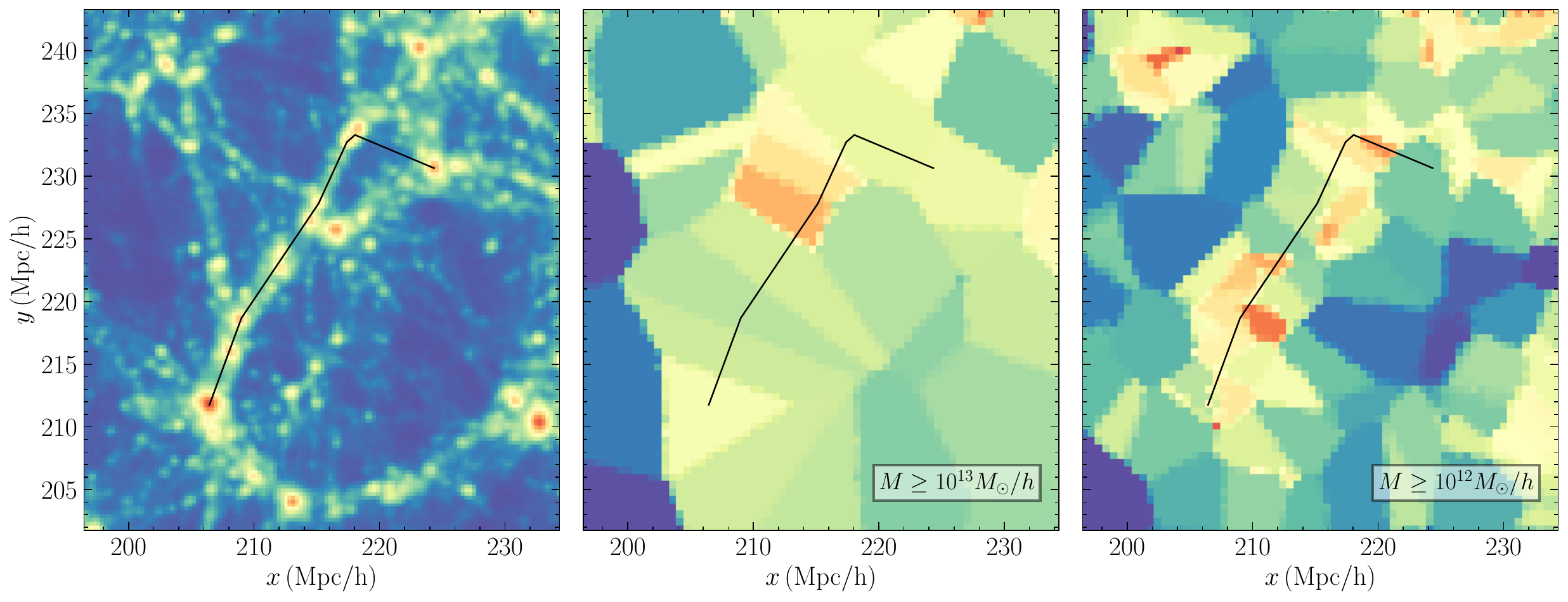}
    \caption{Reconstruction of the prominent filament shown in Figure~\ref{fig:voronoi_filament}. The solid black curve shows the filament identified by \skeletor, overplotted on the same dark matter density slice and Voronoi tessellation. The recovered spine closely follows the visually identified filament.}
    \label{fig:skeletor_filament}
\end{figure}
\begin{figure}
    \centering
    \includegraphics[width=\linewidth]{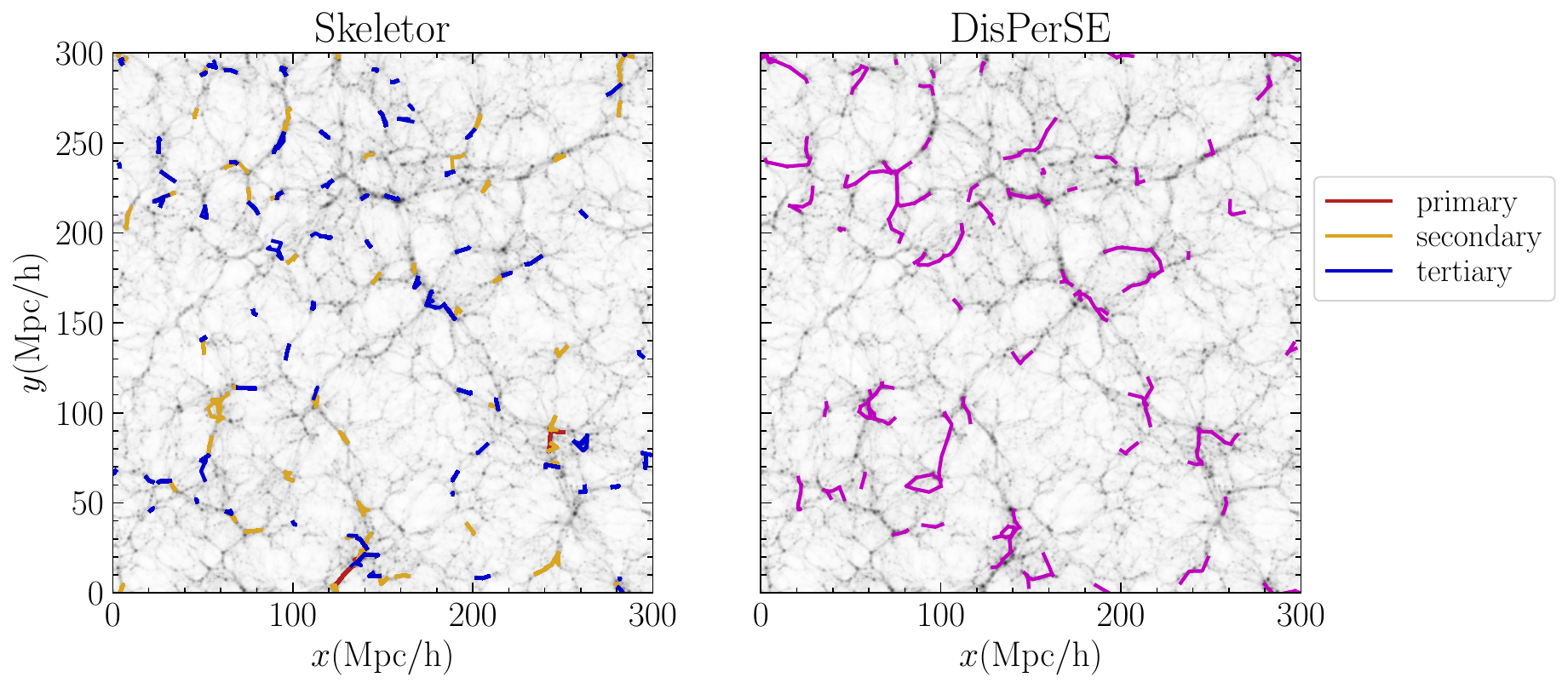}
    \caption{Comparison of filament networks identified by \skeletor\ and \disp. The left panel shows the filament spines recovered by \skeletor, coloured according to their hierarchy level. The right panel shows the corresponding filament network identified by \disp\ using the same halo population. In both panels, the filament spines are overplotted on a dark matter density slice of thickness $10,\Mpch$.}
    \label{fig:skeletor_disp}
\end{figure}
We begin by illustrating the reconstruction obtained with \skeletor\ for a single, visually selected filament. Figure~\ref{fig:skeletor_filament} shows the filament spine recovered by the algorithm overlaid on the same dark matter density slice and Voronoi tessellation presented in Figure~\ref{fig:voronoi_filament}. This filament was chosen because it provides a particularly clear example of the reconstruction. The recovered spine follows the visually identified filament closely, demonstrating that the local anisotropy and connectivity criteria successfully recover the underlying large-scale structure. Having illustrated the reconstruction of an individual filament, we now turn to the properties of the filament network as a whole.

Figure~\ref{fig:skeletor_disp} compares the filament networks recovered by \skeletor\ and \disp. The coloured filaments in the left panel indicate the different hierarchy levels identified by \skeletor, while the grey points show the halo population used in the reconstruction. The dominant large-scale filamentary spine is traced by the highest hierarchy level, with progressively smaller structures appearing at lower levels. Many of the visually apparent filaments in the halo distribution are successfully recovered. The right panel shows the corresponding filament network identified by \disp\ using the same tracer population. Several prominent structures are identified by both methods, for example the filaments located near [120,20] and [180,160], indicating broad agreement on the most significant components of the network. At the same time, noticeable differences are also present. In particular, filaments identified by \skeletor\ often appear more fragmented into shorter segments than those recovered by \disp. At present, it is not clear whether this behaviour reflects a genuine hierarchical subdivision of the filament network or an artifact of the reconstruction procedure, and a more detailed investigation is in progress.\\

\begin{figure}
    \centering
    \includegraphics[width=\linewidth]{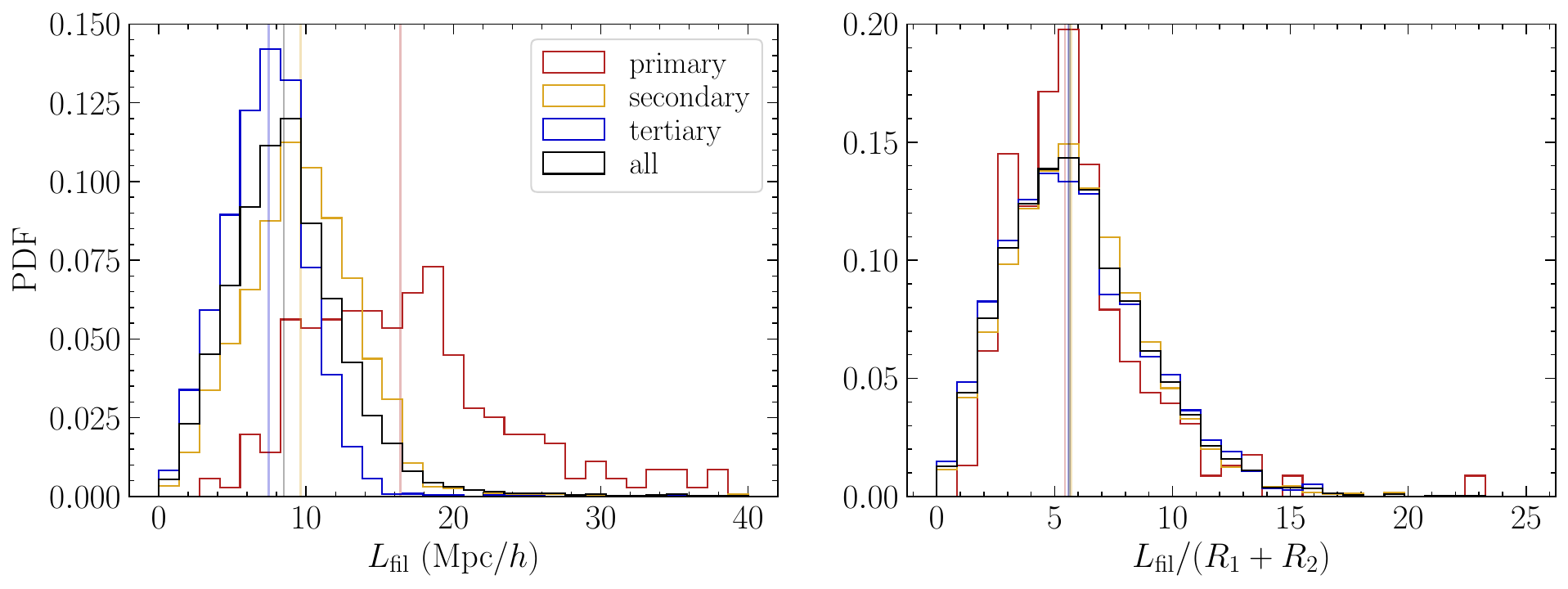}
    \caption{Length distribution of filaments identified by \skeletor. The black curves show the distribution for the full filament population, while the coloured curves correspond to the individual hierarchy levels. The left panel shows the probability distribution function (PDF) of the filament length, $\lfil$. The right panel shows the distribution of filament lengths rescaled by the sum of the radii of the two endpoint nodes. The rescaling substantially reduces the differences between the hierarchy levels, causing the distributions to collapse onto a nearly universal form.}
    \label{fig:lfil_pdf}
\end{figure}
The distribution of filament lengths is shown in Figure~\ref{fig:lfil_pdf}. The left panel displays the probability distribution function of filament lengths for the complete sample as well as for the individual hierarchy levels. The characteristic filament length decreases systematically towards higher hierarchy levels. Primary filaments, which connect the most massive nodes, span the largest distances,hierarchy level is not considered as a primary discriminant of filamentary properties. whereas secondary and tertiary filaments are associated with progressively smaller structures within the cosmic web hierarchy.

A more interesting behaviour emerges when the filament length is rescaled by the sum of the radii of the two endpoint nodes. The right panel of Figure~\ref{fig:lfil_pdf} shows that much of the hierarchy dependence is removed under this rescaling, with the distributions corresponding to different hierarchy levels collapsing onto a nearly universal form. This suggests that the characteristic length of a filament is closely related to the physical sizes of the nodes that it connects. Although the hierarchy levels correspond to very different halo mass scales, the dimensionless quantity $\lfil /(R_1+R_2$) exhibits remarkably similar statistics across the entire filament population. Whether this behaviour persists after accounting for filament fragmentation remains to be investigated. \\
\begin{figure}
    \centering
    \includegraphics[width=\linewidth]{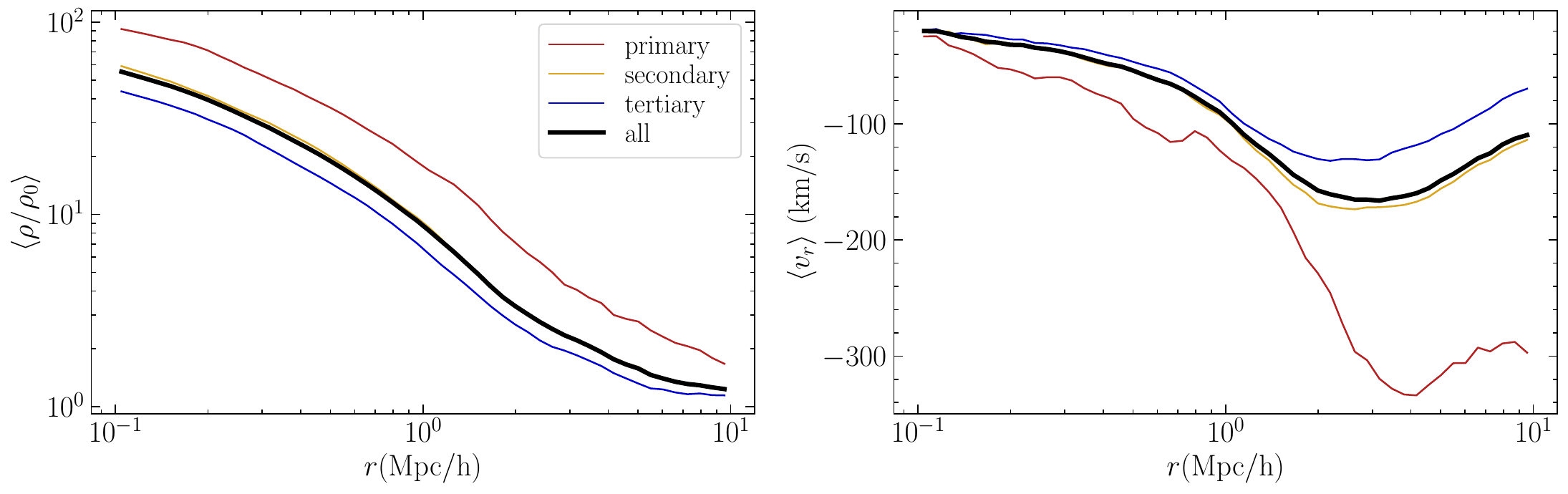}
    \caption{Average radial density (left) and radial velocity (right) profiles of filaments identified by \skeletor. The black curves show the profiles obtained by stacking all filaments, while the coloured curves correspond to individual hierarchy levels. The minima in the radial velocity profiles mark the locations of maximum infall and provide an estimate of the characteristic filament boundaries. Primary filaments exhibit larger characteristic radii, deeper infall velocities, and higher density contrasts. Both the density enhancement and the strength of radial infall decrease systematically from primary to secondary and tertiary filaments, reflecting the hierarchical organization of the filament network.}
    \label{fig:profile_stacks}
\end{figure}
Figure~\ref{fig:profile_stacks} shows the stacked radial density and radial velocity profiles of the recovered filaments. The filaments are weighted by the length over which the profile is measured. The density profiles exhibit the expected ordering with hierarchy, with primary filaments corresponding to the densest structures and lower hierarchy filaments displaying progressively weaker density contrasts. Similar trends are visible in the radial velocity profiles. Primary filaments exhibit stronger radial infall and larger characteristic radii, while secondary and tertiary filaments show progressively weaker accretion signatures. The average profile obtained by stacking all filaments, shown in black, is broadly consistent with previous measurements of dark matter filament profiles, both in amplitude and overall shape (e.g. Figure~2 of \citep{Espinosa+2022}).
The velocity profiles display clear minima corresponding to the locations of largest radial infall. As discussed earlier, these minima provide a physically motivated definition of the filament boundary. The location of the minimum shifts systematically towards smaller radii for lower hierarchy levels, indicating that the characteristic transverse extent of filamentary structures decreases as one moves from the dominant spine of the cosmic web to progressively smaller filaments.

Appendix~\ref{app: node mass and hierarchy} shows that the dominant contribution to the differences between the stacked profiles arises from the masses of the nodes connected by the filaments. Because the hierarchy levels are defined using mass thresholds rather than exclusive mass bins, there is substantial overlap in the node masses associated with different hierarchy levels. As a result, hierarchy is most naturally viewed as a reconstruction tool that allows the filament network to be identified across multiple scales, whereas node mass provides a more physically meaningful variable for comparing filament populations. In the remainder of this thesis, hierarchy level is not considered as a primary discriminant of filamentary properties.
\begin{figure}
    \centering
    \includegraphics[width=\linewidth]{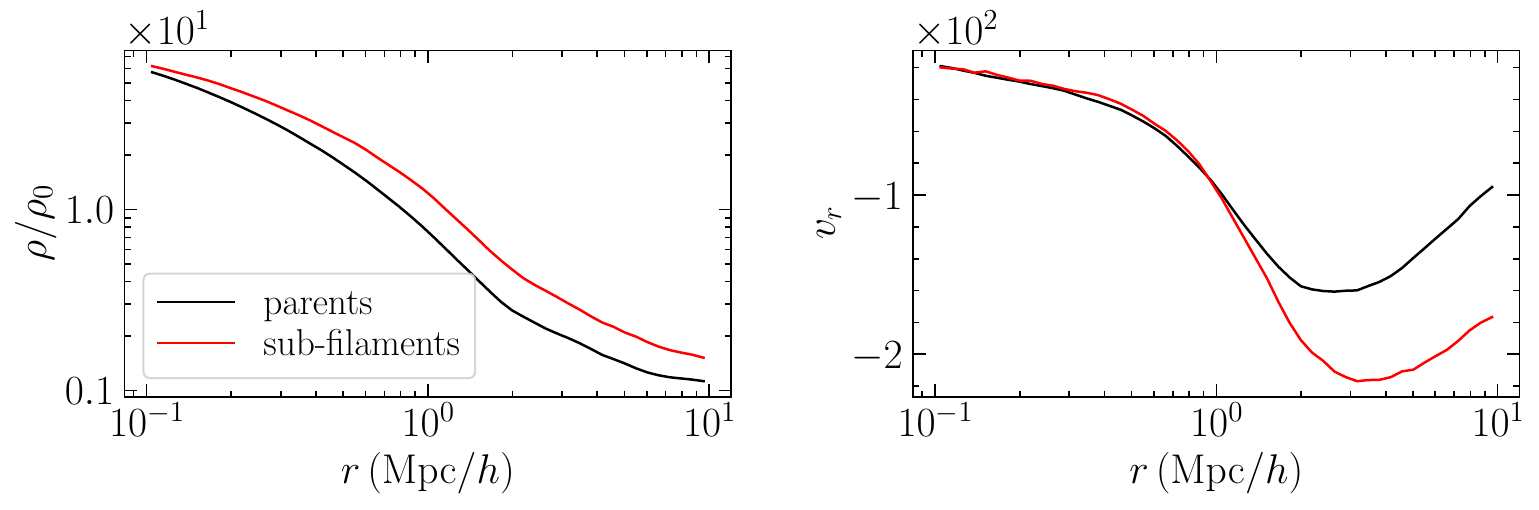}
    \caption{Comparison between the stacked radial profiles of parent (black) and sub-filaments (red). The two represent distinctly different populations.}
    \label{fig:parent_sub_profs}
\end{figure}

Figure~\ref{fig:parent_sub_profs} compares the length-weighed stacked radial density and velocity profiles of the parent and sub-filament populations. The details of how these populations are obtained is presented in the next chapter. The two exhibit systematic differences. Sub-filaments have higher densities at nearly all radii, with the difference becoming more pronounced in the outskirts. They also show stronger radial infall, with deeper velocity minima than the parent filaments. These differences are consistent with sub-filaments residing within the denser environments of larger parent filaments. A more detailed investigation of sub-filaments is presented in the next chapter.

Taken together, these results indicate that \skeletor\ successfully recovers a multiscale filament network whose geometric and dynamical properties are consistent with the expected structure of the cosmic web. The recovered filaments exhibit sensible length distributions, well-defined density and velocity profiles, and physically motivated boundaries inferred from their infall dynamics. The hierarchical reconstruction provides a natural framework for identifying filamentary structures across a wide range of scales, while the resulting filament catalogue forms the basis for the more detailed analysis presented in the following chapter.

\section{Discussion and conclusions}
\label{sec: conclusion}
In this chapter, we introduced \skeletor, a Voronoi-based hierarchical filament finder that reconstructs filamentary networks directly from discrete tracer populations. The method combines local anisotropy measured from Voronoi geometry with a directional connectivity algorithm and a hierarchical reconstruction across multiple mass scales. The resulting filament spines are refined using the surrounding dark matter distribution, and physically motivated filament radii are estimated from radial infall profiles. We further introduced a substructure classifier that identifies parent filaments and embedded subfilaments, providing a framework for studying the hierarchical organization of the cosmic web.

The illustrative results presented in this chapter demonstrate that the recovered filament population exhibits sensible geometric and dynamical properties. The identified filaments trace visually apparent structures in the halo distribution, possess physically reasonable length distributions, and exhibit coherent density and velocity profiles with well-defined infall boundaries. The hierarchy recovered by the algorithm also corresponds broadly to structures of different characteristic scales within the cosmic web.\\

Several caveats should nevertheless be kept in mind. The reconstruction depends on a number of algorithmic choices, most notably the hierarchy thresholds and the connectivity scale \dcut. Although the fiducial choices adopted in this work are physically motivated and yield robust results, they are not unique, and alternative prescriptions may be worth exploring in future work.

A second limitation is the tendency of \skeletor\ to identify relatively short filament segments. Visual inspection suggests that some of these segments may belong to larger coherent structures that are not fully reconstructed by the present connectivity algorithm. One possible way to address this issue would be to incorporate temporal information into the filament reconstruction. Rather than identifying filaments independently in each snapshot, filament segments could be tracked across neighbouring outputs and assigned persistent identities. Segments that repeatedly appear in similar locations and connect the same large-scale structures could then be merged into longer coherent filaments, while transient or poorly resolved structures could be discarded. Such a persistence-based approach would provide information beyond that available from a single snapshot and may help distinguish genuine filament fragmentation from limitations of the reconstruction procedure.

The filament reconstruction itself requires only tracer positions and masses  (or mass proxy values), while the refinement procedure and radius estimation currently depend on information available only in simulations. This suggests that the basic algorithm should be directly applicable to galaxy and group catalogues, with stellar masses/luminosities of individual galaxies and galaxy groups serving as proxies for halo mass. The main challenge in such applications will be accounting for observational effects, particularly incompleteness and redshift-space distortions. Assessing the impact of these effects, and adapting the method to realistic survey data, remains an important direction for future work.

Several other extensions are possible. The hierarchy construction adopted here is based on a discrete set of mass thresholds, whereas the cosmic web is fundamentally continuous and multiscale. Future implementations may benefit from adaptive hierarchy definitions. Similarly, the filament substructure classification introduced in this chapter provides only a first step towards a more complete description of filament assembly and nesting within the cosmic web. A detailed investigation of these hierarchical relationships, and their impact on filament properties, remains an interesting avenue for future study.

A more fundamental extension concerns the definition of the Voronoi tessellation itself. In the present implementation, Voronoi cells are constructed using the standard Euclidean distance between tracers and random points. This choice treats all tracers equally irrespective of their masses. However, the formation and evolution of the cosmic web are governed by gravity, and the influence of a tracer on its surroundings depends strongly on its mass. It may therefore be more natural to replace the Euclidean tessellation with a weighted or tidal Voronoi tessellation in which distances are modified by the masses of tracer halos. The resulting tessellation would no longer partition space according to proximity, but according to gravitational dominance. Since filaments arise from anisotropic gravitational collapse, such a construction may provide a more physically motivated characterization of local filamentary environments than a purely geometric Voronoi tessellation. Exploring whether mass-weighted or tidal tessellations lead to a cleaner separation of filamentary environments, improved spine reconstruction, or more robust identification of substructure represents an interesting direction for future work.\\

The filament catalogue constructed using \skeletor\ forms the basis for the subsequent chapters of this thesis. The next chapter explores the hierarchical nature of the filament network through a detailed study of parent filaments and embedded subfilaments. The following chapter investigates the phase-space structure of filaments, with particular emphasis on the physical interpretation of filament boundaries and the phase-space features associated with filamentary collapse.

    %
%
\let\textcircled=\pgftextcircled
\chapter{subfilaments in the Cosmic Web: Toward a Hierarchical Description of Filamentary Structure}
\label{chapt:sub_filaments}
Cosmic filaments are expected to possess a hierarchical internal structure, with smaller filamentary features embedded within larger filamentary environments. Unlike the well-established distinction between halos and subhalos, however, filament studies have generally treated all filaments as belonging to a single population. In this chapter, we use the hierarchical reconstruction developed in the previous chapter to identify such embedded subfilaments and compare their basic properties with those of the parent filament population. In particular, we examine differences in their radial phase-space profiles and investigate whether subfilaments constitute a statistically distinct class of structures. We also discuss the possible implications of separating parent filaments and subfilaments for future studies of filament statistics and the search for universal filament properties.
\newpage
\section{Introduction}

The previous chapter introduced \skeletor, a hierarchical filament finder that reconstructs the cosmic web across multiple mass scales and identifies filamentary structures embedded within larger filamentary environments. Such embedded structures, which we refer to as \emph{subfilaments}, emerge naturally as the tracer mass threshold is progressively lowered and increasingly smaller scales of the cosmic web are resolved. While hierarchical filamentary structure is an expected consequence of hierarchical structure formation, comparatively little attention has been paid to the physical significance of these subfilaments or to whether they should be regarded as a distinct population.

Most filament-finding algorithms ultimately produce a single catalogue and treat all recovered structures as members of the same statistical population, irrespective of the larger environments in which they reside. This implicitly assumes that an isolated filament and one embedded within a much larger filamentary system should be described by the same physical picture. Such an approach may overlook important environmental effects and obscure physically meaningful trends.

A useful parallel can be drawn with the study of dark matter halos. Hierarchical growth through accretion and mergers produces not only isolated halos but also subhalos embedded within larger host halos. Treating subhalos as a distinct population has proved essential for understanding their structural evolution, tidal stripping, merger histories, and the connection between dark matter halos and galaxy populations.\citep{Springel+2001, Cooray&Sheth2002,Wechsler&Tinker2018}. More generally, recognizing subhalos as a distinct population has led to a more complete description of hierarchical structure formation and has become a standard component of both theoretical and observational studies of galaxy formation.

It is natural to ask whether an analogous distinction should be made for cosmic filaments. A subfilament evolves within the gravitational influence of a larger parent filament and is therefore embedded in a different dynamical environment, experiencing different tidal fields and matter flows than an isolated filament. Even if the two populations have similar geometrical properties, there is little reason to expect their internal phase-space structure to be identical. Mixing them together may therefore obscure physically meaningful trends and contribute to the scatter commonly observed in filament statistics. Conversely, identifying and studying subfilaments separately may reveal cleaner scaling relations and aid the search for universal filament profiles.

The hierarchical reconstruction provided by \skeletor\ makes such a study possible by explicitly identifying the parent and subfilament populations. The definition adopted here is based on geometric overlap between filament spines and the radial extent of larger filaments. This should not be regarded as unique, however, and other definitions incorporating dynamical or evolutionary information may ultimately provide a more physically motivated classification.

The primary goal of this chapter is not to provide a comprehensive characterization of subfilaments but rather to explore whether they exhibit systematic differences from the parent filament population. We compare their node properties and phase-space profiles and discuss the possible implications of these differences for future studies of the cosmic web. More broadly, this chapter serves as an initial exploration of the idea that treating parent filaments and subfilaments as separate populations may lead to a cleaner and more physically motivated statistical description of the filamentary network.

\section{Filament samples}
Here, we use the simulations described in the previous chapter, with a periodic box size of $300~\Mpch$ and particle mass $1.93\times10^9~\Msun/h$. The same halo samples are used, and all analyses are performed at redshift $z=0$. We use five realizations with identical cosmological parameters. The filament samples from all realizations are combined when measuring phase-space profiles, treating the realizations as independent and pooling them together to improve the statistics.

We run \skeletor\ to reconstruct three levels of hierarchy. As described in the previous chapter, the portions of every filament lying within $R_{\mathrm{200b}}$ of the endpoint nodes are removed before any profile measurements are performed. We further discard filaments whose remaining length is smaller than the sum of the radii of the two endpoint nodes. As an additional cleaning step, we discard filaments whose enclosed density never exceeds 10. As seen in the previous chapter, filaments typically reach enclosed densities of several tens of the mean matter density near their centers, with the outskirts of filaments typically reaching $\lesssim 10 \times$ the mean density. Filaments that fail this requirement are therefore unlikely to represent robust detections and are removed from the sample. As a final cleaning step, we discard the $10\%$ of filament segments with the highest curvature. As motivated in the previous chapter, highly curved segments can bias the measured radial profiles and are therefore excluded from the analysis. The dependence of the phase-space structure on filament curvature is examined explicitly in the next chapter. After these cuts, $82\%$ of the original catalogue is retained, corresponding to 27789 filaments. These are then classified into parent filaments and subfilaments.

\section{Parent and subfilament classification}
For every filament identified at a given hierarchy level, we compare its spine with the filament population reconstructed at all previous levels. For each current-level filament, we compute the fraction of its length lying within the estimated radius of previous-level filaments. The radius is taken to be the location of the radial velocity dip, the details of whose determination is described in the next chapter. If more than $50\%$ of the filament lies inside a previously identified filament, it is marked as a candidate subfilament.

Some of these candidates are not physically distinct structures but simply repeated detections of the same filament arising from the use of different tracer populations. To identify such cases, we perform a second comparison using a much smaller matching radius corresponding to the uncertainty in the reconstructed spine position.

This uncertainty is estimated by measuring the mean displacement of the filament spine before and after the mass-dipole minimization procedure. Here, the pre-optimization spine refers to the Fourier-smoothed spine. If more than $80\%$ of a candidate filament lies within three times this characteristic displacement from a previously identified filament, the two are taken to represent the same physical structure, and the shorter resampled filament is removed from the filament catalogue.

The remaining candidates are classified as genuine subfilaments. Throughout this chapter, filaments that are not embedded within larger systems are referred to as \emph{parent filaments}, while the retained embedded population constitutes the \emph{subfilament} catalogue. The focus of this chapter is on comparing these two populations, whereas the next chapter is devoted to a detailed study of the phase-space structure of the parent sample.

The parent-subfilament classification adopted here is based purely on the reconstructed geometry of the filament network. As mentioned before, this is not the only possible definition of filament substructure. Future work could instead incorporate information from the density field, velocity field, or the temporal evolution of filaments. The geometric definition adopted here provides a practical starting point for studying the subfilament population.

\section{Basic properties of the subfilament population}
Having identified the subfilament population, we now compare some of its basic properties with those of the parent filaments. The goal is to determine whether subfilaments represent a statistically distinct subset of the filament population or simply a random sample of all filaments.

\begin{figure}
    \centering
    \includegraphics[width=\linewidth]{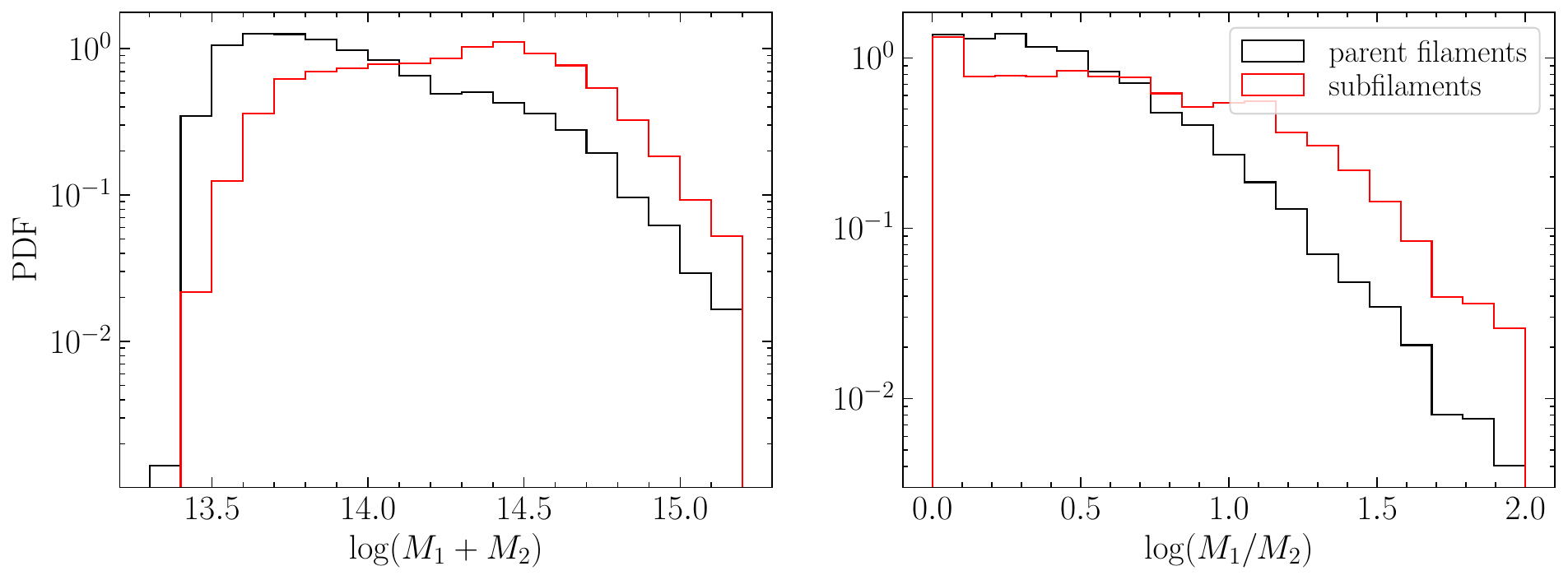}
    \caption{\emph{Left panel:} Distribution of the total node mass, $\log(M_1+M_2)$. \emph{Right panel:} Distribution of the corresponding node mass ratio, $M_1/M_2$. Parent filaments are shown in black and sub-filaments in red. Node masses are reported in units of $\Msun/h$.}
    \label{fig:node_mass_parentsub}
\end{figure}
The left panel of Figure~\ref{fig:node_mass_parentsub} shows the distribution of the total node mass, $\log(M_1+M_2)$. Subfilaments are preferentially associated with larger total node masses, implying that they are more commonly found in the vicinity of massive nodes and in denser environments. This behaviour is expected if hierarchical substructure becomes increasingly abundant around the most prominent filaments of the cosmic web. The higher node masses also suggest that sub-filaments often share a node with their parent filaments.

The right panel of Figure~\ref{fig:node_mass_parentsub} compares the distribution of the node mass ratio, $M_1/M_2$, for parent filaments and subfilaments, where $M_1 \geq M_2$ are the masses of the two nodes connected by a filament. The distribution exhibits a more prominent high-ratio tail for the subfilament population, indicating that subfilaments preferentially connect nodes with more unequal masses. This supports the node-sharing hypothesis mentioned above, where subfilaments are sub-structures of major filaments, connecting to a massive node at one end belonging to the parent filament, while terminating within the body of the parent filament rather than at another comparably massive node. A detailed examination of individual parent--sub-filament pairs to test this interpretation is left for future work.

\begin{figure}
    \centering
    \includegraphics[width=\linewidth]{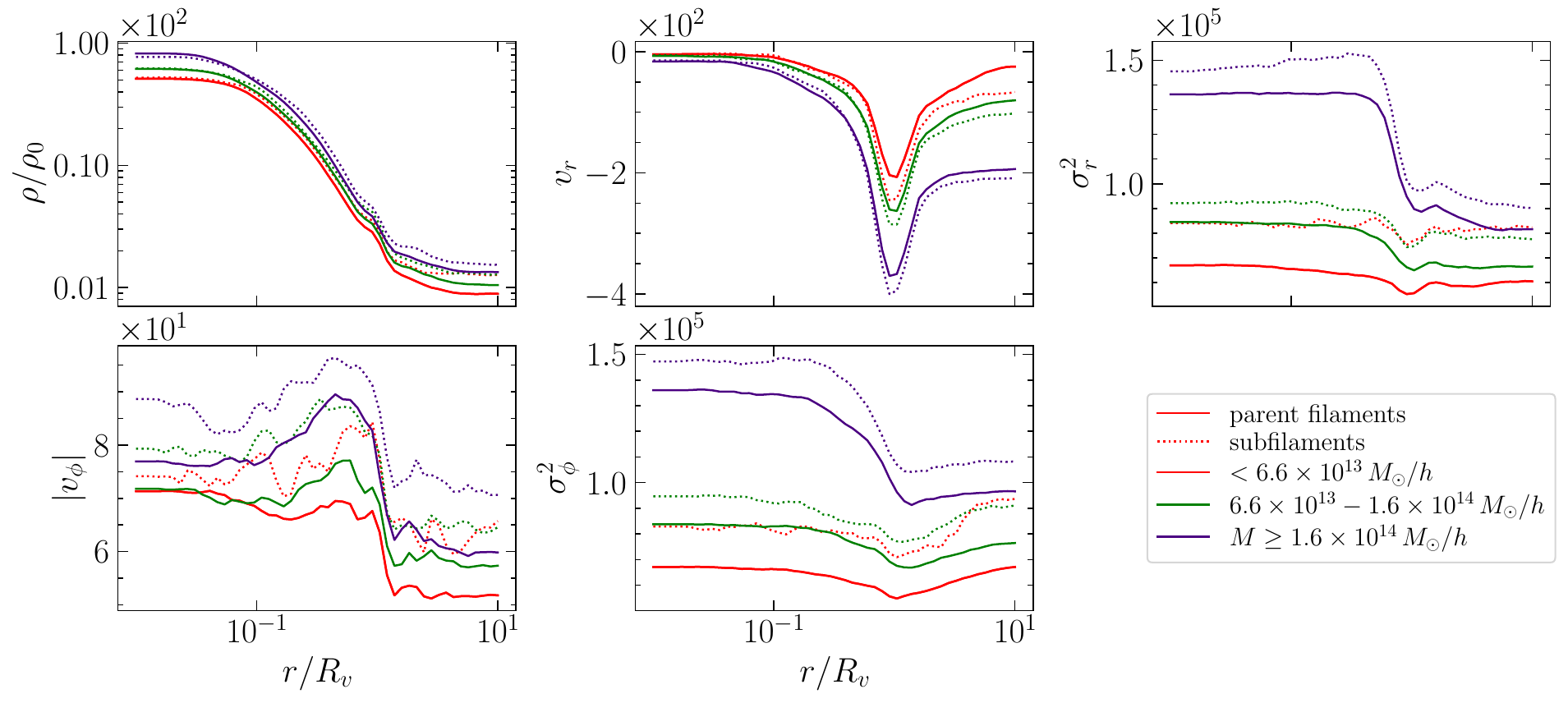}
    \caption{Comparison of the radial density and velocity profiles of parent filaments (solid curves) and subfilaments (dotted curves). Different colours correspond to different bins in the total node mass, defined by the $33.33$ and $66.67$ percentiles of the full filament population. The radial distance is scaled by the radius of the filament,  $R_v$. $v_r, v_\phi$ are given in units of km/s, whereas the corresponding $\sigma^2$ are reported in $(\mathrm{km/s})^2$}
    \label{fig:parent_sub_profs_mnode_split}
\end{figure}
To study whether these environmental differences are reflected in the internal properties of the filaments, we compare radial profiles of the density and velocity fields for parent filaments and subfilaments. We focus on five quantities: the density normalized by the mean background density $\rho/\rho_0$, the radial infall velocity $v_r$, where $r$ is the perpendicular distance to the spine, the absolute value tangential velocity $|v_\phi|$, and the dispersion of both these velocities $\sigma^2_r, \sigma^2_\phi$. When constructing the stacked profiles, each filament is weighted by the length over which its profile is measured. This effectively yields volume-weighted averages, as the area of cross-section is the same for all the filaments in a given radial bin.

Since many filament properties depend strongly on node mass, which will be explored in detail in the next chapter, the comparison is performed within fixed bins of total node mass. The radial distance is scaled by the radius of each filament inferred from the velocity dip, $R_v$, which accounts for most of the filament radius dependence. The filaments are divided into three bins corresponding to the $33.33$ and $66.67$ percentiles of the total node mass distribution. The resulting profiles are shown in Figure~\ref{fig:parent_sub_profs_mnode_split}.

The density profiles show that subfilaments are systematically denser than parent filaments of the same node mass, with the differences becoming particularly pronounced in the outskirts. While the central densities are broadly similar, the density around subfilaments decreases more gradually, indicating that they remain embedded in a denser large-scale environment over a wider radial range.

The radial velocity profiles exhibit a similar trend. Subfilaments show a deeper infall dip than parent filaments, although the location of the minimum remains nearly unchanged. The differences are again most evident in the outskirts, where stronger infall persists out to larger radii for the subfilament population. This suggests that subfilaments continue to reside in anisotropic environments influenced by nearby overdensities and the gravitational potential of larger structures.

The tangential velocity profiles also differ systematically between the two populations. At fixed node mass, subfilaments exhibit larger tangential velocities than parent filaments, implying stronger transverse motions in their surroundings. Likewise, both the radial and tangential velocity dispersions are consistently higher for subfilaments, indicating a dynamically hotter environment.

Overall, these comparisons show that subfilaments differ from parent filaments in several important respects. They preferentially connect more asymmetric and more massive node pairs and exhibit systematically enhanced densities, stronger coherent inflows, larger tangential motions, and higher velocity dispersions even after controlling for node mass and filament radius. These trends suggest that subfilaments constitute a distinct statistical population and that mixing them with parent filaments may contribute to the scatter observed in studies of filament density and velocity profiles.

\section{Discussion and future prospects}
The results presented in this chapter suggest that parent filaments and subfilaments represent physically distinct populations. Even after controlling for node mass, subfilaments exhibit systematically higher densities, stronger radial infall, larger tangential velocities, and higher velocity dispersions. This naturally raises the possibility that treating the two populations separately may provide a more physically meaningful statistical description of the cosmic web.

One possible application of the parent-subfilament classification is in the search for universal filament profiles. Unlike dark matter halos, whose density profiles are well described by simple parametric forms over a wide range of masses, the NFW profile \citep{NFW97}, no such consensus has emerged for cosmic filaments. 
A number of studies have proposed fitting functions or scaling relations for filament density profiles \citep[e.g.][]{Espinosa+2022}, but significant scatter remains and the existence of a universal filament profile is still an open question. It is worth noting, however, that a recent study by \citep{Xu+2026} reports evidence for nearly universal dark matter density profiles after an appropriate rescaling.

Establishing whether filament profiles are universal is important for reasons beyond simply obtaining a convenient fitting function. A universal profile, if it exists, would provide valuable clues about the physical processes that govern the formation and evolution of cosmic filaments and the extent to which these processes are insensitive to the details of the local environment. It would also provide a well-defined baseline against which the effects of different cosmological models, dark matter scenarios, or baryonic physics could be compared in a more robust and systematic manner.

The hierarchical nature of the filament population may itself be one source of the large scatter reported in previous studies. If parent filaments and subfilaments evolve under different physical conditions, combining them into a single sample may wash out otherwise simple scaling relations. A natural next step would therefore be to search for universal profiles within each population separately.\\
\\
The parent–subfilament classification also raises a number of questions about how subfilaments form and evolve. Embedded within larger filamentary systems, they will experience strong directional tidal fields and the coherent mass transport taking place along the parent filament. It is therefore natural to expect their evolution to differ from that of isolated filaments. For example, it is not clear whether subfilaments form independently and are later incorporated into larger structures, or whether they emerge directly as secondary branches during the assembly of the parent filament itself. The subsequent evolution of these structures is equally uncertain. Strong inflows along the parent filament may continuously feed subfilaments, but they may also disrupt them through tidal interactions. Whether subfilaments are long-lived objects with well-defined identities or transient features that appear and disappear over relatively short timescales remains an open question. Understanding the balance between these processes would provide valuable insight into the hierarchical growth of the cosmic web.

The presence of subfilaments may also modify the way matter is transported through the filament network. Rather than flowing directly along a single spine towards a massive node, material may preferentially flow along multiple smaller sub-filaments. It would also be interesting to investigate whether subfilaments exhibit systematically different multistream regions, phase-space caustics, or filament boundaries compared to the parent population. Quantifying this multiscale flow pattern could help build a more complete picture of anisotropic accretion in the cosmic web and clarify how matter is redistributed between different levels of the hierarchy.

Addressing these questions will require tracking individual subfilaments across cosmic time. Constructing such evolutionary histories could clarify whether subfilaments are persistent structures or transient features, how they exchange matter with their parent filaments and are tidally influenced, and what role they play in the hierarchical assembly of the cosmic web.\\
\\

The existence of a subfilament population may also have implications for the connection between galaxies and the cosmic web. In the halo paradigm, distinguishing between central and satellite galaxies has proved essential for accurately describing galaxy clustering and the galaxy-halo connection, reflecting the fact that galaxies residing in different environments evolve differently. A similar situation may arise for filaments. Galaxies embedded within subfilaments are expected to experience a different large-scale environment from those associated with isolated parent filaments, including stronger tidal fields and the coherent mass transport taking place within the surrounding host filament.

Whether these environmental differences leave an imprint on galaxy properties remains an open question. For example, galaxies in subfilaments may exhibit different gas accretion histories, star formation rates, merger histories, or alignments with the local filamentary structure compared to galaxies residing in parent filaments. If such trends exist, explicitly accounting for filament hierarchy could provide a more informative description of the galaxy-cosmic web connection and may complement existing models based primarily on halo properties.\\
\\
It will be interesting to determine whether different classification approaches using extra information such as density field and tidal anisotropy recover the same subfilament population or identify different aspects of the hierarchical organisation of the cosmic web.

Beyond the physical applications discussed above, the overlap algorithm itself may prove useful in other contexts. By requiring substantial geometric agreement between two filament populations, the same framework can be used to match catalogues produced by different filament finders or by the same finder with different parameter choices. Such comparisons could provide a quantitative measure of the robustness of reconstructed filament networks and help establish correspondences between different reconstruction techniques.

The analyses presented in this chapter provide only a first look at the subfilament population. A more detailed study of their abundance, spatial distribution, lifetimes, and evolution will be needed to establish their role in the growth of the cosmic web. These results suggest that treating all filaments as a single population may obscure physically meaningful differences. We hope that this motivates the community to explore hierarchical classifications more systematically in future studies of the cosmic web.

    %
%
\let\textcircled=\pgftextcircled
\chapter{Phase-space Structure of Cosmic Filaments}
\label{chapt: filament_properties}
\initial{C}osmic filaments are dynamically evolving structures produced by anisotropic gravitational collapse and are believed to play an important role in the growth of halos and the redistribution of matter within the cosmic web. While their geometric properties have been studied extensively, much less is known about their internal phase-space structure and the physical processes that determine their extent and evolution. In particular, it remains unclear whether filaments possess genuinely universal density or velocity profiles, how their boundaries should be defined, and whether signatures of shell crossing and multistreaming can be robustly identified in the phase-space structure of filaments measured in cosmological simulations.

This chapter presents a detailed study of the phase-space structure of the parent filament population identified by \skeletor. We examine the dependence of stacked filament profiles on node mass, length, and curvature, with the broader aim of understanding which physical properties have the greatest influence on filament structure and dynamics. We then compare different dynamical and density-based definitions of filament boundaries and search for evidence of multistreaming and caustic-like features in the dark matter distribution, with the aim of developing a more physical understanding of the internal structure of cosmic filaments.
\newpage
\section{Introduction}
The previous chapters introduced \skeletor\ and established a clean sample of parent filaments by removing embedded substructures. With this catalogue in hand, we can now turn to a more detailed study of the internal phase-space structure of parent filaments.

A longstanding question in the study of the cosmic web is whether filaments possess universal internal structure. Although many works have measured stacked density profiles and proposed fitting functions, a generally accepted description has yet to emerge, with significant scatter persisting across different filament populations and reconstruction methods \citep{Colberg_et_al2005, Aragon-calvo+2010a, NEXUS2014}. While the existence of a universal filament profile remains uncertain, recent work has suggested that suitably rescaled dark matter filaments may exhibit remarkably similar density profiles \citep{Xu+2026}, providing fresh motivation for understanding the origin of the remaining scatter.

The previous chapter argued that part of this diversity may arise from mixing parent filaments and subfilaments, motivating the present focus on the parent population alone. Even within this cleaner sample, however, filament properties vary substantially. While the dependence of filament profiles on quantities such as node mass and length has received some attention in the literature, the role of filament curvature remains largely unexplored. A primary goal of this chapter is therefore to examine how the phase-space structure depends on these quantities and to identify which of them has the strongest influence on the filament profiles.

A related problem concerns the definition of filament boundaries. A variety of operational definitions have been proposed, ranging from characteristic overdensity thresholds to splashback-like radii identified from features in the density profile \citep{2pop2020, wang+24}. At present, however, it remains unclear whether cosmic filaments possess a unique physically motivated boundary and, if so, which quantity provides the most appropriate way of identifying it.

The phase-space structure surrounding filaments provides another way of probing the filament boundaries. In Lagrangian descriptions of anisotropic gravitational collapse, shell crossing naturally gives rise to multistream regions and caustics as the initially cold dark matter sheet folds in phase space \citep{Zeldovich1970,Hidding+2014}. Whether such signatures can be identified around filaments reconstructed in cosmological simulations remains largely unexplored. In particular, it is interesting to ask whether they coincide with dynamically meaningful filament boundaries.

In this chapter, we study the internal structure of the parent filament population identified by \skeletor. By comparing stacked phase-space profiles for filaments with different node masses, lengths, and curvatures, we identify which of these properties have the strongest impact on the density and velocity structure. We also compare different definitions of filament boundaries and search for signatures of multistreaming and caustic-like features in the surrounding phase-space distribution.

\section{Filament sample}
The filament catalogue used in this chapter is the parent filament sample introduced in the previous chapter. Briefly, \skeletor\ is run on the five realizations of simulations, and the resulting filament population is cleaned by removing the regions lying within $R_{\mathrm{200b}}$ of the endpoint nodes, rejecting filaments with insufficient remaining length, and applying an enclosed density cut to eliminate spurious detections. Embedded subfilaments are then identified using the substructure classifier and discarded from the analysis while stacking the filament profiles. This is analogous to excluding subhalos when analysing the properties of parent halos.

Following the discussion in Chapter~\ref{chapt:filtools}, segments with the highest $10\%$ curvature are also excluded from the main sample. As we shall show later in this chapter, high curvature regions have systematically different properties and can significantly bias the stacked profiles. Their behaviour is nevertheless qualitatively similar to that of the cleaned sample and is presented separately in Appendix~\ref{app: high kappa fils}. 

The final catalogue contains 21233 parent filaments obtained by combining all five simulation realizations. Throughout this chapter, the realizations are treated as statistically independent and analysed together in order to improve the signal-to-noise of the stacked measurements. While calculating stacks of radial profiles, we weigh each filament by the contributing length, so that all the profiles are equivalently volume weighted.

\section{Dependence of phase-space profiles on filament properties}
We begin by exploring how the phase-space structure of filaments depends on some of the basic properties of filaments. In particular, we consider the effects of filament curvature, length, node mass and velocity-inferred filament radius, and identify which of these quantities has the strongest impact on the measured profiles.

\subsection{Filament curvature}
\begin{figure}
    \centering
    \includegraphics[width=\linewidth]{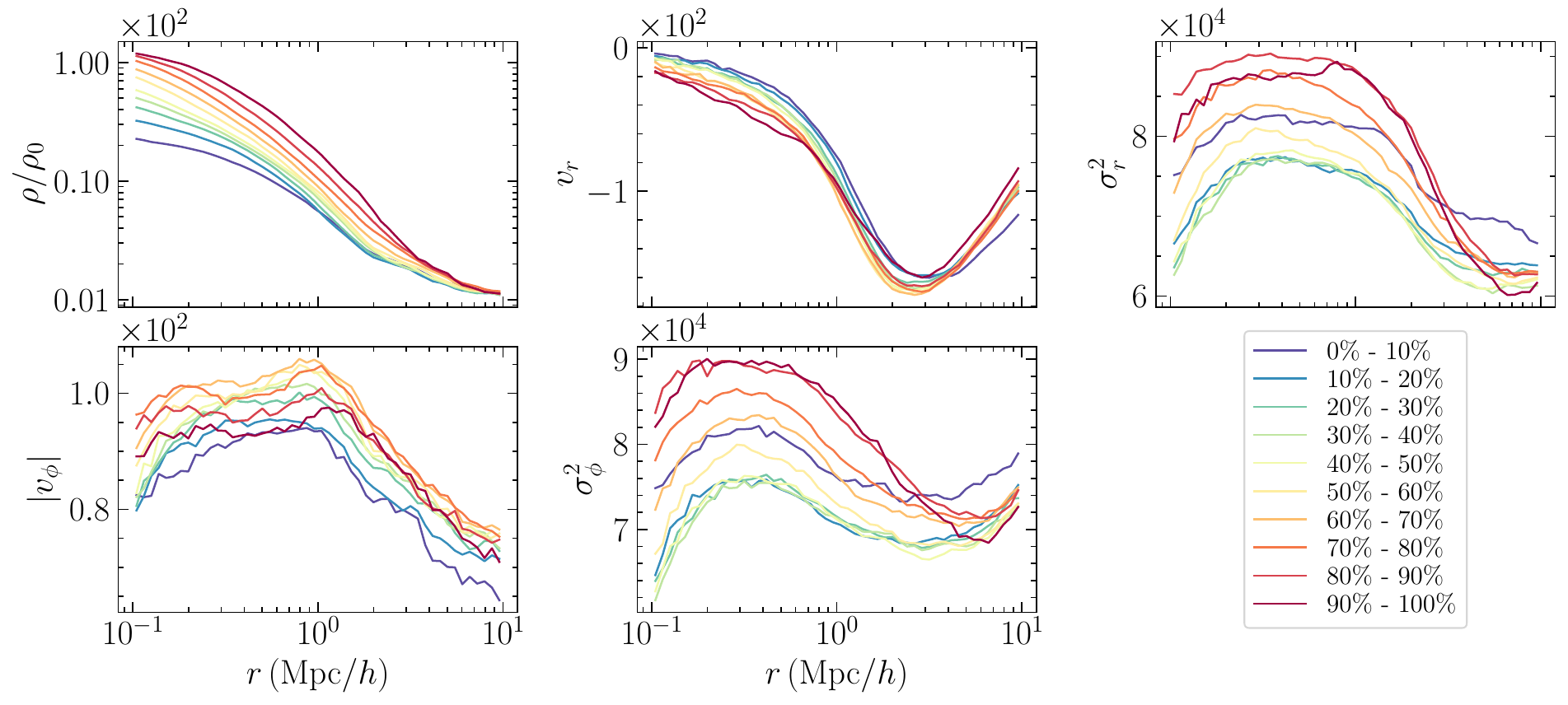}
    \caption{ Stacked radial phase-space profiles in ten percentile bins of filament curvature, $\kappa$. Higher-curvature segments are systematically denser and exhibit stronger radial infall, larger tangential velocities, and higher velocity dispersions, while the radius of maximum infall remains nearly unchanged.  $v_r, v_\phi$ are given in units of km/s, whereas the corresponding $\sigma^2$ are reported in $(\mathrm{km/s})^2.$}
    \label{fig:curvature_split}
\end{figure}
As discussed in Chapter~\ref{chapt:filtools}, filament curvature is a potentially important but largely unexplored property that may influence the measured radial profiles. To study its effect, we compute the local curvature, $\kappa$, along every filament segment following the procedure described in Chapter~\ref{chapt:filtools}. The curvature distribution is determined separately for each simulation realization and divided into ten percentile bins, although the resulting percentile values are nearly identical across realizations. For each filament, all segments belonging to a given curvature bin are grouped together and used to measure the corresponding radial profiles. These profiles are then stacked over all filaments and all five realizations using the contributing filament length as the weight.

The resulting profiles are shown in Figure~\ref{fig:curvature_split}. The dependence on curvature is striking. Higher-curvature segments are systematically denser, exhibit stronger radial infall, larger tangential velocities, and higher velocity dispersions. However, the radius corresponding to the strongest radial infall changes very little with curvature.

The physical origin of these trends is not yet clear. One possibility is that initially straight filaments become bent during their subsequent evolution, causing neighbouring parts of the structure to overlap, enhancing the densities and velocities. An alternative is that the curvature is already imprinted during formation, with strongly curved regions providing a deeper gravitational potential and enhanced accretion. Disentangling these possibilities will require following individual filaments through time and is beyond the scope of the present work.

\begin{figure}
    \centering
    \includegraphics[width=0.7\linewidth]{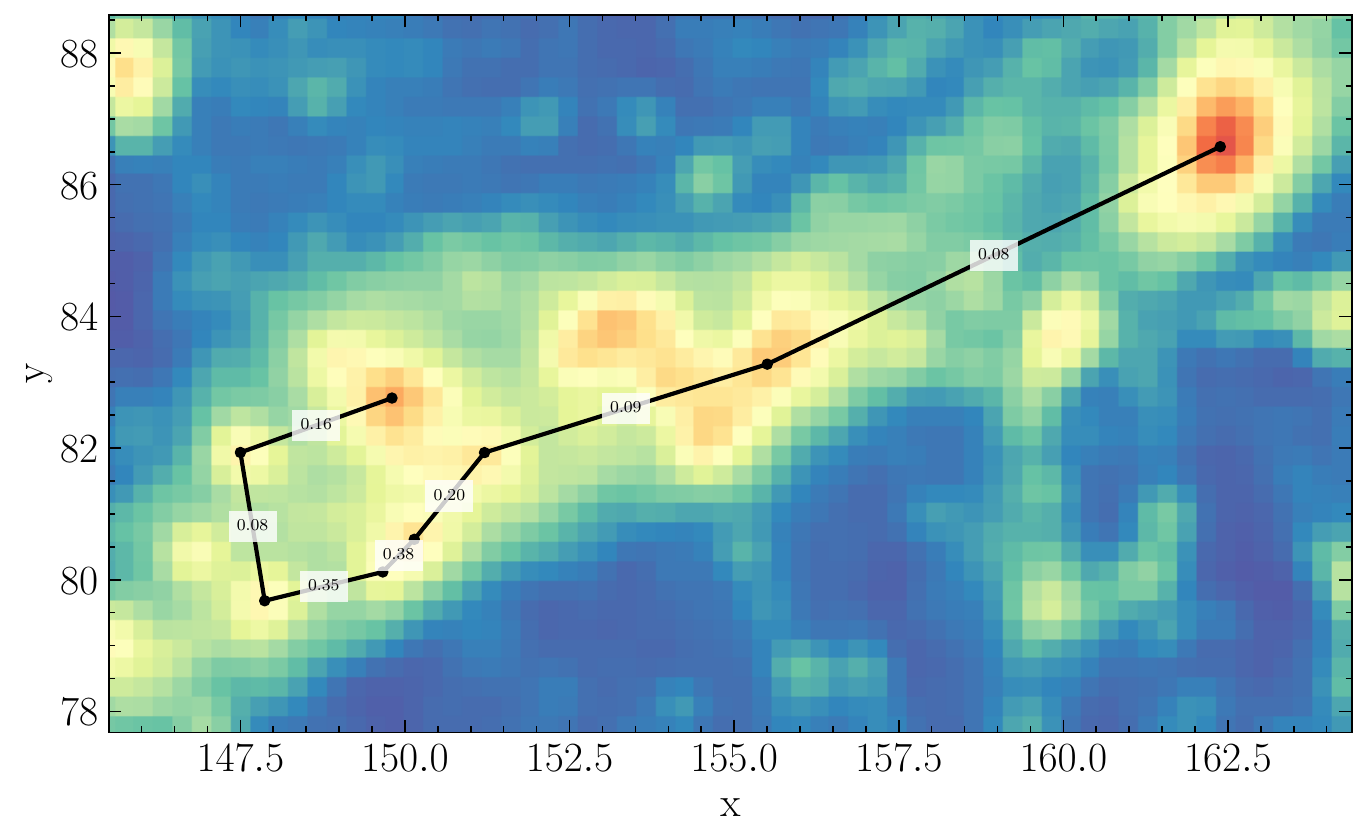}
    \caption{Illustration of filament curvature. The figure shows the projection of a filament identified by \skeletor, overlaid on the corresponding dark matter density slice. The curvature of each filament segment, in units of $h,\mathrm{Mpc}^{-1}$, is annotated along the spine.}
    \label{fig:curvature_visuals}
\end{figure}

Regardless of the underlying mechanism, the trend has an important practical consequence. The most strongly curved regions resemble blobs rather than simple idealized cylindrical filaments and can substantially bias the stacked profiles. This behaviour is illustrated in figure~\ref{fig:curvature_visuals}. Motivated by these results, all subsequent analyses are performed after removing the highest $10\%$ of the curvature distribution. The discarded segments are analysed separately in Appendix~\ref{app: high kappa fils}, where they are found to follow qualitatively similar trends.

We also find that the velocity dispersion increases towards the filament spine, as expected, but then decreases slightly at very small radii, around $r \sim 0.2,\Mpch$. The origin of this feature is not yet understood and will be investigated in future work.

\subsection{Filament length}
\begin{figure}
    \centering
    \includegraphics[width=0.9\linewidth]{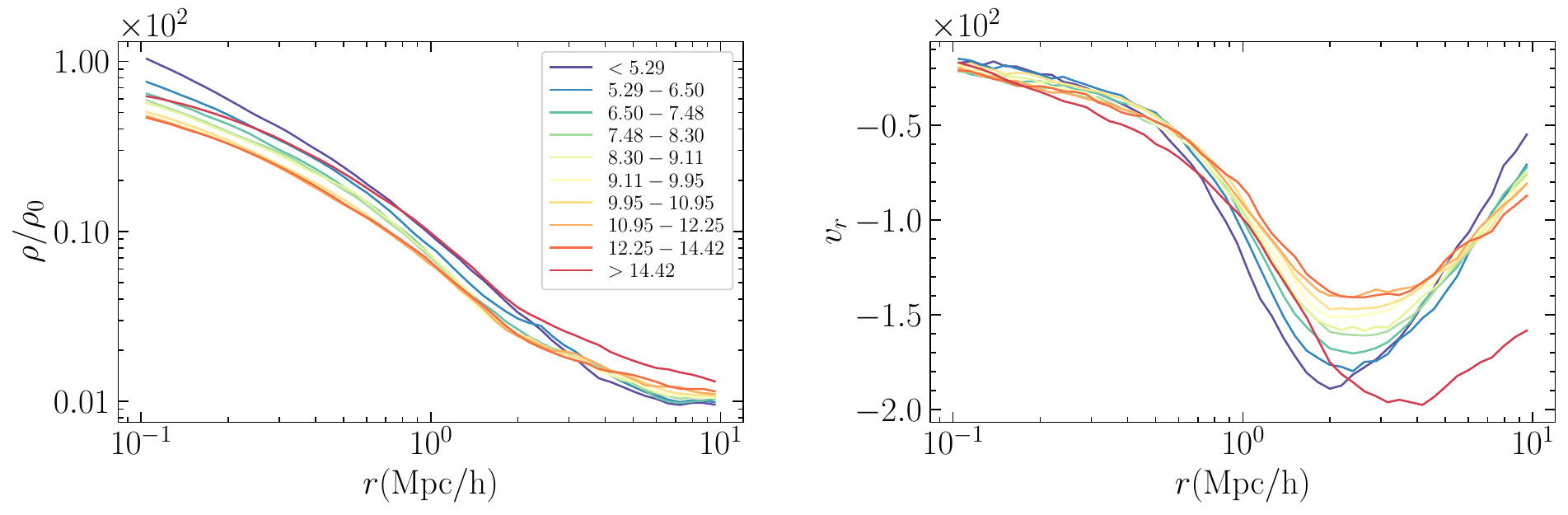}
    \caption{Stacked density and radial velocity profiles in ten percentile bins of filament length. A weak dependence on length is visible, although the interpretation is complicated by potential fragmentation of filaments in the current reconstruction.}
    \label{fig:lfil_dep}
\end{figure}
Studies have reported a dependence of filament profiles on filament length \citep{2pop2020}. We therefore repeat the stacking analysis after splitting the parent filament sample into ten equal percentile bins of filament length, where the percentiles are computed from the full parent population. The resulting density and radial velocity profiles are shown in Figure~\ref{fig:lfil_dep}.

Only the density and radial velocity profiles are shown here. In the current implementation of \skeletor, the measured filament lengths are likely affected by fragmentation, with long physical filaments occasionally reconstructed as multiple shorter segments. The measured lengths should therefore not be interpreted too literally.

Even so, a weak trend is visible. Shorter filaments have somewhat higher central densities, although the ordering reverses beyond $r\gtrsim4~\Mpch$. They also show a deeper radial velocity minimum occurring at slightly smaller radii. Given the uncertainty in the length estimate, however, it is difficult to attach much physical significance to these trends.

\subsection{Node mass}
We next examine the dependence of the phase-space profiles on the total node mass,
\begin{equation}
    M \equiv M_1 + M_2,
\end{equation}
where $M_1$ and $M_2$ are the masses of the two endpoint nodes. The parent filament sample is divided into ten equal percentile bins in $M$, and the profiles are stacked separately for each bin following the same procedure as above. The results are shown in Figure~\ref{fig:mnode_dep}.

The dependence on node mass is very pronounced. The profiles separate cleanly into well-defined sequences over almost the entire radial range. Filaments connecting more massive nodes are systematically denser, exhibit stronger radial infall, larger tangential velocities, and higher velocity dispersions. The radius corresponding to the minimum of the radial velocity profile also increases steadily with node mass.

These trends are consistent with the expectation that more massive nodes are connected by more prominent filamentary structures residing in deeper gravitational potentials. The results also indicate that node mass is one of the primary quantities governing the phase-space structure of filaments and should be taken into account when comparing filament populations or searching for universal radial profiles.
\begin{figure}
    \centering
    \includegraphics[width=\linewidth]{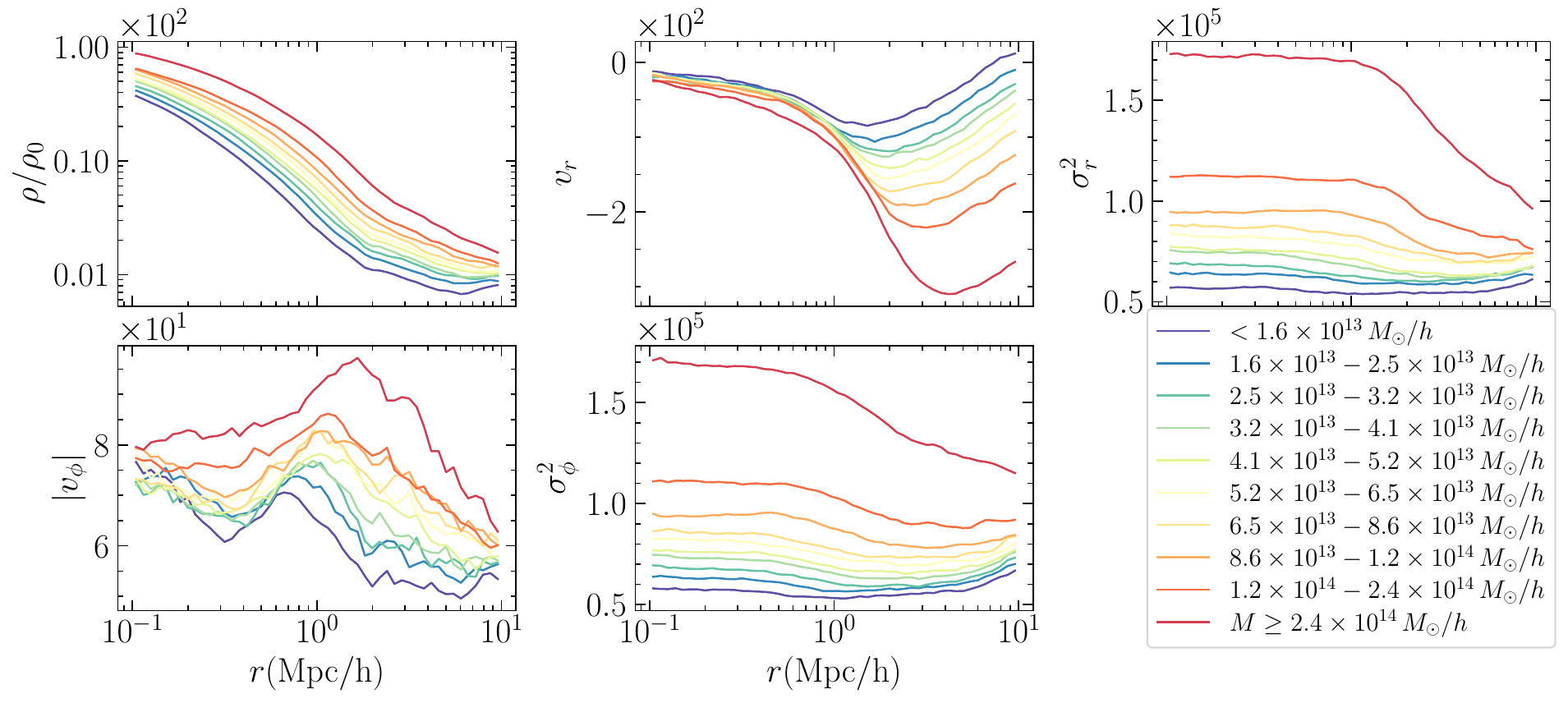}
    \caption{Stacked phase-space profiles in ten percentile bins of the total node mass, $M_1+M_2$. The dependence on node mass is strong: filaments connected to more massive nodes are systematically denser and dynamically hotter, with stronger infall and larger characteristic radii.}
    \label{fig:mnode_dep}
\end{figure}

\subsection{Radius from the radial infall}
\begin{figure}
    \centering
    \includegraphics[width=0.8\linewidth]{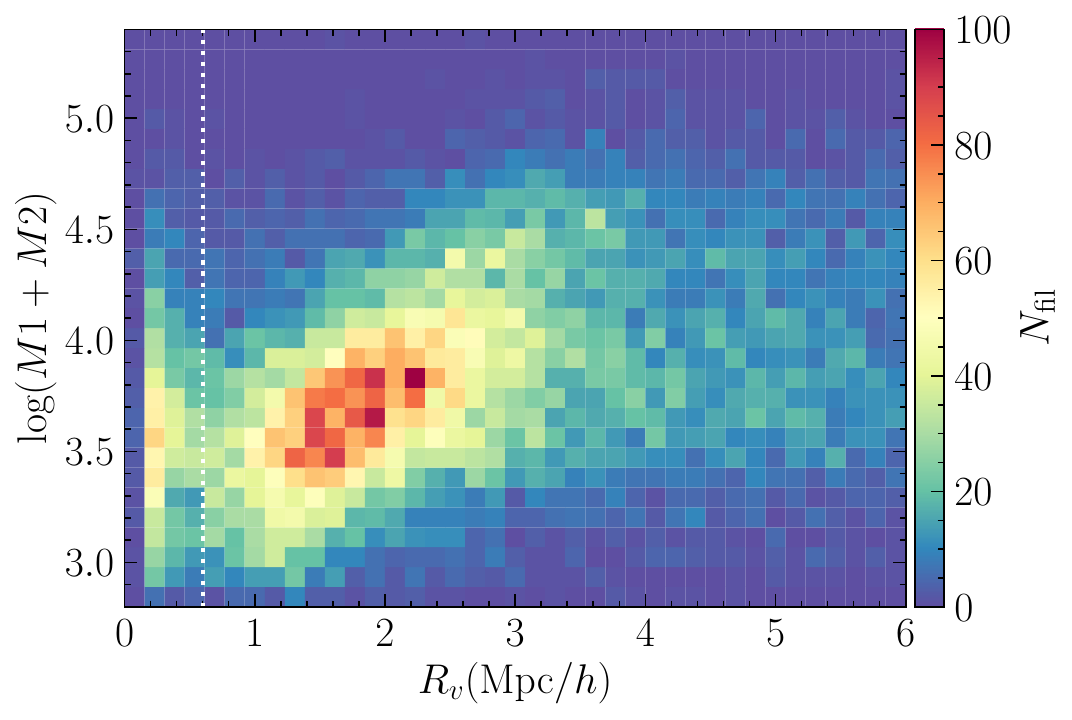}
    \caption{Two-dimensional distribution of the velocity-defined filament radius, $R_v$, and the total node mass, $M_1+M_2$, for the cleaned parent filament sample. The main population shows the expected positive correlation between filament radius and node mass, while a secondary population at low $R_v$ arises from systems where the fitting procedure identifies an unphysical inner minimum in the radial velocity profile. The vertical dashed line marks the cut at $R_v=0.6~\Mpch$ adopted for the remainder of the analysis.}
    \label{fig:rvcut_2d}
\end{figure}

\begin{figure}
    \centering
    \includegraphics[width=\linewidth]{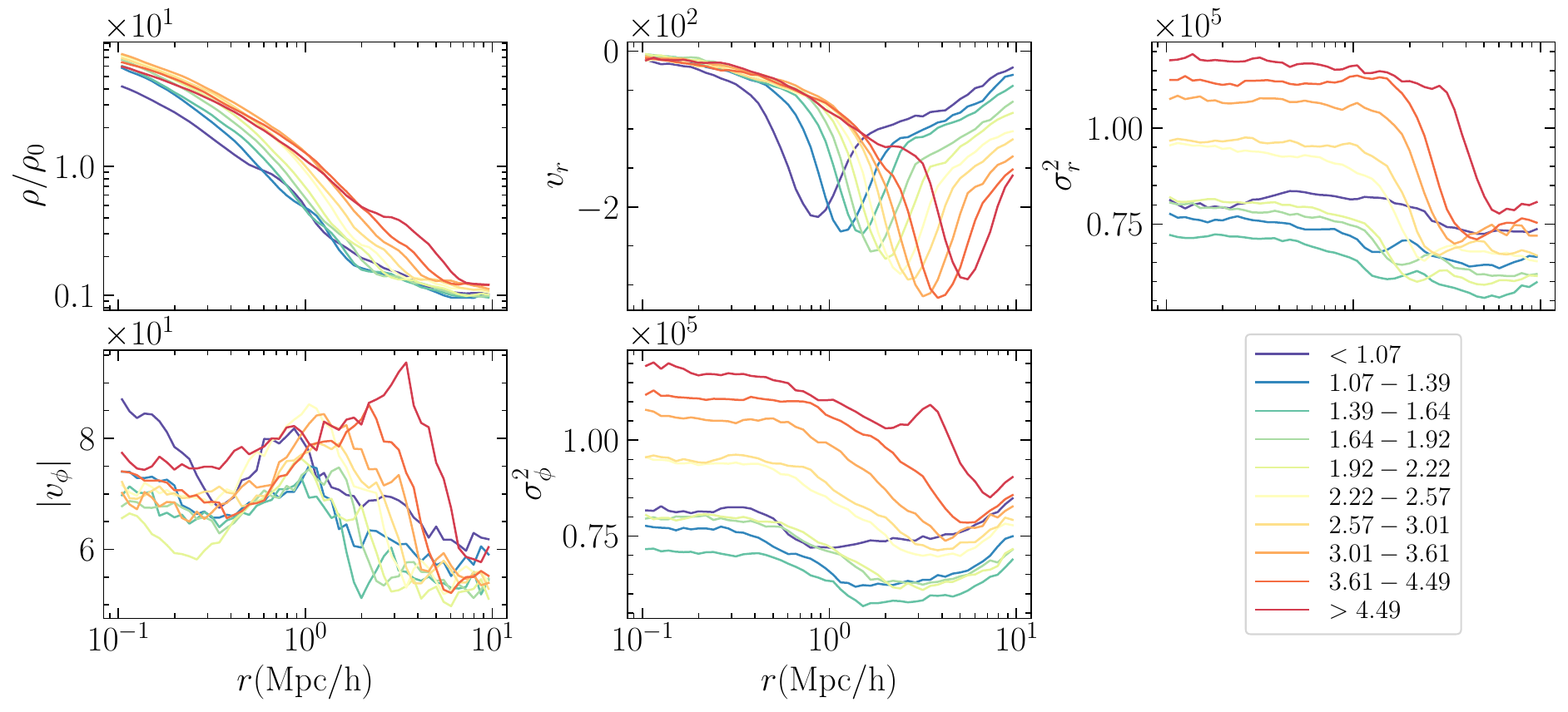}
    \caption{Stacked phase-space profiles in ten percentile bins of the velocity-defined filament radius, $R_v$, after applying the quality cuts described in the text. Filaments with larger $R_v$ are systematically denser and dynamically hotter, exhibiting stronger radial infall, larger velocity dispersions, and slightly enhanced tangential velocities.}
    \label{fig:rv_dep}
\end{figure}

We now turn to the filament radius estimated from the radial velocity profile using the procedure described in Chapter~\ref{chapt:skeletor}. Briefly, the radius is defined as the location of the dip in the radial infall profile, corresponding to the radius of maximum coherent infall.

The radius fitting procedure is not successful for every filament. For around $10\%$ of the sample, the fitting algorithm does not converge within the maximum number of iterations. Some of these profiles have the velocity dip located close to the edge of the measured radial range, leaving too few bins for a reliable quadratic fit. In such cases, the problem is simply that the profiles are not measured out to sufficiently large radii and could be resolved by extending the
fitting range. Around $22\%$ of filaments fail the quality cut on the dip significance, for which we require $\mathrm{SNR}>2$. Roughly $40\%$ of these low-SNR cases also correspond to large fitted radii
and are therefore likely affected by the same limitation of the finite radial range. Most of the unconverged and non-significant dip detected filaments either have very noisy velocity profiles, or strong sub-structure leading to multiple dips. Some also have a broad dip, leading to low SNR. After applying the cleaning criteria, robust velocity-defined radii are obtained for approximately $67\%$ of the cleaned parent filament sample.

Since the previous section established node mass as one of the dominant quantities controlling filament profiles, Figure~\ref{fig:rvcut_2d} shows the relation between the velocity-defined radius, $R_v$, and the total node mass. Two distinct populations are visible. The main population exhibits the expected positive correlation between node mass and filament radius, while a smaller concentration appears at $R_v\lesssim0.6~\Mpch$.

Inspection of these low $R_v$ filaments shows that they probably contain significant substructure, resulting in multiple minima in the radial velocity profile. In such cases, the fitting algorithm can lock onto an inner minimum instead of the physically relevant outer dip. We therefore impose a cut at $R_v=0.6~\Mpch$ and discard filaments below this threshold. The remaining sample contains approximately $65\%$ of the cleaned parent population.

The dependence of the phase-space profiles on $R_v$ is shown in Figure~\ref{fig:rv_dep}, where the surviving filaments are divided into ten equal percentile bins in radius. The radial velocity profiles separate cleanly by construction. Filaments with larger $R_v$ also tend to be denser over most of the radial range, exhibit stronger radial infall, and have larger velocity dispersions. The absolute tangential velocity is similarly enhanced, with its peak shifting systematically to larger radii. Overall, the trends are consistent with the interpretation that the radial velocity dip traces a physically meaningful characteristic scale of the filament.

We also examine the relation between the velocity-defined radius and the enclosed density within that radius. Figure~\ref{fig:rv_rho_2dhist} shows the two-dimensional distribution of $R_v$ and $\rho_{\mathrm{encl}}/\rho_0$.  Over most of the range, the enclosed density depends only weakly on $R_v$, with a slight decrease towards larger radii. At $R_v\lesssim1~\Mpch$, however, the trend becomes noticeably steeper, with thinner filaments reaching significantly larger enclosed densities. The origin of this behaviour is not yet understood and deserves further investigation. The median enclosed density within $R_v$ is $5.4\,\rho_0$, somewhat lower than values often quoted in the literature. For example, \citep{NEXUS2014} report a characteristic overdensity of $\sim10\,\rho_0$ at $z=0$. As discussed in the next section, this difference arises primarily from the adopted definition of the filament radius rather than from an intrinsic physical discrepancy. Nevertheless, in situations where a velocity-based radius cannot be measured directly, the enclosed density of $5.4\rho_0$ provides a useful empirical guide for estimating the characteristic extent of a filament.

\begin{figure}
    \centering
    \includegraphics[width=0.7\linewidth]{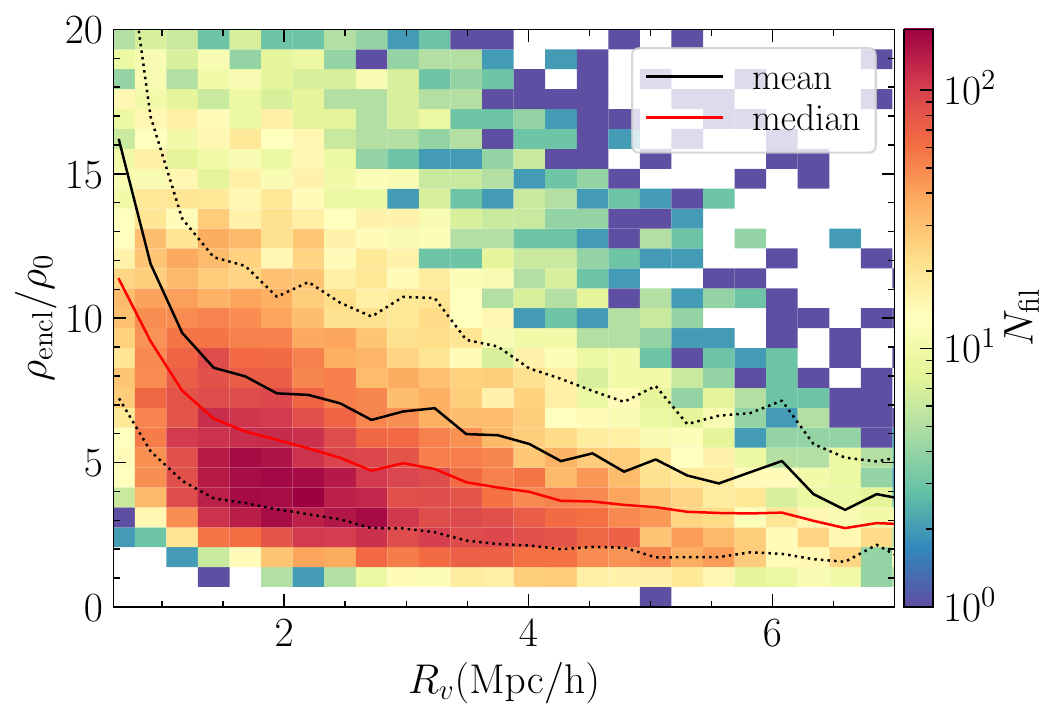}
    \caption{Two-dimensional distribution of the velocity-defined filament radius, $R_v$, and the enclosed overdensity within that radius, $\rho_{\mathrm{encl}}/\rho_0$. The overplotted black and red curves show the mean and median enclosed densities in bins of $R_v$, while the dotted curves indicate the $16^{\mathrm{th}}$ and $84^{\mathrm{th}}$ percentiles. Over most of the range, the enclosed density varies only weakly with filament radius.}
    \label{fig:rv_rho_2dhist}
\end{figure}

The analyses presented in the remainder of this chapter use only the subsample with robustly measured values of $R_v$. In some applications, however, an estimate of the filament radius is required even for objects where the velocity dip cannot be identified reliably. For these filaments, we bin the sample by the total node mass and determine the characteristic radius from the stacked radial velocity profile of each bin. The resulting value is then assigned to all filaments within that node-mass bin and is used wherever an estimate of the filament radius is needed, such as in the subfilament classification described in Chapter~\ref{chapt:skeletor}.

\section{Defining filament boundaries}
\label{sec: filament boundaries}

\begin{figure}
    \centering
    \includegraphics[width=0.8\linewidth]{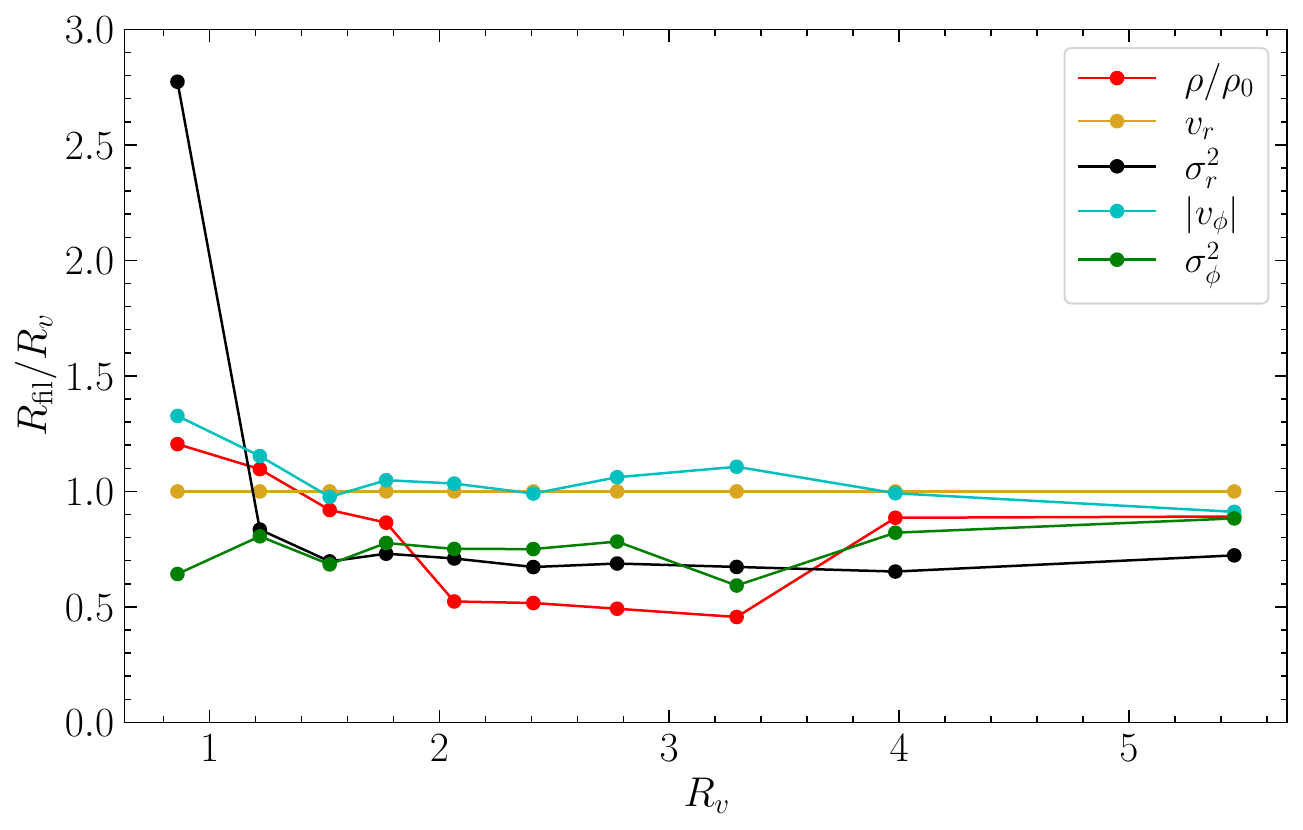}
    \caption{Comparison of different filament radius definitions relative to the velocity-defined radius $R_v$, shown as a function of $R_v$. Radii are obtained from stacked profiles in bins of $R_v$, using density gradients, radial velocity minima, tangential velocity structure, and velocity dispersion gradients. With the exception of the $|v_\phi|$-based estimate, all definitions systematically lie within $R_v$ across most of the range. }
    \label{fig:rfil_comparison}
\end{figure}

Unlike dark matter halos, cosmic filaments do not possess a well-accepted physical boundary, and different studies therefore adopt different operational definitions of the filament radius. Having identified a characteristic scale from the radial velocity field, it is natural to ask whether other phase-space quantities select the same scale or probe distinct regions of the filament.

To address this, we again divide the sample into ten $R_v$ percentile bins as in the previous section and analyze the corresponding stacked profiles. For each stack, we define a characteristic radius from five different phase-space quantities. For the density profile, the radius is taken as the location where $d\log\rho/d\log r$ reaches its minimum, corresponding to the steepest decline. For the radial velocity profile, the characteristic scale is defined by the minimum of $v_r$. For the tangential velocity, we use the location of the maximum in $d|v_\phi|/d\log r$. For the radial and tangential velocity dispersions, the radius is defined as the point where $d\sigma_i^2/d\log r$ is most negative, with $i \in \{r,\phi\}$.

In all cases involving a minimum, a quadratic polynomial is fitted locally using the point of interest and five neighbouring points on either side, and the analytic minimum of the fitted function is taken as the radius estimate. The corresponding stacked profiles and derivative-based estimate of filament radii $R_{\mathrm{fil}}$ are shown in Appendix~\ref{app: diff rfil def}. The comparison across definitions is shown in Figure~\ref{fig:rfil_comparison}, where the ratio of each radius estimate to $R_v$ in each filament radius bin is plotted as a function of $R_v$, where $R_v$ is now the fitted minimum of the stack in the given bin. With the exception of the radius inferred from $|v_\phi|$, all definitions lie systematically within $R_v$ over most of the range. The density-based radius can be as small as $\sim 50\%$ of $R_v$, which directly explains why the enclosed overdensity measured within $R_v$ is lower than values commonly quoted in the literature, where filament radii are typically defined using density-based criteria. The dispersion-based radii show little variation in the ratio $R_{\mathrm{fil}}/R_v$, while the density-based definition exhibits a mild but noticeable scale dependence.
\begin{figure}
    \centering
    \includegraphics[width=\linewidth]{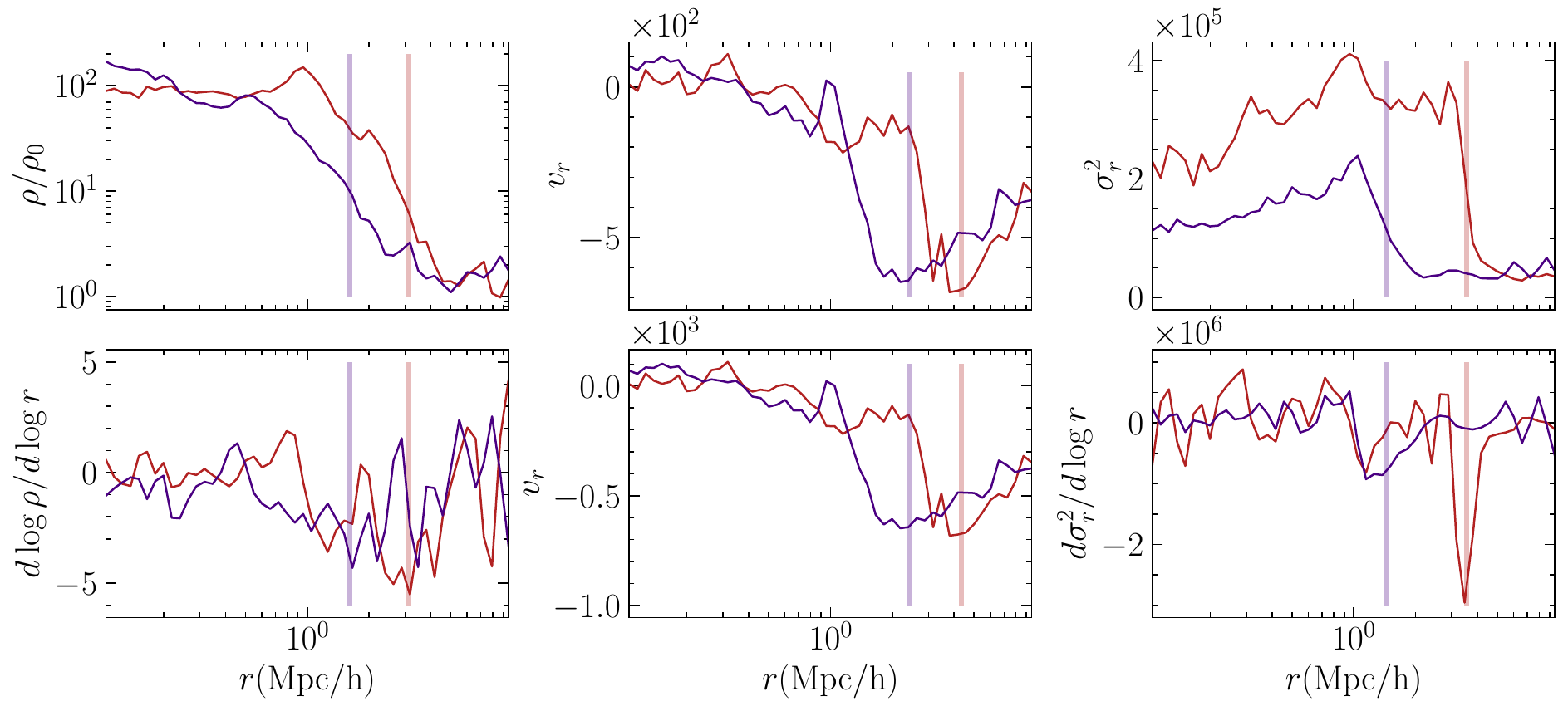}
    \caption{Individual filament profiles for two representative primary filaments (red and blue). Top panels show the radial profiles of density, radial velocity, and radial velocity dispersion. Bottom panels show the corresponding derived quantities used for radius estimation: $d\log\rho/d\log r$, $v_r$, and $d\sigma_r^2/d\log r$. Vertical lines mark the inferred characteristic radii from each estimator.}
    \label{fig:individual_rfil}
\end{figure}

Figure~\ref{fig:individual_rfil} illustrates similar results for two individual primary filaments, highlighting the radii inferred from $\rho$, $v_r$, and $\sigma_r^2$. The behaviour of $\sigma_\phi^2$ is qualitatively similar to that of $\sigma_r^2$, whereas the estimator based on $v_\phi$ is not robust at the level of individual filaments due to noise. The top panels show the profiles, while the bottom panels show the quantities used for the radius estimation, with vertical lines marking the inferred radii. We find that in some cases where the minimum in $v_r$ is poorly constrained, the dispersion-based estimator still provides a stable measurement, while in other cases the reverse is true. A systematic combination of these estimators is left for future work, particularly for improving robustness at the level of individual filaments.

The systematic differences between different filament radius definitions suggest that the various definitions probe different stages of the infall process rather than a single sharp boundary. The minimum of the radial velocity profile traces the location of maximum coherent inflow toward the filament spine, whereas the density gradient and velocity dispersion are more sensitive to the inner region where infalling material begins to accumulate and multistreaming sets in. Under this interpretation, it is natural that the extrema in density and dispersion occur at smaller radii than the velocity-defined scale. The behaviour of the stacked profiles is consistent with this picture.

The characteristic radii discussed above are mainly inferred from stacked profiles. To better understand their physical meaning, we next turn to the phase-space structure of individual filaments and search for signatures of multistreaming and caustic formation.
\section{Multistreaming and caustics around filaments}

\begin{figure}[ht]
    \centering
    \begin{minipage}[c]{0.3\textwidth}
        \centering
        \includegraphics[width=\linewidth, height=5cm, keepaspectratio]{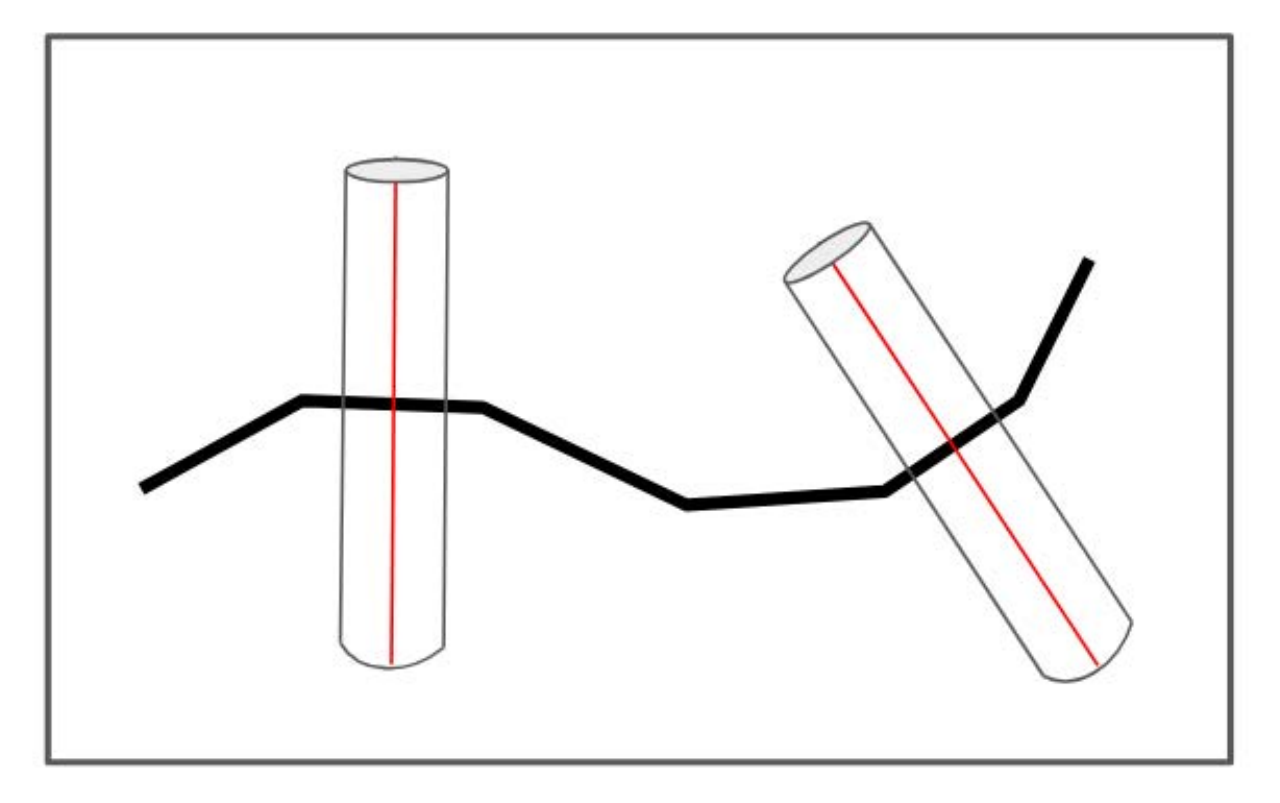}
    \end{minipage}
    \hfill
    \begin{minipage}[c]{0.69\textwidth}
        \centering
        \includegraphics[width=\linewidth, height=5cm, keepaspectratio]{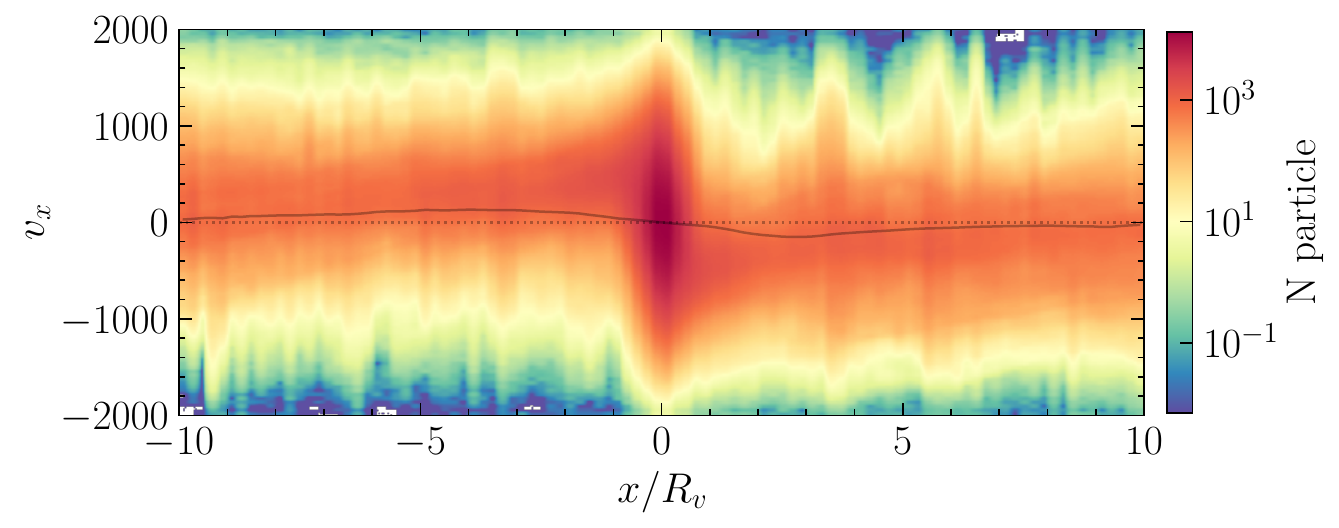}
    \end{minipage}

    \caption{(\emph{left panel:}) Schematic illustrating the construction of the filament phase-space distribution. The solid black line represents the filament spine made up of discrete segments, with cylinders of radius $R_v$ centred on individual spine segments and oriented perpendicular to the local filament direction. For clarity, only two representative cylinders are shown, although the procedure is repeated for every segment along the filament. (\emph{right panel:}) Stacked $(x,v_x)$ phase-space distribution obtained by combining all parent filaments in one simulation realization. The black curve shows the median velocity at each position, and black dotted line marks $v_x=0$ for reference. At large distances from the filament spine, the median velocity approaches zero, while closer to the spine coherent infall becomes apparent through the correlation between the sign of $x$ and $v_x$. Near the spine, the rapid increase in the velocity dispersion is consistent with the onset of multistreaming.}
    \label{fig:avg_caustic}
\end{figure}


\begin{figure}
    \centering
    \includegraphics[width=0.75\linewidth]{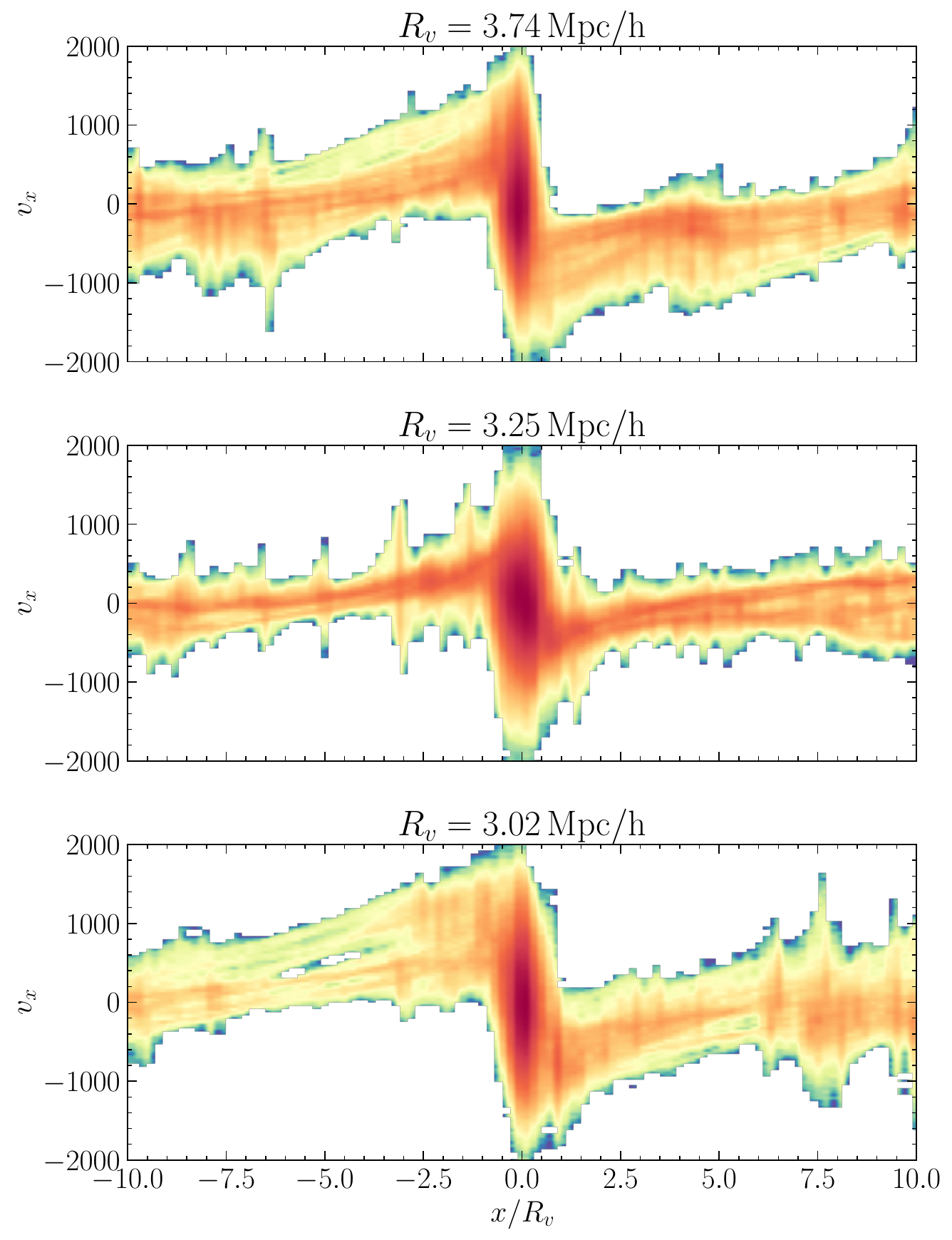}
    \caption{Examples of $(x,v_x)$ phase-space diagrams for three individual parent filaments. In all cases, coherent infall towards the filament spine is followed by multistreaming, and eventually a sharp broadening of the velocity distribution in the central regions. Localized enhancements in the dispersion correspond to dark matter halos intersecting the cylindrical volume used in the analysis.}
    \label{fig:caustic_example}
\end{figure}
The comparison of different filament radius definitions in the previous section suggests that different observables become prominent at different distances from the filament spine. A natural question is whether these transitions correspond to the onset of shell crossing and multistreaming. To investigate this, we move beyond stacked radial profiles and examine the phase-space structure of individual filaments.

For each parent filament with a well-defined velocity radius, $R_v$, we construct local phase-space diagrams by passing a cylinder of radius $R_v$ through the center of every filament segment. The cylinder axis is chosen to be perpendicular to the direction of the segment, which is the local tangent to the filament spine, and defines the coordinate $x$. The left panel of Figure~\ref{fig:avg_caustic} illustrates this construction. For all dark matter particles inside the cylinder, we measure the component of the velocity along the cylinder axis, $v_x$, and construct the two-dimensional distribution in $(x,v_x)$. The distributions from all segments belonging to a given filament are combined before further analysis.

Right panel in figure~\ref{fig:avg_caustic} shows the resulting stacked phase-space diagram obtained by combining all parent filaments in one simulation realization. At large distances from the spine, the median velocity approaches zero, as expected for material that has not yet been significantly affected by the filament potential. Closer to the spine, particles on opposite sides acquire coherent velocities directed towards the filament, producing the characteristic pattern in which negative $x$ is associated with positive $v_x$ and positive $x$ with negative $v_x$. This behaviour directly reflects the large-scale infall responsible for filament growth.

Near the filament spine, however, the velocity distribution broadens rapidly. Instead of a single cold stream, particles occupy a wide range of velocities, indicating that multiple streams overlap in the same region of configuration space. Such behaviour is expected once shell crossing occurs and is qualitatively consistent with the formation of caustics predicted by Lagrangian descriptions of anisotropic gravitational collapse.

The behaviour becomes even clearer in individual systems. Figure~\ref{fig:caustic_example} shows three representative examples. In each case, the coherent infall towards the filament is followed by  multiple streams manifesting as a striated pattern of the particle distribution, and eventually a rapid broadening of the velocity distribution close to the spine. Small localized spikes in the dispersion are also visible at intermediate positions and are associated with dark matter halos intersecting the cylindrical volume.

\begin{figure}
    \centering
    \includegraphics[width=\linewidth]{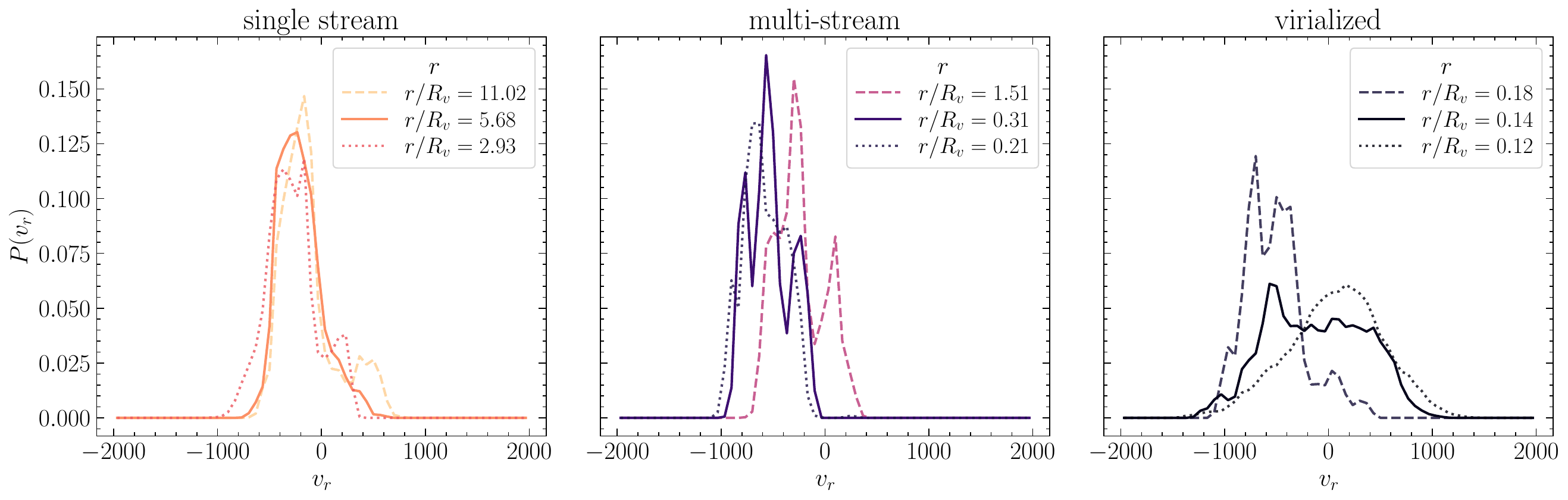}
    \caption{Radial velocity distributions, $P(v_r)$, at three neighbouring cylindrical radii for a representative parent filament. The left, centre, and right panels correspond to regions dominated by coherent infall, multistreaming, and the central dynamically mixed region, respectively. Within each panel, the solid line shows the distribution at the central radius, while the dashed and dotted lines correspond to slightly larger and smaller radii. The appearance of multiple peaks in the middle panel indicates the coexistence of several velocity streams at the same radius.}
    \label{fig:multistreaming}
\end{figure}

Motivated by the striated patterns indicative of multistreaming, we adopt an alternative approach to probe its onset by examining the distribution of radial velocities in narrow cylindrical shells around the filament spine. Figure~\ref{fig:multistreaming} shows one such example for a representative filament, where we take concentric cylinders around every segment in the filament, and combine shells with the same radius together to get the distribution of radial velocity at a given $r$. At large radii, the distribution is dominated by a single peak at negative $v_r$, indicating coherent infall towards the spine. Moving inward, additional peaks begin to emerge, signalling the coexistence of multiple streams at the same radius. Towards the filament spine, these components broaden and overlap until the distinct peaks are no longer resolved, giving rise to a broad velocity distribution characteristic of the central region.

The exact appearance of the phase-space distribution, including the number of streams, varies from one filament to another, as expected from differences in their formation history and local environment. Nevertheless, the overall sequence remains qualitatively similar: a single-stream infall region is followed by the onset of multistreaming and a rapid increase in velocity dispersion towards the filament centre. The transition is also visible in the $(x,v_x)$ diagrams, where the initially narrow velocity distribution striates and then broadens substantially near the spine.

These features are consistent with the picture of anisotropic gravitational collapse, in which matter first converges towards the filament and subsequently undergoes shell crossing, leading to multiple overlapping streams in phase space. While multistream regions have been discussed extensively in Lagrangian descriptions of the cosmic web, direct visualizations around filaments identified in cosmological simulations have received comparatively little attention. The examples presented here demonstrate that such signatures can be identified in individual systems and motivate a more systematic study of their statistical properties. 

Several interesting questions remain open. In particular, it would be useful to investigate how the onset of multistreaming is related to the velocity-defined radius, $R_v$, and whether the ratio between these two characteristic scales depends systematically on quantities such as node mass, filament curvature, or environment. Such studies may help clarify whether the onset of shell crossing provides a more physically motivated definition of the filament boundary. It would also be interesting to compare these phase-space signatures across different cosmologies and dark matter models, where changes in the nonlinear collapse history may leave measurable imprints on the internal dynamics of filaments. The \Sahyadri\, simulation suite discussed in Chapter~\ref{chap:sahyadri}, with its systematic cosmological parameter variations, is well suited to such studies.

\section{Discussion}

The results presented in this chapter indicate that the internal structure of cosmic filaments depends strongly on the physical properties of the filament itself. Even after restricting the analysis to the parent population, quantities such as node mass, filament radius, and local curvature produce systematic changes in the measured phase-space profiles. This suggests that caution is needed when searching for universal filament profiles, since stacking together filaments with very different properties is likely to increase the observed scatter.

One result that deserves particular attention is the dependence on curvature. While the effects of node mass have been discussed in earlier work, curvature has remained largely unexplored. The trends found here show that highly curved regions are systematically denser and dynamically hotter than straighter segments. It is not yet clear whether this reflects the formation of intrinsically curved filaments or the subsequent bending of initially straighter structures. Resolving this question will require following individual filaments through multiple simulation snapshots, and is left for future work.

An interesting result is the systematic offset between the different definitions of filament radius. The characteristic scales obtained from density, velocity, and velocity dispersion do not coincide, suggesting that each captures a different aspect of the underlying dynamics. Rather than identifying a unique filament edge, these quantities may instead mark successive stages in the transition from coherent inflow to multistreamed regions.

The study of individual phase-space profiles provides further support for this interpretation. The examples shown in this chapter exhibit a clear progression from coherent infall to multistreaming and finally to a broad central distribution where the individual streams are no longer easily distinguished. Although shell crossing and caustic formation are expected from theoretical descriptions of anisotropic collapse, they have rarely been examined directly in reconstructed filament populations. The results presented here show that these signatures can be identified in cosmological simulations and deserve a more systematic investigation.

An obvious next step is to study the relation between the onset of multistreaming and the velocity-defined filament radius introduced in this work. It would be interesting to determine whether the two are connected by a simple scaling and whether that relation depends on quantities such as node mass or curvature. More generally, extending the same analysis to different cosmologies or dark matter models may reveal how changes in the growth history are reflected in the internal dynamics of cosmic filaments.
    %
%
\let\textcircled=\pgftextcircled
\chapter{Conclusions and Future directions}
\label{chapt: conclusion}

\initial{N}umerical simulations and robust methods for identifying and characterizing cosmic structures have become indispensable ingredients in modern studies of the nonlinear Universe. This thesis has contributed to both aspects through the development of new simulation resources, reconstruction techniques, and analyses of the internal structure and dynamics of cosmic filaments.

This concluding chapter summarizes the principal results of the thesis, discusses their broader implications for studies of the Large-Scale Structure, and outlines several directions in which the ideas developed here may be extended in the future.

\newpage

The coming decade of cosmology will be defined by the wealth of high-precision observations delivered by surveys such as DESI \citep{DESICollaboration2016a, DESICollaboration2016b}, LSST \citep{Ivezic2019} and Euclid \citep{EuclidCollaboration2022}. These experiments will map the Large-Scale Structure of the Universe over unprecedented volumes and redshift ranges, placing increasingly stringent demands on theoretical modelling. While  perturbation theories provide a successful description on large scales \citep{Bernardeauetal02}, much of the available cosmological information resides in the nonlinear regime, where gravitational collapse generates halos, filaments, walls, and voids with highly non-Gaussian statistics. Extracting this information in a robust and physically interpretable manner remains one of the central challenges of modern cosmology.

This thesis approaches the nonlinear Universe from several complementary directions. It develops new numerical resources, introduces methodologies for reconstructing the cosmic web, and uses these tools to investigate the physical properties of the filamentary structures. The common thread throughout is the development of reliable tools for probing the nonlinear Universe and extracting physical information from the cosmic web.

One component of this work is the development of the \Sahyadri\ simulation suite. The suite was designed to explore controlled variations of cosmological parameters around a fiducial model while maintaining high mass resolution over cosmological volumes. It therefore provides a useful resource for studying nonlinear structure formation in the low-redshift Universe and its cosmology dependence. Beyond traditional analyses of halo populations and clustering statistics, such simulations can be used to develop and test new summary statistics, perform Fisher forecasts for upcoming surveys, and investigate a variety of problems in nonlinear Large-Scale Structure. To illustrate these possibilities, we examine the dependence of several observables on $\Om$, including the Voronoi Volume Function (VVF), $k$-nearest neighbour (KNN) statistics, and the environmental dependence of halo properties, in addition to conventional quantities such as the matter power spectrum and halo mass function. These examples serve as a demonstration of the wider range of cosmological applications that the simulation suite is intended to support.

A significant part of the work presented in this thesis concerns the development, calibration, and physical interpretation of methods for reconstructing the filamentary cosmic web and understanding the properties of the resulting filament population. Unlike halos, filaments do not possess a unique definition, and their inferred properties depend on the methodology used to identify them \citep{Libeskind_et_al2018}. To address this, the thesis introduces two publicly available tools: \filgen, which generates controlled mock filament catalogues with known properties, and \filapt, which provides a framework for analysing filament populations and quantifying reconstruction biases. Using these tools, we study the effects of smoothing, spine reconstruction, tracer selection, and geometric systematics on recovered filament profiles, providing a clearer picture of the uncertainties and biases associated with filament reconstruction and a practical framework for calibrating filament finders. We also develop robust Fourier-space techniques for processing filament spines and identify filament curvature as an important, previously unexplored quantity influencing measured radial profiles.

A major contribution of this thesis is the development of \skeletor, a hierarchical filament finder based on Voronoi tessellations and local anisotropy in the tracer distribution. The algorithm identifies filamentary environments directly from discrete tracers, reconstructs connected skeletons through a hierarchical procedure, and classifies embedded subfilaments separately from their parent structures. The reconstruction is then refined using the underlying dark matter distribution, and characteristic filament radii are estimated from a physically motivated feature in the radial velocity profile. The resulting catalogue retains information about the hierarchical organization of the cosmic web while providing a physically motivated description of the reconstructed filament population.

The hierarchical nature of the reconstruction also opens up a new way in which filament populations can be studied. Embedded subfilaments are found to differ systematically from their parent structures in their phase-space properties, indicating that they should be treated as a distinct population rather than combined into a single sample. This distinction has practical consequences for statistical studies of filaments and suggests that at least part of the diversity reported in the literature may arise from combining physically different classes of objects into a single sample.

The analysis of parent filaments also reveals a strong dependence of their internal structure on quantities such as node mass, filament radius, and local curvature. While the influence of node mass has been explored in previous studies, curvature is shown here to be an equally important parameter despite receiving little attention in the literature. These results suggest that any attempt to identify universal filament profiles must first account for the broad range of filament properties. At the same time, they raise the possibility that much of the observed diversity reflects a superposition of different filament populations, and that appropriate rescaling or subdivision may uncover simpler underlying relations.

The thesis also revisits the question of what constitutes the boundary of a cosmic filament \citep[e.g.][]{wang+24}. Different observables identify systematically different characteristic radii, implying that no single profile captures the entire transition between the surrounding environment and the filament interior. The velocity-defined radius introduced here provides one physically motivated scale tied directly to coherent infall, but its comparison with density-based and dispersion-based definitions suggests that these quantities trace different stages of the nonlinear collapse process rather than a unique geometric edge.

Another novel aspect of this thesis is the direct examination of the phase-space structure of individual filaments. The analysis reveals coherent infall towards the filament spine, followed by the appearance of multistreaming and caustic-like features as the dark matter distribution becomes increasingly mixed. While such behaviour is expected from Lagrangian descriptions of anisotropic collapse \citep{Zeldovich1970, Arnold+1982}, it has received relatively little attention in studies of reconstructed filament populations. The results presented here show that the full phase-space distribution contains information beyond conventional stacked density profiles and may provide a useful new window into the dynamics of filament formation.\\
\\

Several natural extensions of this work remain. The \Sahyadri\ simulations provide a useful framework for developing new cosmological observables and quantifying their constraining power through Fisher forecasts. The hierarchical reconstruction implemented in \skeletor\ can also be adapted for application to galaxy surveys, although this will require accounting for survey geometry, selection effects, redshift-space distortions, and the absence of direct halo mass estimates. One possible approach is to replace halo masses with observable proxies such as stellar mass or galaxy luminosity. The analysis of filament velocity fields and phase-space structure presented in this thesis may also help in understanding how filaments appear in redshift space and in developing models that incorporate their peculiar velocity field.

On the theoretical side, a detailed understanding of filament structure and dynamics should help guide the development of semi-analytical models of cosmic filaments and their evolution. It will also be interesting to investigate whether controlling for quantities such as node mass, curvature, and hierarchical structure leads to genuinely universal filament profiles or reveals simpler scaling relations hidden by the diversity of the full filament population.

Filaments may themselves become useful cosmological probes. Their abundance, geometry, and internal dynamics are expected to depend on the underlying growth of structure and could therefore provide complementary constraints on cosmological parameters or alternative dark matter models \citep{Sousbie+2008, Codis+2018}. Likewise, a better understanding of the relation between coherent infall, shell crossing, and multistreaming may eventually lead to a more physically motivated definition of filament boundaries than those currently in use. More broadly, improved knowledge of filament dynamics and the cosmic web may prove useful for developing environment-dependent models of redshift-space distortions and extracting additional cosmological information from galaxy surveys.

As cosmology enters an era of increasingly precise observations, understanding the nonlinear Universe will become progressively more important. The work presented in this thesis contributes to that effort through new simulations, publicly available tools for studying the cosmic web, and a detailed study of the reconstruction and dynamics of cosmic filaments. The developments presented in this thesis help lay the groundwork for future studies aimed at connecting the geometry and dynamics of the cosmic web with the underlying physics of structure formation.

\else
\fi

\clearemptydoublepage
%
%
\appendix
%
%

\chapter{Appendix for chapter 2}
\label{app:app02}
\newpage
\section{Numerical specifications for Sahyadri simulations}
\label{App: sahyadri numerical spcifications}
We employed the massively parallel tree-PM code \textsc{gadget-4} \citep{Gadget4}\footnote{\url{https://wwwmpa.mpa-garching.mpg.de/gadget4/}} with the ``lean'' memory configuration. The simulations adopted a comoving Plummer-equivalent force softening length of $\epsilon = L_{\mathrm box}/(30 \times 2048) \approx 3.26\,h^{-1}\,{\mathrm kpc}$, corresponding to $1/30$ of the mean inter-particle spacing. Long-range gravitational forces were computed using a particle-mesh (PM) algorithm on a $4096^3$ grid, ensuring Nyquist sampling of the particle distribution.

Initial conditions were generated at $z = 49$ using second-order Lagrangian perturbation theory \citep[2LPT;][]{Scoccimarro1998} via the \textsc{N-GenIC} code integrated into \textsc{gadget-4}. The linear matter power spectrum was computed using the Boltzmann code \textsc{class} \citep{Blas2011,class-II}.\footnote{\url{https://lesgourg.github.io/class_public/class.html}} To enable controlled cosmological derivatives, all parameter-varied simulations employed identical random number seeds (i.e., matching initial phases) as the fiducial run, ensuring that differences arise solely from the cosmological parameter changes. Although this is not sufficient to control the gravitationally induced stochasticity due to phase-mixing at small scales, which affects statistical probes involving halos \citep[e.g.,][]{coulton+23}, such effects can be addressed using additional post-processing optimization strategies as described in \cite{Fisher_stabilization2026}. This allows \Sahyadri\ to be used reliably for estimating derivatives of small-scale observables with respect to cosmological parameters.

Each simulation generated 101 snapshots between $z=12$ and $z=0$, uniformly spaced in scale factor $a = (1+z)^{-1}$ with $\Delta a = 0.01$. While the full snapshot sequence is available, the analyses presented in this work focus primarily on $z=0$ and $z=1$. Each simulation was executed on 12 compute nodes with 32 cores each, utilizing approximately 250 GB of memory per node and requiring $\sim 0.43$ million CPU hours per run.

\section{Halo environment catalogs}
\label{App: sahyadri vahc}
Here, we briefly describe the environment variables calculated for the value added halo catalogs.
\begin{itemize}
\item Eigen-values $\{\lambda_1,\lambda_2,\lambda_3\}$ of the halo-centric tidal tensor $T_{ij}(\xx_h;R)$ smoothed at scales $R \in \{2,4,6,8\}\times R_{\mathrm 200b}$. (Here $\xx_h$ denotes the spatial location of halo $h$.)
\item Eigen-values $\{\lambda_1,\lambda_2,\lambda_3\}$ of $T_{ij}(\xx_h;R)$ smoothed at the fixed scales $R\in\{2,3,5\}$ \Mpch.
\item Eigen-values $\{h_1,h_2,h_3\}$ of the halo-centric density Hessian $H_{ij}(\xx_h;R)$ smoothed at the fixed scales $R\in\{2,3,5\}$ \Mpch.
\item Halo-by-halo bias $b_1(h)$.
\end{itemize}

The technique for calculating the halo-centric tidal tensor and density Hessian is the same as described in \cite{phs18}. Briefly, the smoothed tidal tensor $T_{ij}(\xx;R)=\partial_i\partial_j\psi(\xx;R)$ is estimated by first inverting the unsmoothed Poisson equation $\nabla^2\psi=\delta$ in Fourier space and multiplying by the Fourier transform $W(kR)$ of the relevant (Gaussian) smoothing kernel, before inverse Fourier transforming. Here, $\delta$ is the matter density contrast estimated using cloud-in-cell (CIC) smoothing on a $1024^3$ grid, with Fourier transform $\delta_{\kk}$, so that
\be
T_{ij}(\kk)=\frac{k_ik_j}{k^2}\delta_{\kk}\,\longrightarrow T_{ij}(\xx;R) = {\mathrm F.T.}\left[T_{ij}(\kk)W(kR)\right]\,.
\ee
For the halo-scaled smoothing radius values $R\propto R_{\mathrm 200b}$, we first estimate the tensor on a range of fixed smoothing scales and then interpolate to the scale of interest. For the fixed smoothing scales $\{2,3,5\}$ \Mpch, the tensor is calculated by directly smoothing at the respective scale. 
In each case, $T_{ij}(\xx;R)$ is finally interpolated to the halo locations $\xx\to\{\xx_h\}$.

Below, we will showcase \Sahyadri\ results using the \emph{tidal anisotropy} variable $\alpha$ introduced by \cite{phs18} and defined as
\begin{equation}
\alpha \equiv \sqrt{q^2}/(1+\delta)\,,
\label{eq:alpha-def}
\end{equation}
where $q^2=\frac12\left[(\lambda_1-\lambda_2)^2+(\lambda_1-\lambda_3)^2+(\lambda_2-\lambda_3)^2\right]$ is the halo-centric tidal shear and $\delta=\lambda_1+\lambda_2+\lambda_3$ is the halo-centric density contrast, with $\{\lambda_i\}$ in this case being the tidal tensor eigen-values evaluated at smoothing scale $R=4R_{\mathrm 200b}$.

The density Hessian $H_{ij}$ is constructed similarly to the tidal tensor in Fourier space at the three fixed smoothing scales and then inverse Fourier transformed and interpolated to the halo locations:
\be
H_{ij}(\xx;R)=\partial_i\partial_j\delta(\xx;R)={\mathrm F.T.}\left[-k_ik_j\delta_{\kk}W(kR)\right]\,\longrightarrow H_{ij}(\xx_h;R)\,.
\ee
The halo-by-halo bias $b_1(h)$ is calculated as a suitably weighted, sharp-$k$ filtered halo-centric density using the method described by \cite{VVF2020}, which is an improved version of the technique originally introduced by \cite{phs18}. 
Schematically, for halo $h$,
\be
b_{1}(h)=\sum_{{\mathrm low-}k}N_k\avg{\e{i\kk\cdot\xx_h}\delta^\ast_{\kk}}_k\,/\,\sum_{{\mathrm low-}k}N_k\,P_{\mathrm mm}(k)\,,
\ee
where $\e{i\kk\cdot\xx_h}$ denotes an inverse-CIC weighted phase factor centered on the halo location $\xx_h$, $P_{\mathrm mm}(k)$ is the matter auto power spectrum, the angular brackets indicate an angle-average over all \kk\ modes in a bin of $k=|\kk|$, $N_k$ is the number of these modes and the sums are over low-$k$ modes, typically restricted to $k<0.1\hMpc$. We refer the reader to \cite{phs18,VVF2020} for further details of the procedure, including tests and convergence studies. The quantity $b_1(h)$ has the property that its arithmetic mean over a sample of halos is the same as the traditional definition of linear halo bias of the sample as a ratio of the halo-matter cross power spectrum and the matter auto power spectrum.

\section{Comparison of \Sahyadri\, results with literature fitting functions}
\label{App:sahyadri fitfuncs}
\begin{figure}[h!]
    \centering
    \includegraphics[width=\linewidth]{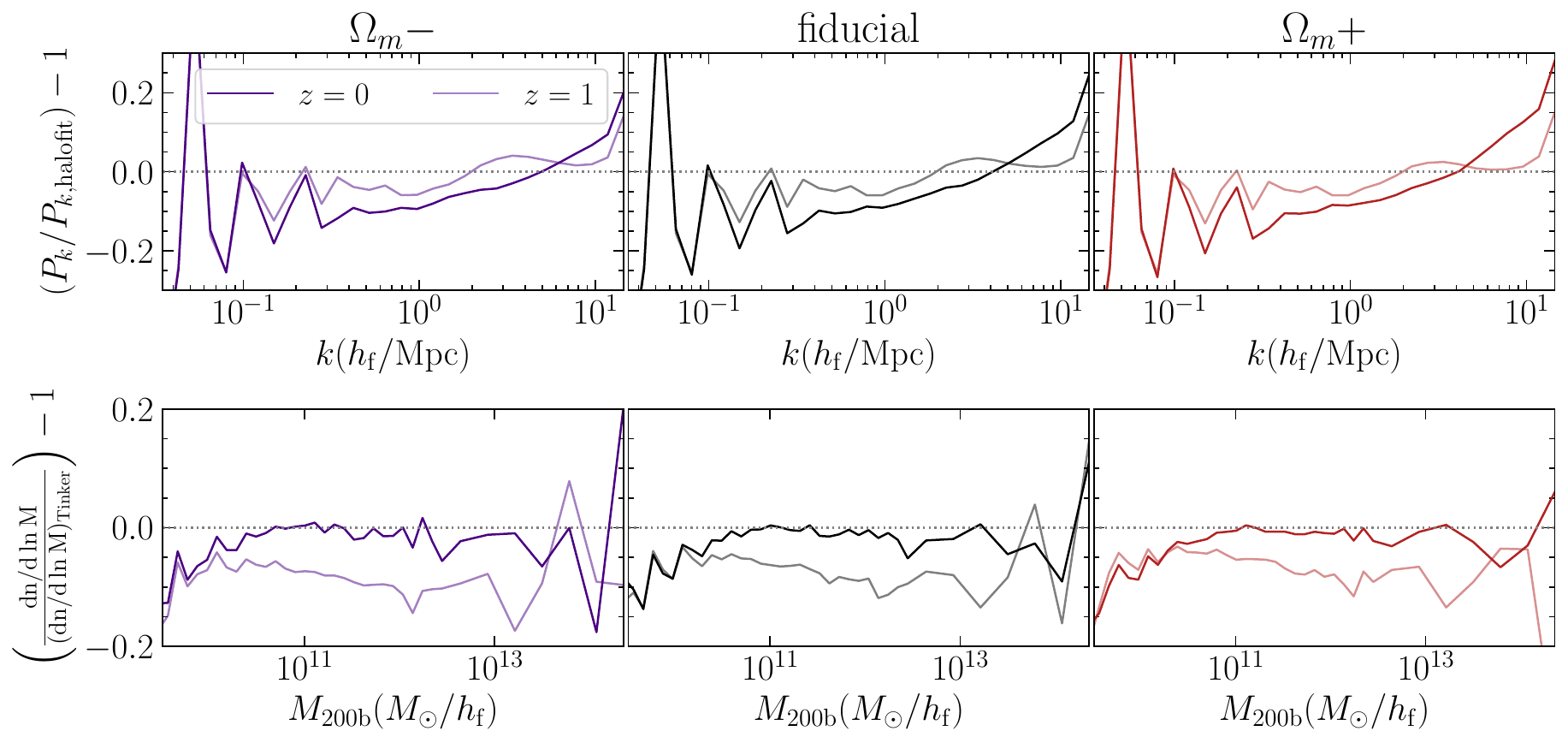}
    \caption{Comparison of the simulation results with expectations from literature. Top (bottom) panels show the ratios of simulation matter power spectrum (halo mass function) with the corresponding halofit (Tinker) expectations.}
    \label{fig:fitfuncs}
\end{figure}

As a consistency check, we compare the matter power spectrum and halo mass function measured from our simulations with widely used fitting functions from the literature, namely \textsc{halofit} \citep{Takahashi2012} for the matter power spectrum and the Tinker mass function \citep{Tinker+2008}. Figure~\ref{fig:fitfuncs} shows these comparisons, with the top panels displaying the matter power spectrum and the bottom panels showing the halo mass function. The three columns correspond to the three cosmologies considered in this work, and results are shown at $z=0$ and $z=1$.

The halo mass function shows excellent agreement with the Tinker fit at $z=0$, with deviations typically below $\lesssim 5\%$ across the resolved mass range. At $z=1$, the level of agreement is slightly weaker, with deviations reaching up to $\sim 15\%$, which is consistent with expectations given the increased uncertainty in fitting functions at higher redshifts.

For the matter power spectrum, we find that at $z=0$ the measured power is systematically lower than \textsc{halofit} on large scales and higher on small scales, with deviations reaching up to $\sim 20\%$. At $z=1$, the discrepancies are reduced, remaining at the level of $\lesssim 10\%$ over the range of scales shown. These trends are consistent with previous studies, e.g. see figure~3 in  \citep{Klypin2016} for halo mass function and figure~1 in \citep{Daalen+2011}.

\section{Length scales in VVF}
\label{App: sahdyadri vvf-scale}
\begin{figure}[h!]
    \centering
    \includegraphics[width=0.9\linewidth]{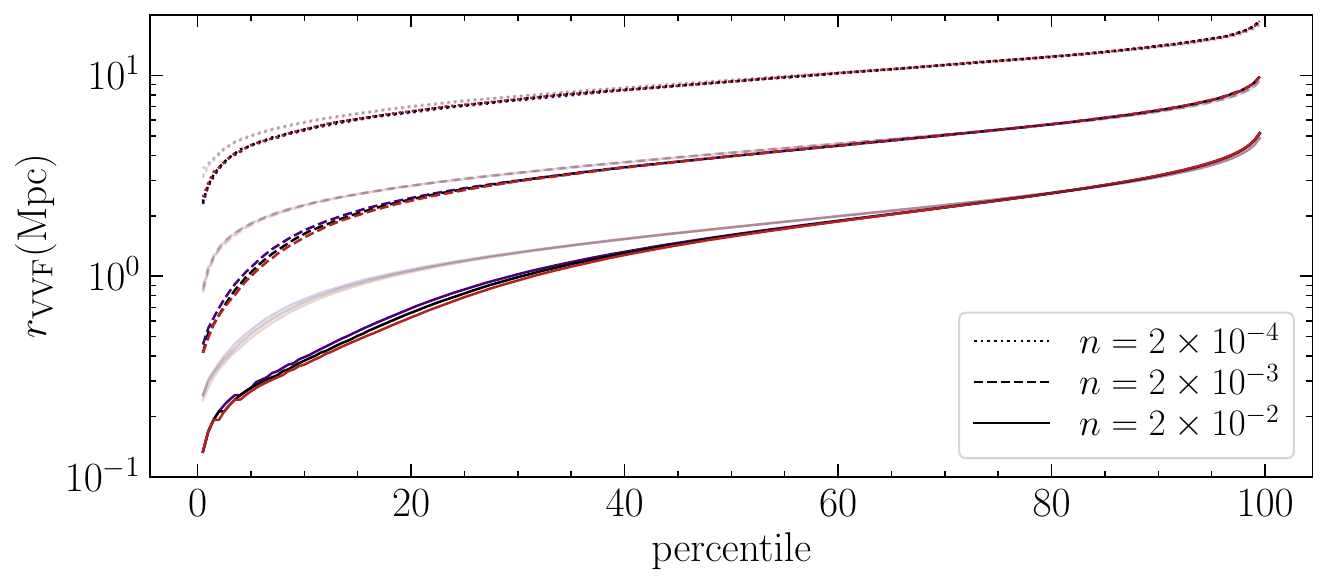}
    \caption{Effective length scales corresponding to VVF percentiles for all tracer number densities, \Om\, variations, and redshifts considered in this work. The effective scale $r(p)$ is obtained by associating each Voronoi cell volume at percentile $p$ with the radius of a sphere of equal volume. Dark (pale) curves correspond to $z=0$ ($z=1$). The color and linestyle scheme follows that of Figure~\ref{fig:vvf_100}.
    }
    \label{fig:rVVF}
\end{figure}

The VVF is a relatively new statistic, and this work presents its first application to high-density tracer samples. It is therefore useful to identify the physical length scales probed by the VVF. Since the statistic is normalized by the mean tracer number density, it does not explicitly retain absolute spatial scales. These can, however, be recovered by associating VVF percentiles with effective length scales.\\
The VVF at percentile $p$ is defined as
\begin{align}
\mathrm{VVF}(p) &= \frac{V(p)}{\langle V \rangle} = nV(p),
\end{align}
where $V(p)$ is the Voronoi cell volume at percentile $p$, $\langle V \rangle$ is the mean cell volume, and $n$ is the tracer number density. Defining an effective spherical radius as:
\begin{align}
\frac{4}{3}\pi r^3(p) &= V(p), \
r(p) = \left(\frac{3}{4\pi}V(p)\right)^{1/3},
\end{align}
gives a characteristic length scale corresponding to each VVF percentile.

Figure~\ref{fig:rVVF} shows these length scales for all tracer number densities, cosmological variations, and for both redshifts considered in this work. Dark (pale) curves correspond to $z=0$ ($z=1$), with the color and linestyle scheme matching Figure~\ref{fig:vvf_100}. The VVF probes scales ranging from $\sim 0.1,\mathrm{Mpc}$ to $\sim 20,\mathrm{Mpc}$, with the range depending on tracer number density and redshift. Higher number densities probe systematically smaller scales, as expected. We also find a clear redshift dependence of the effective length scales associated with the VVF percentiles. At fixed tracer number density, percentiles below $\sim 80$ correspond to systematically larger effective radii, and hence larger absolute Voronoi cell volumes, at $z=1$ compared to $z=0$; while the highest percentiles are largely unchanged. This behaviour reflects the less clustered nature of the tracer distribution at higher redshift, which leads to a narrower distribution of Voronoi volumes with fewer extreme small cells associated with dense environments.

Comparison with Figure~\ref{fig:vvf_100} shows that, for the highest-density tracer sample, the strongest response to changes in $\Om$ occurs at scales $\lesssim 1,\mathrm{Mpc}$. This indicates that the VVF is sensitive to information from the non-linear regime.

\section{Volume dependence of $b_1$}
\label{App: sahyadri b1-voldep}

\begin{table*}[h!]
\centering
\begin{tabular}{lcccc}
\hline\hline
Simulation & Cosmology & $L_{\mathrm box}$ & $N_{\mathrm part}$  \\
 &  & $[h^{-1}\,{\mathrm Mpc}]$ &  \\
\hline
A & C1 & 150 & $1024^3$\\
B & Planck18 & 200 & $256^3$\\
C & C1 & 300 & $1024^3$\\
D & Planck18 & 400 & $512^3$\\
E & C1 & 600 & $1024^3$\\

\hline
\end{tabular}
\caption{Summary of simulations used to study the volume dependence of halo-by-halo bias $b_1$. The default \Sahyadri\, cosmology is Planck18, while C1 corresponds $\Om=0.276, \Ob=0.045, h=0.7, \ns=0.961, \sigma_8=0.811$.} 
\label{tab:diffvol_sims}
\end{table*}

\begin{figure}
    \centering
    \includegraphics[width=\linewidth]{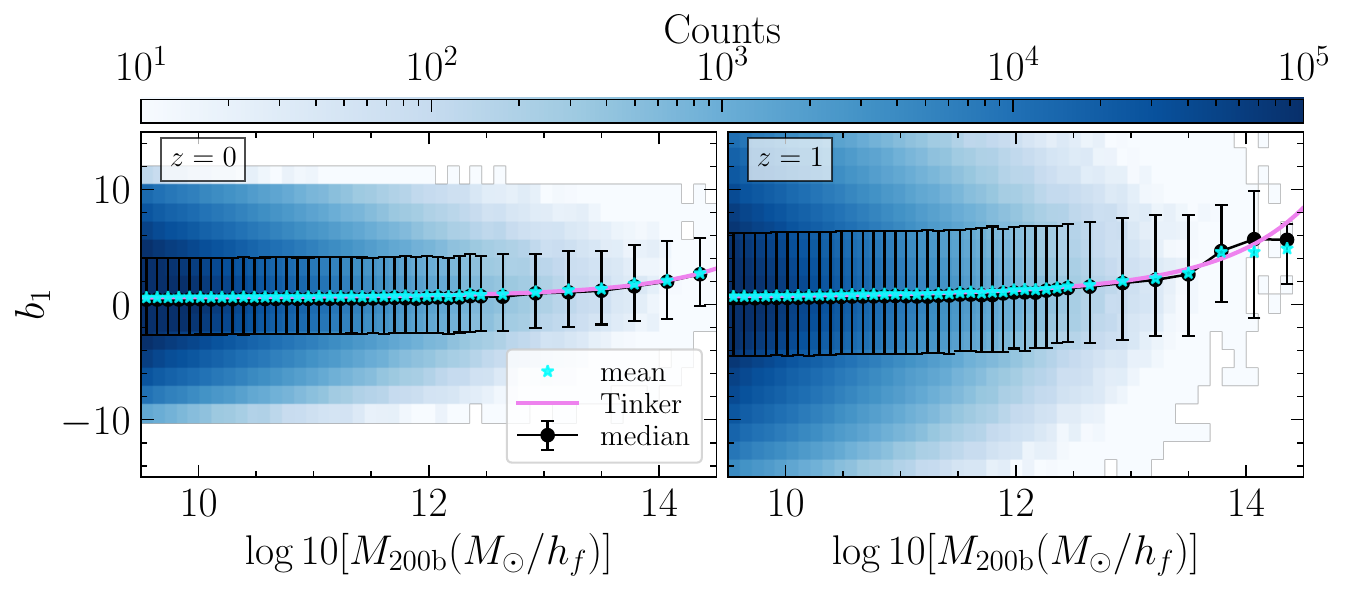}
    \caption{Distribution of halo-by-halo bias $b_1$ as a function of halo mass at $z=0$ (left) and $z=1$ (right). The background shows the 2D histogram of halo counts in the $(M_{\mathrm 200},b_1)$ plane.
    Black points with error bars indicate the median and the $16$–$84$ percentile range in each mass bin. Cyan stars denote the mean bias, while the violet curve shows the Tinker prediction. }
    \label{fig:b1_2d}
\end{figure}

\begin{figure}
    \centering
    \includegraphics[width=\linewidth]{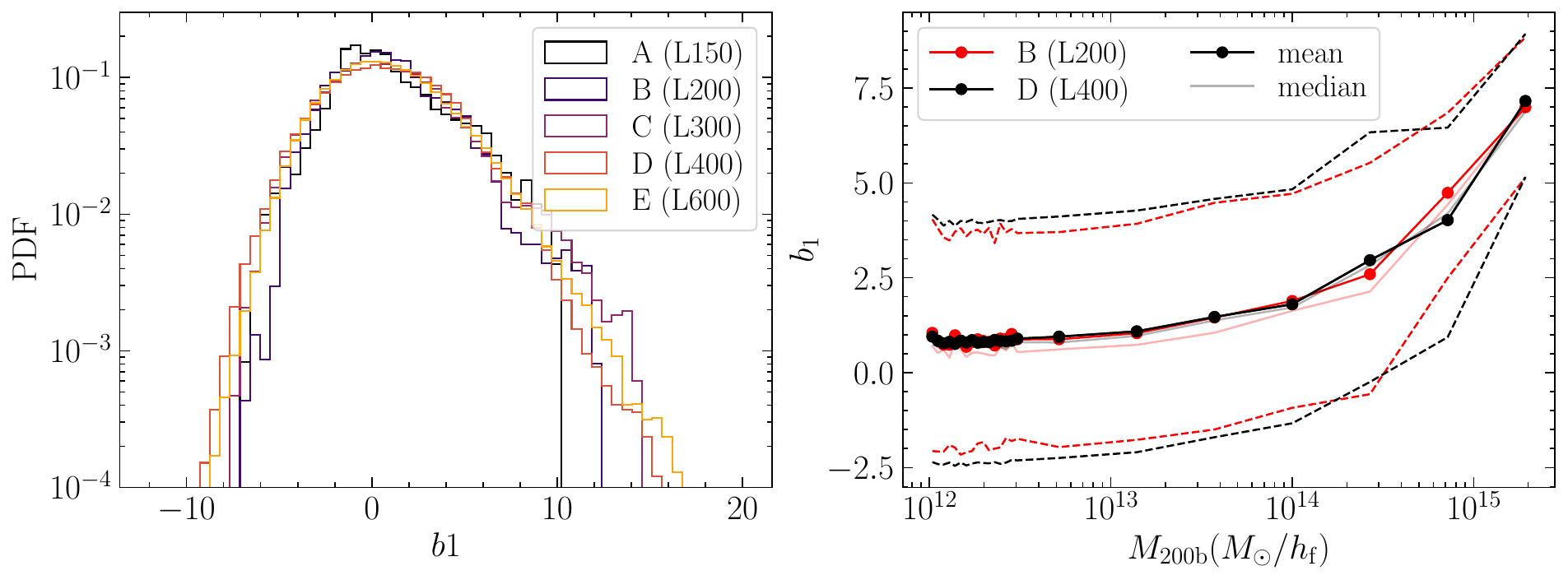}
    \caption{\emph{Left panel:} Distribution of $b_1$ for five simulations with varying box sizes (box size in brackets, in units of $\mathrm{Mpc}/h$). Smaller volumes exhibit truncated tails in $b_1$, indicating the absence of extreme environments. \emph{Right panel:}Comparison of $b_1$ as a function of mass between two simulations with identical cosmology and resolution, but different volumes. Dark (pale) curves show the mean (median) bias in each mass bin, while dashed lines indicate the $16$–$84$ percentile range. 
    }
    \label{fig:b1_vol}
\end{figure}

The halo-by-halo bias $b_1$ plays an important role in the environment-based analyses presented in this work. Since $b_1$ is sensitive to the surrounding large-scale density field, its measurement can be affected by finite-volume effects, particularly in the tails of the bias distribution. In this Appendix, we examine how its distribution and mass dependence varies with simulation volume.

Figure~\ref{fig:b1_2d} shows the distribution of $b_1$ in bins of halo mass for $z=0, 1$, using halos cleaned with the QE criterion and restricted to parent halos in the default \Sahyadri\, simulation. The mean values of $b_1$ agree well with the Tinker expectation \citep{Tinker+2010}. However, we see a sharp truncation at very low and high values of $b_1$, at $z=0$. This is because the bias, being influenced by the large-scale environment, may be sensitive to finite-volume effects which lead to missing long-wavelength modes in the simulation volume.

To study this, we consider five different simulations spanning a range of volumes. Simulations B, D have the same cosmology as our fiducial one, whereas simulations A, C, E have a slightly different cosmology as summarized in Table~\ref{tab:diffvol_sims}. Simulations A, C and E are the same as those used by \citep{NPaul+2019}, whereas B is the default, low-resolution \Sinhagad\ simulation. We clean the halos based on the QE criteion, and retain only parent halos with $M_{\mathrm 200}\geq 10^{12} \Msun/h$. Figure~\ref{fig:b1_vol} illustrates the volume dependence of $b_1$. The \emph{left panel} shows the distribution of $b_1$ for these simulations. We see that the tails of the distribution are progressively suppressed in lower volumes, reflecting the absence of rare overdense and underdense environments in smaller simulation boxes.

The \emph{right panel} of the Figure shows a comparison between the mass dependence of $b_1$ for the default \Sinhagad\, simulation (labeled B), and a simulation with identical cosmology and mass resolution but eight times larger volume (labeled D). Dark (pale) curves represent the mean (median) value of $b_1$ in each mass bin, while the dashed curves show the $16$th and $84$th percentiles. The median bias shows a systematic dependence on volume, with smaller volume consistently under-estimating the median $b_1$ by $\Delta b\sim0.2$. The spread $\sigma_b$ of $b_1$ distribution is also under-estimated in the smaller simulation box $\Delta\sigma_b\sim0.7$. In contrast, the mean value remains unbiased. 
Overall, we find that finite-volume effects mainly alter the distribution of halo-by-halo bias by truncating its tails, especially at high $b_1$, while while leaving the mean mass–bias relation largely unchanged. Because our analysis is based on Spearman rank correlations, these effects will reduce the dynamic range of $b_1$ and may slightly weaken the inferred correlations, without qualitatively changing the trends.

%
%

\chapter{Appendix for chapter 3}
\label{app:app03}
\section{Notations}
\label{App: filtools Notations}
Throughout chapter~3, $r$ refers to the perpendicular distance from the $z$-axis in the cylindrical coordinate system. $z, \phi$ are the axial coordinate and the azimuthal angle in the cylindrical coordinate system respectively. The spherical polar coordinates (radius, polar angle, azimuthal angle) will be denoted by $(r_{sp}, \theta, \phi_{sp})$, and the Cartesian coordinates by $(x, y, z)$. The axis of the cylindrical coordinate system will always be aligned with the $z$-axis of the Cartesian coordinate system. The cubical box that is simulated has a side length $L$, and all the particles are assumed to have the same mass ($m_{\mathrm p}$). The expression $x_{N} \sim f(\alpha_i)$ means that $N$ independent values of the coordinate $x$ are drawn randomly from the distribution $f$, which is a function of the parameters $\alpha_i$. When there is no subscript $N$, it means that a single value is drawn. A uniform distribution between $[a,b]$ will be represented by $\mathcal{U}(a,b)$, and a Gaussian with mean $\mu$ and standard deviation $\sigma$ will be represented by $\mathcal{N}(\mu, \sigma)$.
The information generated by the filament finder as an output and the one accepted by the profile estimator as its input is in the Cartesian coordinate system. The units of length and velocities are \Mpch\ and \kms, respectively, although throughout the discussion, the units are irrelevant.

\section{Technical details of \filgen}
\label{App: filtools filgen}
Here, we discuss in detail the steps followed to generate the particles in various regions of the box, namely the filament, the nodes at its ends, the node outskirts and a uniform background.\\
\begin{figure}
    \centering
    \includegraphics[width=\textwidth]{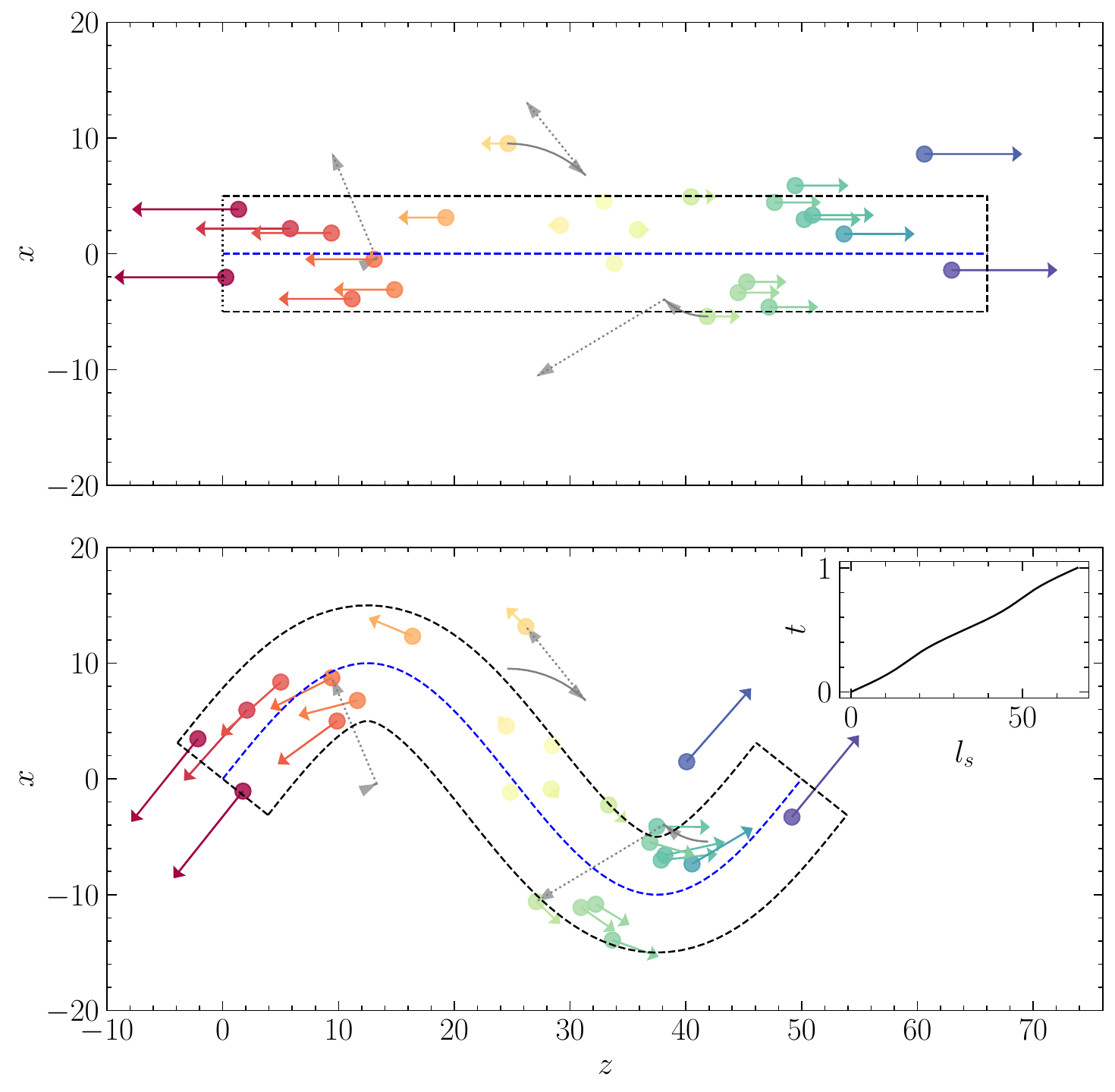}
    \caption{An illustration to demonstrate the generation of a filament. The two panels show the $y$-projection of the filament, and the filament spine in both cases has its $y$-coordinate to be identically zero. Top (bottom) panel shows the straight (curved) filament. The dashed blue line shows the spine of the filament, and the dashed black lines show the region at a radial distance of $r_f$ away from the spine. The markers show the position of the particles generated from a cylindrical Gaussian density profile. They are coloured by their $z$-coordinates before curving, so as to match the particles across the panels. The arrows show the velocity of the particles, which is assumed to have only longitudinal component for easy interpretation. The rotation and translation that curve the filament are shown for three of the particles using solid and dotted grey lines, respectively, the arrows indicating their direction. The bottom panel illustrates that the transformations are valid, retaining the radial and longitudinal profiles. The mapping between the length along the spine $l_s$ and the parameter $t$ used to define the spine of this filament is shown in the top right corner of the bottom panel. This mapping is necessary while curving a filament in \filgen. }
    \label{fig:cartoon}
\end{figure}
To model the filament, we assume that the filament density profile depends only on the perpendicular distance from the spine ($\rho(\vec{r})=\rho(r)$). The radial profile is assumed to be a truncated Gaussian:
\begin{align}
\rho(r) &= c_f\rho_0\exp{\left(-\frac{r^2}{2r_f^2}\right)} \,;\quad r\leq R_f\,,
\label{eq:Gaussian}\\
R_f & \equiv \sqrt{2r_f^2 \ln{(c_f)}} \,,
\label{eq:Gaussian_Rfdef}
\end{align}
where $r_f$ is the radius of the filament, and $c_f$ is the filament concentration defined to be the ratio of the density at the centre of the filament to the mean background density ($\rho_0$). The Gaussian profile is truncated at $R_f$, the radius at which the Gaussian density equals the background density. The cumulative distribution function (CDF) corresponding to this density profile is:
\begin{equation}
\label{eq: Gaussian_cdf}
{\mathrm{CDF}}(r) = \frac{ 1-\e{ -r^2/(2r_f^2}) }{1-\e{ -R_f^2/(2r_f^2)}}\,;\quad r \leq R_f\,.
\end{equation}
The velocity components $(v_r, v_z, v_\phi)$ are assumed to have Gaussian profiles, whose mean and standard deviation have to be specified by the user as functions of $\vec{r}$. One needs to describe the spine of the filament by a continuously differentiable parametric curve $\vec{P}(t)$ parameterized by $t \in [0,1]$. The tangent to this curve is given by $\vec{T}(t) \equiv d\vec{P}(t)/dl_s =( d\vec{P}/dt) (dt/dl_s)$, where $l_s$ is the length along the curve.

The process of creating the filament can be split into two broad tasks: (a) creating a straight filament with the given radial and longitudinal profiles, and (b) curving the filament according to the given spine; both of which we shall describe briefly. 
\begin{itemize}
    \item \textbf{Creating a straight filament}
    \begin{enumerate}
        \item Based on the input mean background density $\rho_0$, mass of the particles $m_{\mathrm p}$, and filament parameters $r_f$, $c_f$,  the number of particles $N_f$ making up the filament are calculated. From the given spine, the length of the filament $l_f$ is also calculated.
        \item The azimuthal and axial coordinates are generated as $\phi_{N_f} \sim \mathcal{U}(0, 2\pi)$ and $z_{N_f} \sim \mathcal{U}(0, l_f)$.
        \item To generate the coordinate $r$, uniform random numbers are generated between $[0,1]$, and these CDF values are converted to the corresponding $r$ values by inverting the CDF given in \eqn{eq: Gaussian_cdf}. 
        \item In case the velocity information is required, the means ($\mu_{i} (\vec{r})$) and standard deviations ($\sigma_{i}(\vec{r})$) of the Gaussians describing the velocity components at the position ($\vec{r}$) of each particle are evaluated from the input functions, where $i \in \{r, \phi, z\}$. 
        Then the velocity components are generated as $v_{i}(\vec{r}) \sim \mathcal{N}(\mu_i(\vec{r}), \sigma_i(\vec{r}))$. The set of cylindrical polar basis vectors form a local orthonormal basis; thus the velocity components are drawn from independent Gaussians with different dispersions.
        \item Appropriate coordinate transformations are used to get the position and velocity components in Cartesian coordinate system. This completes the generation of a straight filament.
    \end{enumerate}
    \item \textbf{Curving the filament}
    \begin{enumerate}
        \item A one-to-one mapping between the length along the spine ($l_s$), and the parameter $t$ is generated. Since initially the filament is aligned with the $z$-axis, the $z$-coordinate of the particle is the same as $l_s$. Thus, for any given $z_1$, there exists a unique $t_1$ and position on the spine ($\vec{P}(t_1)$) associated with it. Note that this mapping between the straight and curved spine preserves the length of the spine by construction.
        \item To curve the filament, two operations are performed on the position vector of every particle. First, a rotation $\mathcal{R}(t_1)$  centered on $[0,0,z_1]$ that takes the unit vector $\hat{z}$ to $\hat{T}(t_1) \equiv \vec{T}(t_1)/\lVert \vec{T} (t_1) \rVert$, which is the unit vector along the required tangent at $z_1$. Second, a translation $\mathcal{T}(t_1)$ that takes the point $[0,0,z_1]$ to $\vec{P}(t_1)$. 
        Note that the exact same operations need to be performed on all the particles lying in a plane perpendicular to the spine at any given value of $z$, and hence $\mathcal{R}, \mathcal{T}$ are a function of the $z$-coordinate of the particle alone (and not $x, y$). The velocity vectors only need to be rotated, and not translated.
    \end{enumerate}
\end{itemize}
Finally, the curved filament is translated to the specified starting point inside the box. Figure~\ref{fig:cartoon} illustrates this procedure for a toy filament. The rotation and translation is shown for three random points using grey curves. The mapping between $l_s$ and $t$ for this filament is shown in the top right corner of the bottom panel. Note that although we have restricted ourselves to a Gaussian radial density profile here, it is trivial to change this to an arbitrary radial profile, as long as the CDF is well-defined. Imagine the initial straight filament to be made up of thin discs placed adjacent to each other, which are to be
rotated in accordance with the local direction of the final curved spine in order to obtain the curved filament. When the curvature is low, the discs do not affect each others’ profiles substantially after rotation, thus retaining the radial and longitudinal profiles. This is exactly what is being done in our case.

We assume that the nodes at the ends of the filament are NFW halos truncated at $R_{200c}$, where $R_{200c}$ is the radius at which the mean enclosed density is $200$ times the critical density $\rho_{\mathrm{crit}}$ of the Universe. 
The density profile is given by:
\begin{align}
\rho_{\mathrm{NFW}}(\rsp) \propto & \, r_{\mathrm{sp}}^{-1}\,(r_{\mathrm{sp}}+r_{\mathrm s})^{-2}\,,
\label{eq:rho_NFW}
\end{align}
where $r_{\mathrm s}$ is the scale radius (with the usual concentration parameter being $c\equiv R_{\mathrm{200c}}/r_{\mathrm s}$), and the normalization is chosen so as to enclose a density of $200\rho_{\mathrm{crit}}$ inside the radius $R_{\mathrm{200c}}$. 
To generate the position coordinates of the particles of a given halo, the number of particles $N_h$ are calculated from the given $m_{\mathrm p}$, $R_\mathrm{200c}$ and $\Om$. Then, the angular coordinates are generated as $N_h$ realizations of $\phi_{\mathrm{sp}} \sim \mathcal{U}(0, 2\pi)$ and $\mu   \sim \mathcal{U}(-1,1)$, 
where $\mu \equiv \cos(\theta)$. The radial coordinates \rsp\ are generated using inversion on the tabulated CDF of the NFW profile.

To get the velocity of the particles, we follow the approach described in the appendix of  \cite{Shethetal2001}, modifying the equations to include a constant velocity anisotropy $\beta$ \cite{Binney&Tremaine1987}, where we solve the Jeans equation for an NFW halo:
\begin{equation}
    \frac{\partial
    (\rho_{\mathrm{NFW}}\sigma_{\rsp}^2 )}{\partial \rsp}+ \frac{2 \rho_{\mathrm{NFW}}}{\rsp}\beta \sigma_{\rsp}^2+\rho_{\mathrm{NFW}}\frac{G M(< \rsp)}{\rsp^2}=0 \,,
\end{equation}
to obtain the radial and tangential velocity dispersion $\sigma_{\rsp}, \sigma_t$ respectively.  Here $M(<\rsp)$ is the mass enclosed within the radius $\rsp$. We assume $\sigma_\theta^2=\sigma_{\phi_{\mathrm{sp}}}^2 = \sigma_t^2/2$, and $\sigma_t^2 = 2\sigma_{\rsp}^2(1-\beta)$. Here, $\beta$ quantifies the velocity anisotropy of the halo, and can take values between $(-\infty, 1]$, with $\beta=0$ representing isotropic velocity dispersion. We assume that $\beta$ is independent of $\vec{r}$, and one needs to specify its value as an input. In principle, this can be extended to include realistic anisotropy profiles that might be relevant for alignment statistics of halos with filaments (see, e.g., \cite{Catelan_et_al1996, Jeeson-Daniel_et_al2011, Ramakrishnan+2019}), but we do not pursue this here. Similarly, we leave the modelling of halo shapes and their tidal alignment, which are potentially relevant for weak-lensing studies (see, e.g., \cite{Blazek_et_al2011, Maion_et_al2023}), to future work. The differential equation is solved numerically with the boundary condition $\sigma_{\rsp}^2=0$ at $\rsp= R_{200c}$, to obtain $\sigma_{\rsp}$ as a function of $\rsp$. The velocity components for each particle with position $\vec{r}$ is generated as $v_i \sim \mathcal{N}(0, \sigma_i)$, where $i \in \{\rsp, \theta, \phi_{\mathrm{sp}} \}$. The position and velocity coordinates are transformed into Cartesian coordinates, and the halos are translated so that their centres coincide with the end points of the filament.

We assume that the outskirts of a given halo are defined between $R_{200c} < \rsp \leq 4R_{200m}$,\footnote{$\sim 4R_{200m}$  defines the tidal environment of a halo, as motivated by \cite{phs18, Ramakrishnan+2019}. } where $R_{200m}$ is the radius at which the mean enclosed density is $200$ times the mean density $\rho_0$. The generation of density and velocity coordinates of the particles in this region is in principle similar to that of the generation of the NFW halos. The CDFs corresponding to the density profiles in these regions have to be specified as an input, along with the velocity mean and dispersion profiles as functions of $\vec{r}$. One also needs to specify the mean density in this region. First, the number of particles in the halo outskirts is calculated from the mean density. The position coordinates are generated following the steps used in NFW halos, except the CDF is different. The velocity profiles are assumed to be Gaussians with given means and standard deviations. The velocity components are generated accordingly. The rest of the steps are the same as in the case of the NFW halos. After the creation of the halos and their outskirts, the parts of the filament lying inside the halos are deleted, so that the filament ends at the edges of the halos; however, we retain the filament sections lying inside the halo outskirts. The choices for the density and velocity profiles of halo outskirts in the present work are arbitrary, since accurate characterizations of these, while correctly accounting for mid- to large-scale density and velocity correlations, do not yet exist (although see \cite{Tinker_et_al2005, Bosch_et_al2013}; in halo model language, this is the so-called 1-halo to 2-halo transition regime). In future work, we plan to characterise these regions using halo environments measured in simulations in order to improve this part of \filgen. 

Finally, a background is added. We assume that the background particles are distributed uniformly with a density $\rho_0$, and have isotropic velocities. The position and velocity Cartesian coordinates of the particles are generated as $x_{N_b}, y_{N_b}, z_{N_b} \sim \mathcal{U}(0, L)$ and $v_{x,N_b}, v_{y,N_b}, v_{z, N_b} \sim \mathcal{N}(0, \sigma_b)$, where $N_b= \rho_0 L^3/m_p$, and $\sigma_b$ is the $1$-dimensional isotropic background velocity dispersion. All the background particles lying within a distance $R$ from the centres of the two halos are deleted. Here, $R=R_{200c}$ of the respective halos if the halo outskirts are not modelled, otherwise $R=4R_{200m}$ of the respective halos. All the background particles lying within a perpendicular distance less than the truncation radius $R_f$ (see equation~\ref{eq:Gaussian_Rfdef})  from the filament spine are also deleted. To calculate the distance to the spine, the spine is very finely sampled, and the distance for each background particle to all these sample points is calculated. The minimum of these distances is taken to be the distance of the particle to the spine. The deletion of the particles is to ensure that the filament, halos and their outskirts have the expected profiles (e.g. NFW in the case of halos), and no extra constant component.

Now that all the regions have been modelled, the position and velocity information of all the particles in the box is combined and returned at the output. For illustrations, see Figure~\ref{fig:filament_illustration}, which shows the density projections along the three Cartesian axes of a filament generated using this tool. Here, we have set the node masses and radii to zero (i.e., no node halos), and modelled the halo outskirts to be identical to the uniform background (i.e., no halo outskirts). 

\section{Technical details of \filapt}
\label{App: filtools filapt}
Here, we assume that the filament is made up of tiny straight cylinders whose axes are the line segments joining consecutive points of the discretely sampled spine. This is a good approximation as long as the length of the line segment at a point is substantially smaller than the radius of curvature $R_\kappa=\kappa^{-1}$ of the spine at that point. We shall elaborate on this point in section~\ref{sec:results}. In case the number of particles in the box is high, we provide a way to sort them into smaller sub-boxes; the code then processes only a subset of particles that lie in the sub-boxes close to the filament spine, resulting in a substantial speed-up. Once the phase space information of all the particles in the box along with the filament spine are provided to \filapt, the following operations are performed by the module \filapt\texttt{.ExtractProfiles}:
\begin{enumerate}
    \item The code starts by defining radial bins (logarithmically spaced by default), based on the user-defined parameters. The minimum and maximum radii of interest are $r_{\mathrm{min}}, r_{\mathrm{max}}$ respectively. Then, it goes over each line segment making up the spine one at a time, and performs the same operations as follows accounting for the boundary conditions when necessary.
    \item A rotation is performed on both the positions and velocities of all the particles, which takes the vector directed along the segment to the $z$-direction. Then, the coordinate system is translated so that the starting point of the segment lies at the origin. In the case of periodic boundary conditions, the particles are ``unfolded'' accordingly.
    \item All the particles with $r_{\mathrm{min}} \leq r \equiv \sqrt{x^2+y^2} < r_{\mathrm{max}}$ and $0\leq z < d_s$ (where $d_s$ is the length of the segment) are selected. These form a cylinder around the segment. The cylinder is translated so that its starting point coincides with the end point of the previous segment.
    \item When all the segments are processed, one is left with a straightened filament in the form of a cylinder, with its axis lying along the $z$-direction. This is chopped into 3D bins of $(r, z, \phi)$. The density $\rho$, mean radial and longitudinal velocities $\langle v_r \rangle, \langle v_z\rangle$, and their corresponding dispersions $\sigma_{r}, \sigma_{z}$ are calculated for each bin. This gives the 3D profiles for the filament. Note that the number of longitudinal bins is independent of the initial sampling of the spine, as the bins are created after straightening the filament. 
    \item In order to calculate the 1D radial (longitudinal) profiles, all the particles in the given $r$ ($z$)-bin are considered irrespective of their $\phi$ and $z$ ($r$) coordinates. The statistics of interest are evaluated from the data of all the particles in that bin.
    \item When splitting by curvature is required, the $\kappa$ values for all the segments are calculated at the beginning using equations~\eqref{eq:kappa_f}-\eqref{eq:kappa_final} by the module \filapt\texttt{.ComputeCurvature}. The segments are split into appropriate groups selected by $\kappa$ and the 1D radial profiles are calculated for each of the groups separately. This is useful for studying the effect of curvature on the radial profiles, as we discuss in section~\ref{sec:effect of radius}. 
\end{enumerate}

\section{Curvature $\kappa$ of a curve}
\label{App: filtools curvature}
Here, we first give the standard mathematical definition of curvature $\kappa$ for a continuous curve. Based on this definition, we give a formula to calculate $\kappa$ for a discretely sampled curve, and show that this is in good agreement with the analytical definition.

\subsection{Continuous curves}
\label{App:filtools curvature_continuous}

Curvature quantifies how much a given curve deviates from a straight line, or by how much the tangent to the curve changes its direction per unit length along the curve. In calculus, curvature $\kappa$ for a curve $\mathcal{C}$ is defined to be \cite[Chapter~14]{apostolcalculus}:
\begin{equation}
\label{eq:kappa_basic}
    \kappa \equiv \left\lVert \frac{ d \hat{T}}{dl_s}\right\rVert \,,
\end{equation}
where $\hat{T}$ is the unit tangent, and $l_s$ is the length along the curve. For a general parametrization of the curve $\mathcal{C} = \vec{C(}t)$, the curvature can be written as:
\begin{equation}
\label{eq:kappa_parametric}
    \kappa (t)= \frac{\lVert \hat{T}'(t) \rVert}{\lVert \vec{C}'(t)\rVert}= \frac{\lVert \vec{C}''(t) \times \vec{C}'(t)\rVert}{\lVert \vec{C}'(t) \rVert^3}\,,
\end{equation}
where $' \equiv d/dt,\, '' \equiv d^2/dt^2$, and the second equality follows from using chain rule and the fact that $\hat{T} \perp \hat{T}'$ (\cite{apostolcalculus}, equation~14.22). It is seen from equation~\eqref{eq:kappa_basic} that $\kappa$ has units of inverse length. For a circle, $\kappa$ is constant, and is equal to the inverse of its radius, whereas it is identically zero for a straight line. For an arbitrary curve, the inverse of $\kappa$ at a given point is called the radius of curvature $R_\kappa=\kappa^{-1}$.

\subsection{Discretely sampled curves}
\label{App: filtools curvature_discrete}
\begin{figure}[h!]
    \centering
    \includegraphics[width=0.7\linewidth]{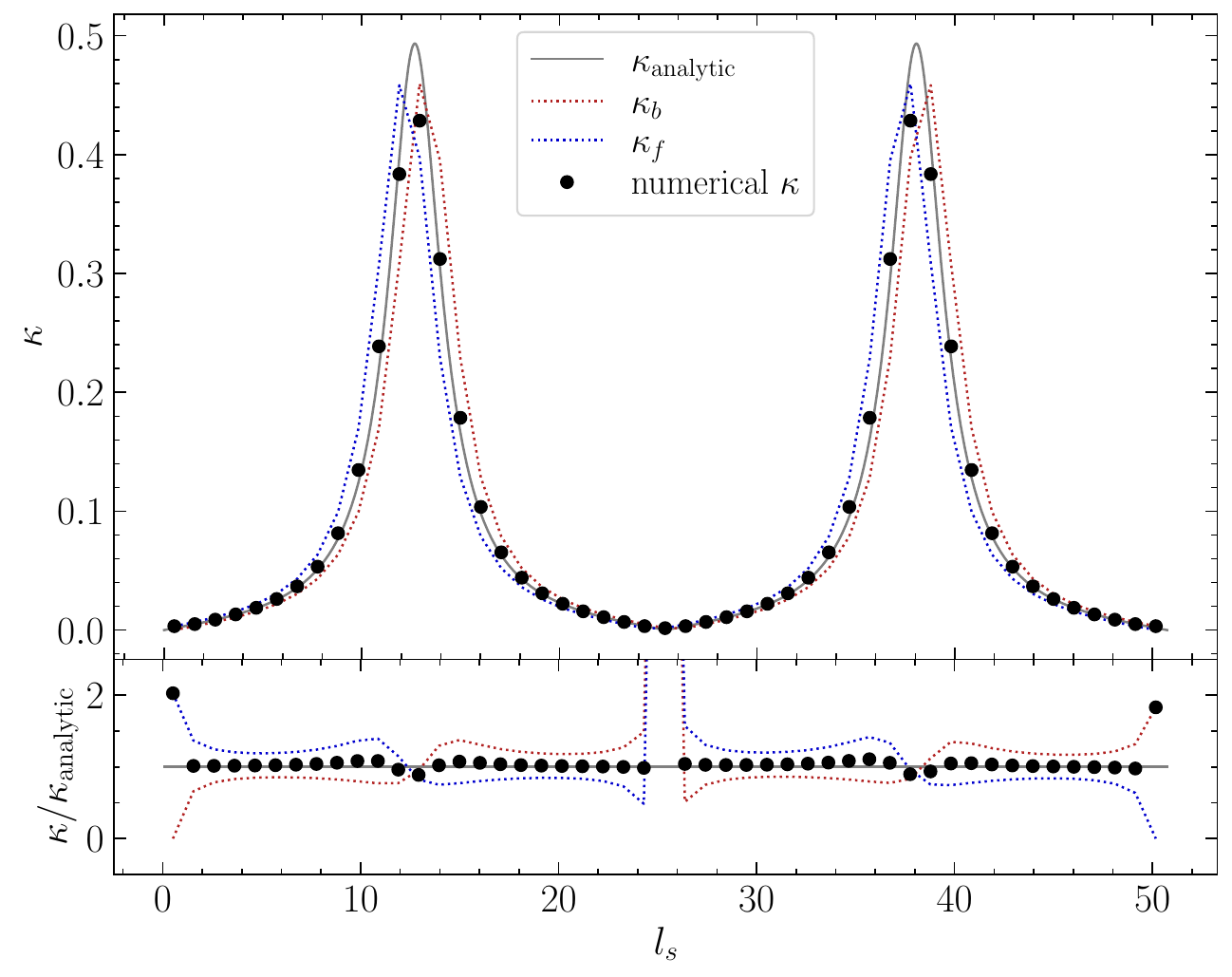}
    \caption{Comparison between the analytical curvature of a curve and the one obtained using \eqns{eq:kappa_f}-\eqref{eq:kappa_final}. The curve used is the spine of the filament described in section~\ref{sec:fiducial}. In the top panel, the curvature $\kappa$ is plotted as a function of length along the curve $l_s$. Grey solid curve shows the analytical value $\kappa_{\mathrm{ analytic}}$. Blue and red dotted curves show the curvature calculated for the discretely sampled curve using the directions of outgoing and incoming segments, respectively. Black markers show the final curvature values calculated using equation~\eqref{eq:kappa_final}. 
    Bottom panel shows a ratio between the calculated curvature and $\kappa_{\mathrm{analytic}}$. The ratio blows up near the centre because the denominator is approaching zero. We see that there is an excellent agreement between the black points and the analytical curvature.}
    \label{fig:curvature}
\end{figure}
We have introduced the standard definition of curvature above, which defines $\kappa$ for a continuously differentiable curve parametrized by a single parameter $t$. However, we want to find the curvature of filaments found by filament finders, which usually define the filament spine to be a set of points rather than an analytical curve. Here, we define the curvature of each segment of the spine based on the basic definition that $\kappa$ is the rate of change of the angle of the tangent per unit length (equation~\ref{eq:kappa_basic}): \begin{equation}
\label{eq:kappa_i}
    \kappa_i \equiv \frac{\lvert \cos^{-1}(\vec{p}_{1\,i} \cdot \vec{p}_{2\,i})\rvert}{l_i}\,,
\end{equation}
where the subscript $i$ refers to a particular segment, $l_i$ is the length of the segment, and $\vec{p}_{1,2\,i}$ are the unit vectors specifying the direction of the tangent to the curve at the start and end points of the segment respectively. Now, the direction of the tangent at a point in this discretely sampled curve can either be taken to be the direction of the incoming segment at the point, or that of the outgoing segment. Thus, one can define two curvature values, corresponding to the two choices of $\vec{p}_{1,2\,i}$ used.  Let the $i$th segment join the points $\vec{r}_i,\, \vec{r}_{i+1}$, with $\vec{l}_i \equiv \vec{r}_{i+1}-\vec{r}_i$ and $l_i \equiv \lVert \vec{l}_i \rVert$. Then the two curvature values for this segment are given by:
\begin{align}
    \kappa_{f,i} & \equiv \frac{1}{l_i}{\left| \cos^{-1} \left(\frac{\vec{l}_i \cdot \vec{l}_{i+1}}{l_i \, l_{i+1}} \right)\right|}\,,
    \label{eq:kappa_f}\\
    \kappa_{b,i} & \equiv \frac{1}{l_i}{\left| \cos^{-1} \left(\frac{\vec{l}_{i-1} \cdot \vec{l}_{i}}{l_i \, l_{i-1}} \right)\right|}\,,
    \label{eq:kappa_b}
\end{align}
Here, $\kappa_f, \kappa_b$ are the curvatures calculated using the directions of the outgoing (or `forward') and incoming (or `backward') segments, respectively. We define the final curvature value for the segment to be the average of these two quantities, accounting for edge effects at the ends of the filament:
\begin{equation}
\label{eq:kappa_final}
    \kappa_i \equiv 
     \begin{cases}
       (\kappa_{f,i}+\kappa_{b,i})/{2} &\,; \quad 1< i < N\\
       \kappa_{f,i} &\,; \quad i=1\\
       \kappa_{b,i} &\,; \quad i=N\\
     \end{cases}
\end{equation}
where $N$ is the total number of segments in the given filament. Figure~\ref{fig:curvature} shows a plot of $\kappa_f, \kappa_b$ and the average $\kappa$ given in the above equation along with the analytical expectation for a curve. The final values of $\kappa$ calculated using our approach (represented using black markers) are in good agreement with the analytical values, showing that this is a valid and reliable method to calculate curvatures of discretely sampled curves.

\section{Smoothing the filament spine}
\label{App: filtools spine_smoothing}
Here, we briefly describe two of the smoothing techniques used to process the spines obtained using a filament finder: the widely used neighbour smoothing (which is usually used to process \disp\, spines), and Fourier smoothing, which we introduce in this work. We also describe in detail a way to optimize the smoothing parameters for any given filament.

\begin{figure}
    \centering
    \includegraphics[width=\textwidth]{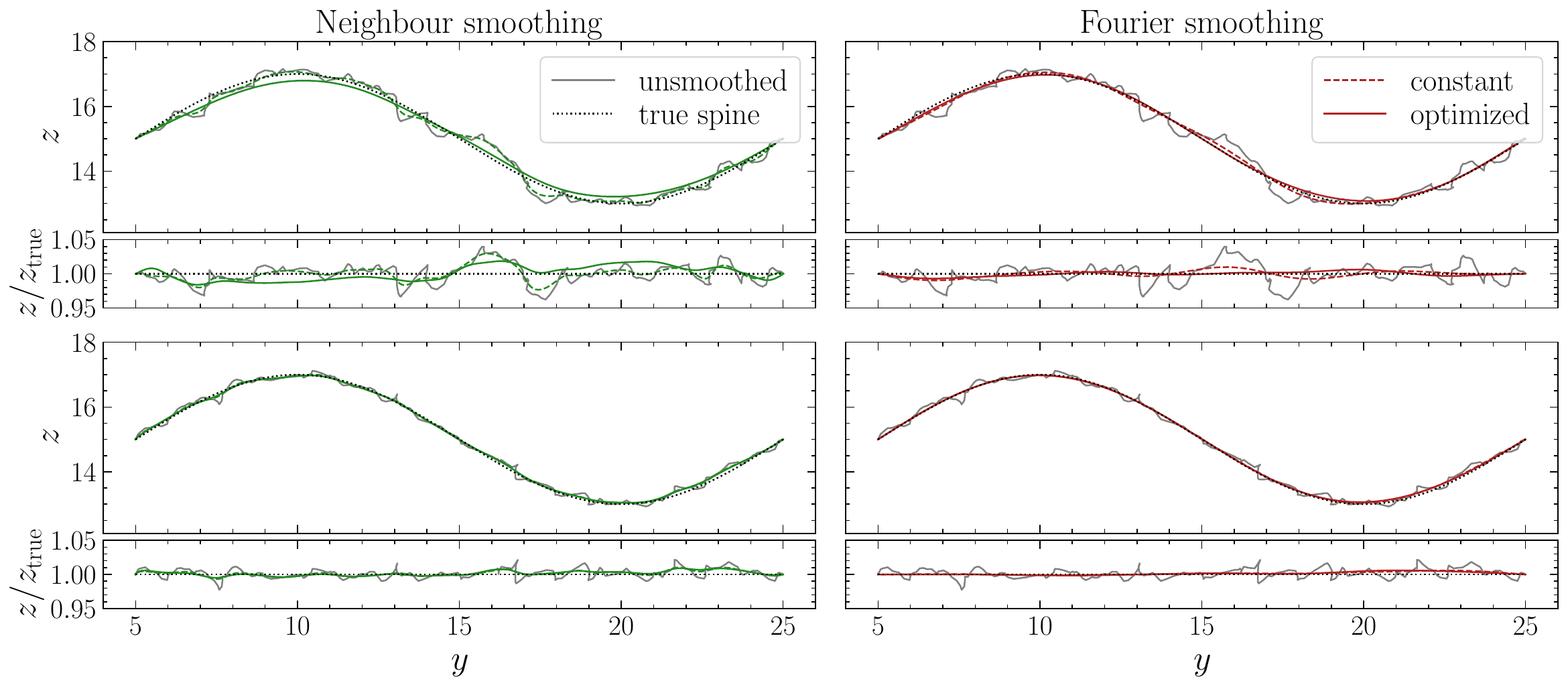}
    \caption{Comparison between different kinds of smoothings applied to two filaments with the same input spine and similar $c_f, \rho_0$, but different $r_f$. The main panels show the $x$-projections of the spines of two filaments. In the narrow panels below each main panel, the $z$-coordinate is divided by the corresponding $z$-coordinate of the true spine, to better illustrate the deviations. The filament in the top panels has approximately twice the $r_f$ as compared to the one shown in the lower panels. The black dotted curves represent the input spine, whereas the unsmoothed \disp\ spine is shown using solid grey curves. Left (right) panels show neighbour (Fourier) smoothing. In each case, optimized and constant smoothings are represented using solid and dashed curves, respectively. We see that the optimized Fourier smoothing gives excellent results in recovering the spines of both the filaments.}
    \label{fig:smoothing_comparison}
\end{figure}

\subsection{Neighbour smoothing}
\label{App: filtools neighbour_smo}
In the neighbour smoothing, the end points of the filament are fixed, and the smoothed positions of all the other points are given by:
\begin{equation}
\label{eq:neighbour_smoothing}
\vec{r}_{i, \, s}= \frac{1}{4}\left(\vec{r}_{i-1}+2\vec{r}_i+\vec{r}_{i+1} \right) \,
\end{equation}
where $\vec{r}_i$ is the position of the point to be smoothed, and $\vec{r}_{i \pm 1}$ are those of its immediate neighbours. The smoothing is performed \Nsm\ number of times to give the smoothed filament.

The green dashed curves in the first column in Figure~\ref{fig:smoothing_comparison} show the application of this smoothing to two filaments similar in all regards except for $r_f$, which differs by a factor $\approx 2$ (see the Figure caption for details), for $\Nsm=30$. We see that this choice of smoothing works for the filament with smaller $r_f$ (bottom two panels), but the thick filament (top two panels) remains under-smoothed. 

\subsection{Fourier smoothing}
\label{App:filtools fourier_smo}
The idea here is to smooth the filaments in Fourier space using a low-pass filter. This is a valid approach, as the noise will contribute to high wave numbers, and removing these should lead to recovery of the actual spine. First, we calculate the length along the spine $l_s$, so that the curve (or spine) is parameterized by this length. Then, we interpolate to get equi-spaced points on the spine. The three Cartesian coordinates of all the points tracing the curve are separately Fourier transformed with respect to $l_s$. A rounded top hat \cite{Bond_et_al1991} low pass filter (RTHF) with the same parameters is applied to all three Fourier spectra. The RTHF is defined as a convolution of a sharp-$k$ low pass filter and a Gaussian in Fourier space:
\begin{equation}
\label{eq:RTHF}
    {\mathrm{RTHF}}(k, k_0, q) = \frac{1}{2}\left[ \erf{\frac{k+k_0}{\sqrt{2q^2}}}- \erf{\frac{k-k_0}{\sqrt{2q^2}}}\right]\,.
\end{equation}
Here, $k$ is the Fourier space variable, $k_0$ is the cutoff wave number of the low-pass filter, $q$ is the standard deviation of the apodizing Gaussian, and $\erf{x}$ is the error function. We do not use a sharp-$k$ filter, as it leads to ringing in real space. We have checked that the smoothed spine is relatively independent of the choice of $q$, and a value $q=k_0/5$ works well enough for any kind of spine or $k_0$. Thus, the only free parameter is the cutoff wave number $k_0$. There are undesirable edge effects if one directly uses the Fourier transform of the input spine for smoothing. Instead, we first pad the spine on either side with an odd padding of length equal to the length of the spine. This padded curve is Fourier transformed, the filter is applied, and an inverse Fourier transform is performed to get back to the real space. This curve is cut off on either side to remove the initial padding, and get the smoothed spine. We choose odd padding as it fixes the end points. However, one can also opt for constant or even padding, and only small differences are seen at the end points of the smoothed filament, those too when $q$ is very low.

The second column in Figure~\ref{fig:smoothing_comparison} shows Fourier smoothing applied to two different filaments. We have chosen a constant $k_0=0.1\,\hMpc ; \, q=k_0/5$, and the smoothed spines are shown using dashed red curves. This choice of parameters works well for the thinner filament (bottom two panels). Although the smoothed spine of the thicker filament (top two panels) shows some deviations, the performance of constant Fourier smoothing for the thick filament is much better than the constant neighbour smoothing. This is because although the noise in this case has higher amplitude, it still contributes to higher wave numbers in Fourier space, which are killed using the low pass filter. Another advantage of Fourier space smoothing is that the cutoff wave number is bounded; the mean spacing between the points in the input spine provides an upper bound on $k_0$, and the length of the spine provides a lower bound. There is no such upper bound on \Nsm\ in neighbour smoothing. Although Fourier smoothing offers an improvement, some longer wavelength noise modes are retained, and the choice of smoothing parameters is still subjective. We address this issue in the next section.

\subsection{Smoothing optimization}
\label{App: filtools smoothing_optimization}
As seen from Figure~\ref{fig:smoothing_comparison}, different types of filaments lead to different noise properties of the spines obtained using a filament finder (in this case, \disp). Thus, a particular choice of smoothing that works for one filament might not work for another; and there is a need to optimize the smoothing parameters separately for individual filaments. Here, we introduce a physically motivated way to achieve this. 

One expects the filaments to be densest at their core, and the density is supposed to drop with increasing $r$, to reach a constant value at large $r$. This constant value might not be equal to the mean density $\rho_0$, since the filaments could be embedded in different environments like a sheet or a void. When the spine of the filament is not robustly estimated (under- or over-smoothed), the radial density profile of the filament will get smoothed out (inner, denser regions contributing to larger $r$ while estimating the profiles and vice versa), increasing the width of the density profile. Therefore, it is logical to assume that the choice of smoothing parameters that gives rise to the narrowest radial density profile is the optimum choice. 

Here, we use filaments generated using \filgen, which have (by construction) a Gaussian radial density profile. Thus, there is a unique way to estimate the width of the profile, which is the standard deviation of the Gaussian, and is equal to $r_f$. For any given choice of smoothing, the smoothed spine is used to estimate the radial density profile, and the errors are taken to be Poissonian. The profile is fitted with a Gaussian (see \eqn{eq:Gaussian}), with two free parameters $(c_f,  r_f)$. $\rho_0$ is fixed, as it is completely degenerate with $c_f$. Note that here, we choose to fit the complete, un-truncated Gaussian. This procedure is repeated for different values of smoothings, till a converged minimum $r_f$ is reached. This corresponds to the optimum smoothing. We use the publicly available code \texttt{PICASA} \cite{picasa} to implement anisotropic simulated annealing (ASA), which is a fast and robust method to achieve the optimum. In future work, we will extend the technique so as to estimate the filament thickness using more flexible functions than a single Gaussian.

The results of implementing this optimization using neighbour (Fourier) smoothing are illustrated in the left (right) panels of Figure~\ref{fig:smoothing_comparison}, using solid green (red) curves. There are no visual differences between the constant and optimized smoothing for the thin filament (lower panels). For the thick filament (upper panels), optimized neighbour smoothing does provide an improvement. Although the smoothed spine does not coincide with the input spine completely, a substantial amount of the noise has been removed, and the smoothed spine agrees better with the input as compared to the case of constant neighbour smoothing. We see that optimized Fourier smoothing gives an excellent performance in recovering the spines of both the filaments. Overall, it is evident that optimization (especially using Fourier smoothing) gives a substantial improvement over the widely used constant neighbour smoothing, along with eliminating the ambiguity associated with the smoothing parameters.

\section{\disp\ filament identification}
\label{app: filtools filament spine identification}
To obtain the spines of the filaments explained in sections~\ref{sec:fiducial} and \ref{sec:optimized_smoothing}, we first add two NFW halos at the ends of these filaments, with radii $R_{200c}=0.4, \, 0.6\, \Mpch$ and concentrations $c=8, 5$ respectively. The filament identification is not sensitive to these choices, one only requires the presence of density maxima at the ends of the filament. The density field is estimated using the Delaunay Tessellation Field Estimator (DTFE), which is implemented using the \texttt{delaunay\_3D} function of \disp. This DTFE density is smoothed using the \texttt{netconv} function with the parameter \texttt{nsmo=5}. The filament identification is done using the \texttt{mse} function, with \texttt{nsig=4}. Finally, the binary output is converted to human readable form using \texttt{skelconv}, with the keyword \texttt{-breakdown}, which merges overlapping filament segments.

Along with the spine of the filament, \disp\ also produces some nonphysical filaments, and the output needs to be cleaned to remove these artifacts. First, we remove any loops formed between two critical points that are not maxima. We also remove any saddle or bifurcation point that is linked to a single filament, and the filament associated with it. This is done recursively, until no freely hanging filament (one which ends at a point that is not a maximum) remains. Now, starting from one node, we loop over all the remaining filaments, joining those which share a common end point, until the other node is reached. This gives a single connected filament, which is used to estimate the profiles in section~\ref{sec:error on the spine}. Note that since here we have a single physical filament, this choice of the filament joining the two nodes is unique. However, in cosmological simulations, one might have multiple filaments joining at a point that is not a maximum (node). Then, the joining procedure is not so trivial, and one will have to make a choice (say choosing the longest or the shortest path joining two nodes). 

%
%

\chapter{Appendix for chapter 4}
\label{app:app04}
\section{Adaptive dip locator}
\label{app:skeletor adaptmin}

The filament radius in this work is defined by the location of maximum radial infall around the filament spine. In practice, however, radial velocity profiles measured around individual filaments can be noisy, particularly for low-mass filaments. Simply identifying the minimum velocity bin often leads to unstable radius estimates that are sensitive to fluctuations in the profile. To obtain a more robust estimate, we developed an adaptive dip locator.

The procedure begins with an initial estimate of the filament radius obtained from the global minimum of the noisy data. A fitting window centred on this initial estimate is then selected and the radial velocity profile within the window is fitted with a second-order polynomial,
\begin{equation}
v_r(r)=ax^2+bx+c,
\end{equation}
where $x\equiv \log{r}$. The fit is performed using seven neighbouring radial bins centred on the current estimate of the minimum, five on either side. This provides enough points to obtain a stable quadratic fit while remaining sufficiently local to follow the shape of the velocity dip. The minimum of the fitted polynomial is used as an estimate of the location of maximum infall.

A fit is considered acceptable only if the inferred minimum lies sufficiently far from the boundaries of the fitting window. This concept is inspired by the implementation in \textsc{PICASA}\citep{picasa}. Specifically, we define an edge region occupying a fraction $f_{\mathrm{edge}}$ of the window width on either side. Throughout this work we adopt $f_{\mathrm{edge}}=0.3$. If the fitted minimum falls within one of these edge regions, the fitting window is shifted in the corresponding direction and the fit is repeated. This prevents the algorithm from locking onto minima that are only partially contained within the fitting interval.

The magnitude of the shift is adjusted adaptively. If successive fits indicate that the minimum lies in the same direction, the shift size is increased. If the direction changes between iterations, indicating that the search has overshot the dip, the shift size is reduced. The procedure continues until the fitted minimum lies comfortably within the fitting window or a maximum number of iterations is reached.



In practice, the procedure converges within a few iterations for the vast majority of filaments and provides significantly more stable radius estimates than a direct search for the minimum velocity bin.

\section{Sensitivity to mass bins}
\label{app: mass sensitivity}
\begin{figure}[h!]
    \centering
    \includegraphics[width=0.6\linewidth]{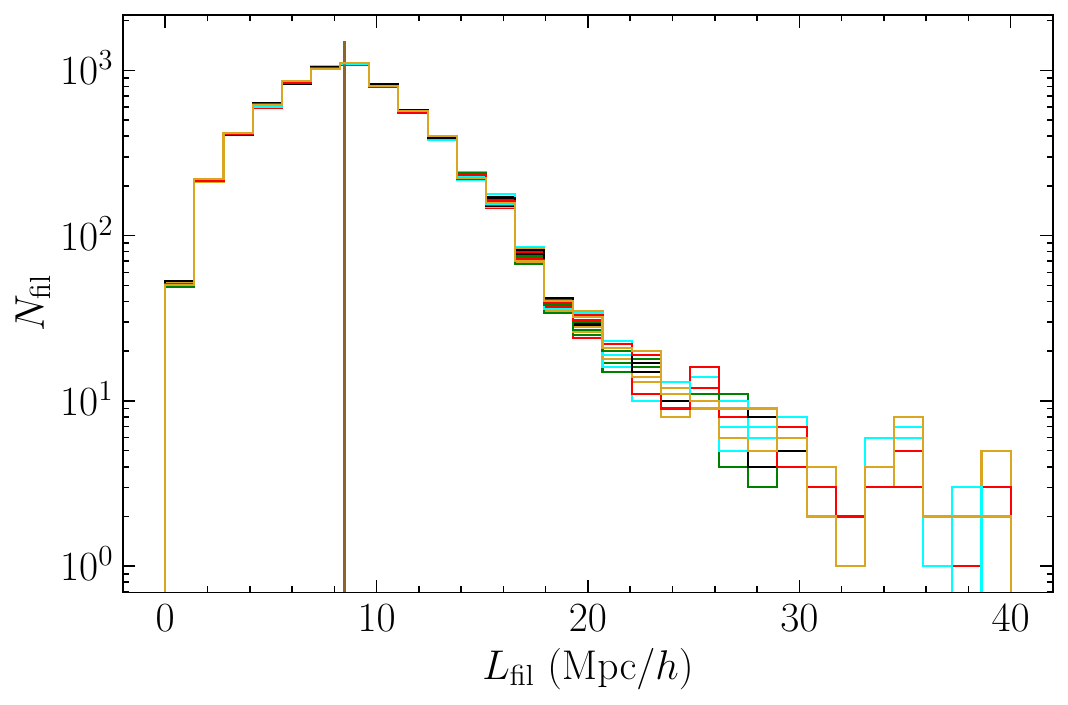}
    \caption{Dependence of the filament length distribution on the hierarchy mass thresholds. The curves show the filament populations recovered from all 27 combinations of threshold variations around the fiducial reconstruction, discussed in the text. The resulting distributions exhibit very little scatter, demonstrating the stability of the recovered filament population against reasonable variations in the mass cuts.}
    \label{fig:mcut_dep}
\end{figure}
The hierarchical reconstruction implemented in \skeletor\ requires the specification of a set of mass thresholds that define the node and tracer populations at each hierarchy level. Since these thresholds are not uniquely determined, it is important to assess the sensitivity of the recovered filament population to their precise values.

To quantify this, we vary the mass thresholds around the fiducial values adopted in the main analysis. Let $N$ denote the number of filament-like tracers lying above a given mass threshold. For each tracer threshold, we construct two additional thresholds by varying the number of filament-like tracers by $\pm 2\sqrt{N}$ relative to the fiducial value. Since $\sqrt{N}$ represents the Poisson uncertainty in the number counts, this corresponds to exploring threshold variations at roughly twice the Poisson error. The corresponding halo masses are then used as alternative tracer thresholds. A similar procedure is applied to the highest mass threshold defining the primary nodes, where $N$ now denotes the number of halos satisfying $M>M_1$. The last mass threshold defining the tracers of the tertiary filaments is kept the same.

Applying these variations independently to the three thresholds used in the fiducial reconstruction produces $3^3=27$ distinct realizations of the filament catalogue. Each realization is processed through the full \skeletor\ pipeline, yielding a corresponding population of reconstructed filaments.

Figure~\ref{fig:mcut_dep} compares the filament length distributions obtained for all 27 realizations. The distributions are nearly indistinguishable across the full range of filament lengths probed by the catalogue. No systematic shifts in the abundance of either short or long filaments are observed, and the scatter between the realizations remains small.

This behaviour indicates that the filament population identified by \skeletor\ is not strongly sensitive to modest variations in the adopted mass thresholds. While the precise membership of individual filaments can change as tracers enter or leave the reconstruction, the statistical properties of the recovered filament network remain stable. The chosen thresholds therefore appear to lie in a regime where the reconstruction is dominated by the underlying filamentary structure rather than by the specific details of the threshold selection.

\section{Sensitivity to $\alpha_{\mathrm{th}}$}
\label{app:alpha_sensitivity}
\begin{figure}[h!]
    \centering
    \includegraphics[width=\linewidth]{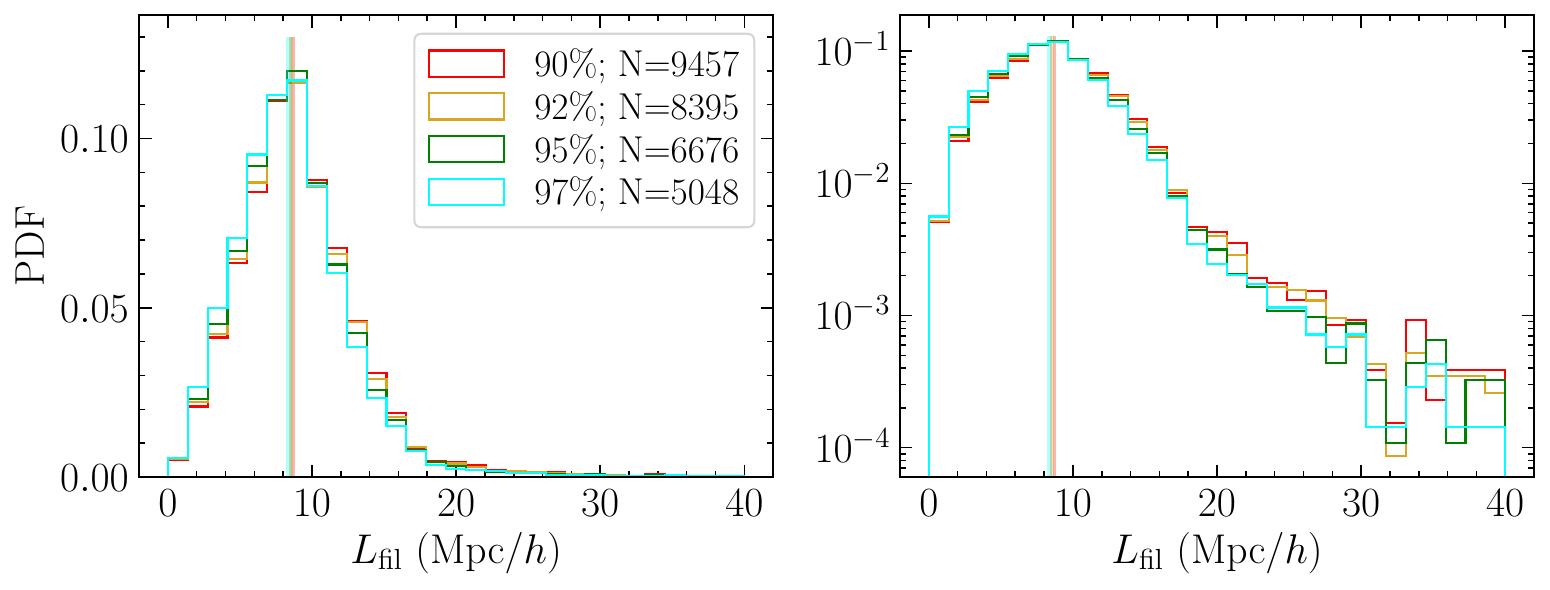}
    \caption{Filament length distributions obtained for different anisotropy thresholds $\alpha_{\mathrm{th}}$. Left (right) panles show the y-axis on linear (log) scales. The thresholds correspond to the $90$th, $92$nd, $95$th and $97$th percentiles of the random-catalogue $\alpha$ distribution. The number of filaments recovered in each case is indicated in the figure.}
    \label{fig:alpha_dep}
\end{figure}
The identification of filament-like cells in \skeletor\ relies on the anisotropy threshold $\alpha_{\mathrm{th}}$. Cells with $\alpha\geq\alpha_{\mathrm{th}}$ are considered sufficiently anisotropic to be potential members of the filamentary skeleton. Since the choice of $\alpha_{\mathrm{th}}$ is not unique, it is important to assess the impact of this parameter on the recovered filament population.

In the fiducial reconstruction, $\alpha_{\mathrm{th}}$ is chosen to correspond to the $95$ th percentile of the anisotropy distribution measured from a uniform random catalogue. To test the sensitivity of the reconstruction, we repeat the filament finding procedure using thresholds corresponding to the $90$th, $92$nd, $95$th and $97$th  percentiles.

Figure~\ref{fig:alpha_dep} compares the resulting filament length distributions. The shape of the distribution remains remarkably stable across the full range of thresholds considered. The probability density functions are nearly identical, indicating that the characteristic length scales of the recovered filament population are largely insensitive to the precise value of $\alpha_{\mathrm{th}}$.

The total number of recovered filaments, however, shows a stronger dependence on the threshold. Lower values of $\alpha_{\mathrm{th}}$ admit a larger fraction of Voronoi cells into the filament candidate population, leading to a larger number of reconstructed filaments. Conversely, increasing the threshold restricts the reconstruction to only the most anisotropic cells and reduces the number of identified filaments.

The stability of the length distribution suggests that the anisotropy threshold primarily affects the completeness of the filament sample rather than the properties of the recovered filaments themselves. The fiducial choice of the $95$th percentile therefore represents a compromise between retaining a sufficiently large filament population and restricting the reconstruction to strongly anisotropic environments.

\section{Sensitivity to \dcut}
\label{app:dcut_sensitivity}
\begin{figure}[h!]
    \centering
    \includegraphics[width=0.6\linewidth]{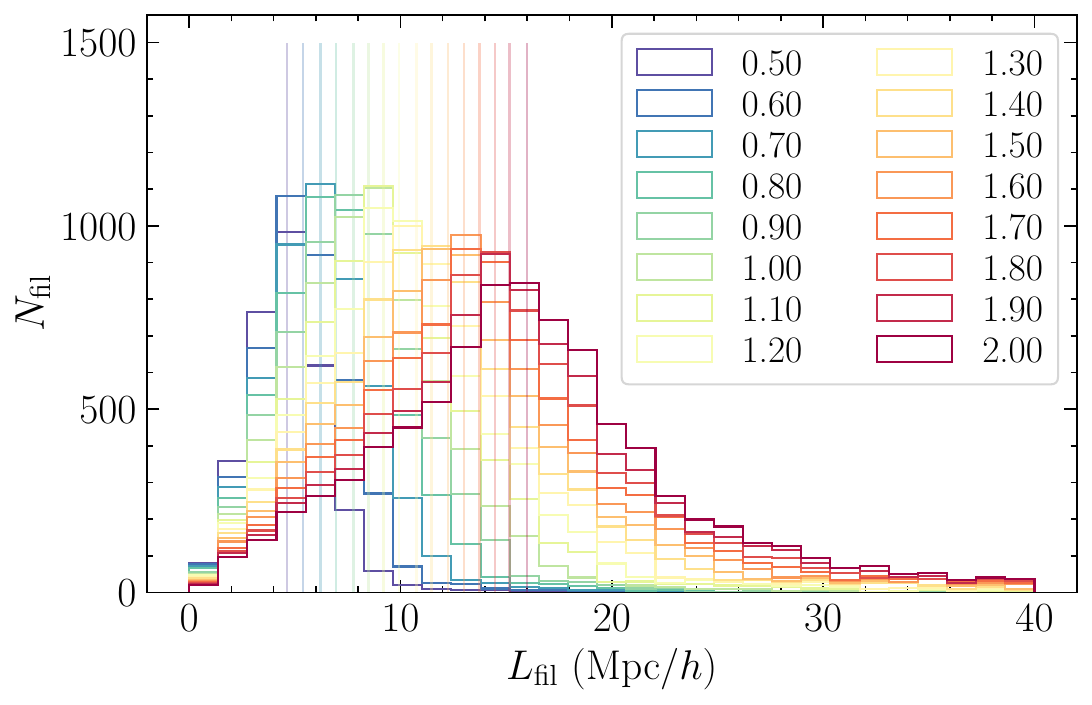}
    \caption{Dependence of the filament length distribution on the connectivity parameter \dcut. The reconstruction is repeated using $\dcut=fl_{\mathrm{mean}}$. Increasing \dcut systematically shifts the filament population towards larger lengths, producing a more extended high-length tail. This behaviour reflects the increased ability of the algorithm to connect neighbouring filament segments across larger gaps.}
    \label{fig:dcut_dep}
\end{figure}

The parameter \dcut defines the maximum distance over which filament-like cells can be connected during the reconstruction. In the fiducial implementation, \dcut is chosen to be equal to the mean inter-tracer separation at the corresponding hierarchy level, $l_{\mathrm{mean}}$. Since filaments are typically denser than the average tracer population, this provides a physically motivated upper limit on plausible filament connections.

To assess the impact of this choice, we repeat the reconstruction using values of \dcut ranging from $0.5\times l_{\mathrm{mean}}$ to $2\times l_{\mathrm{mean}}$. Figure~\ref{fig:dcut_dep} shows the resulting filament length distributions. Unlike the mass thresholds, the filament population exhibits a clear dependence on \dcut. Increasing the connection length systematically shifts the distribution towards larger filament lengths, increases the abundance of long filaments, and produces a more extended high-length tail. At the same time, the total number of recovered filaments also increases. Conversely, smaller values of \dcut lead to a more fragmented filament network dominated by shorter structures. This behaviour is expected: larger values of \dcut allow filament-like cells to be linked across larger gaps, increasing the likelihood that neighbouring segments are connected into a single filament. Smaller values impose a stricter connectivity requirement and therefore tend to break the network into shorter pieces. The choice $\dcut= l_{\mathrm{mean}}$ provides a natural compromise between excessive fragmentation at small values and overly permissive connections at large values.

\begin{figure}
    \centering
    \includegraphics[width=\linewidth]{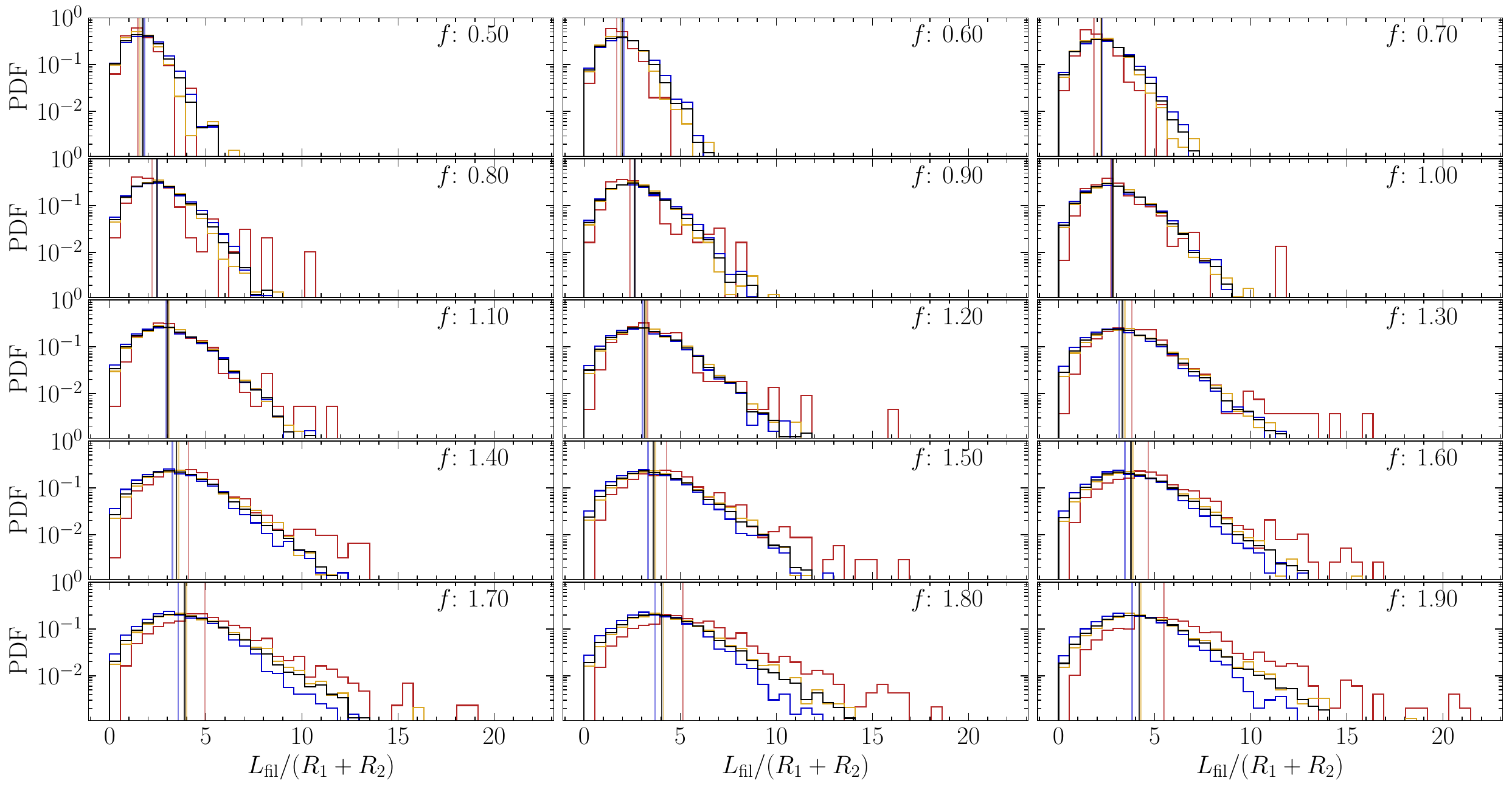}
    \caption{Comparison of the rescaled filament length distributions, $\lfil/(R_1+R_2)$, for different values of the connectivity parameter \dcut. Each panel corresponds to a different value of $\dcut/l_{\mathrm{mean}}$, indicated in the upper-right corner, and shows the three hierarchy levels together with the full filament population. Red, yellow, blue colours correspond to the primary, secondary and teriary levels respectively, wheras black colour shows the full population. The approximate collapse of the distributions is recovered only near the fiducial choice, $\dcut=l_{\mathrm{mean}}$.}
    \label{fig:dcut_universal}
\end{figure}
An independent argument in favour of the fiducial choice is provided by the rescaled filament length distribution discussed in figure~\ref{fig:lfil_pdf}. There we showed that the distributions of $\lfil/(R_1+R_2)$ for different hierarchy levels collapse onto an approximately universal form. Figure~\ref{fig:dcut_universal} examines whether this behaviour persists for different values of \dcut. The collapse is found to be strongest near $\dcut=l_{\mathrm{mean}}$. For larger values of \dcut, the primary filaments develop a systematically extended high-length tail relative to the secondary and tertiary levels, whereas the opposite trend is seen for smaller values of \dcut. The approximate universality of the rescaled length distribution is therefore recovered only close to the fiducial choice, providing additional support for adopting $\dcut=l_{\mathrm{mean}}$.

\section{Hierarchy and node mass}
\label{app: node mass and hierarchy}
\begin{figure}[h!]
    \centering
    \includegraphics[width=\linewidth]{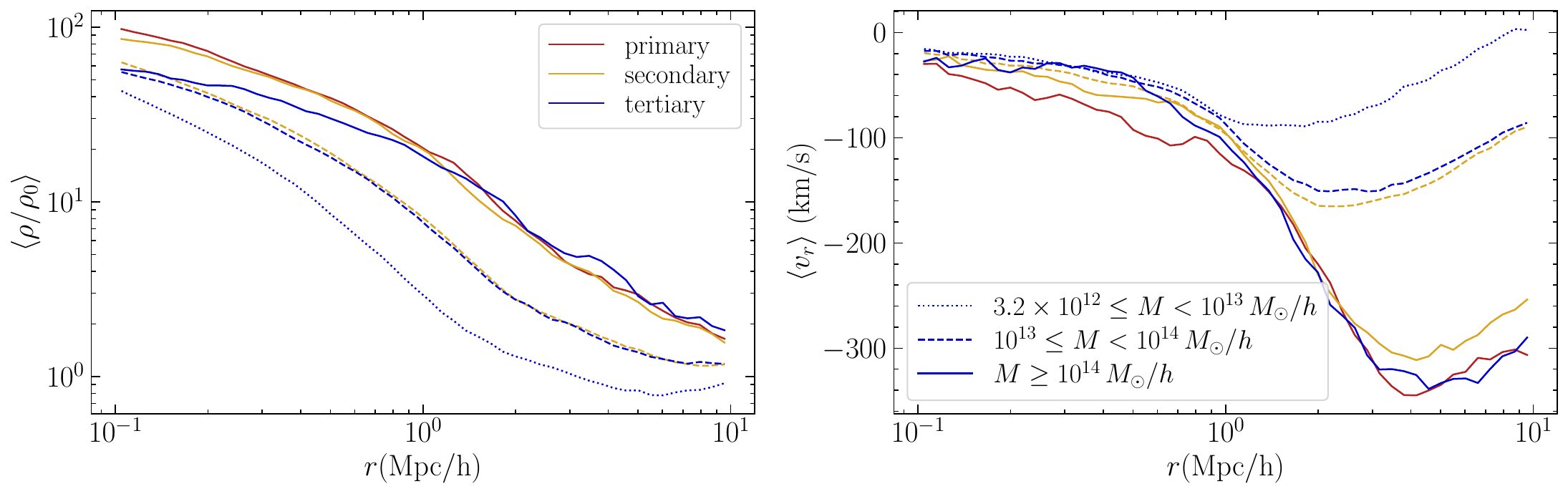}
    \caption{Density and velocity profiles, analogous to Figure~\ref{fig:profile_stacks}, but with the filament sample further divided by the average node mass. Colours denote the hierarchy level, while line styles correspond to different node-mass bins. Much of the hierarchy dependence seen in Figure~\ref{fig:profile_stacks} is reduced once the node mass is controlled for, although small differences between hierarchy levels remain.}
    \label{fig:hier_mass}
\end{figure}
The differences between the stacked profiles at different hierarchy levels could arise either because the filaments belong to different levels or because they connect nodes of different masses. To separate these effects, we repeat the stacking after controlling for the node mass. For each filament, we define the characteristic node mass, $M$, as the average mass of its two endpoint nodes, and divide the sample into bins with edges given by the mass thresholds defining the hierarchy.

Figure~\ref{fig:hier_mass} shows the radial density and velocity profiles for the three hierarchy levels in these node-mass bins. Different colours denote the hierarchy level, while different line styles correspond to different node-mass bins. Once the node mass is fixed, the differences between the hierarchy levels are substantially reduced, although small systematic differences remain. This indicates that the trends seen in Figure~\ref{fig:profile_stacks} are driven primarily by the node mass, with the hierarchy level contributing a weaker secondary effect.
%
%

\chapter{Appendix for chapter 6}
\label{app:app06}
\section{High curvature regions of filaments}
\label{app: high kappa fils}
\begin{figure}
    \centering
    \includegraphics[width=\linewidth]{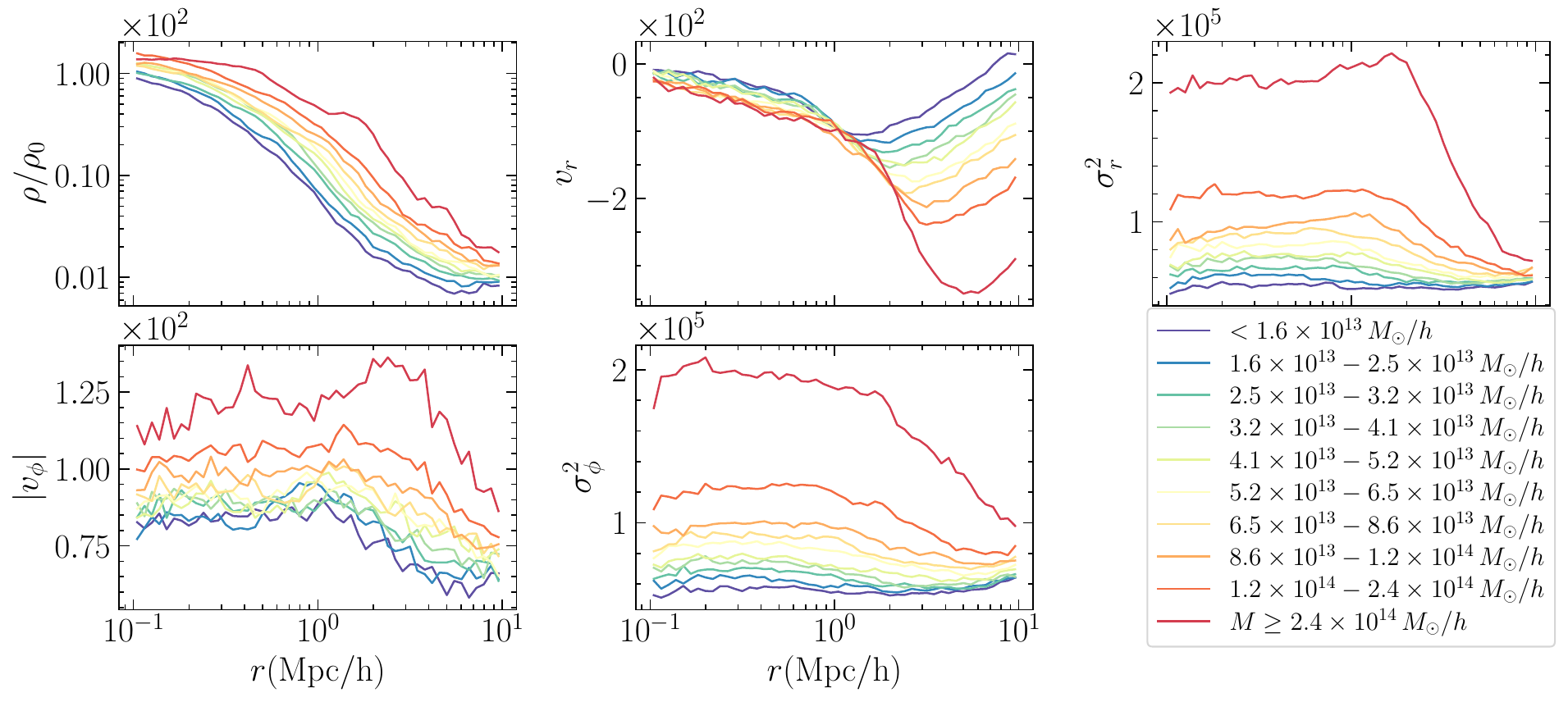}
    \caption{Node mass dependence of radial profiles of high curvature regions in the filament. The figure is similar to figure~\ref{fig:mnode_dep}, but for the segments with top $10$ percentiles in curvature.}
    \label{fig:high_kappa}
\end{figure}
In the main text, we studied the dependence of stacked filament profiles on velocity-defined radius $R_v$, node mass, and filament length $L_{\mathrm{fil}}$ after removing segments belonging to the highest $10\%$ in curvature. This was done to isolate the behaviour of relatively straight filament segments and to reduce geometric biases associated with strongly bent structures.

In this appendix, we instead focus directly on the high-curvature population. Specifically, we repeat the analysis for filament segments belonging to the top $10\%$ in curvature, while keeping the same binning in node mass as used in the main text. The goal is to assess whether strongly curved filaments exhibit systematically different phase-space behaviour compared to the full and curvature-cleaned samples.

Figure~\ref{fig:high_kappa} shows the stacked profiles in bins of total node mass for this high-curvature subset. The qualitative trends are broadly consistent with those seen in the main text: higher node mass systems are denser, exhibit stronger radial infall, and show enhanced velocity dispersion relative to lower-mass systems. However, the signals are noticeably noisier, reflecting the reduced sample size. No qualitatively new trends are introduced by restricting to high-curvature regions. The observed behaviour remains consistent with the combined interpretation of curvature and node-mass dependence discussed in the main analysis. In particular, curvature primarily modulates the scatter and clarity of the profiles rather than introducing a distinct physical trend.

Similar behaviour is observed when binning in $R_v$ and $L_{\mathrm{fil}}$: the overall ordering of profiles remains unchanged, and the dependence on these quantities is consistent with that reported in the main text. For brevity, we show only the node-mass dependence here, as it provides the clearest comparison with the curvature-cleaned case.

\section{Different definitions of filament radius}
\label{app: diff rfil def}
\begin{figure}
    \centering
    \includegraphics[width=\linewidth]{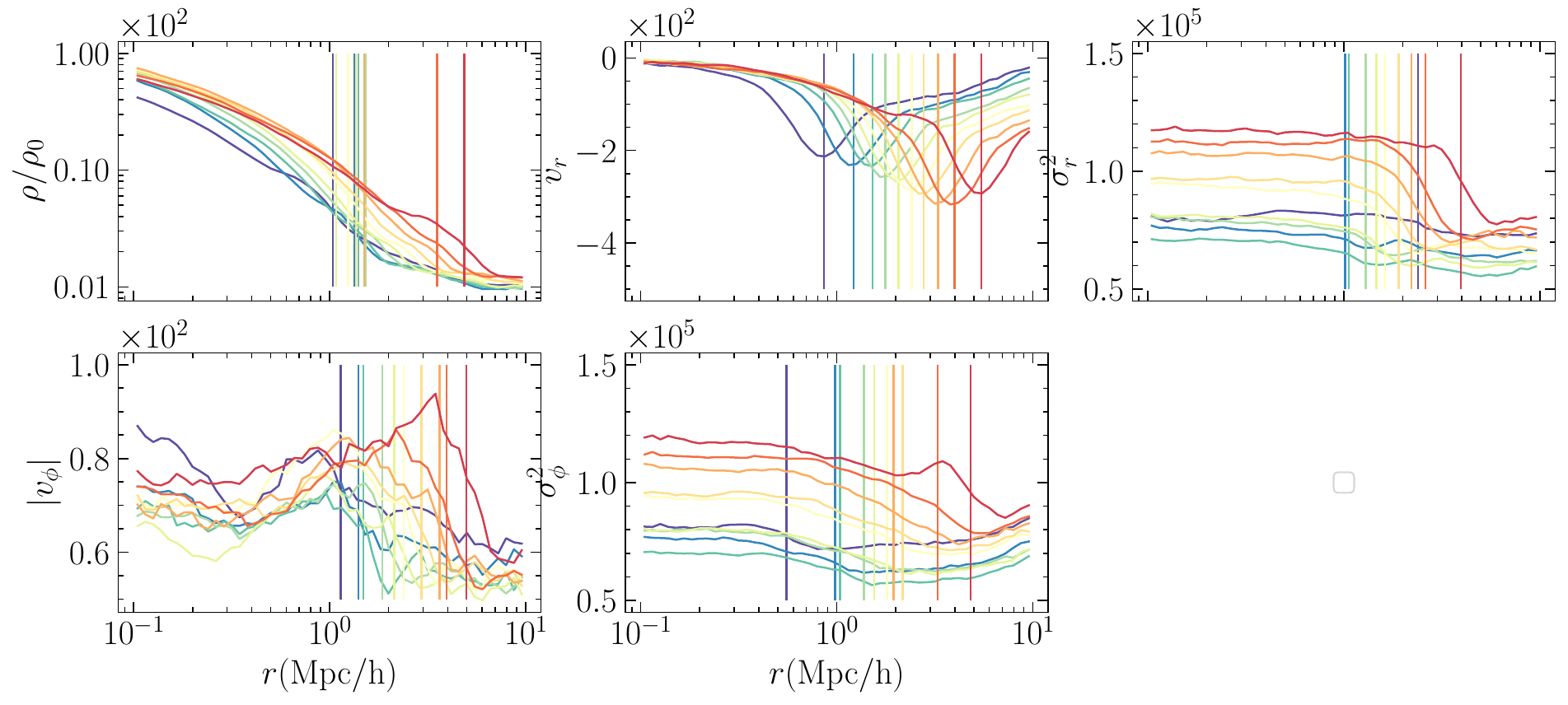}
     \caption{Same as Fig.~\ref{fig:rv_dep}, but with vertical lines marking the estimates of $R_{\mathrm{fil}}$ obtained from the corresponding radial profiles in each $R_v$ bin.}
    \label{fig:rfil_stacks}
\end{figure}

\begin{figure}
    \centering
    \includegraphics[width=\linewidth]{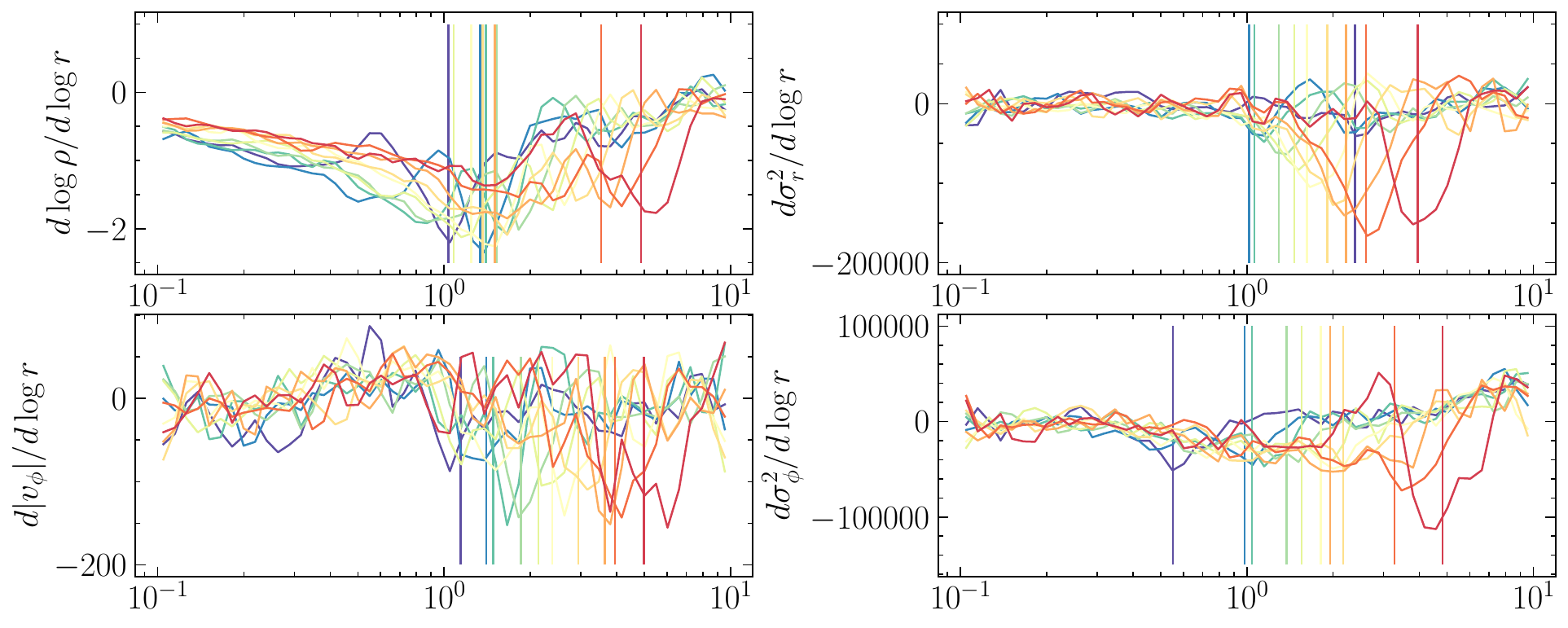}
    \caption{Derivatives of the stacked radial profiles used to estimate the filament radii. The vertical lines indicate the locations of the corresponding minima used in the radius definition.}
    \label{fig:rfil_stacks_deriv}
\end{figure}
This appendix presents the derivatives of stacked profiles used to obtain characteristic radius estimates in bins of $R_v$, as discussed in Section~\ref{sec: filament boundaries}. 

For each $R_v$ percentile bin, we compute stacked profiles of density, radial and tangential velocity, and velocity dispersion, together with their corresponding logarithmic derivatives, except for $v_r$. Characteristic radii are then defined from the minima of these quantities: $d\log\rho/d\log r$, $v_r$, $d|v_\phi|/d\log r$, and $d\sigma_i^2/d\log r$ with $i \in \{r,\phi\}$. Each minimum is estimated using a local quadratic fit over five neighbouring bins.

Figure~\ref{fig:rfil_stacks} shows the stacked profiles with the inferred radii indicated by vertical lines. Figure~\ref{fig:rfil_stacks_deriv} shows the corresponding derivative profiles used to locate these extrema. The radial velocity profile itself is not shown in the derivative panel since its minimum is directly used for the radius estimate. The vertical lines mark the characteristic radii obtained from each definition.

\clearemptydoublepage
\backmatter
\bibliographystyle{siam}
\refstepcounter{chapter}
\bibliography{superbib}

@preamble{ "\newcommand\aap{A\&A} \let\astap=\aap
           \newcommand\aapr{A\&ARv} \newcommand\aaps{A\&AS}
           \newcommand\actaa{Acta Astron.} \newcommand\afz{Afz}
           \newcommand\aj{AJ} \newcommand\ao{Appl. Opt.}
           \let\applopt=\ao \newcommand\aplett{Astrophys.~Lett.}
           \newcommand\apj{ApJ} \newcommand\apjl{ApJ} \let\apjlett=\apjl
           \newcommand\apjs{ApJS} \let\apjsupp=\apjs
           \newcommand\apss{Ap\&SS} \newcommand\araa{ARA\&A}
           \newcommand\arep{Astron. Rep.} \newcommand\aspc{ASP Conf.
           Ser.} \newcommand\azh{Azh} \newcommand\baas{BAAS}
           \newcommand\bac{Bull. Astron. Inst. Czechoslovakia}
           \newcommand\bain{Bull. Astron. Inst. Netherlands}
           \newcommand\caa{Chinese Astron. Astrophys.}
           \newcommand\cjaa{Chinese J.~Astron. Astrophys.}
           \newcommand\fcp{Fundamentals Cosmic Phys.}
           \newcommand\gca{Geochimica Cosmochimica Acta}
           \newcommand\grl{Geophys. Res. Lett.}
           \newcommand\iaucirc{IAU~Circ.} \newcommand\icarus{Icarus}
           \newcommand\japa{J.~Astrophys. Astron.}
           \newcommand\jcap{J.~Cosmology Astropart. Phys.}
           \newcommand\jcp{J.~Chem.~Phys.}
           \newcommand\jgr{J.~Geophys.~Res.} \newcommand\jqsrt{J.~Quant.
           Spectrosc. Radiative Transfer}
           \newcommand\jrasc{J.~R.~Astron. Soc. Canada}
           \newcommand\mnassa{MNASSA} \newcommand\mnras{MNRAS}
           \newcommand\memras{Mem.~RAS} \newcommand\memsai{Mem. Soc.
           Astron. Italiana} \newcommand\nat{Nature}
           \newcommand\na{New~Astron.} \newcommand\nar{New~Astron.~Rev.}
           \newcommand\nphysa{Nuclear Phys.~A} \newcommand\pra{Phys.
           Rev.~A} \newcommand\prb{Phys. Rev.~B} \newcommand\prd{Phys.
           Rev.~D} \newcommand\pre{Phys. } \newcommand\prl{Phys.
           Rev.~Lett.} \newcommand\pasa{Publ. Astron. Soc. Australia}
           \newcommand\pasp{PASP} \newcommand\pasj{PASJ}
           \newcommand\physrep{Phys.~Rep.}
           \newcommand\physscr{Phys.~Scr.} \newcommand\planss{Planet.
           Space~Sci.} \newcommand\rmxaa{Rev. Mex. Astron. Astrofis.}
           \newcommand\qjras{QJRAS} \newcommand\sci{Science}
           \newcommand\skytel{Sky \& Telesc.}
           \newcommand\solphys{Sol.~Phys.}
           \newcommand\sovast{Soviet~Ast.} \newcommand\ssr{Space Sci.
           Rev.} \newcommand\zap{Z.~Astrophys.} "
}

@article{Springel2005,
  author =        {Springel, Volker and White, Simon D. M. and
                   Jenkins, Adrian and others},
  journal =       {Nature},
  number =        {7042},
  pages =         {629-636},
  title =         {Simulations of the formation, evolution and
                   clustering of galaxies and quasars},
  volume =        {435},
  year =          {2005},
  doi =           {10.1038/nature03597},
}

@article{Angulo+2008,
  author =        {{Angulo}, R.~E. and {Baugh}, C.~M. and {Frenk}, C.~S. and
                   {Lacey}, C.~G.},
  journal =       {\mnras},
  month =         jan,
  number =        {2},
  pages =         {755-776},
  title =         {{The detectability of baryonic acoustic oscillations
                   in future galaxy surveys}},
  volume =        {383},
  year =          {2008},
  doi =           {10.1111/j.1365-2966.2007.12587.x},
}

@article{Heitmann+2014,
  author =        {{Heitmann}, Katrin and {Lawrence}, Earl and
                   {Kwan}, Juliana and {Habib}, Salman and
                   {Higdon}, David},
  journal =       {\apj},
  month =         jan,
  number =        {1},
  pages =         {111},
  title =         {{The Coyote Universe Extended: Precision Emulation of
                   the Matter Power Spectrum}},
  volume =        {780},
  year =          {2014},
  doi =           {10.1088/0004-637X/780/1/111},
  eid =           {111},
}

@article{TNS,
  author =        {{Taruya}, Atsushi and {Nishimichi}, Takahiro and
                   {Saito}, Shun},
  journal =       {\prd},
  month =         {September},
  number =        {6},
  pages =         {063522},
  title =         {{Baryon acoustic oscillations in 2D: Modeling
                   redshift-space power spectrum from perturbation
                   theory}},
  volume =        {82},
  year =          {2010},
  doi =           {10.1103/PhysRevD.82.063522},
  eid =           {063522},
}

@article{Cooray&Sheth2002,
  author =        {{Cooray}, Asantha and {Sheth}, Ravi},
  journal =       {\physrep},
  month =         dec,
  number =        {1},
  pages =         {1-129},
  title =         {{Halo models of large scale structure}},
  volume =        {372},
  year =          {2002},
  doi =           {10.1016/S0370-1573(02)00276-4},
}

@article{Zheng+2005,
  author =        {{Zheng}, Zheng and {Berlind}, Andreas A. and
                   {Weinberg}, David H. and {Benson}, Andrew J. and
                   {Baugh}, Carlton M. and {Cole}, Shaun and
                   {Dav{\'e}}, Romeel and {Frenk}, Carlos S. and
                   {Katz}, Neal and {Lacey}, Cedric G.},
  journal =       {\apj},
  month =         nov,
  number =        {2},
  pages =         {791-809},
  title =         {{Theoretical Models of the Halo Occupation
                   Distribution: Separating Central and Satellite
                   Galaxies}},
  volume =        {633},
  year =          {2005},
  doi =           {10.1086/466510},
}

@article{DESICollaboration2016a,
  author =        {{DESI Collaboration} and Aghamousa, A. and
                   Aguilar, J. and Ahlen, S. and others},
  journal =       {arXiv e-prints},
  title =         {The DESI Experiment Part I: Science, Targeting, and
                   Survey Design},
  year =          {2016},
}

@article{DESICollaboration2016b,
  author =        {{DESI Collaboration} and Aghamousa, A. and
                   Aguilar, J. and Ahlen, S. and others},
  journal =       {arXiv e-prints},
  title =         {The DESI Experiment Part II: Instrument Design},
  year =          {2016},
}

@article{EuclidCollaboration2022,
  author =        {{Euclid Collaboration} and Mellier, Y. and
                   Abdalla, F. B. and others},
  journal =       {arXiv e-prints},
  title =         {Euclid: Overview of the mission and survey},
  year =          {2022},
}

@article{Ivezic2019,
  author =        {Ivezi{\'c}, {\v{Z}}eljko and Kahn, Steven M. and
                   Tyson, J. Anthony and others},
  journal =       {The Astrophysical Journal},
  number =        {2},
  pages =         {111},
  title =         {LSST: From Science Drivers to Reference Design and
                   Anticipated Data Products},
  volume =        {873},
  year =          {2019},
  doi =           {10.3847/1538-4357/ab042c},
}

@article{Wechsler&Tinker2018,
  author =        {{Wechsler}, Risa H. and {Tinker}, Jeremy L.},
  journal =       {\araa},
  month =         sep,
  pages =         {435-487},
  title =         {{The Connection Between Galaxies and Their Dark
                   Matter Halos}},
  volume =        {56},
  year =          {2018},
  doi =           {10.1146/annurev-astro-081817-051756},
}

@article{Joachimi+2015,
  author =        {{Joachimi}, Benjamin and {Cacciato}, Marcello and
                   {Kitching}, Thomas D. and {Leonard}, Adrienne and
                   {Mandelbaum}, Rachel and
                   {Sch{\"a}fer}, Bj{\"o}rn Malte and
                   {Sif{\'o}n}, Crist{\'o}bal and {Hoekstra}, Henk and
                   {Kiessling}, Alina and {Kirk}, Donnacha and
                   {Rassat}, Anais},
  journal =       {\ssr},
  month =         nov,
  number =        {1-4},
  pages =         {1-65},
  title =         {{Galaxy Alignments: An Overview}},
  volume =        {193},
  year =          {2015},
  doi =           {10.1007/s11214-015-0177-4},
}

@article{Percival+2011,
  author =        {{Percival}, W.~J. and {Samushia}, L. and
                   {Ross}, A.~J. and {Shapiro}, C. and {Raccanelli}, A.},
  journal =       {Philosophical Transactions of the Royal Society of
                   London Series A},
  month =         dec,
  number =        {1957},
  pages =         {5058-5067},
  title =         {{Redshift-space distortions}},
  volume =        {369},
  year =          {2011},
  doi =           {10.1098/rsta.2011.0370},
}

@article{Daalen+2011,
  author =        {{van Daalen}, Marcel P. and {Schaye}, Joop and
                   {Booth}, C.~M. and {Dalla Vecchia}, Claudio},
  journal =       {\mnras},
  month =         {August},
  number =        {4},
  pages =         {3649-3665},
  title =         {{The effects of galaxy formation on the matter power
                   spectrum: a challenge for precision cosmology}},
  volume =        {415},
  year =          {2011},
  doi =           {10.1111/j.1365-2966.2011.18981.x},
}

@article{Semboloni+2011,
  author =        {{Semboloni}, Elisabetta and {Hoekstra}, Henk and
                   {Schaye}, Joop and {van Daalen}, Marcel P. and
                   {McCarthy}, Ian G.},
  journal =       {\mnras},
  month =         nov,
  number =        {3},
  pages =         {2020-2035},
  title =         {{Quantifying the effect of baryon physics on weak
                   lensing tomography}},
  volume =        {417},
  year =          {2011},
  doi =           {10.1111/j.1365-2966.2011.19385.x},
}

@article{GADGET2005,
  author =        {{Springel}, Volker},
  journal =       {\mnras},
  month =         {December},
  number =        {4},
  pages =         {1105-1134},
  title =         {{The cosmological simulation code GADGET-2}},
  volume =        {364},
  year =          {2005},
  doi =           {10.1111/j.1365-2966.2005.09655.x},
}

@article{Gadget4,
  author =        {{Springel}, Volker and {Pakmor}, R{\"u}diger and
                   {Zier}, Oliver and {Reinecke}, Martin},
  journal =       {\mnras},
  month =         {September},
  number =        {2},
  pages =         {2871-2949},
  title =         {{Simulating cosmic structure formation with the
                   GADGET-4 code}},
  volume =        {506},
  year =          {2021},
  doi =           {10.1093/mnras/stab1855},
}

@article{AREPO2010,
  author =        {{Springel}, Volker},
  journal =       {\mnras},
  month =         jan,
  number =        {2},
  pages =         {791-851},
  title =         {{E pur si muove: Galilean-invariant cosmological
                   hydrodynamical simulations on a moving mesh}},
  volume =        {401},
  year =          {2010},
  doi =           {10.1111/j.1365-2966.2009.15715.x},
}

@article{RAMSES2002,
  author =        {{Teyssier}, R.},
  journal =       {\aap},
  month =         apr,
  pages =         {337-364},
  title =         {{Cosmological hydrodynamics with adaptive mesh
                   refinement. A new high resolution code called
                   RAMSES}},
  volume =        {385},
  year =          {2002},
  doi =           {10.1051/0004-6361:20011817},
}

@phdthesis{PKDGRAV2001,
  author =        {{Stadel}, Joachim Gerhard},
  month =         jan,
  school =        {University of Washington, Seattle},
  title =         {{Cosmological N-body simulations and their analysis}},
  year =          {2001},
}

@ARTICLE{Potter+2017,
       author = {{Potter}, Douglas and {Stadel}, Joachim and {Teyssier}, Romain},
        title = "{PKDGRAV3: beyond trillion particle cosmological simulations for the next era of galaxy surveys}",
      journal = {Computational Astrophysics and Cosmology},
         year = 2017,
        month = may,
       volume = {4},
       number = {1},
          eid = {2},
        pages = {2},
          doi = {10.1186/s40668-017-0021-1},
archivePrefix = {arXiv},
       eprint = {1609.08621},
 primaryClass = {astro-ph.IM},
       adsurl = {https://ui.adsabs.harvard.edu/abs/2017ComAC...4....2P}
}

@article{ABACUS_code2021,
  author =        {{Garrison}, Lehman H. and {Eisenstein}, Daniel J. and
                   {Ferrer}, Douglas and {Maksimova}, Nina A. and
                   {Pinto}, Philip A.},
  journal =       {\mnras},
  month =         nov,
  number =        {1},
  pages =         {575-596},
  title =         {{The ABACUS cosmological N-body code}},
  volume =        {508},
  year =          {2021},
  doi =           {10.1093/mnras/stab2482},
}

@article{CUBE3PM,
  author =        {{Harnois-D{\'e}raps}, Joachim and {Pen}, Ue-Li and
                   {Iliev}, Ilian T. and {Merz}, Hugh and
                   {Emberson}, J.~D. and {Desjacques}, Vincent},
  journal =       {\mnras},
  month =         nov,
  number =        {1},
  pages =         {540-559},
  title =         {{High-performance P$^{3}$M N-body code:
                   CUBEP$^{3}$M}},
  volume =        {436},
  year =          {2013},
  doi =           {10.1093/mnras/stt1591},
}

@article{HACC2016,
  author =        {{Habib}, Salman and {Pope}, Adrian and {Finkel}, Hal and
                   {Frontiere}, Nicholas and {Heitmann}, Katrin and
                   {Daniel}, David and {Fasel}, Patricia and
                   {Morozov}, Vitali and {Zagaris}, George and
                   {Peterka}, Tom and {Vishwanath}, Venkatram and
                   {Luki{\'c}}, Zarija and {Sehrish}, Saba and
                   {Liao}, Wei-keng},
  journal =       {\na},
  month =         jan,
  pages =         {49-65},
  title =         {{HACC: Simulating sky surveys on state-of-the-art
                   supercomputing architectures}},
  volume =        {42},
  year =          {2016},
  doi =           {10.1016/j.newast.2015.06.003},
}

@article{SWIFT2024,
  author =        {{Schaller}, Matthieu and {Borrow}, Josh and
                   {Draper}, Peter W. and {Ivkovic}, Mladen and
                   {McAlpine}, Stuart and {Vandenbroucke}, Bert and
                   {Bah{\'e}}, Yannick and {Chaikin}, Evgenii and
                   {Chalk}, Aidan B.~G. and {Chan}, Tsang Keung and
                   {Correa}, Camila and {van Daalen}, Marcel and
                   {Elbers}, Willem and {Gonnet}, Pedro and
                   {Hausammann}, Lo{\"\i}c and {Helly}, John and
                   {Hu{\v{s}}ko}, Filip and {Kegerreis}, Jacob A. and
                   {Nobels}, Folkert S.~J. and {Ploeckinger}, Sylvia and
                   {Revaz}, Yves and {Roper}, William J. and
                   {Ruiz-Bonilla}, Sergio and {Sandnes}, Thomas D. and
                   {Uyttenhove}, Yolan and {Willis}, James S. and
                   {Xiang}, Zhen},
  journal =       {\mnras},
  month =         may,
  number =        {2},
  pages =         {2378-2419},
  title =         {{SWIFT: A modern highly-parallel gravity and smoothed
                   particle hydrodynamics solver for astrophysical and
                   cosmological applications}},
  volume =        {530},
  year =          {2024},
  doi =           {10.1093/mnras/stae922},
}

@article{Davis+1985,
  author =        {{Davis}, M. and {Efstathiou}, G. and {Frenk}, C.~S. and
                   {White}, S.~D.~M.},
  journal =       {\apj},
  month =         may,
  pages =         {371-394},
  title =         {{The evolution of large-scale structure in a universe
                   dominated by cold dark matter}},
  volume =        {292},
  year =          {1985},
  doi =           {10.1086/163168},
}

@article{Springel+2001,
  author =        {{Springel}, Volker and {White}, Simon D.~M. and
                   {Tormen}, Giuseppe and {Kauffmann}, Guinevere},
  journal =       {\mnras},
  month =         dec,
  number =        {3},
  pages =         {726-750},
  title =         {{Populating a cluster of galaxies - I. Results at
                   z=0}},
  volume =        {328},
  year =          {2001},
  doi =           {10.1046/j.1365-8711.2001.04912.x},
}

@article{Behroozi2019,
  author =        {Behroozi, Peter and Wechsler, Risa H. and
                   Hearin, Andrew P. and Conroy, Charlie},
  journal =       {Monthly Notices of the Royal Astronomical Society},
  number =        {3},
  pages =         {3143-3194},
  title =         {UniverseMachine: The correlation between galaxy
                   growth and dark matter halo assembly from z = 0-10},
  volume =        {488},
  year =          {2019},
  doi =           {10.1093/mnras/stz1182},
}

@article{Nelson2019,
  author =        {Nelson, Dylan and Pillepich, Annalisa and
                   Springel, Volker and others},
  journal =       {Computational Astrophysics and Cosmology},
  number =        {1},
  pages =         {2},
  title =         {The IllustrisTNG simulations: public data release},
  volume =        {6},
  year =          {2019},
  doi =           {10.1186/s40668-019-0028-x},
}

@article{Schaye2015,
  author =        {Schaye, Joop and Crain, Robert A. and
                   Bower, Richard G. and others},
  journal =       {Monthly Notices of the Royal Astronomical Society},
  number =        {1},
  pages =         {521-554},
  title =         {The EAGLE project: simulating the evolution and
                   assembly of galaxies and their environments},
  volume =        {446},
  year =          {2015},
  doi =           {10.1093/mnras/stu2058},
}

@article{Dave2019,
  author =        {Dav{\'e}, Romeel and Angl{\'e}s-Alc{\'a}zar, Daniel and
                   Narayanan, Desika and others},
  journal =       {Monthly Notices of the Royal Astronomical Society},
  number =        {2},
  pages =         {2827-2849},
  title =         {SIMBA: Cosmological simulations with black hole
                   growth and feedback},
  volume =        {486},
  year =          {2019},
  doi =           {10.1093/mnras/stz937},
}

@article{Chisari+2019,
  author =        {{Chisari}, Nora Elisa and {Mead}, Alexander J. and
                   {Joudaki}, Shahab and {Ferreira}, Pedro G. and
                   {Schneider}, Aurel and {Mohr}, Joseph and
                   {Tr{\"o}ster}, Tilman and {Alonso}, David and
                   {McCarthy}, Ian G. and {Martin-Alvarez}, Sergio and
                   {Devriendt}, Julien and {Slyz}, Adrianne and
                   {van Daalen}, Marcel P.},
  journal =       {The Open Journal of Astrophysics},
  month =         jun,
  number =        {1},
  pages =         {4},
  title =         {{Modelling baryonic feedback for survey cosmology}},
  volume =        {2},
  year =          {2019},
  doi =           {10.21105/astro.1905.06082},
  eid =           {4},
}

@article{4MOST2019,
  author =        {de Jong, R. S. and Agertz, O. and Berbel, A. Agudo and
                   others},
  journal =       {The Messenger},
  pages =         {3-11},
  title =         {4MOST: Project overview and information for the First
                   Call for Proposals},
  volume =        {175},
  year =          {2019},
  doi =           {10.18727/0722-6691/5117},
}

@article{Takada2014,
  author =        {Takada, Masahiro and Ellis, Richard S. and
                   Chiba, Masashi and others},
  journal =       {Publications of the Astronomical Society of Japan},
  number =        {1},
  pages =         {R1},
  title =         {Extragalactic science, cosmology, and Galactic
                   archaeology with the Subaru Prime Focus Spectrograph},
  volume =        {66},
  year =          {2014},
  doi =           {10.1093/pasj/pst019},
}

@article{Ishiyama2021,
  author =        {Ishiyama, Tomoaki and Prada, Francisco and
                   Klypin, Anatoly A. and others},
  journal =       {Monthly Notices of the Royal Astronomical Society},
  number =        {3},
  pages =         {4210-4231},
  title =         {The Uchuu simulations: Data Release 1 and dark matter
                   halo concentrations},
  volume =        {506},
  year =          {2021},
  doi =           {10.1093/mnras/stab1755},
}

@article{Vogelsberger2014,
  author =        {Vogelsberger, Mark and Genel, Shy and
                   Springel, Volker and others},
  journal =       {Monthly Notices of the Royal Astronomical Society},
  number =        {2},
  pages =         {1518-1547},
  title =         {Introducing the Illustris Project: simulating the
                   coevolution of dark and visible matter in the
                   Universe},
  volume =        {444},
  year =          {2014},
  doi =           {10.1093/mnras/stu1536},
}

@article{Pillepich2018,
  author =        {Pillepich, Annalisa and Nelson, Dylan and
                   Hernquist, Lars and others},
  journal =       {Monthly Notices of the Royal Astronomical Society},
  number =        {1},
  pages =         {648-675},
  title =         {First results from the IllustrisTNG simulations: the
                   galaxy colour bimodality},
  volume =        {475},
  year =          {2018},
  doi =           {10.1093/mnras/stx3112},
}

@article{Boylan-Kolchin2009,
  author =        {Boylan-Kolchin, Michael and Springel, Volker and
                   White, Simon D. M. and Jenkins, Adrian and
                   Lemson, Gerard},
  journal =       {Monthly Notices of the Royal Astronomical Society},
  number =        {3},
  pages =         {1150-1164},
  title =         {Resolving cosmic structure formation with the
                   Millennium-II Simulation},
  volume =        {398},
  year =          {2009},
  doi =           {10.1111/j.1365-2966.2009.15191.x},
}

@article{Klypin2011,
  author =        {Klypin, Anatoly A. and Trujillo-Gomez, Sebastian and
                   Primack, Joel},
  journal =       {The Astrophysical Journal},
  number =        {2},
  pages =         {102},
  title =         {Dark Matter Halos in the Standard Cosmological Model:
                   Results from the Bolshoi Simulation},
  volume =        {740},
  year =          {2011},
  doi =           {10.1088/0004-637X/740/2/102},
}

@article{Klypin2016,
  author =        {Klypin, Anatoly and Yepes, Gustavo and
                   Gottl{\"o}ber, Stefan and Prada, Francisco and
                   He{\ss}, Steffen},
  journal =       {Monthly Notices of the Royal Astronomical Society},
  number =        {4},
  pages =         {4340-4359},
  title =         {MultiDark simulations: the story of dark matter halo
                   concentrations and density profiles},
  volume =        {457},
  year =          {2016},
  doi =           {10.1093/mnras/stw248},
}

@article{camels-2021,
  author =        {{Villaescusa-Navarro}, Francisco and
                   {Angl{\'e}s-Alc{\'a}zar}, Daniel and {Genel}, Shy and
                   {Spergel}, David N. and {Somerville}, Rachel S. and
                   {Dave}, Romeel and {Pillepich}, Annalisa and
                   {Hernquist}, Lars and {Nelson}, Dylan and
                   {Torrey}, Paul and {Narayanan}, Desika and {Li}, Yin and
                   {Philcox}, Oliver and {La Torre}, Valentina and
                   {Maria Delgado}, Ana and {Ho}, Shirley and
                   {Hassan}, Sultan and {Burkhart}, Blakesley and
                   {Wadekar}, Digvijay and {Battaglia}, Nicholas and
                   {Contardo}, Gabriella and {Bryan}, Greg L.},
  journal =       {\apj},
  month =         {July},
  number =        {1},
  pages =         {71},
  title =         {{The CAMELS Project: Cosmology and Astrophysics with
                   Machine-learning Simulations}},
  volume =        {915},
  year =          {2021},
  doi =           {10.3847/1538-4357/abf7ba},
  eid =           {71},
}

@article{Villaescusa-Navarro2020,
  author =        {Villaescusa-Navarro, Francisco and Hahn, ChangHoon and
                   Massara, Elena and others},
  journal =       {The Astrophysical Journal Supplement Series},
  number =        {1},
  pages =         {2},
  title =         {The Quijote simulations},
  volume =        {250},
  year =          {2020},
  doi =           {10.3847/1538-4365/ab9d82},
}

@article{DeRose2019,
  author =        {DeRose, Joseph and Wechsler, Risa H. and
                   Tinker, Jeremy L. and others},
  journal =       {The Astrophysical Journal},
  number =        {1},
  pages =         {69},
  title =         {The Aemulus Project. II. Emulating the Halo Mass
                   Function},
  volume =        {875},
  year =          {2019},
  doi =           {10.3847/1538-4357/ab1085},
}

@article{McClintock2019,
  author =        {McClintock, Thomas and Varga, T. N. and Gruen, D. and
                   others},
  journal =       {Monthly Notices of the Royal Astronomical Society},
  number =        {1},
  pages =         {1352-1378},
  title =         {Dark Energy Survey Year 1 Results: Weak Lensing Mass
                   Calibration of redMaPPer Galaxy Clusters},
  volume =        {482},
  year =          {2019},
  doi =           {10.1093/mnras/sty2711},
}

@article{Nishimichi2019,
  author =        {Nishimichi, Takahiro and Shirata, Akihito and
                   Taruya, Atsushi and others},
  journal =       {The Astrophysical Journal},
  number =        {1},
  pages =         {29},
  title =         {Dark Quest. I. Fast and Accurate Emulation of Halo
                   Clustering Statistics and Its Application to Galaxy
                   Clustering},
  volume =        {884},
  year =          {2019},
  doi =           {10.3847/1538-4357/ab3719},
}

@article{Maksimova2021,
  author =        {Maksimova, Nina A. and Garrison, Lehman H. and
                   Hadzhiyska, Boryana and others},
  journal =       {Monthly Notices of the Royal Astronomical Society},
  number =        {3},
  pages =         {4017-4037},
  title =         {AbacusSummit: a massive set of high-accuracy,
                   high-resolution N-body simulations},
  volume =        {508},
  year =          {2021},
  doi =           {10.1093/mnras/stab2484},
}

@article{Alam_2021,
  author =        {Alam, Shadab and Peacock, John A and Farrow, Daniel J and
                   Loveday, J and Hopkins, A M},
  journal =       {Monthly Notices of the Royal Astronomical Society},
  month =         {February},
  number =        {1},
  pages =         {59–76},
  publisher =     {Oxford University Press (OUP)},
  title =         {Using GAMA to probe the impact of small-scale galaxy
                   physics on nonlinear redshift-space distortions},
  volume =        {503},
  year =          {2021},
  doi =           {10.1093/mnras/stab409},
  issn =          {1365-2966},
  url =           {http://dx.doi.org/10.1093/mnras/stab409},
}

@article{Hahn2023,
  author =        {Hahn, ChangHoon and Wilson, Michael J. and
                   Ruiz-Macias, Omar and others},
  journal =       {The Astronomical Journal},
  number =        {6},
  pages =         {253},
  title =         {The DESI Bright Galaxy Survey: Final Target
                   Selection, Design, and Validation},
  volume =        {165},
  year =          {2023},
  doi =           {10.3847/1538-3881/accff8},
}

@article{Yildiz2020,
  author =        {Yildiz, Eleonora and Biagetti, Matteo and
                   Seljak, Uro{\v{s}} and others},
  journal =       {Monthly Notices of the Royal Astronomical Society},
  number =        {4},
  pages =         {4783-4815},
  title =         {Towards testing the theory of gravity with DESI:
                   summary statistics, model predictions and future
                   simulation requirements},
  volume =        {493},
  year =          {2020},
  doi =           {10.1093/mnras/staa421},
}

@article{Baldry2012,
  author =        {Baldry, I. K. and Driver, S. P. and Loveday, J. and
                   others},
  journal =       {Monthly Notices of the Royal Astronomical Society},
  number =        {1},
  pages =         {621-634},
  title =         {Galaxy And Mass Assembly (GAMA): the galaxy stellar
                   mass function at z < 0.06},
  volume =        {421},
  year =          {2012},
  doi =           {10.1111/j.1365-2966.2012.20340.x},
}

@article{Wetzel2013,
  author =        {Wetzel, Andrew R. and Tinker, Jeremy L. and
                   Conroy, Charlie and van den Bosch, Frank C.},
  journal =       {Monthly Notices of the Royal Astronomical Society},
  number =        {1},
  pages =         {336-358},
  title =         {Galaxy Evolution in Groups and Clusters: Satellite
                   Star Formation Histories and Quenching Time-scales in
                   a Hierarchical Universe},
  volume =        {432},
  year =          {2013},
  doi =           {10.1093/mnras/stt469},
}

@article{Reines2013,
  author =        {Reines, Amy E. and Greene, Jenny E. and Geha, Marla},
  journal =       {The Astrophysical Journal},
  number =        {2},
  pages =         {116},
  title =         {Dwarf Galaxies with Optical Signatures of Active
                   Massive Black Holes},
  volume =        {775},
  year =          {2013},
  doi =           {10.1088/0004-637X/775/2/116},
}

@article{Riess2022,
  author =        {Riess, Adam G. and Yuan, Wenlong and Macri, Lucas M. and
                   others},
  journal =       {The Astrophysical Journal Letters},
  number =        {1},
  pages =         {L7},
  title =         {A Comprehensive Measurement of the Local Value of the
                   Hubble Constant with 1 km s$^{-1}$ Mpc$^{-1}$
                   Uncertainty from the Hubble Space Telescope and the
                   SH0ES Team},
  volume =        {934},
  year =          {2022},
  doi =           {10.3847/2041-8213/ac5c5b},
}

@article{Planck18-VI-cosmoparam,
  author =        {{Planck Collaboration} and Aghanim, N. and Akrami, Y. and
                   others},
  journal =       {Astronomy \& Astrophysics},
  pages =         {A6},
  title =         {Planck 2018 results. VI. Cosmological parameters},
  volume =        {641},
  year =          {2020},
  doi =           {10.1051/0004-6361/201833910},
}

@article{DESIDR1_cosmo,
  author =        {{DESI Collaboration} and {Adame}, A.~G. and
                   {Aguilar}, J. and {Ahlen}, S. and {Alam}, S. and
                   {Alexander}, D.~M. and {Alvarez}, M. and {Alves}, O. and
                   {Anand}, A. and {Andrade}, U. and {Armengaud}, E. and
                   {Avila}, S. and {Aviles}, A. and {Awan}, H. and
                   {Bahr-Kalus}, B. and {Bailey}, S. and {Baltay}, C. and
                   {Bault}, A. and {Behera}, J. and {BenZvi}, S. and
                   {Bera}, A. and {Beutler}, F. and {Bianchi}, D. and
                   {Blake}, C. and {Blum}, R. and {Brieden}, S. and
                   {Brodzeller}, A. and {Brooks}, D. and
                   {Buckley-Geer}, E. and {Burtin}, E. and
                   {Calderon}, R. and {Canning}, R. and
                   {Carnero Rosell}, A. and {Cereskaite}, R. and
                   {Cervantes-Cota}, J.~L. and {Chabanier}, S. and
                   {Chaussidon}, E. and {Chaves-Montero}, J. and
                   {Chen}, S. and {Chen}, X. and {Claybaugh}, T. and
                   {Cole}, S. and {Cuceu}, A. and {Davis}, T.~M. and
                   {Dawson}, K. and {de la Macorra}, A. and
                   {de Mattia}, A. and {Deiosso}, N. and {Dey}, A. and
                   {Dey}, B. and {Ding}, Z. and {Doel}, P. and
                   {Edelstein}, J. and {Eftekharzadeh}, S. and
                   {Eisenstein}, D.~J. and {Elliott}, A. and
                   {Fagrelius}, P. and {Fanning}, K. and {Ferraro}, S. and
                   {Ereza}, J. and {Findlay}, N. and {Flaugher}, B. and
                   {Font-Ribera}, A. and {Forero-S{\'a}nchez}, D. and
                   {Forero-Romero}, J.~E. and {Frenk}, C.~S. and
                   {Garcia-Quintero}, C. and {Gazta{\~n}aga}, E. and
                   {Gil-Mar{\'\i}n}, H. and {Gontcho}, S. Gontcho A and
                   {Gonzalez-Morales}, A.~X. and {Gonzalez-Perez}, V. and
                   {Gordon}, C. and {Green}, D. and {Gruen}, D. and
                   {Gsponer}, R. and {Gutierrez}, G. and {Guy}, J. and
                   {Hadzhiyska}, B. and {Hahn}, C. and {Hanif}, M.~M. S and
                   {Herrera-Alcantar}, H.~K. and {Honscheid}, K. and
                   {Howlett}, C. and {Huterer}, D. and
                   {Ir{\v{s}}i{\v{c}}}, V. and {Ishak}, M. and
                   {Juneau}, S. and {Kara{\c{c}}ayl{\i}}, N.~G. and
                   {Kehoe}, R. and {Kent}, S. and {Kirkby}, D. and
                   {Kremin}, A. and {Krolewski}, A. and {Lai}, Y. and
                   {Lan}, T. -W. and {Landriau}, M. and {Lang}, D. and
                   {Lasker}, J. and {Le Goff}, J.~M. and
                   {Le Guillou}, L. and {Leauthaud}, A. and
                   {Levi}, M.~E. and {Li}, T.~S. and {Linder}, E. and
                   {Lodha}, K. and {Magneville}, C. and {Manera}, M. and
                   {Margala}, D. and {Martini}, P. and {Maus}, M. and
                   {McDonald}, P. and {Medina-Varela}, L. and
                   {Meisner}, A. and {Mena-Fern{\'a}ndez}, J. and
                   {Miquel}, R. and {Moon}, J. and {Moore}, S. and
                   {Moustakas}, J. and {Mudur}, N. and {Mueller}, E. and
                   {Mu{\~n}oz-Guti{\'e}rrez}, A. and {Myers}, A.~D. and
                   {Nadathur}, S. and {Napolitano}, L. and {Neveux}, R. and
                   {Newman}, J.~A. and {Nguyen}, N.~M. and {Nie}, J. and
                   {Niz}, G. and {Noriega}, H.~E. and {Padmanabhan}, N. and
                   {Paillas}, E. and {Palanque-Delabrouille}, N. and
                   {Pan}, J. and {Penmetsa}, S. and {Percival}, W.~J. and
                   {Pieri}, M.~M. and {Pinon}, M. and {Poppett}, C. and
                   {Porredon}, A. and {Prada}, F. and
                   {P{\'e}rez-Fern{\'a}ndez}, A. and
                   {P{\'e}rez-R{\`a}fols}, I. and {Rabinowitz}, D. and
                   {Raichoor}, A. and {Ram{\'\i}rez-P{\'e}rez}, C. and
                   {Ramirez-Solano}, S. and {Ravoux}, C. and
                   {Rashkovetskyi}, M. and {Rezaie}, M. and {Rich}, J. and
                   {Rocher}, A. and {Rockosi}, C. and {Roe}, N.~A. and
                   {Rosado-Marin}, A. and {Ross}, A.~J. and {Rossi}, G. and
                   {Ruggeri}, R. and {Ruhlmann-Kleider}, V. and
                   {Samushia}, L. and {Sanchez}, E. and {Saulder}, C. and
                   {Schlafly}, E.~F. and {Schlegel}, D. and
                   {Schubnell}, M. and {Seo}, H. and {Shafieloo}, A. and
                   {Sharples}, R. and {Silber}, J. and {Slosar}, A. and
                   {Smith}, A. and {Sprayberry}, D. and {Tan}, T. and
                   {Tarl{\'e}}, G. and {Taylor}, P. and {Trusov}, S. and
                   {Ure{\~n}a-L{\'o}pez}, L.~A. and {Vaisakh}, R. and
                   {Valcin}, D. and {Valdes}, F. and
                   {Vargas-Maga{\~n}a}, M. and {Verde}, L. and
                   {Walther}, M. and {Wang}, B. and {Wang}, M.~S. and
                   {Weaver}, B.~A. and {Weaverdyck}, N. and
                   {Wechsler}, R.~H. and {Weinberg}, D.~H. and
                   {White}, M. and {Yu}, J. and {Yu}, Y. and {Yuan}, S. and
                   {Y{\`e}che}, C. and {Zaborowski}, E.~A. and
                   {Zarrouk}, P. and {Zhang}, H. and {Zhao}, C. and
                   {Zhao}, R. and {Zhou}, R. and {Zhuang}, T. and
                   {Zou}, H.},
  journal =       {arXiv e-prints},
  month =         {April},
  pages =         {arXiv:2404.03002},
  title =         {{DESI 2024 VI: Cosmological Constraints from the
                   Measurements of Baryon Acoustic Oscillations}},
  year =          {2024},
  doi =           {10.48550/arXiv.2404.03002},
  eid =           {arXiv:2404.03002},
}

@article{DESIDR2_cosmo,
  author =        {Abdul Karim, M. and Aguilar, J. and Ahlen, S. and
                   Alam, S. and Allen, L. and Prieto, C. Allende and
                   Alves, O. and Anand, A. and Andrade, U. and
                   Armengaud, E. and Aviles, A. and Bailey, S. and
                   Baltay, C. and Bansal, P. and Bault, A. and
                   Behera, J. and BenZvi, S. and Bianchi, D. and
                   Blake, C. and Brieden, S. and Brodzeller, A. and
                   Brooks, D. and Buckley-Geer, E. and Burtin, E. and
                   Calderon, R. and Canning, R. and Rosell, A. Carnero and
                   Carrilho, P. and Casas, L. and Castander, F. J. and
                   Charles, M. and Chaussidon, E. and Chaves-Montero, J. and
                   Chebat, D. and Chen, X. and Claybaugh, T. and
                   Cole, S. and Cooper, A. P. and Cuceu, A. and
                   Dawson, K. S. and de la Macorra, A. and
                   de Mattia, A. and Deiosso, N. and Della Costa, J. and
                   Demina, R. and Dey, A. and Dey, B. and Ding, Z. and
                   Doel, P. and Edelstein, J. and Eisenstein, D. J. and
                   Elbers, W. and Fagrelius, P. and Fanning, K. and
                   Fernández-García, E. and Ferraro, S. and
                   Font-Ribera, A. and Forero-Romero, J. E. and
                   Frenk, C. S. and Garcia-Quintero, C. and
                   Garrison, L. H. and Gaztañaga, E. and
                   Gil-Marín, H. and Gontcho, S. Gontcho A. and
                   Gonzalez, D. and Gonzalez-Morales, A. X. and
                   Gordon, C. and Green, D. and Gutierrez, G. and
                   Guy, J. and Hadzhiyska, B. and Hahn, C. and He, S. and
                   Herbold, M. and Herrera-Alcantar, H. K. and
                   Ho, M.-F. and Honscheid, K. and Howlett, C. and
                   Huterer, D. and Ishak, M. and Juneau, S. and
                   Kamble, N. V. and Karaçayl, N. G. and
                   Kehoe, R. and Kent, S. and Kim, A. G. and
                   Kirkby, D. and Kisner, T. and Koposov, S. E. and
                   Kremin, A. and Krolewski, A. and Lahav, O. and
                   Lamman, C. and Landriau, M. and Lang, D. and
                   Lasker, J. and Le Goff, J. M. and Le Guillou, L. and
                   Leauthaud, A. and Levi, M. E. and Li, Q. and
                   Li, T. S. and Lodha, K. and Lokken, M. and
                   Lozano-Rodríguez, F. and Magneville, C. and
                   Manera, M. and Martini, P. and Matthewson, W. L. and
                   Meisner, A. and Mena-Fernández, J. and Menegas, A. and
                   Mergulhão, T. and Miquel, R. and Moustakas, J. and
                   Muñoz-Gutiérrez, A. and Muñoz-Santos, D. and
                   Myers, A. D. and Nadathur, S. and Naidoo, K. and
                   Napolitano, L. and Newman, J. A. and Niz, G. and
                   Noriega, H. E. and Paillas, E. and
                   Palanque-Delabrouille, N. and Pan, J. and
                   Peacock, J. A. and Ibanez, M. P. and
                   Percival, W. J. and Pérez-Fernández, A. and
                   Pérez-Ràfols, I. and Pieri, M. M. and Poppett, C. and
                   Prada, F. and Rabinowitz, D. and Raichoor, A. and
                   Ramírez-Pérez, C. and Rashkovetskyi, M. and
                   Ravoux, C. and Rich, J. and Rocher, A. and
                   Rockosi, C. and Rohlf, J. and Román-Herrera, J. O. and
                   Ross, A. J. and Rossi, G. and Ruggeri, R. and
                   Ruhlmann-Kleider, V. and Samushia, L. and Sanchez, E. and
                   Sanders, N. and Schlegel, D. and Schubnell, M. and
                   Seo, H. and Shafieloo, A. and Sharples, R. and
                   Silber, J. and Sinigaglia, F. and Sprayberry, D. and
                   Tan, T. and Tarlé, G. and Taylor, P. and Turner, W. and
                   Ureña-López, L. A. and Vaisakh, R. and Valdes, F. and
                   Valogiannis, G. and Vargas-Magaña, M. and Verde, L. and
                   Walther, M. and Weaver, B. A. and Weinberg, D. H. and
                   White, M. and Wolfson, M. and Yèche, C. and Yu, J. and
                   Zaborowski, E. A. and Zarrouk, P. and Zhai, Z. and
                   Zhang, H. and Zhao, C. and Zhao, G. B. and Zhou, R. and
                   Zou, H.},
  journal =       {Physical Review D},
  month =         {October},
  number =        {8},
  publisher =     {American Physical Society (APS)},
  title =         {DESI DR2 results. II. Measurements of baryon acoustic
                   oscillations and cosmological constraints},
  volume =        {112},
  year =          {2025},
  doi =           {10.1103/tr6y-kpc6},
  issn =          {2470-0029},
  url =           {http://dx.doi.org/10.1103/tr6y-kpc6},
}

@article{Wadekar2020,
  author =        {Wadekar, Digvijay and Scoccimarro, Rom{\'a}n},
  journal =       {Physical Review D},
  number =        {12},
  pages =         {123521},
  title =         {Galaxy power spectrum multipoles covariance in
                   perturbation theory},
  volume =        {102},
  year =          {2020},
  doi =           {10.1103/PhysRevD.102.123521},
}

@article{Hang_2020,
  author =        {Hang, Qianjun and Alam, Shadab and Peacock, John A and
                   Cai, Yan-Chuan},
  journal =       {Monthly Notices of the Royal Astronomical Society},
  month =         {December},
  number =        {1},
  pages =         {1481–1498},
  publisher =     {Oxford University Press (OUP)},
  title =         {Galaxy clustering in the DESI Legacy Survey and its
                   imprint on the CMB},
  volume =        {501},
  year =          {2020},
  doi =           {10.1093/mnras/staa3738},
  issn =          {1365-2966},
  url =           {http://dx.doi.org/10.1093/mnras/staa3738},
}

@article{Ishiyama2024,
  author =        {Ishiyama, Tomoaki and Prada, Francisco and
                   Klypin, Anatoly A.},
  journal =       {Phys. Rev. D},
  month =         {Aug},
  pages =         {043504},
  publisher =     {American Physical Society},
  title =         {Evolution of clustering in cosmological models with
                   time-varying dark energy},
  volume =        {112},
  year =          {2025},
  doi =           {10.1103/4k5f-gyrx},
  url =           {https://link.aps.org/doi/10.1103/4k5f-gyrx},
}

@article{Rockstar2013,
  author =        {{Behroozi}, Peter S. and {Wechsler}, Risa H. and
                   {Wu}, Hao-Yi},
  journal =       {\apj},
  month =         {January},
  number =        {2},
  pages =         {109},
  title =         {{The ROCKSTAR Phase-space Temporal Halo Finder and
                   the Velocity Offsets of Cluster Cores}},
  volume =        {762},
  year =          {2013},
  doi =           {10.1088/0004-637X/762/2/109},
  eid =           {109},
}

@article{Consistent_trees2013,
  author =        {{Behroozi}, Peter S. and {Wechsler}, Risa H. and
                   {Wu}, Hao-Yi and {Busha}, Michael T. and
                   {Klypin}, Anatoly A. and {Primack}, Joel R.},
  journal =       {\apj},
  month =         {January},
  number =        {1},
  pages =         {18},
  title =         {{Gravitationally Consistent Halo Catalogs and Merger
                   Trees for Precision Cosmology}},
  volume =        {763},
  year =          {2013},
  doi =           {10.1088/0004-637X/763/1/18},
  eid =           {18},
}

@article{phs18,
  author =        {{Paranjape}, A. and {Hahn}, O. and {Sheth}, R.~K.},
  journal =       {\mnras},
  month =         {May},
  pages =         {3631-3647},
  title =         {{Halo assembly bias and the tidal anisotropy of the
                   local halo environment}},
  volume =        {476},
  year =          {2018},
  doi =           {10.1093/mnras/sty496},
}

@article{VVF2020,
  author =        {{Paranjape}, Aseem and {Alam}, Shadab},
  journal =       {\mnras},
  month =         {July},
  number =        {3},
  pages =         {3233-3251},
  title =         {{Voronoi volume function: a new probe of cosmology
                   and galaxy evolution}},
  volume =        {495},
  year =          {2020},
  doi =           {10.1093/mnras/staa1379},
}

@article{Bett+2007,
  author =        {{Bett}, Philip and {Eke}, Vincent and
                   {Frenk}, Carlos S. and {Jenkins}, Adrian and
                   {Helly}, John and {Navarro}, Julio},
  journal =       {\mnras},
  month =         {March},
  number =        {1},
  pages =         {215-232},
  title =         {{The spin and shape of dark matter haloes in the
                   Millennium simulation of a {\ensuremath{\Lambda}}
                   cold dark matter universe}},
  volume =        {376},
  year =          {2007},
  doi =           {10.1111/j.1365-2966.2007.11432.x},
}

@article{Reddick+2013,
  author =        {{Reddick}, Rachel M. and {Wechsler}, Risa H. and
                   {Tinker}, Jeremy L. and {Behroozi}, Peter S.},
  journal =       {\apj},
  month =         {July},
  number =        {1},
  pages =         {30},
  title =         {{The Connection between Galaxies and Dark Matter
                   Structures in the Local Universe}},
  volume =        {771},
  year =          {2013},
  doi =           {10.1088/0004-637X/771/1/30},
  eid =           {30},
}

@article{SHAM_2015,
  author =        {{Carretero}, J. and {Castander}, F.~J. and
                   {Gazta{\~n}aga}, E. and {Crocce}, M. and
                   {Fosalba}, P.},
  journal =       {\mnras},
  month =         {February},
  number =        {1},
  pages =         {646-670},
  title =         {{An algorithm to build mock galaxy catalogues using
                   MICE simulations}},
  volume =        {447},
  year =          {2015},
  doi =           {10.1093/mnras/stu2402},
}

@article{SHAM2016,
  author =        {{Chaves-Montero}, Jon{\'a}s and {Angulo}, Raul E. and
                   {Schaye}, Joop and {Schaller}, Matthieu and
                   {Crain}, Robert A. and {Furlong}, Michelle and
                   {Theuns}, Tom},
  journal =       {\mnras},
  month =         {August},
  number =        {3},
  pages =         {3100-3118},
  title =         {{Subhalo abundance matching and assembly bias in the
                   EAGLE simulation}},
  volume =        {460},
  year =          {2016},
  doi =           {10.1093/mnras/stw1225},
}

@article{SHAM_RSD2022,
  author =        {{DeRose}, Joseph and {Becker}, Matthew R. and
                   {Wechsler}, Risa H.},
  journal =       {\apj},
  month =         {November},
  number =        {1},
  pages =         {13},
  title =         {{Modeling Redshift-space Clustering with Abundance
                   Matching}},
  volume =        {940},
  year =          {2022},
  doi =           {10.3847/1538-4357/ac9968},
  eid =           {13},
}

@article{Takahashi2012,
  author =        {{Takahashi}, Ryuichi and {Sato}, Masanori and
                   {Nishimichi}, Takahiro and {Taruya}, Atsushi and
                   {Oguri}, Masamune},
  journal =       {\apj},
  month =         {December},
  number =        {2},
  pages =         {152},
  title =         {{Revising the Halofit Model for the Nonlinear Matter
                   Power Spectrum}},
  volume =        {761},
  year =          {2012},
  doi =           {10.1088/0004-637X/761/2/152},
  eid =           {152},
}

@article{Tinker+2008,
  author =        {{Tinker}, Jeremy and {Kravtsov}, Andrey V. and
                   {Klypin}, Anatoly and {Abazajian}, Kevork and
                   {Warren}, Michael and {Yepes}, Gustavo and
                   {Gottl{\"o}ber}, Stefan and {Holz}, Daniel E.},
  journal =       {\apj},
  month =         {December},
  number =        {2},
  pages =         {709-728},
  title =         {{Toward a Halo Mass Function for Precision Cosmology:
                   The Limits of Universality}},
  volume =        {688},
  year =          {2008},
  doi =           {10.1086/591439},
}

@article{Voronoi1908,
  author =        {Voronoi, Georges},
  journal =       {Journal für die reine und angewandte Mathematik},
  pages =         {198-287},
  title =         {Nouvelles applications des paramètres continus à la
                   théorie des formes quadratiques. Deuxième mémoire.
                   Recherches sur les parallélloèdres primitifs.},
  volume =        {134},
  year =          {1908},
  url =           {http://eudml.org/doc/149291},
}

@article{Banerjee:2020,
  author =        {Banerjee, Arka and Abel, Tom},
  journal =       {Mon. Not. Roy. Astron. Soc.},
  number =        {4},
  pages =         {5479--5499},
  title =         {{Nearest neighbour distributions: New statistical
                   measures for cosmological clustering}},
  volume =        {500},
  year =          {2020},
  doi =           {10.1093/mnras/staa3604},
}

@article{Banerjee:2021,
  author =        {Banerjee, Arka and Abel, Tom},
  journal =       {Mon. Not. Roy. Astron. Soc.},
  number =        {2},
  pages =         {2911--2923},
  title =         {{Cosmological cross-correlations and nearest
                   neighbour distributions}},
  volume =        {504},
  year =          {2021},
  doi =           {10.1093/mnras/stab961},
}

@article{Banerjee:2022,
  author =        {Banerjee, Arka and Abel, Tom},
  journal =       {Mon. Not. Roy. Astron. Soc.},
  number =        {4},
  pages =         {4856--4868},
  title =         {{Tracer-field cross-correlations with k-nearest
                   neighbour distributions}},
  volume =        {519},
  year =          {2023},
  doi =           {10.1093/mnras/stac3813},
}

@article{Gangopadhyay:2025,
  author =        {Gangopadhyay, Kwanit and Banerjee, Arka and
                   Abel, Tom},
  journal =       {Mon. Not. Roy. Astron. Soc.},
  pages =         {3427},
  title =         {{Geometric Interpretations of the $k$-Nearest
                   Neighbour Distributions}},
  volume =        {3409},
  year =          {2025},
  doi =           {10.1093/mnras/staf1637},
}

@article{Banerjee:2021cmi,
  author =        {Banerjee, Arka and Kokron, Nickolas and Abel, Tom},
  journal =       {Mon. Not. Roy. Astron. Soc.},
  number =        {2},
  pages =         {2765--2781},
  title =         {{Modelling nearest neighbour distributions of biased
                   tracers using hybrid effective field theory}},
  volume =        {511},
  year =          {2022},
  doi =           {10.1093/mnras/stac193},
}

@article{Ramakrishnan+2019,
  author =        {{Ramakrishnan}, Sujatha and {Paranjape}, Aseem and
                   {Hahn}, Oliver and {Sheth}, Ravi K.},
  journal =       {\mnras},
  month =         {November},
  number =        {3},
  pages =         {2977-2996},
  title =         {{Cosmic web anisotropy is the primary indicator of
                   halo assembly bias}},
  volume =        {489},
  year =          {2019},
  doi =           {10.1093/mnras/stz2344},
}

@article{Ramakrishnan+2025,
  author =        {{Ramakrishnan}, Sujatha and {Gonzalez-Perez}, Violeta and
                   {Parimbelli}, Gabriele and {Yepes}, Gustavo},
  journal =       {\aap},
  month =         {May},
  pages =         {A70},
  title =         {{The multi-dimensional halo assembly bias can be
                   preserved when enhancing halo properties with
                   HALOSCOPE}},
  volume =        {697},
  year =          {2025},
  doi =           {10.1051/0004-6361/202453030},
  eid =           {A70},
}

@article{RamakrishnanVelmani2022,
  author =        {{Ramakrishnan}, Sujatha and {Velmani}, Premvijay},
  journal =       {\mnras},
  month =         {November},
  number =        {4},
  pages =         {5849-5862},
  title =         {{Properties beyond mass for unresolved haloes across
                   redshift and cosmology using correlations with local
                   halo environment}},
  volume =        {516},
  year =          {2022},
  doi =           {10.1093/mnras/stac2605},
}

@article{RPS2021,
  author =        {{Ramakrishnan}, Sujatha and {Paranjape}, Aseem and
                   {Sheth}, Ravi K.},
  journal =       {\mnras},
  month =         {May},
  number =        {2},
  pages =         {2053-2064},
  title =         {{Mock halo catalogues: assigning unresolved halo
                   properties using correlations with local halo
                   environment}},
  volume =        {503},
  year =          {2021},
  doi =           {10.1093/mnras/stab541},
}

@article{Scoccimarro1998,
  author =        {{Scoccimarro}, Roman},
  journal =       {\mnras},
  month =         {October},
  number =        {4},
  pages =         {1097-1118},
  title =         {{Transients from initial conditions: a perturbative
                   analysis}},
  volume =        {299},
  year =          {1998},
  doi =           {10.1046/j.1365-8711.1998.01845.x},
}

@article{Blas2011,
  author =        {Blas, Diego and Lesgourgues, Julien and Tram, Thomas},
  journal =       {Journal of Cosmology and Astroparticle Physics},
  number =        {07},
  pages =         {034},
  title =         {The Cosmic Linear Anisotropy Solving System (CLASS).
                   Part II: Approximation schemes},
  volume =        {2011},
  year =          {2011},
  doi =           {10.1088/1475-7516/2011/07/034},
}

@article{class-II,
  author =        {Lesgourgues, Julien},
  journal =       {arXiv e-prints},
  title =         {The Cosmic Linear Anisotropy Solving System (CLASS)
                   I: Overview},
  year =          {2011},
}

@article{coulton+23,
  author =        {{Coulton}, William R. and
                   {Villaescusa-Navarro}, Francisco and {Jamieson}, Drew and
                   {Baldi}, Marco and {Jung}, Gabriel and
                   {Karagiannis}, Dionysios and {Liguori}, Michele and
                   {Verde}, Licia and {Wandelt}, Benjamin D.},
  journal =       {\apj},
  month =         {February},
  number =        {2},
  pages =         {178},
  title =         {{Quijote-PNG: The Information Content of the Halo
                   Power Spectrum and Bispectrum}},
  volume =        {943},
  year =          {2023},
  doi =           {10.3847/1538-4357/aca7c1},
  eid =           {178},
}

@article{AbacusSummit,
  author =        {Maksimova, Nina A. and Garrison, Lehman H. and
                   Hadzhiyska, Boryana and Bose, Sownak and
                   Eisenstein, Daniel J.},
  journal =       {Monthly Notices of the Royal Astronomical Society},
  number =        {3},
  pages =         {4017-4037},
  title =         {AbacusSummit: a massive set of high-accuracy,
                   high-resolution N-body simulations},
  volume =        {508},
  year =          {2021},
  doi =           {10.1093/mnras/stab2484},
}

@article{Tinker+2010,
  author =        {{Tinker}, Jeremy L. and {Robertson}, Brant E. and
                   {Kravtsov}, Andrey V. and {Klypin}, Anatoly and
                   {Warren}, Michael S. and {Yepes}, Gustavo and
                   {Gottl{\"o}ber}, Stefan},
  journal =       {\apj},
  month =         {December},
  number =        {2},
  pages =         {878-886},
  title =         {{The Large-scale Bias of Dark Matter Halos: Numerical
                   Calibration and Model Tests}},
  volume =        {724},
  year =          {2010},
  doi =           {10.1088/0004-637X/724/2/878},
}

@article{NPaul+2019,
  author =        {{Paul}, Niladri and {Pahwa}, Isha and
                   {Paranjape}, Aseem},
  journal =       {\mnras},
  month =         {September},
  number =        {1},
  pages =         {1220-1234},
  title =         {{Global analysis of luminosity- and colour-dependent
                   galaxy clustering in the Sloan Digital Sky Survey}},
  volume =        {488},
  year =          {2019},
  doi =           {10.1093/mnras/stz1764},
}

@article{Abacus_cosmo2018,
 adsurl = {https://ui.adsabs.harvard.edu/abs/2018ApJS..236...43G},
 archiveprefix = {arXiv},
 author = {{Garrison}, Lehman H. and {Eisenstein}, Daniel J. and {Ferrer}, Douglas and {Tinker}, Jeremy L. and {Pinto}, Philip A. and {Weinberg}, David H.},
 doi = {10.3847/1538-4365/aabfd3},
 eid = {43},
 eprint = {1712.05768},
 journal = {\apjs},
 month = {June},
 number = {2},
 pages = {43},
 primaryclass = {astro-ph.CO},
 title = {{The Abacus Cosmos: A Suite of Cosmological N-body Simulations}},
 volume = {236},
 year = {2018}
}

@ARTICLE{Crain+2023,
       author = {{Crain}, Robert A. and {van de Voort}, Freeke},
        title = "{Hydrodynamical Simulations of the Galaxy Population: Enduring Successes and Outstanding Challenges}",
      journal = {\araa},
         year = 2023,
        month = aug,
       volume = {61},
        pages = {473-515},
          doi = {10.1146/annurev-astro-041923-043618},
archivePrefix = {arXiv},
       eprint = {2309.17075},
 primaryClass = {astro-ph.GA},
       adsurl = {https://ui.adsabs.harvard.edu/abs/2023ARA&A..61..473C}
}

@ARTICLE{Vogelsberger+2020,
       author = {{Vogelsberger}, Mark and {Marinacci}, Federico and {Torrey}, Paul and {Puchwein}, Ewald},
        title = "{Cosmological simulations of galaxy formation}",
      journal = {Nature Reviews Physics},
         year = 2020,
        month = jan,
       volume = {2},
       number = {1},
        pages = {42-66},
          doi = {10.1038/s42254-019-0127-2},
archivePrefix = {arXiv},
       eprint = {1909.07976},
 primaryClass = {astro-ph.GA},
       adsurl = {https://ui.adsabs.harvard.edu/abs/2020NatRP...2...42V}
}

@book{Dodelson&Schmidt,
 author = {{Dodelson}, Scott and {Schmidt}, Fabian},
 edition = {2},
 isbn = {9780128159484},
 pages = {336-339},
 publisher = {Elsevier Science},
 title = {{Modern Cosmology}},
 year = {2020}
}

@ARTICLE{SDSS_2017,
       author = {{Alam}, Shadab and {Ata}, Metin and {Bailey}, Stephen and {Beutler}, Florian and {Bizyaev}, Dmitry and {Blazek}, Jonathan A. and {Bolton}, Adam S. and {Brownstein}, Joel R. and {Burden}, Angela and {Chuang}, Chia-Hsun and {Comparat}, Johan and {Cuesta}, Antonio J. and {Dawson}, Kyle S. and {Eisenstein}, Daniel J. and {Escoffier}, Stephanie and {Gil-Mar{\'\i}n}, H{\'e}ctor and {Grieb}, Jan Niklas and {Hand}, Nick and {Ho}, Shirley and {Kinemuchi}, Karen and {Kirkby}, David and {Kitaura}, Francisco and {Malanushenko}, Elena and {Malanushenko}, Viktor and {Maraston}, Claudia and {McBride}, Cameron K. and {Nichol}, Robert C. and {Olmstead}, Matthew D. and {Oravetz}, Daniel and {Padmanabhan}, Nikhil and {Palanque-Delabrouille}, Nathalie and {Pan}, Kaike and {Pellejero-Ibanez}, Marcos and {Percival}, Will J. and {Petitjean}, Patrick and {Prada}, Francisco and {Price-Whelan}, Adrian M. and {Reid}, Beth A. and {Rodr{\'\i}guez-Torres}, Sergio A. and {Roe}, Natalie A. and {Ross}, Ashley J. and {Ross}, Nicholas P. and {Rossi}, Graziano and {Rubi{\~n}o-Mart{\'\i}n}, Jose Alberto and {Saito}, Shun and {Salazar-Albornoz}, Salvador and {Samushia}, Lado and {S{\'a}nchez}, Ariel G. and {Satpathy}, Siddharth and {Schlegel}, David J. and {Schneider}, Donald P. and {Sc{\'o}ccola}, Claudia G. and {Seo}, Hee-Jong and {Sheldon}, Erin S. and {Simmons}, Audrey and {Slosar}, An{\v{z}}e and {Strauss}, Michael A. and {Swanson}, Molly E.~C. and {Thomas}, Daniel and {Tinker}, Jeremy L. and {Tojeiro}, Rita and {Maga{\~n}a}, Mariana Vargas and {Vazquez}, Jose Alberto and {Verde}, Licia and {Wake}, David A. and {Wang}, Yuting and {Weinberg}, David H. and {White}, Martin and {Wood-Vasey}, W. Michael and {Y{\`e}che}, Christophe and {Zehavi}, Idit and {Zhai}, Zhongxu and {Zhao}, Gong-Bo},
        title = "{The clustering of galaxies in the completed SDSS-III Baryon Oscillation Spectroscopic Survey: cosmological analysis of the DR12 galaxy sample}",
      journal = {\mnras},
         year = 2017,
        month = sep,
       volume = {470},
       number = {3},
        pages = {2617-2652},
          doi = {10.1093/mnras/stx721},
archivePrefix = {arXiv},
       eprint = {1607.03155},
 primaryClass = {astro-ph.CO},
       adsurl = {https://ui.adsabs.harvard.edu/abs/2017MNRAS.470.2617A}
}

@ARTICLE{Gao&White2007,
       author = {{Gao}, Liang and {White}, Simon D.~M.},
        title = "{Assembly bias in the clustering of dark matter haloes}",
      journal = {\mnras},
         year = 2007,
        month = apr,
       volume = {377},
       number = {1},
        pages = {L5-L9},
          doi = {10.1111/j.1745-3933.2007.00292.x},
archivePrefix = {arXiv},
       eprint = {astro-ph/0611921},
 primaryClass = {astro-ph},
       adsurl = {https://ui.adsabs.harvard.edu/abs/2007MNRAS.377L...5G}
}

@ARTICLE{SDSS2000,
       author = {{York}, Donald G. and {Adelman}, J. and {Anderson}, Jr., John E. and {Anderson}, Scott F. and {Annis}, James and {Bahcall}, Neta A. and {Bakken}, J.~A. and {Barkhouser}, Robert and {Bastian}, Steven and {Berman}, Eileen and {Boroski}, William N. and {Bracker}, Steve and {Briegel}, Charlie and {Briggs}, John W. and {Brinkmann}, J. and {Brunner}, Robert and {Burles}, Scott and {Carey}, Larry and {Carr}, Michael A. and {Castander}, Francisco J. and {Chen}, Bing and {Colestock}, Patrick L. and {Connolly}, A.~J. and {Crocker}, J.~H. and {Csabai}, Istv{\'a}n and {Czarapata}, Paul C. and {Davis}, John Eric and {Doi}, Mamoru and {Dombeck}, Tom and {Eisenstein}, Daniel and {Ellman}, Nancy and {Elms}, Brian R. and {Evans}, Michael L. and {Fan}, Xiaohui and {Federwitz}, Glenn R. and {Fiscelli}, Larry and {Friedman}, Scott and {Frieman}, Joshua A. and {Fukugita}, Masataka and {Gillespie}, Bruce and {Gunn}, James E. and {Gurbani}, Vijay K. and {de Haas}, Ernst and {Haldeman}, Merle and {Harris}, Frederick H. and {Hayes}, J. and {Heckman}, Timothy M. and {Hennessy}, G.~S. and {Hindsley}, Robert B. and {Holm}, Scott and {Holmgren}, Donald J. and {Huang}, Chi-hao and {Hull}, Charles and {Husby}, Don and {Ichikawa}, Shin-Ichi and {Ichikawa}, Takashi and {Ivezi{\'c}}, {\v{Z}}eljko and {Kent}, Stephen and {Kim}, Rita S.~J. and {Kinney}, E. and {Klaene}, Mark and {Kleinman}, A.~N. and {Kleinman}, S. and {Knapp}, G.~R. and {Korienek}, John and {Kron}, Richard G. and {Kunszt}, Peter Z. and {Lamb}, D.~Q. and {Lee}, B. and {Leger}, R. French and {Limmongkol}, Siriluk and {Lindenmeyer}, Carl and {Long}, Daniel C. and {Loomis}, Craig and {Loveday}, Jon and {Lucinio}, Rich and {Lupton}, Robert H. and {MacKinnon}, Bryan and {Mannery}, Edward J. and {Mantsch}, P.~M. and {Margon}, Bruce and {McGehee}, Peregrine and {McKay}, Timothy A. and {Meiksin}, Avery and {Merelli}, Aronne and {Monet}, David G. and {Munn}, Jeffrey A. and {Narayanan}, Vijay K. and {Nash}, Thomas and {Neilsen}, Eric and {Neswold}, Rich and {Newberg}, Heidi Jo and {Nichol}, R.~C. and {Nicinski}, Tom and {Nonino}, Mario and {Okada}, Norio and {Okamura}, Sadanori and {Ostriker}, Jeremiah P. and {Owen}, Russell and {Pauls}, A. George and {Peoples}, John and {Peterson}, R.~L. and {Petravick}, Donald and {Pier}, Jeffrey R. and {Pope}, Adrian and {Pordes}, Ruth and {Prosapio}, Angela and {Rechenmacher}, Ron and {Quinn}, Thomas R. and {Richards}, Gordon T. and {Richmond}, Michael W. and {Rivetta}, Claudio H. and {Rockosi}, Constance M. and {Ruthmansdorfer}, Kurt and {Sandford}, Dale and {Schlegel}, David J. and {Schneider}, Donald P. and {Sekiguchi}, Maki and {Sergey}, Gary and {Shimasaku}, Kazuhiro and {Siegmund}, Walter A. and {Smee}, Stephen and {Smith}, J. Allyn and {Snedden}, S. and {Stone}, R. and {Stoughton}, Chris and {Strauss}, Michael A. and {Stubbs}, Christopher and {SubbaRao}, Mark and {Szalay}, Alexander S. and {Szapudi}, Istvan and {Szokoly}, Gyula P. and {Thakar}, Anirudda R. and {Tremonti}, Christy and {Tucker}, Douglas L. and {Uomoto}, Alan and {Vanden Berk}, Dan and {Vogeley}, Michael S. and {Waddell}, Patrick and {Wang}, Shu-i. and {Watanabe}, Masaru and {Weinberg}, David H. and {Yanny}, Brian and {Yasuda}, Naoki and {SDSS Collaboration}},
        title = "{The Sloan Digital Sky Survey: Technical Summary}",
      journal = {\aj},
         year = 2000,
        month = sep,
       volume = {120},
       number = {3},
        pages = {1579-1587},
          doi = {10.1086/301513},
archivePrefix = {arXiv},
       eprint = {astro-ph/0006396},
 primaryClass = {astro-ph},
       adsurl = {https://ui.adsabs.harvard.edu/abs/2000AJ....120.1579Y}
}

@ARTICLE{SDSS_BAO_2005,
       author = {{Eisenstein}, Daniel J. and {Zehavi}, Idit and {Hogg}, David W. and {Scoccimarro}, Roman and {Blanton}, Michael R. and {Nichol}, Robert C. and {Scranton}, Ryan and {Seo}, Hee-Jong and {Tegmark}, Max and {Zheng}, Zheng and {Anderson}, Scott F. and {Annis}, Jim and {Bahcall}, Neta and {Brinkmann}, Jon and {Burles}, Scott and {Castander}, Francisco J. and {Connolly}, Andrew and {Csabai}, Istvan and {Doi}, Mamoru and {Fukugita}, Masataka and {Frieman}, Joshua A. and {Glazebrook}, Karl and {Gunn}, James E. and {Hendry}, John S. and {Hennessy}, Gregory and {Ivezi{\'c}}, Zeljko and {Kent}, Stephen and {Knapp}, Gillian R. and {Lin}, Huan and {Loh}, Yeong-Shang and {Lupton}, Robert H. and {Margon}, Bruce and {McKay}, Timothy A. and {Meiksin}, Avery and {Munn}, Jeffery A. and {Pope}, Adrian and {Richmond}, Michael W. and {Schlegel}, David and {Schneider}, Donald P. and {Shimasaku}, Kazuhiro and {Stoughton}, Christopher and {Strauss}, Michael A. and {SubbaRao}, Mark and {Szalay}, Alexander S. and {Szapudi}, Istv{\'a}n and {Tucker}, Douglas L. and {Yanny}, Brian and {York}, Donald G.},
        title = "{Detection of the Baryon Acoustic Peak in the Large-Scale Correlation Function of SDSS Luminous Red Galaxies}",
      journal = {\apj},
         year = 2005,
        month = nov,
       volume = {633},
       number = {2},
        pages = {560-574},
          doi = {10.1086/466512},
archivePrefix = {arXiv},
       eprint = {astro-ph/0501171},
 primaryClass = {astro-ph},
       adsurl = {https://ui.adsabs.harvard.edu/abs/2005ApJ...633..560E}
}

@ARTICLE{SKA2009,
       author = {{Dewdney}, P.~E. and {Hall}, P.~J. and {Schilizzi}, R.~T. and {Lazio}, T.~J.~L.~W.},
        title = "{The Square Kilometre Array}",
      journal = {IEEE Proceedings},
         year = 2009,
        month = aug,
       volume = {97},
       number = {8},
        pages = {1482-1496},
          doi = {10.1109/JPROC.2009.2021005},
       adsurl = {https://ui.adsabs.harvard.edu/abs/2009IEEEP..97.1482D}
}

@ARTICLE{Weinberg+2013,
       author = {{Weinberg}, David H. and {Mortonson}, Michael J. and {Eisenstein}, Daniel J. and {Hirata}, Christopher and {Riess}, Adam G. and {Rozo}, Eduardo},
        title = "{Observational probes of cosmic acceleration}",
      journal = {\physrep},
         year = 2013,
        month = sep,
       volume = {530},
       number = {2},
        pages = {87-255},
          doi = {10.1016/j.physrep.2013.05.001},
archivePrefix = {arXiv},
       eprint = {1201.2434},
 primaryClass = {astro-ph.CO},
       adsurl = {https://ui.adsabs.harvard.edu/abs/2013PhR...530...87W}
}

@article{Libeskind_et_al2018,
 adsurl = {https://ui.adsabs.harvard.edu/abs/2018MNRAS.473.1195L},
 archiveprefix = {arXiv},
 author = {{Libeskind}, Noam I. and {van de Weygaert}, Rien and {Cautun}, Marius and {Falck}, Bridget and {Tempel}, Elmo and {Abel}, Tom and {Alpaslan}, Mehmet and {Arag{\'o}n-Calvo}, Miguel A. and {Forero-Romero}, Jaime E. and {Gonzalez}, Roberto and {Gottl{\"o}ber}, Stefan and {Hahn}, Oliver and {Hellwing}, Wojciech A. and {Hoffman}, Yehuda and {Jones}, Bernard J.~T. and {Kitaura}, Francisco and {Knebe}, Alexander and {Manti}, Serena and {Neyrinck}, Mark and {Nuza}, Sebasti{\'a}n E. and {Padilla}, Nelson and {Platen}, Erwin and {Ramachandra}, Nesar and {Robotham}, Aaron and {Saar}, Enn and {Shandarin}, Sergei and {Steinmetz}, Matthias and {Stoica}, Radu S. and {Sousbie}, Thierry and {Yepes}, Gustavo},
 doi = {10.1093/mnras/stx1976},
 eprint = {1705.03021},
 journal = {\mnras},
 month = {January},
 number = {1},
 pages = {1195-1217},
 primaryclass = {astro-ph.CO},
 title = {{Tracing the cosmic web}},
 volume = {473},
 year = {2018}
}

@ARTICLE{Sheth&Weygaert,
       author = {{Sheth}, Ravi K. and {van de Weygaert}, Rien},
        title = "{A hierarchy of voids: much ado about nothing}",
      journal = {\mnras},
         year = 2004,
        month = may,
       volume = {350},
       number = {2},
        pages = {517-538},
          doi = {10.1111/j.1365-2966.2004.07661.x},
archivePrefix = {arXiv},
       eprint = {astro-ph/0311260},
 primaryClass = {astro-ph},
       adsurl = {https://ui.adsabs.harvard.edu/abs/2004MNRAS.350..517S}
}

@ARTICLE{Bond+1996,
       author = {{Bond}, J. Richard and {Kofman}, Lev and {Pogosyan}, Dmitry},
        title = "{How filaments of galaxies are woven into the cosmic web}",
      journal = {\nat},
         year = 1996,
        month = apr,
       volume = {380},
       number = {6575},
        pages = {603-606},
          doi = {10.1038/380603a0},
archivePrefix = {arXiv},
       eprint = {astro-ph/9512141},
 primaryClass = {astro-ph},
       adsurl = {https://ui.adsabs.harvard.edu/abs/1996Natur.380..603B}
}

@article{Zeldovich1970,
 adsurl = {https://ui.adsabs.harvard.edu/abs/1970A&A.....5...84Z},
 author = {{Zel'dovich}, Ya. B.},
 journal = {\aap},
 month = {March},
 pages = {84-89},
 title = {{Gravitational instability: An approximate theory for large density perturbations.}},
 volume = {5},
 year = {1970}
}

@article{NEXUS2013,
 adsurl = {https://ui.adsabs.harvard.edu/abs/2013MNRAS.429.1286C},
 archiveprefix = {arXiv},
 author = {{Cautun}, Marius and {van de Weygaert}, Rien and {Jones}, Bernard J.~T.},
 doi = {10.1093/mnras/sts416},
 eprint = {1209.2043},
 journal = {\mnras},
 month = {February},
 number = {2},
 pages = {1286-1308},
 primaryclass = {astro-ph.CO},
 title = {{NEXUS: tracing the cosmic web connection}},
 volume = {429},
 year = {2013}
}

@article{NEXUS2014,
 adsurl = {https://ui.adsabs.harvard.edu/abs/2014MNRAS.441.2923C},
 archiveprefix = {arXiv},
 author = {{Cautun}, Marius and {van de Weygaert}, Rien and {Jones}, Bernard J.~T. and {Frenk}, Carlos S.},
 doi = {10.1093/mnras/stu768},
 eprint = {1401.7866},
 journal = {\mnras},
 month = {July},
 number = {4},
 pages = {2923-2973},
 primaryclass = {astro-ph.CO},
 title = {{Evolution of the cosmic web}},
 volume = {441},
 year = {2014}
}

@article{NFW97,
 adsurl = {https://ui.adsabs.harvard.edu/abs/1997ApJ...490..493N},
 archiveprefix = {arXiv},
 author = {{Navarro}, Julio F. and {Frenk}, Carlos S. and {White}, Simon D.~M.},
 doi = {10.1086/304888},
 eprint = {astro-ph/9611107},
 journal = {\apj},
 month = {December},
 number = {2},
 pages = {493-508},
 primaryclass = {astro-ph},
 title = {{A Universal Density Profile from Hierarchical Clustering}},
 volume = {490},
 year = {1997}
}

@article{2pop2020,
 adsurl = {https://ui.adsabs.harvard.edu/abs/2020A&A...641A.173G},
 archiveprefix = {arXiv},
 author = {{Gal{\'a}rraga-Espinosa}, Daniela and {Aghanim}, Nabila and {Langer}, Mathieu and {Gouin}, C{\'e}line and {Malavasi}, Nicola},
 doi = {10.1051/0004-6361/202037986},
 eid = {A173},
 eprint = {2003.09697},
 journal = {\aap},
 month = {September},
 pages = {A173},
 primaryclass = {astro-ph.CO},
 title = {{Populations of filaments from the distribution of galaxies in numerical simulations}},
 volume = {641},
 year = {2020}
}

@article{Malavasi_et_al2020,
 adsurl = {https://ui.adsabs.harvard.edu/abs/2020A&A...642A..19M},
 archiveprefix = {arXiv},
 author = {{Malavasi}, Nicola and {Aghanim}, Nabila and {Douspis}, Marian and {Tanimura}, Hideki and {Bonjean}, Victor},
 doi = {10.1051/0004-6361/202037647},
 eid = {A19},
 eprint = {2002.01486},
 journal = {\aap},
 month = {October},
 pages = {A19},
 primaryclass = {astro-ph.CO},
 title = {{Characterising filaments in the SDSS volume from the galaxy distribution}},
 volume = {642},
 year = {2020}
}

@article{Martizzi_et_al2019,
 adsurl = {https://ui.adsabs.harvard.edu/abs/2019MNRAS.486.3766M},
 archiveprefix = {arXiv},
 author = {{Martizzi}, Davide and {Vogelsberger}, Mark and {Artale}, Maria Celeste and {Haider}, Markus and {Torrey}, Paul and {Marinacci}, Federico and {Nelson}, Dylan and {Pillepich}, Annalisa and {Weinberger}, Rainer and {Hernquist}, Lars and {Naiman}, Jill and {Springel}, Volker},
 doi = {10.1093/mnras/stz1106},
 eprint = {1810.01883},
 journal = {\mnras},
 month = {July},
 number = {3},
 pages = {3766-3787},
 primaryclass = {astro-ph.CO},
 title = {{Baryons in the Cosmic Web of IllustrisTNG - I: gas in knots, filaments, sheets, and voids}},
 volume = {486},
 year = {2019}
}

@article{Aragon-calvo+2010a,
 adsurl = {https://ui.adsabs.harvard.edu/abs/2010MNRAS.408.2163A},
 archiveprefix = {arXiv},
 author = {{Arag{\'o}n-Calvo}, Miguel A. and {van de Weygaert}, Rien and {Jones}, Bernard J.~T.},
 doi = {10.1111/j.1365-2966.2010.17263.x},
 eprint = {1007.0742},
 journal = {\mnras},
 month = {November},
 number = {4},
 pages = {2163-2187},
 primaryclass = {astro-ph.CO},
 title = {{Multiscale phenomenology of the cosmic web}},
 volume = {408},
 year = {2010}
}

@ARTICLE{Dressler1980,
       author = {{Dressler}, A.},
        title = "{Galaxy morphology in rich clusters: implications for the formation and evolution of galaxies.}",
      journal = {\apj},
         year = 1980,
        month = mar,
       volume = {236},
        pages = {351-365},
          doi = {10.1086/157753},
       adsurl = {https://ui.adsabs.harvard.edu/abs/1980ApJ...236..351D}
}

@ARTICLE{Blanton&Moustakas2009,
       author = {{Blanton}, Michael R. and {Moustakas}, John},
        title = "{Physical Properties and Environments of Nearby Galaxies}",
      journal = {\araa},
         year = 2009,
        month = sep,
       volume = {47},
       number = {1},
        pages = {159-210},
          doi = {10.1146/annurev-astro-082708-101734},
archivePrefix = {arXiv},
       eprint = {0908.3017},
 primaryClass = {astro-ph.GA},
       adsurl = {https://ui.adsabs.harvard.edu/abs/2009ARA&A..47..159B}
}

@ARTICLE{Codis+2012,
       author = {{Codis}, Sandrine and {Pichon}, Christophe and {Devriendt}, Julien and {Slyz}, Adrianne and {Pogosyan}, Dmitry and {Dubois}, Yohan and {Sousbie}, Thierry},
        title = "{Connecting the cosmic web to the spin of dark haloes: implications for galaxy formation}",
      journal = {\mnras},
         year = 2012,
        month = dec,
       volume = {427},
       number = {4},
        pages = {3320-3336},
          doi = {10.1111/j.1365-2966.2012.21636.x},
archivePrefix = {arXiv},
       eprint = {1201.5794},
 primaryClass = {astro-ph.CO},
       adsurl = {https://ui.adsabs.harvard.edu/abs/2012MNRAS.427.3320C}
}

@ARTICLE{Laigle+2015,
       author = {{Laigle}, C. and {Pichon}, C. and {Codis}, S. and {Dubois}, Y. and {Le Borgne}, D. and {Pogosyan}, D. and {Devriendt}, J. and {Peirani}, S. and {Prunet}, S. and {Rouberol}, S. and {Slyz}, A. and {Sousbie}, T.},
        title = "{Swirling around filaments: are large-scale structure vortices spinning up dark haloes?}",
      journal = {\mnras},
         year = 2015,
        month = jan,
       volume = {446},
       number = {3},
        pages = {2744-2759},
          doi = {10.1093/mnras/stu2289},
archivePrefix = {arXiv},
       eprint = {1310.3801},
 primaryClass = {astro-ph.CO},
       adsurl = {https://ui.adsabs.harvard.edu/abs/2015MNRAS.446.2744L}
}

@article{Kaiser87,
 adsurl = {https://ui.adsabs.harvard.edu/abs/1987MNRAS.227....1K},
 author = {{Kaiser}, Nick},
 doi = {10.1093/mnras/227.1.1},
 journal = {\mnras},
 month = {July},
 pages = {1-21},
 title = {{Clustering in real space and in redshift space}},
 volume = {227},
 year = {1987}
}

@ARTICLE{Scoccimarro2004,
       author = {{Scoccimarro}, Rom{\'a}n},
        title = "{Redshift-space distortions, pairwise velocities, and nonlinearities}",
      journal = {\prd},
         year = 2004,
        month = oct,
       volume = {70},
       number = {8},
          eid = {083007},
        pages = {083007},
          doi = {10.1103/PhysRevD.70.083007},
archivePrefix = {arXiv},
       eprint = {astro-ph/0407214},
 primaryClass = {astro-ph},
       adsurl = {https://ui.adsabs.harvard.edu/abs/2004PhRvD..70h3007S}
}

@INPROCEEDINGS{Hamilton1998,
       author = {{Hamilton}, A.~J.~S.},
        title = "{Linear Redshift Distortions: a Review}",
    booktitle = {The Evolving Universe},
         year = 1998,
       editor = {{Hamilton}, Donald},
       series = {Astrophysics and Space Science Library},
       volume = {231},
        month = jan,
        pages = {185},
          doi = {10.1007/978-94-011-4960-0_17},
archivePrefix = {arXiv},
       eprint = {astro-ph/9708102},
 primaryClass = {astro-ph},
       adsurl = {https://ui.adsabs.harvard.edu/abs/1998ASSL..231..185H}
}

@ARTICLE{Guzzo+2008,
       author = {{Guzzo}, L. and {Pierleoni}, M. and {Meneux}, B. and {Branchini}, E. and {Le F{\`e}vre}, O. and {Marinoni}, C. and {Garilli}, B. and {Blaizot}, J. and {De Lucia}, G. and {Pollo}, A. and {McCracken}, H.~J. and {Bottini}, D. and {Le Brun}, V. and {Maccagni}, D. and {Picat}, J.~P. and {Scaramella}, R. and {Scodeggio}, M. and {Tresse}, L. and {Vettolani}, G. and {Zanichelli}, A. and {Adami}, C. and {Arnouts}, S. and {Bardelli}, S. and {Bolzonella}, M. and {Bongiorno}, A. and {Cappi}, A. and {Charlot}, S. and {Ciliegi}, P. and {Contini}, T. and {Cucciati}, O. and {de la Torre}, S. and {Dolag}, K. and {Foucaud}, S. and {Franzetti}, P. and {Gavignaud}, I. and {Ilbert}, O. and {Iovino}, A. and {Lamareille}, F. and {Marano}, B. and {Mazure}, A. and {Memeo}, P. and {Merighi}, R. and {Moscardini}, L. and {Paltani}, S. and {Pell{\`o}}, R. and {Perez-Montero}, E. and {Pozzetti}, L. and {Radovich}, M. and {Vergani}, D. and {Zamorani}, G. and {Zucca}, E.},
        title = "{A test of the nature of cosmic acceleration using galaxy redshift distortions}",
      journal = {\nat},
         year = 2008,
        month = jan,
       volume = {451},
       number = {7178},
        pages = {541-544},
          doi = {10.1038/nature06555},
archivePrefix = {arXiv},
       eprint = {0802.1944},
 primaryClass = {astro-ph},
       adsurl = {https://ui.adsabs.harvard.edu/abs/2008Natur.451..541G}
}

@article{Bernardeauetal02,
 adsurl = {https://ui.adsabs.harvard.edu/abs/2002PhR...367....1B},
 archiveprefix = {arXiv},
 author = {{Bernardeau}, F. and {Colombi}, S. and {Gazta{\~n}aga}, E. and {Scoccimarro}, R.},
 doi = {10.1016/S0370-1573(02)00135-7},
 eprint = {astro-ph/0112551},
 journal = {\physrep},
 month = {September},
 primaryclass = {astro-ph},
 title = {{Large-scale structure of the Universe and cosmological perturbation theory}},
 volume = {367},
 year = {2002}
}

@ARTICLE{Carlson+2009,
       author = {{Carlson}, Jordan and {White}, Martin and {Padmanabhan}, Nikhil},
        title = "{Critical look at cosmological perturbation theory techniques}",
      journal = {\prd},
         year = 2009,
        month = aug,
       volume = {80},
       number = {4},
          eid = {043531},
        pages = {043531},
          doi = {10.1103/PhysRevD.80.043531},
archivePrefix = {arXiv},
       eprint = {0905.0479},
 primaryClass = {astro-ph.CO},
       adsurl = {https://ui.adsabs.harvard.edu/abs/2009PhRvD..80d3531C}
}

@ARTICLE{Baumann+2012,
       author = {{Baumann}, Daniel and {Nicolis}, Alberto and {Senatore}, Leonardo and {Zaldarriaga}, Matias},
        title = "{Cosmological non-linearities as an effective fluid}",
      journal = {\jcap},
         year = 2012,
        month = jul,
       volume = {2012},
       number = {7},
          eid = {051},
        pages = {051},
          doi = {10.1088/1475-7516/2012/07/051},
archivePrefix = {arXiv},
       eprint = {1004.2488},
 primaryClass = {astro-ph.CO},
       adsurl = {https://ui.adsabs.harvard.edu/abs/2012JCAP...07..051B}
}

@ARTICLE{Carrasco+2012,
       author = {{Carrasco}, John Joseph M. and {Hertzberg}, Mark P. and {Senatore}, Leonardo},
        title = "{The effective field theory of cosmological large scale structures}",
      journal = {Journal of High Energy Physics},
         year = 2012,
        month = sep,
       volume = {2012},
          eid = {82},
        pages = {82},
          doi = {10.1007/JHEP09(2012)082},
archivePrefix = {arXiv},
       eprint = {1206.2926},
 primaryClass = {astro-ph.CO},
       adsurl = {https://ui.adsabs.harvard.edu/abs/2012JHEP...09..082C}
}

@article{Lapparent+1986,
 adsurl = {https://ui.adsabs.harvard.edu/abs/1986ApJ...302L...1D},
 author = {{de Lapparent}, V. and {Geller}, M.~J. and {Huchra}, J.~P.},
 doi = {10.1086/184625},
 journal = {\apjl},
 month = {March},
 pages = {L1},
 title = {{A Slice of the Universe}},
 volume = {302},
 year = {1986}
}

@article{Colberg_et_al2005,
 adsurl = {https://ui.adsabs.harvard.edu/abs/2005MNRAS.359..272C},
 archiveprefix = {arXiv},
 author = {{Colberg}, J{\"o}rg M. and {Krughoff}, K. Simon and {Connolly}, Andrew J.},
 doi = {10.1111/j.1365-2966.2005.08897.x},
 eprint = {astro-ph/0406665},
 journal = {\mnras},
 month = {May},
 number = {1},
 pages = {272-282},
 primaryclass = {astro-ph},
 title = {{Intercluster filaments in a {\ensuremath{\Lambda}}CDM Universe}},
 volume = {359},
 year = {2005}
}

@article{aragon-calvo24,
 adsurl = {https://ui.adsabs.harvard.edu/abs/2023arXiv230816186A},
 archiveprefix = {arXiv},
 author = {{Aragon-Calvo}, M.~A.},
 doi = {10.48550/arXiv.2308.16186},
 eid = {arXiv:2308.16186},
 eprint = {2308.16186},
 journal = {arXiv e-prints},
 month = {August},
 pages = {arXiv:2308.16186},
 primaryclass = {astro-ph.CO},
 title = {{Hierarchical Reconstruction of the Cosmic Web, The H-Spine method}},
 year = {2023}
}

@article{Hahn_et_al2007b,
 adsurl = {https://ui.adsabs.harvard.edu/abs/2007MNRAS.375..489H},
 archiveprefix = {arXiv},
 author = {{Hahn}, Oliver and {Porciani}, Cristiano and {Carollo}, C. Marcella and {Dekel}, Avishai},
 doi = {10.1111/j.1365-2966.2006.11318.x},
 eprint = {astro-ph/0610280},
 journal = {\mnras},
 month = {February},
 number = {2},
 pages = {489-499},
 primaryclass = {astro-ph},
 title = {{Properties of dark matter haloes in clusters, filaments, sheets and voids}},
 volume = {375},
 year = {2007}
}

@article{Kraljic_et_al2018,
 adsurl = {https://ui.adsabs.harvard.edu/abs/2018MNRAS.474..547K},
 archiveprefix = {arXiv},
 author = {{Kraljic}, K. and {Arnouts}, S. and {Pichon}, C. and {Laigle}, C. and {de la Torre}, S. and {Vibert}, D. and {Cadiou}, C. and {Dubois}, Y. and {Treyer}, M. and {Schimd}, C. and {Codis}, S. and {de Lapparent}, V. and {Devriendt}, J. and {Hwang}, H.~S. and {Le Borgne}, D. and {Malavasi}, N. and {Milliard}, B. and {Musso}, M. and {Pogosyan}, D. and {Alpaslan}, M. and {Bland-Hawthorn}, J. and {Wright}, A.~H.},
 doi = {10.1093/mnras/stx2638},
 eprint = {1710.02676},
 journal = {\mnras},
 month = {February},
 number = {1},
 pages = {547-571},
 primaryclass = {astro-ph.GA},
 title = {{Galaxy evolution in the metric of the cosmic web}},
 volume = {474},
 year = {2018}
}

@article{Dubois_et_al2014,
 adsurl = {https://ui.adsabs.harvard.edu/abs/2014MNRAS.444.1453D},
 archiveprefix = {arXiv},
 author = {{Dubois}, Y. and {Pichon}, C. and {Welker}, C. and {Le Borgne}, D. and {Devriendt}, J. and {Laigle}, C. and {Codis}, S. and {Pogosyan}, D. and {Arnouts}, S. and {Benabed}, K. and {Bertin}, E. and {Blaizot}, J. and {Bouchet}, F. and {Cardoso}, J. -F. and {Colombi}, S. and {de Lapparent}, V. and {Desjacques}, V. and {Gavazzi}, R. and {Kassin}, S. and {Kimm}, T. and {McCracken}, H. and {Milliard}, B. and {Peirani}, S. and {Prunet}, S. and {Rouberol}, S. and {Silk}, J. and {Slyz}, A. and {Sousbie}, T. and {Teyssier}, R. and {Tresse}, L. and {Treyer}, M. and {Vibert}, D. and {Volonteri}, M.},
 doi = {10.1093/mnras/stu1227},
 eprint = {1402.1165},
 journal = {\mnras},
 month = {October},
 number = {2},
 pages = {1453-1468},
 primaryclass = {astro-ph.CO},
 title = {{Dancing in the dark: galactic properties trace spin swings along the cosmic web}},
 volume = {444},
 year = {2014}
}

@article{hahn+09,
 adsurl = {https://ui.adsabs.harvard.edu/abs/2009MNRAS.398.1742H},
 archiveprefix = {arXiv},
 author = {{Hahn}, Oliver and {Porciani}, Cristiano and {Dekel}, Avishai and {Carollo}, C. Marcella},
 doi = {10.1111/j.1365-2966.2009.15271.x},
 eprint = {0803.4211},
 journal = {\mnras},
 month = {October},
 number = {4},
 pages = {1742-1756},
 primaryclass = {astro-ph},
 title = {{Tidal effects and the environment dependence of halo assembly}},
 volume = {398},
 year = {2009}
}

@article{musso+18,
 adsurl = {https://ui.adsabs.harvard.edu/abs/2018MNRAS.476.4877M},
 archiveprefix = {arXiv},
 author = {{Musso}, M. and {Cadiou}, C. and {Pichon}, C. and {Codis}, S. and {Kraljic}, K. and {Dubois}, Y.},
 doi = {10.1093/mnras/sty191},
 eprint = {1709.00834},
 journal = {\mnras},
 month = {June},
 number = {4},
 pages = {4877-4906},
 primaryclass = {astro-ph.CO},
 title = {{How does the cosmic web impact assembly bias?}},
 volume = {476},
 year = {2018}
}

@article{zomg-I,
 adsurl = {https://ui.adsabs.harvard.edu/abs/2017MNRAS.469..594B},
 archiveprefix = {arXiv},
 author = {{Borzyszkowski}, Mikolaj and {Porciani}, Cristiano and {Romano-D{\'\i}az}, Emilio and {Garaldi}, Enrico},
 doi = {10.1093/mnras/stx873},
 eprint = {1610.04231},
 journal = {\mnras},
 month = {July},
 number = {1},
 pages = {594-611},
 primaryclass = {astro-ph.CO},
 title = {{ZOMG - I. How the cosmic web inhibits halo growth and generates assembly bias}},
 volume = {469},
 year = {2017}
}

@article{Keres_et_al2005,
 adsurl = {https://ui.adsabs.harvard.edu/abs/2005MNRAS.363....2K},
 archiveprefix = {arXiv},
 author = {{Kere{\v{s}}}, Du{\v{s}}an and {Katz}, Neal and {Weinberg}, David H. and {Dav{\'e}}, Romeel},
 doi = {10.1111/j.1365-2966.2005.09451.x},
 eprint = {astro-ph/0407095},
 journal = {\mnras},
 month = {October},
 number = {1},
 pages = {2-28},
 primaryclass = {astro-ph},
 title = {{How do galaxies get their gas?}},
 volume = {363},
 year = {2005}
}

@inproceedings{Raychaudhury&Porter2005,
 adsurl = {https://ui.adsabs.harvard.edu/abs/2005AAS...20717715R},
 author = {{Raychaudhury}, S. and {Porter}, S.~C.},
 booktitle = {American Astronomical Society Meeting Abstracts},
 eid = {177.15},
 month = {December},
 pages = {177.15},
 series = {American Astronomical Society Meeting Abstracts},
 title = {{Star formation properties of galaxies in Supercluster filaments}},
 volume = {207},
 year = {2005}
}

@article{Kirk_et_al2013,
 adsurl = {https://ui.adsabs.harvard.edu/abs/2013ApJ...766..115K},
 archiveprefix = {arXiv},
 author = {{Kirk}, Helen and {Myers}, Philip C. and {Bourke}, Tyler L. and {Gutermuth}, Robert A. and {Hedden}, Abigail and {Wilson}, Grant W.},
 doi = {10.1088/0004-637X/766/2/115},
 eid = {115},
 eprint = {1301.6792},
 journal = {\apj},
 month = {April},
 number = {2},
 pages = {115},
 primaryclass = {astro-ph.GA},
 title = {{Filamentary Accretion Flows in the Embedded Serpens South Protocluster}},
 volume = {766},
 year = {2013}
}

@article{Seth&Raychaudhury2020,
 adsurl = {https://ui.adsabs.harvard.edu/abs/2020MNRAS.497..466S},
 archiveprefix = {arXiv},
 author = {{Seth}, Ruchika and {Raychaudhury}, Somak},
 doi = {10.1093/mnras/staa1779},
 eprint = {2006.09898},
 journal = {\mnras},
 month = {September},
 number = {1},
 pages = {466-481},
 primaryclass = {astro-ph.GA},
 title = {{Evolution of galaxies in groups in the Coma Supercluster}},
 volume = {497},
 year = {2020}
}

@article{Konyves_et_al2015,
 adsurl = {https://ui.adsabs.harvard.edu/abs/2015A&A...584A..91K},
 archiveprefix = {arXiv},
 author = {{K{\"o}nyves}, V. and {Andr{\'e}}, Ph. and {Men'shchikov}, A. and {Palmeirim}, P. and {Arzoumanian}, D. and {Schneider}, N. and {Roy}, A. and {Didelon}, P. and {Maury}, A. and {Shimajiri}, Y. and {Di Francesco}, J. and {Bontemps}, S. and {Peretto}, N. and {Benedettini}, M. and {Bernard}, J. -Ph. and {Elia}, D. and {Griffin}, M.~J. and {Hill}, T. and {Kirk}, J. and {Ladjelate}, B. and {Marsh}, K. and {Martin}, P.~G. and {Motte}, F. and {Nguy{\^e}n Luong}, Q. and {Pezzuto}, S. and {Roussel}, H. and {Rygl}, K.~L.~J. and {Sadavoy}, S.~I. and {Schisano}, E. and {Spinoglio}, L. and {Ward-Thompson}, D. and {White}, G.~J.},
 doi = {10.1051/0004-6361/201525861},
 eid = {A91},
 eprint = {1507.05926},
 journal = {\aap},
 month = {December},
 pages = {A91},
 primaryclass = {astro-ph.GA},
 title = {{A census of dense cores in the Aquila cloud complex: SPIRE/PACS observations from the Herschel Gould Belt survey}},
 volume = {584},
 year = {2015}
}

@article{Zhu_et_al2021,
 adsurl = {https://ui.adsabs.harvard.edu/abs/2021ApJ...920....2Z},
 archiveprefix = {arXiv},
 author = {{Zhu}, Weishan and {Zhang}, Fupeng and {Feng}, Long-Long},
 doi = {10.3847/1538-4357/ac15f1},
 eid = {2},
 eprint = {2107.08663},
 journal = {\apj},
 month = {October},
 number = {1},
 pages = {2},
 primaryclass = {astro-ph.CO},
 title = {{Profiles of Cosmic Filaments Since z = 4.0 in Cosmological Hydrodynamical Simulation}},
 volume = {920},
 year = {2021}
}

@article{wang+24,
 adsurl = {https://ui.adsabs.harvard.edu/abs/2024arXiv240211678W},
 archiveprefix = {arXiv},
 author = {{Wang}, Wei and {Wang}, Peng and {Guo}, Hong and {Kang}, Xi and {Libeskind}, Noam I. and {Galarraga-Espinosa}, Daniela and {Springel}, Volker and {Kannan}, Rahul and {Hernquist}, Lars and {Pakmor}, Rudiger and {Yu}, Haoran and {Bose}, Sownak and {Guo}, Quan and {Yu}, Luo and {Hernandez-Aguayo}, Cesar},
 eid = {arXiv:2402.11678},
 eprint = {2402.11678},
 journal = {arXiv e-prints},
 month = {February},
 pages = {arXiv:2402.11678},
 primaryclass = {astro-ph.CO},
 title = {{The boundary of cosmic filaments}},
 year = {2024}
}

@article{Einasto1965,
 adsurl = {https://ui.adsabs.harvard.edu/abs/1965TrAlm...5...87E},
 author = {{Einasto}, J.},
 journal = {Trudy Astrofizicheskogo Instituta Alma-Ata},
 month = {January},
 pages = {87-100},
 title = {{On the Construction of a Composite Model for the Galaxy and on the Determination of the System of Galactic Parameters}},
 volume = {5},
 year = {1965}
}

@article{merritt+06,
 adsurl = {https://ui.adsabs.harvard.edu/abs/2006AJ....132.2685M},
 archiveprefix = {arXiv},
 author = {{Merritt}, David and {Graham}, Alister W. and {Moore}, Ben and {Diemand}, J{\"u}rg and {Terzi{\'c}}, Bal{\v{s}}a},
 doi = {10.1086/508988},
 eprint = {astro-ph/0509417},
 journal = {\aj},
 month = {December},
 number = {6},
 pages = {2685-2700},
 primaryclass = {astro-ph},
 title = {{Empirical Models for Dark Matter Halos. I. Nonparametric Construction of Density Profiles and Comparison with Parametric Models}},
 volume = {132},
 year = {2006}
}

@article{Pan_et_al2012,
 adsurl = {https://ui.adsabs.harvard.edu/abs/2012MNRAS.421..926P},
 archiveprefix = {arXiv},
 author = {{Pan}, Danny C. and {Vogeley}, Michael S. and {Hoyle}, Fiona and {Choi}, Yun-Young and {Park}, Changbom},
 doi = {10.1111/j.1365-2966.2011.20197.x},
 eprint = {1103.4156},
 journal = {\mnras},
 month = {April},
 number = {2},
 pages = {926-934},
 primaryclass = {astro-ph.CO},
 title = {{Cosmic voids in Sloan Digital Sky Survey Data Release 7}},
 volume = {421},
 year = {2012}
}

@article{Nadathur_et_al2015,
 adsurl = {https://ui.adsabs.harvard.edu/abs/2015MNRAS.449.3997N},
 archiveprefix = {arXiv},
 author = {{Nadathur}, S. and {Hotchkiss}, S. and {Diego}, J.~M. and {Iliev}, I.~T. and {Gottl{\"o}ber}, S. and {Watson}, W.~A. and {Yepes}, G.},
 doi = {10.1093/mnras/stv513},
 eprint = {1407.1295},
 journal = {\mnras},
 month = {June},
 number = {4},
 pages = {3997-4009},
 primaryclass = {astro-ph.CO},
 title = {{Self-similarity and universality of void density profiles in simulation and SDSS data}},
 volume = {449},
 year = {2015}
}

@article{Hamaus_et_al2014,
 adsurl = {https://ui.adsabs.harvard.edu/abs/2014PhRvL.112y1302H},
 archiveprefix = {arXiv},
 author = {{Hamaus}, Nico and {Sutter}, P.~M. and {Wandelt}, Benjamin D.},
 doi = {10.1103/PhysRevLett.112.251302},
 eid = {251302},
 eprint = {1403.5499},
 journal = {\prl},
 month = {June},
 number = {25},
 pages = {251302},
 primaryclass = {astro-ph.CO},
 title = {{Universal Density Profile for Cosmic Voids}},
 volume = {112},
 year = {2014}
}

@article{Yang_et_al2022,
 adsurl = {https://ui.adsabs.harvard.edu/abs/2022MNRAS.516.6041Y},
 archiveprefix = {arXiv},
 author = {{Yang}, Tianyi and {Hudson}, Michael J. and {Afshordi}, Niayesh},
 doi = {10.1093/mnras/stac2564},
 eprint = {2203.16170},
 journal = {\mnras},
 month = {November},
 number = {4},
 pages = {6041-6054},
 primaryclass = {astro-ph.CO},
 title = {{A universal profile for stacked filaments from cold dark matter simulations}},
 volume = {516},
 year = {2022}
}

@ARTICLE{Xu+2026,
       author = {{Xu}, Peng and {Jiang}, Fangzhou and {Hasan}, Farhanul and {Woo}, Joanna and {Hellinger}, Douglas and {Primack}, Joel R. and {Faber}, Sandra M. and {Koo}, David},
        title = "{Universal Dark-matter Density Profiles of Cosmic Filaments}",
      journal = {arXiv e-prints},
         year = 2026,
        month = apr,
          eid = {arXiv:2604.05033},
        pages = {arXiv:2604.05033},
          doi = {10.48550/arXiv.2604.05033},
archivePrefix = {arXiv},
       eprint = {2604.05033},
 primaryClass = {astro-ph.CO},
       adsurl = {https://ui.adsabs.harvard.edu/abs/2026arXiv260405033X}
}

@ARTICLE{Hoffman+2012,
       author = {{Hoffman}, Yehuda and {Metuki}, Ofer and {Yepes}, Gustavo and {Gottl{\"o}ber}, Stefan and {Forero-Romero}, Jaime E. and {Libeskind}, Noam I. and {Knebe}, Alexander},
        title = "{A kinematic classification of the cosmic web}",
      journal = {\mnras},
         year = 2012,
        month = sep,
       volume = {425},
       number = {3},
        pages = {2049-2057},
          doi = {10.1111/j.1365-2966.2012.21553.x},
archivePrefix = {arXiv},
       eprint = {1201.3367},
 primaryClass = {astro-ph.CO},
       adsurl = {https://ui.adsabs.harvard.edu/abs/2012MNRAS.425.2049H}
}

@article{Disperse_illustration,
 adsurl = {https://ui.adsabs.harvard.edu/abs/2011MNRAS.414..384S},
 archiveprefix = {arXiv},
 author = {{Sousbie}, T. and {Pichon}, C. and {Kawahara}, H.},
 doi = {10.1111/j.1365-2966.2011.18395.x},
 eprint = {1009.4014},
 journal = {\mnras},
 month = {June},
 number = {1},
 pages = {384-403},
 primaryclass = {astro-ph.CO},
 title = {{The persistent cosmic web and its filamentary structure - II. Illustrations}},
 volume = {414},
 year = {2011}
}

@article{Disperse_theory,
 adsurl = {https://ui.adsabs.harvard.edu/abs/2011MNRAS.414..350S},
 archiveprefix = {arXiv},
 author = {{Sousbie}, T.},
 doi = {10.1111/j.1365-2966.2011.18394.x},
 eprint = {1009.4015},
 journal = {\mnras},
 month = {June},
 number = {1},
 pages = {350-383},
 primaryclass = {astro-ph.CO},
 title = {{The persistent cosmic web and its filamentary structure - I. Theory and implementation}},
 volume = {414},
 year = {2011}
}

@ARTICLE{Barrow+1985,
       author = {{Barrow}, J.~D. and {Bhavsar}, S.~P. and {Sonoda}, D.~H.},
        title = "{Minimal spanning trees, filaments and galaxy clustering}",
      journal = {\mnras},
         year = 1985,
        month = sep,
       volume = {216},
        pages = {17-35},
          doi = {10.1093/mnras/216.1.17},
       adsurl = {https://ui.adsabs.harvard.edu/abs/1985MNRAS.216...17B}
}

@ARTICLE{Pereyra+2020,
       author = {{Pereyra}, Luis A. and {Sgr{\'o}}, Mario A. and {Merch{\'a}n}, Manuel E. and {Stasyszyn}, Federico A. and {Paz}, Dante J.},
        title = "{Detection and analysis of cluster-cluster filaments}",
      journal = {\mnras},
         year = 2020,
        month = dec,
       volume = {499},
       number = {4},
        pages = {4876-4886},
          doi = {10.1093/mnras/staa3112},
archivePrefix = {arXiv},
       eprint = {1911.06768},
 primaryClass = {astro-ph.CO},
       adsurl = {https://ui.adsabs.harvard.edu/abs/2020MNRAS.499.4876P}
}

@ARTICLE{Abel+2012,
       author = {{Abel}, Tom and {Hahn}, Oliver and {Kaehler}, Ralf},
        title = "{Tracing the dark matter sheet in phase space}",
      journal = {\mnras},
         year = 2012,
        month = nov,
       volume = {427},
       number = {1},
        pages = {61-76},
          doi = {10.1111/j.1365-2966.2012.21754.x},
archivePrefix = {arXiv},
       eprint = {1111.3944},
 primaryClass = {astro-ph.CO},
       adsurl = {https://ui.adsabs.harvard.edu/abs/2012MNRAS.427...61A}
}

@ARTICLE{Shandarin+2012,
       author = {{Shandarin}, Sergei and {Habib}, Salman and {Heitmann}, Katrin},
        title = "{Cosmic web, multistream flows, and tessellations}",
      journal = {\prd},
         year = 2012,
        month = apr,
       volume = {85},
       number = {8},
          eid = {083005},
        pages = {083005},
          doi = {10.1103/PhysRevD.85.083005},
archivePrefix = {arXiv},
       eprint = {1111.2366},
 primaryClass = {astro-ph.CO},
       adsurl = {https://ui.adsabs.harvard.edu/abs/2012PhRvD..85h3005S}
}

@ARTICLE{Sahyadri2026,
       author = {{Dhawalikar}, Saee and {Alam}, Shadab and {Paranjape}, Aseem and {Banerjee}, Arka},
        title = "{Sahyadri: A simulation suite for the cosmology dependence of the Cosmic Web}",
      journal = {arXiv e-prints},
         year = 2026,
        month = jan,
          eid = {arXiv:2601.07924},
        pages = {arXiv:2601.07924},
          doi = {10.48550/arXiv.2601.07924},
archivePrefix = {arXiv},
       eprint = {2601.07924},
 primaryClass = {astro-ph.CO},
       adsurl = {https://ui.adsabs.harvard.edu/abs/2026arXiv260107924D}
}

@article{ramsoy+21,
 adsurl = {https://ui.adsabs.harvard.edu/abs/2021MNRAS.502..351R},
 archiveprefix = {arXiv},
 author = {{Rams{\o}y}, Marius and {Slyz}, Adrianne and {Devriendt}, Julien and {Laigle}, Clotilde and {Dubois}, Yohan},
 doi = {10.1093/mnras/stab015},
 eprint = {2101.00844},
 journal = {\mnras},
 month = {March},
 number = {1},
 pages = {351-368},
 primaryclass = {astro-ph.GA},
 title = {{Rivers of gas - I. Unveiling the properties of high redshift filaments}},
 volume = {502},
 year = {2021}
}

@article{Flows_around_galaxiesI,
 adsurl = {https://ui.adsabs.harvard.edu/abs/2023A&A...671A.160G},
 archiveprefix = {arXiv},
 author = {{Gal{\'a}rraga-Espinosa}, Daniela and {Garaldi}, Enrico and {Kauffmann}, Guinevere},
 doi = {10.1051/0004-6361/202244935},
 eid = {A160},
 eprint = {2209.05495},
 journal = {\aap},
 month = {March},
 pages = {A160},
 primaryclass = {astro-ph.GA},
 title = {{Flows around galaxies. I. The dependence of galaxy connectivity on cosmic environments and effects on the star formation rate}},
 volume = {671},
 year = {2023}
}

@article{Rost_et_al2024,
 adsurl = {https://ui.adsabs.harvard.edu/abs/2024MNRAS.527.1301R},
 archiveprefix = {arXiv},
 author = {{Rost}, Agust{\'\i}n M. and {Nuza}, Sebasti{\'a}n E. and {Stasyszyn}, Federico and {Kuchner}, Ulrike and {Hoeft}, Matthias and {Welker}, Charlotte and {Pearce}, Frazer and {Gray}, Meghan and {Knebe}, Alexander and {Cui}, Weiguang and {Yepes}, Gustavo},
 doi = {10.1093/mnras/stad3208},
 eprint = {2310.12245},
 journal = {\mnras},
 month = {January},
 number = {1},
 pages = {1301-1316},
 primaryclass = {astro-ph.CO},
 title = {{The three hundred project: thermodynamical properties, shocks, and gas dynamics in simulated galaxy cluster filaments and their surroundings}},
 volume = {527},
 year = {2024}
}

@article{Muru_tempel2021,
 adsurl = {https://ui.adsabs.harvard.edu/abs/2021A&A...649A.108M},
 archiveprefix = {arXiv},
 author = {{Muru}, Moorits Mihkel and {Tempel}, Elmo},
 doi = {10.1051/0004-6361/202039169},
 eid = {A108},
 eprint = {2103.06619},
 journal = {\aap},
 month = {May},
 pages = {A108},
 primaryclass = {astro-ph.CO},
 title = {{Assessing the reliability of the Bisous filament finder}},
 volume = {649},
 year = {2021}
}

@article{bl24,
 adsurl = {https://ui.adsabs.harvard.edu/abs/2024arXiv240204837B},
 archiveprefix = {arXiv},
 author = {{Boldrini}, Pierre and {Laigle}, Clotilde},
 doi = {10.48550/arXiv.2402.04837},
 eid = {arXiv:2402.04837},
 eprint = {2402.04837},
 journal = {arXiv e-prints},
 month = {February},
 pages = {arXiv:2402.04837},
 primaryclass = {astro-ph.CO},
 title = {{Distinguish dark matter theories with the cosmic web and next-generation surveys I: an alternative theory of gravity}},
 year = {2024}
}

@article{Matsubara2008,
  title = {Resumming cosmological perturbations via the Lagrangian picture: One-loop results in real space and in redshift space},
  author = {Matsubara, Takahiko},
  journal = {Phys. Rev. D},
  volume = {77},
  issue = {6},
  pages = {063530},
  numpages = {19},
  year = {2008},
  month = {Mar},
  publisher = {American Physical Society},
  doi = {10.1103/PhysRevD.77.063530},
  url = {https://link.aps.org/doi/10.1103/PhysRevD.77.063530}
}

@ARTICLE{Fisher_stabilization2026,
       author = {{Dhawalikar}, Saee and {Paranjape}, Aseem and {Alam}, Shadab},
        title = "{Stabilizing simulation-based cosmological Fisher forecasts: a case study using the Voronoi volume function}",
      journal = {\jcap},
         year = 2026,
        month = mar,
       volume = {2026},
       number = {3},
          eid = {040},
        pages = {040},
          doi = {10.1088/1475-7516/2026/03/040},
archivePrefix = {arXiv},
       eprint = {2506.16408},
 primaryClass = {astro-ph.CO},
       adsurl = {https://ui.adsabs.harvard.edu/abs/2026JCAP...03..040D}
}

@article{Galarraga_Espinosa_et_al2023a,
 adsurl = {https://ui.adsabs.harvard.edu/abs/2023arXiv230908659G},
 archiveprefix = {arXiv},
 author = {{Gal{\'a}rraga-Espinosa}, Daniela and {Cadiou}, Corentin and {Gouin}, C{\'e}line and {White}, Simon D.~M. and {Springel}, Volker and {Pakmor}, R{\"u}diger and {Hadzhiyska}, Boryana and {Bose}, Sownak and {Ferlito}, Fulvio and {Hernquist}, Lars and {Kannan}, Rahul and {Barrera}, Monica and {Delgado}, Ana Maria and {Hern{\'a}ndez-Aguayo}, C{\'e}sar},
 doi = {10.48550/arXiv.2309.08659},
 eid = {arXiv:2309.08659},
 eprint = {2309.08659},
 journal = {arXiv e-prints},
 month = {September},
 pages = {arXiv:2309.08659},
 primaryclass = {astro-ph.CO},
 title = {{Evolution of cosmic filaments in the MTNG simulation}},
 year = {2023}
}

@article{ps12,
 adsurl = {https://ui.adsabs.harvard.edu/abs/2012MNRAS.426.2789P},
 archiveprefix = {arXiv},
 author = {{Paranjape}, Aseem and {Sheth}, Ravi K.},
 doi = {10.1111/j.1365-2966.2012.21911.x},
 eprint = {1206.3506},
 journal = {\mnras},
 month = {November},
 number = {4},
 pages = {2789-2796},
 primaryclass = {astro-ph.CO},
 title = {{Peaks theory and the excursion set approach}},
 volume = {426},
 year = {2012}
}

@article{Shethetal2001,
 adsurl = {https://ui.adsabs.harvard.edu/abs/2001MNRAS.325.1288S},
 archiveprefix = {arXiv},
 author = {{Sheth}, Ravi K. and {Hui}, Lam and {Diaferio}, Antonaldo and {Scoccimarro}, Rom{\'a}n},
 doi = {10.1046/j.1365-8711.2001.04222.x},
 eprint = {astro-ph/0009167},
 journal = {\mnras},
 month = {August},
 number = {4},
 pages = {1288-1302},
 primaryclass = {astro-ph},
 title = {{Linear and non-linear contributions to pairwise peculiar velocities}},
 volume = {325},
 year = {2001}
}

@book{Binney&Tremaine1987,
 adsurl = {https://ui.adsabs.harvard.edu/abs/1987gady.book.....B},
 author = {{Binney}, James and {Tremaine}, Scott},
 title = {{Galactic dynamics}},
 year = {1987}
}

@article{Catelan_et_al1996,
 adsurl = {https://ui.adsabs.harvard.edu/abs/1996MNRAS.282..436C},
 archiveprefix = {arXiv},
 author = {{Catelan}, Paolo and {Theuns}, Tom},
 doi = {10.1093/mnras/282.2.436},
 eprint = {astro-ph/9604077},
 journal = {\mnras},
 month = {September},
 number = {2},
 pages = {436-454},
 primaryclass = {astro-ph},
 title = {{Evolution of the angular momentum of protogalaxies from tidal torques: Zel'dovich approximation}},
 volume = {282},
 year = {1996}
}

@article{Jeeson-Daniel_et_al2011,
 adsurl = {https://ui.adsabs.harvard.edu/abs/2011MNRAS.415L..69J},
 archiveprefix = {arXiv},
 author = {{Jeeson-Daniel}, Akila and {Dalla Vecchia}, Claudio and {Haas}, Marcel R. and {Schaye}, Joop},
 doi = {10.1111/j.1745-3933.2011.01081.x},
 eprint = {1103.5467},
 journal = {\mnras},
 month = {July},
 number = {1},
 pages = {L69-L73},
 primaryclass = {astro-ph.CO},
 title = {{The correlation structure of dark matter halo properties}},
 volume = {415},
 year = {2011}
}

@article{Blazek_et_al2011,
 adsurl = {https://ui.adsabs.harvard.edu/abs/2011JCAP...05..010B},
 archiveprefix = {arXiv},
 author = {{Blazek}, Jonathan and {McQuinn}, Matthew and {Seljak}, Uro{\v{s}}},
 doi = {10.1088/1475-7516/2011/05/010},
 eid = {010},
 eprint = {1101.4017},
 journal = {\jcap},
 month = {May},
 number = {5},
 pages = {010},
 primaryclass = {astro-ph.CO},
 title = {{Testing the tidal alignment model of galaxy intrinsic alignment}},
 volume = {2011},
 year = {2011}
}

@article{Maion_et_al2023,
 adsurl = {https://ui.adsabs.harvard.edu/abs/2023arXiv230713754M},
 archiveprefix = {arXiv},
 author = {{Maion}, Francisco and {Angulo}, Raul E. and {Bakx}, Thomas and {Chisari}, Nora Elisa and {Kurita}, Toshiki and {Pellejero-Ib{\'a}{\~n}ez}, Marcos},
 doi = {10.48550/arXiv.2307.13754},
 eid = {arXiv:2307.13754},
 eprint = {2307.13754},
 journal = {arXiv e-prints},
 month = {July},
 pages = {arXiv:2307.13754},
 primaryclass = {astro-ph.CO},
 title = {{HYMALAIA: A Hybrid Lagrangian Model for Intrinsic Alignments}},
 year = {2023}
}

@article{Tinker_et_al2005,
 adsurl = {https://ui.adsabs.harvard.edu/abs/2005ApJ...631...41T},
 archiveprefix = {arXiv},
 author = {{Tinker}, Jeremy L. and {Weinberg}, David H. and {Zheng}, Zheng and {Zehavi}, Idit},
 doi = {10.1086/432084},
 eprint = {astro-ph/0411777},
 journal = {\apj},
 month = {September},
 number = {1},
 pages = {41-58},
 primaryclass = {astro-ph},
 title = {{On the Mass-to-Light Ratio of Large-Scale Structure}},
 volume = {631},
 year = {2005}
}

@article{Bosch_et_al2013,
 adsurl = {https://ui.adsabs.harvard.edu/abs/2013MNRAS.430..725V},
 archiveprefix = {arXiv},
 author = {{van den Bosch}, Frank C. and {More}, Surhud and {Cacciato}, Marcello and {Mo}, Houjun and {Yang}, Xiaohu},
 doi = {10.1093/mnras/sts006},
 eprint = {1206.6890},
 journal = {\mnras},
 month = {April},
 number = {2},
 pages = {725-746},
 primaryclass = {astro-ph.CO},
 title = {{Cosmological constraints from a combination of galaxy clustering and lensing - I. Theoretical framework}},
 volume = {430},
 year = {2013}
}

@book{apostolcalculus,
 author = {Apostol, T.M.},
 isbn = {9788126515196},
 publisher = {Wiley India Pvt. Limited},
 title = {Calculus, Volume I, 2nd Ed One-variable Calculus, with an Introduction to Linear Algebra},
 url = {https://books.google.co.in/books?id=vTpbq0UPDaQC},
 year = {2007}
}

@article{Bond_et_al1991,
 adsurl = {https://ui.adsabs.harvard.edu/abs/1991ApJ...379..440B},
 author = {{Bond}, J.~R. and {Cole}, S. and {Efstathiou}, G. and {Kaiser}, N.},
 doi = {10.1086/170520},
 journal = {\apj},
 month = {October},
 pages = {440},
 title = {{Excursion Set Mass Functions for Hierarchical Gaussian Fluctuations}},
 volume = {379},
 year = {1991}
}

@article{picasa,
 adsurl = {https://ui.adsabs.harvard.edu/abs/2022arXiv220507906P},
 archiveprefix = {arXiv},
 author = {{Paranjape}, Aseem},
 doi = {10.48550/arXiv.2205.07906},
 eid = {arXiv:2205.07906},
 eprint = {2205.07906},
 journal = {arXiv e-prints},
 month = {May},
 pages = {arXiv:2205.07906},
 primaryclass = {astro-ph.CO},
 title = {{A simulated annealing approach to parameter inference with expensive likelihoods}},
 year = {2022}
}

@ARTICLE{Gao+2004,
       author = {{Gao}, L. and {White}, S.~D.~M. and {Jenkins}, A. and {Stoehr}, F. and {Springel}, V.},
        title = "{The subhalo populations of {\ensuremath{\Lambda}}CDM dark haloes}",
      journal = {\mnras},
         year = 2004,
        month = dec,
       volume = {355},
       number = {3},
        pages = {819-834},
          doi = {10.1111/j.1365-2966.2004.08360.x},
archivePrefix = {arXiv},
       eprint = {astro-ph/0404589},
 primaryClass = {astro-ph},
       adsurl = {https://ui.adsabs.harvard.edu/abs/2004MNRAS.355..819G}
}

@ARTICLE{ZOBOV2008,
       author = {{Neyrinck}, Mark C.},
        title = "{ZOBOV: a parameter-free void-finding algorithm}",
      journal = {\mnras},
         year = 2008,
        month = jun,
       volume = {386},
       number = {4},
        pages = {2101-2109},
          doi = {10.1111/j.1365-2966.2008.13180.x},
archivePrefix = {arXiv},
       eprint = {0712.3049},
 primaryClass = {astro-ph},
       adsurl = {https://ui.adsabs.harvard.edu/abs/2008MNRAS.386.2101N}
}

@ARTICLE{Platen+2007,
       author = {{Platen}, Erwin and {van de Weygaert}, Rien and {Jones}, Bernard J.~T.},
        title = "{A cosmic watershed: the WVF void detection technique}",
      journal = {\mnras},
         year = 2007,
        month = sep,
       volume = {380},
       number = {2},
        pages = {551-570},
          doi = {10.1111/j.1365-2966.2007.12125.x},
archivePrefix = {arXiv},
       eprint = {0706.2788},
 primaryClass = {astro-ph},
       adsurl = {https://ui.adsabs.harvard.edu/abs/2007MNRAS.380..551P}
}

@ARTICLE{Sutter+2015,
       author = {{Sutter}, P.~M. and {Lavaux}, G. and {Hamaus}, N. and {Pisani}, A. and {Wandelt}, B.~D. and {Warren}, M. and {Villaescusa-Navarro}, F. and {Zivick}, P. and {Mao}, Q. and {Thompson}, B.~B.},
        title = "{VIDE: The Void IDentification and Examination toolkit}",
      journal = {Astronomy and Computing},
         year = 2015,
        month = mar,
       volume = {9},
        pages = {1-9},
          doi = {10.1016/j.ascom.2014.10.002},
archivePrefix = {arXiv},
       eprint = {1406.1191},
 primaryClass = {astro-ph.CO},
       adsurl = {https://ui.adsabs.harvard.edu/abs/2015A&C.....9....1S}
}

@PHDTHESIS{Schaap2007,
       author = {{Schaap}, Willem Egbert},
        title = "{DTFE: The Delaunay tessellation field estimator}",
       school = {University of Groningen, Netherlands},
         year = 2007,
        month = jan,
       adsurl = {https://ui.adsabs.harvard.edu/abs/2007PhDT.......486S}
}

@ARTICLE{Schaap+2000,
       author = {{Schaap}, W.~E. and {van de Weygaert}, R.},
        title = "{Continuous fields and discrete samples: reconstruction through Delaunay tessellations}",
      journal = {\aap},
         year = 2000,
        month = nov,
       volume = {363},
        pages = {L29-L32},
          doi = {10.48550/arXiv.astro-ph/0011007},
archivePrefix = {arXiv},
       eprint = {astro-ph/0011007},
 primaryClass = {astro-ph},
       adsurl = {https://ui.adsabs.harvard.edu/abs/2000A&A...363L..29S}
}

@ARTICLE{Diaz+2007,
       author = {{Romano-D{\'\i}az}, Emilio and {van de Weygaert}, Rien},
        title = "{Delaunay Tessellation Field Estimator analysis of the PSCz local Universe: density field and cosmic flow}",
      journal = {\mnras},
         year = 2007,
        month = nov,
       volume = {382},
       number = {1},
        pages = {2-28},
          doi = {10.1111/j.1365-2966.2007.12190.x},
       adsurl = {https://ui.adsabs.harvard.edu/abs/2007MNRAS.382....2R}
}

@ARTICLE{vdWeygaert+2009,
       author = {{van de Weygaert}, Rien and {Aragon-Calvo}, Miguel A. and {Jones}, Bernard J.~T. and {Platen}, Erwin},
        title = "{Geometry and Morphology of the Cosmic Web: Analyzing Spatial Patterns in the Universe}",
      journal = {arXiv e-prints},
         year = 2009,
        month = dec,
          eid = {arXiv:0912.3448},
        pages = {arXiv:0912.3448},
          doi = {10.48550/arXiv.0912.3448},
archivePrefix = {arXiv},
       eprint = {0912.3448},
 primaryClass = {astro-ph.IM},
       adsurl = {https://ui.adsabs.harvard.edu/abs/2009arXiv0912.3448V}
}

@ARTICLE{Espinosa+2022,
       author = {{Gal{\'a}rraga-Espinosa}, Daniela and {Langer}, Mathieu and {Aghanim}, Nabila},
        title = "{Relative distribution of dark matter, gas, and stars around cosmic filaments in the IllustrisTNG simulation}",
      journal = {\aap},
         year = 2022,
        month = may,
       volume = {661},
          eid = {A115},
        pages = {A115},
          doi = {10.1051/0004-6361/202141974},
archivePrefix = {arXiv},
       eprint = {2109.06198},
 primaryClass = {astro-ph.CO},
       adsurl = {https://ui.adsabs.harvard.edu/abs/2022A&A...661A.115G}
}

@ARTICLE{Hidding+2014,
       author = {{Hidding}, Johan and {Shandarin}, Sergei F. and {van de Weygaert}, Rien},
        title = "{The Zel'dovich approximation: key to understanding cosmic web complexity}",
      journal = {\mnras},
         year = 2014,
        month = feb,
       volume = {437},
       number = {4},
        pages = {3442-3472},
          doi = {10.1093/mnras/stt2142},
archivePrefix = {arXiv},
       eprint = {1311.7134},
 primaryClass = {astro-ph.CO},
       adsurl = {https://ui.adsabs.harvard.edu/abs/2014MNRAS.437.3442H}
}

@ARTICLE{Sousbie+2008,
       author = {{Sousbie}, T. and {Pichon}, C. and {Colombi}, S. and {Novikov}, D. and {Pogosyan}, D.},
        title = "{The 3D skeleton: tracing the filamentary structure of the Universe}",
      journal = {\mnras},
         year = 2008,
        month = feb,
       volume = {383},
       number = {4},
        pages = {1655-1670},
          doi = {10.1111/j.1365-2966.2007.12685.x},
archivePrefix = {arXiv},
       eprint = {0707.3123},
 primaryClass = {astro-ph},
       adsurl = {https://ui.adsabs.harvard.edu/abs/2008MNRAS.383.1655S}
}

@ARTICLE{Codis+2018,
       author = {{Codis}, Sandrine and {Pogosyan}, Dmitri and {Pichon}, Christophe},
        title = "{On the connectivity of the cosmic web: theory and implications for cosmology and galaxy formation}",
      journal = {\mnras},
         year = 2018,
        month = sep,
       volume = {479},
       number = {1},
        pages = {973-993},
          doi = {10.1093/mnras/sty1643},
archivePrefix = {arXiv},
       eprint = {1803.11477},
 primaryClass = {astro-ph.CO},
       adsurl = {https://ui.adsabs.harvard.edu/abs/2018MNRAS.479..973C}
}

@ARTICLE{Sheth&Tormen2004,
       author = {{Sheth}, Ravi K. and {Tormen}, Giuseppe},
        title = "{On the environmental dependence of halo formation}",
      journal = {\mnras},
         year = 2004,
        month = jun,
       volume = {350},
       number = {4},
        pages = {1385-1390},
          doi = {10.1111/j.1365-2966.2004.07733.x},
archivePrefix = {arXiv},
       eprint = {astro-ph/0402237},
 primaryClass = {astro-ph},
       adsurl = {https://ui.adsabs.harvard.edu/abs/2004MNRAS.350.1385S}
}

@ARTICLE{Crocce&Scoccimarro2006a,
       author = {{Crocce}, Mart{\'\i}n and {Scoccimarro}, Rom{\'a}n},
        title = "{Renormalized cosmological perturbation theory}",
      journal = {\prd},
         year = 2006,
        month = mar,
       volume = {73},
       number = {6},
          eid = {063519},
        pages = {063519},
          doi = {10.1103/PhysRevD.73.063519},
archivePrefix = {arXiv},
       eprint = {astro-ph/0509418},
 primaryClass = {astro-ph},
       adsurl = {https://ui.adsabs.harvard.edu/abs/2006PhRvD..73f3519C}
}

@ARTICLE{Crocce&Scoccimarro2006b,
       author = {{Crocce}, Mart{\'\i}n and {Scoccimarro}, Rom{\'a}n},
        title = "{Memory of initial conditions in gravitational clustering}",
      journal = {\prd},
         year = 2006,
        month = mar,
       volume = {73},
       number = {6},
          eid = {063520},
        pages = {063520},
          doi = {10.1103/PhysRevD.73.063520},
archivePrefix = {arXiv},
       eprint = {astro-ph/0509419},
 primaryClass = {astro-ph},
       adsurl = {https://ui.adsabs.harvard.edu/abs/2006PhRvD..73f3520C}
}

@ARTICLE{Crocce&Scoccimarro2008,
       author = {{Crocce}, Mart{\'\i}n and {Scoccimarro}, Rom{\'a}n},
        title = "{Nonlinear evolution of baryon acoustic oscillations}",
      journal = {\prd},
         year = 2008,
        month = jan,
       volume = {77},
       number = {2},
          eid = {023533},
        pages = {023533},
          doi = {10.1103/PhysRevD.77.023533},
archivePrefix = {arXiv},
       eprint = {0704.2783},
 primaryClass = {astro-ph},
       adsurl = {https://ui.adsabs.harvard.edu/abs/2008PhRvD..77b3533C}
}

@ARTICLE{Lemaitre1927,
       author = {{Lema{\^\i}tre}, G.},
        title = "{Un Univers homog{\`e}ne de masse constante et de rayon croissant rendant compte de la vitesse radiale des n{\'e}buleuses extra-galactiques}",
      journal = {Annales de la Soci{\'e}t{\'e} Scientifique de Bruxelles},
         year = 1927,
        month = jan,
       volume = {47},
        pages = {49-59},
       adsurl = {https://ui.adsabs.harvard.edu/abs/1927ASSB...47...49L}
}

@ARTICLE{Hubble1929,
       author = {{Hubble}, Edwin},
        title = "{A Relation between Distance and Radial Velocity among Extra-Galactic Nebulae}",
      journal = {Proceedings of the National Academy of Science},
         year = 1929,
        month = mar,
       volume = {15},
       number = {3},
        pages = {168-173},
          doi = {10.1073/pnas.15.3.168},
       adsurl = {https://ui.adsabs.harvard.edu/abs/1929PNAS...15..168H}
}

@BOOK{Peebles_cosmology,
       author = {{Peebles}, P.~J.~E.},
        title = "{Principles of Physical Cosmology}",
         year = 1993,
          doi = {10.1515/9780691206721},
       adsurl = {https://ui.adsabs.harvard.edu/abs/1993ppc..book.....P}
}

@BOOK{Peacock_cosmology,
       author = {{Peacock}, John A.},
        title = "{Cosmological Physics}",
         year = 1999,
       adsurl = {https://ui.adsabs.harvard.edu/abs/1999coph.book.....P}
}

@ARTICLE{Manera+2013,
       author = {{Manera}, Marc and {Scoccimarro}, Roman and {Percival}, Will J. and {Samushia}, Lado and {McBride}, Cameron K. and {Ross}, Ashley J. and {Sheth}, Ravi K. and {White}, Martin and {Reid}, Beth A. and {S{\'a}nchez}, Ariel G. and {de Putter}, Roland and {Xu}, Xiaoying and {Berlind}, Andreas A. and {Brinkmann}, Jonathan and {Maraston}, Claudia and {Nichol}, Bob and {Montesano}, Francesco and {Padmanabhan}, Nikhil and {Skibba}, Ramin A. and {Tojeiro}, Rita and {Weaver}, Benjamin A.},
        title = "{The clustering of galaxies in the SDSS-III Baryon Oscillation Spectroscopic Survey: a large sample of mock galaxy catalogues}",
      journal = {\mnras},
         year = 2013,
        month = jan,
       volume = {428},
       number = {2},
        pages = {1036-1054},
          doi = {10.1093/mnras/sts084},
archivePrefix = {arXiv},
       eprint = {1203.6609},
 primaryClass = {astro-ph.CO},
       adsurl = {https://ui.adsabs.harvard.edu/abs/2013MNRAS.428.1036M}
}

@ARTICLE{DES_mocks2018,
       author = {{Avila}, S. and {Crocce}, M. and {Ross}, A.~J. and {Garc{\'\i}a-Bellido}, J. and {Percival}, W.~J. and {Banik}, N. and {Camacho}, H. and {Kokron}, N. and {Chan}, K.~C. and {Andrade-Oliveira}, F. and {Gomes}, R. and {Gomes}, D. and {Lima}, M. and {Rosenfeld}, R. and {Salvador}, A.~I. and {Friedrich}, O. and {Abdalla}, F.~B. and {Annis}, J. and {Benoit-L{\'e}vy}, A. and {Bertin}, E. and {Brooks}, D. and {Carrasco Kind}, M. and {Carretero}, J. and {Castander}, F.~J. and {Cunha}, C.~E. and {da Costa}, L.~N. and {Davis}, C. and {De Vicente}, J. and {Doel}, P. and {Fosalba}, P. and {Frieman}, J. and {Gerdes}, D.~W. and {Gruen}, D. and {Gruendl}, R.~A. and {Gutierrez}, G. and {Hartley}, W.~G. and {Hollowood}, D. and {Honscheid}, K. and {James}, D.~J. and {Kuehn}, K. and {Kuropatkin}, N. and {Miquel}, R. and {Plazas}, A.~A. and {Sanchez}, E. and {Scarpine}, V. and {Schindler}, R. and {Schubnell}, M. and {Sevilla-Noarbe}, I. and {Smith}, M. and {Sobreira}, F. and {Suchyta}, E. and {Swanson}, M.~E.~C. and {Tarle}, G. and {Thomas}, D. and {Walker}, A.~R. and {Dark Energy Survey Collaboration}},
        title = "{Dark Energy Survey Year-1 results: galaxy mock catalogues for BAO}",
      journal = {\mnras},
         year = 2018,
        month = sep,
       volume = {479},
       number = {1},
        pages = {94-110},
          doi = {10.1093/mnras/sty1389},
archivePrefix = {arXiv},
       eprint = {1712.06232},
 primaryClass = {astro-ph.CO},
       adsurl = {https://ui.adsabs.harvard.edu/abs/2018MNRAS.479...94A}
}

@ARTICLE{Blot+2019,
       author = {{Blot}, Linda and {Crocce}, Martin and {Sefusatti}, Emiliano and {Lippich}, Martha and {S{\'a}nchez}, Ariel G. and {Colavincenzo}, Manuel and {Monaco}, Pierluigi and {Alvarez}, Marcelo A. and {Agrawal}, Aniket and {Avila}, Santiago and {Balaguera-Antol{\'\i}nez}, Andr{\'e}s and {Bond}, Richard and {Codis}, Sandrine and {Dalla Vecchia}, Claudio and {Dorta}, Antonio and {Fosalba}, Pablo and {Izard}, Albert and {Kitaura}, Francisco-Shu and {Pellejero-Ibanez}, Marcos and {Stein}, George and {Vakili}, Mohammadjavad and {Yepes}, Gustavo},
        title = "{Comparing approximate methods for mock catalogues and covariance matrices II: power spectrum multipoles}",
      journal = {\mnras},
         year = 2019,
        month = may,
       volume = {485},
       number = {2},
        pages = {2806-2824},
          doi = {10.1093/mnras/stz507},
archivePrefix = {arXiv},
       eprint = {1806.09497},
 primaryClass = {astro-ph.CO},
       adsurl = {https://ui.adsabs.harvard.edu/abs/2019MNRAS.485.2806B}
}

@ARTICLE{Cranmer+2020,
       author = {{Cranmer}, Kyle and {Brehmer}, Johann and {Louppe}, Gilles},
        title = "{The frontier of simulation-based inference}",
      journal = {Proceedings of the National Academy of Science},
         year = 2020,
        month = dec,
       volume = {117},
       number = {48},
        pages = {30055-30062},
          doi = {10.1073/pnas.1912789117},
archivePrefix = {arXiv},
       eprint = {1911.01429},
 primaryClass = {stat.ML},
       adsurl = {https://ui.adsabs.harvard.edu/abs/2020PNAS..11730055C}
}

@ARTICLE{Angulo&Hahn2022,
       author = {{Angulo}, Raul E. and {Hahn}, Oliver},
        title = "{Large-scale dark matter simulations}",
      journal = {Living Reviews in Computational Astrophysics},
         year = 2022,
        month = dec,
       volume = {8},
       number = {1},
          eid = {1},
        pages = {1},
          doi = {10.1007/s41115-021-00013-z},
archivePrefix = {arXiv},
       eprint = {2112.05165},
 primaryClass = {astro-ph.CO},
       adsurl = {https://ui.adsabs.harvard.edu/abs/2022LRCA....8....1A}
}

@ARTICLE{Grove+2022,
       author = {{Grove}, Cameron and {Chuang}, Chia-Hsun and {Devi}, Ningombam Chandrachani and {Garrison}, Lehman and {L'Huillier}, Benjamin and {Feng}, Yu and {Helly}, John and {Hern{\'a}ndez-Aguayo}, C{\'e}sar and {Alam}, Shadab and {Zhang}, Hanyu and {Yu}, Yu and {Cole}, Shaun and {Eisenstein}, Daniel and {Norberg}, Peder and {Wechsler}, Risa and {Brooks}, David and {Dawson}, Kyle and {Landriau}, Martin and {Meisner}, Aaron and {Poppett}, Claire and {Tarl{\'e}}, Gregory and {Valenzuela}, Octavio},
        title = "{The DESI N-body simulation project - I. Testing the robustness of simulations for the DESI dark time survey}",
      journal = {\mnras},
         year = 2022,
        month = sep,
       volume = {515},
       number = {2},
        pages = {1854-1870},
          doi = {10.1093/mnras/stac1947},
archivePrefix = {arXiv},
       eprint = {2112.09138},
 primaryClass = {astro-ph.CO},
       adsurl = {https://ui.adsabs.harvard.edu/abs/2022MNRAS.515.1854G}
}

@ARTICLE{Smith+2024,
       author = {{Smith}, A. and {Grove}, C. and {Cole}, S. and {Norberg}, P. and {Zarrouk}, P. and {Yuan}, S. and {Aguilar}, J. and {Ahlen}, S. and {Brooks}, D. and {Claybaugh}, T. and {de la Macorra}, A. and {Doel}, P. and {Forero-Romero}, J.~E. and {Gazta{\~n}aga}, E. and {Gontcho}, S. Gontcho A. and {Hahn}, C. and {Kehoe}, R. and {Kremin}, A. and {Levi}, M.~E. and {Manera}, M. and {Meisner}, A. and {Miquel}, R. and {Moustakas}, J. and {Nie}, J. and {Percival}, W.~J. and {Rezaie}, M. and {Rossi}, G. and {Sanchez}, E. and {Seo}, H. and {Tarl{\'e}}, G. and {Zhou}, Z.},
        title = "{Generating mock galaxy catalogues for flux-limited samples like the DESI Bright Galaxy Survey}",
      journal = {\mnras},
         year = 2024,
        month = jul,
       volume = {532},
       number = {1},
        pages = {903-919},
          doi = {10.1093/mnras/stae1503},
archivePrefix = {arXiv},
       eprint = {2312.08792},
 primaryClass = {astro-ph.CO},
       adsurl = {https://ui.adsabs.harvard.edu/abs/2024MNRAS.532..903S}
}

@ARTICLE{Zehavi+2005,
       author = {{Zehavi}, Idit and {Zheng}, Zheng and {Weinberg}, David H. and {Frieman}, Joshua A. and {Berlind}, Andreas A. and {Blanton}, Michael R. and {Scoccimarro}, Roman and {Sheth}, Ravi K. and {Strauss}, Michael A. and {Kayo}, Issha and {Suto}, Yasushi and {Fukugita}, Masataka and {Nakamura}, Osamu and {Bahcall}, Neta A. and {Brinkmann}, Jon and {Gunn}, James E. and {Hennessy}, Greg S. and {Ivezi{\'c}}, {\v{Z}}eljko and {Knapp}, Gillian R. and {Loveday}, Jon and {Meiksin}, Avery and {Schlegel}, David J. and {Schneider}, Donald P. and {Szapudi}, Istvan and {Tegmark}, Max and {Vogeley}, Michael S. and {York}, Donald G. and {SDSS Collaboration}},
        title = "{The Luminosity and Color Dependence of the Galaxy Correlation Function}",
      journal = {\apj},
         year = 2005,
        month = sep,
       volume = {630},
       number = {1},
        pages = {1-27},
          doi = {10.1086/431891},
archivePrefix = {arXiv},
       eprint = {astro-ph/0408569},
 primaryClass = {astro-ph},
       adsurl = {https://ui.adsabs.harvard.edu/abs/2005ApJ...630....1Z}
}

@ARTICLE{Zehavi+2011,
       author = {{Zehavi}, Idit and {Zheng}, Zheng and {Weinberg}, David H. and {Blanton}, Michael R. and {Bahcall}, Neta A. and {Berlind}, Andreas A. and {Brinkmann}, Jon and {Frieman}, Joshua A. and {Gunn}, James E. and {Lupton}, Robert H. and {Nichol}, Robert C. and {Percival}, Will J. and {Schneider}, Donald P. and {Skibba}, Ramin A. and {Strauss}, Michael A. and {Tegmark}, Max and {York}, Donald G.},
        title = "{Galaxy Clustering in the Completed SDSS Redshift Survey: The Dependence on Color and Luminosity}",
      journal = {\apj},
         year = 2011,
        month = jul,
       volume = {736},
       number = {1},
          eid = {59},
        pages = {59},
          doi = {10.1088/0004-637X/736/1/59},
archivePrefix = {arXiv},
       eprint = {1005.2413},
 primaryClass = {astro-ph.CO},
       adsurl = {https://ui.adsabs.harvard.edu/abs/2011ApJ...736...59Z}
}

@software{Crocce_2LPTIC,
       author = {{Crocce}, M. and {Pueblas}, S. and {Scoccimarro}, R.},
        title = "{2LPTIC: 2nd-order Lagrangian Perturbation Theory Initial Conditions}",
 howpublished = {Astrophysics Source Code Library, record ascl:1201.005},
         year = 2012,
        month = jan,
          eid = {ascl:1201.005},
archivePrefix = {ascl},
       eprint = {1201.005},
       adsurl = {https://ui.adsabs.harvard.edu/abs/2012ascl.soft01005C}
}

@software{Hahn&Abel_MUSIC,
       author = {{Hahn}, Oliver and {Abel}, Tom},
        title = "{MUSIC: MUlti-Scale Initial Conditions}",
 howpublished = {Astrophysics Source Code Library, record ascl:1311.011},
         year = 2013,
        month = nov,
          eid = {ascl:1311.011},
archivePrefix = {ascl},
       eprint = {1311.011},
       adsurl = {https://ui.adsabs.harvard.edu/abs/2013ascl.soft11011H}
}

@ARTICLE{White1994,
       author = {{White}, Simon D.~M.},
        title = "{Formation and Evolution of Galaxies: Les Houches Lectures}",
      journal = {arXiv e-prints},
         year = 1994,
        month = oct,
          eid = {astro-ph/9410043},
        pages = {astro-ph/9410043},
          doi = {10.48550/arXiv.astro-ph/9410043},
archivePrefix = {arXiv},
       eprint = {astro-ph/9410043},
 primaryClass = {astro-ph},
       adsurl = {https://ui.adsabs.harvard.edu/abs/1994astro.ph.10043W}
}

@ARTICLE{Barnes&Hut1986,
       author = {{Barnes}, Josh and {Hut}, Piet},
        title = "{A hierarchical O(N log N) force-calculation algorithm}",
      journal = {\nat},
         year = 1986,
        month = dec,
       volume = {324},
       number = {6096},
        pages = {446-449},
          doi = {10.1038/324446a0},
       adsurl = {https://ui.adsabs.harvard.edu/abs/1986Natur.324..446B}
}

@ARTICLE{Bagla2002,
       author = {{Bagla}, J.~S.},
        title = "{TreePM: A Code for Cosmological N-Body Simulations}",
      journal = {Journal of Astrophysics and Astronomy},
         year = 2002,
        month = dec,
       volume = {23},
       number = {3-4},
        pages = {185-196},
          doi = {10.1007/BF02702282},
archivePrefix = {arXiv},
       eprint = {astro-ph/9911025},
 primaryClass = {astro-ph},
       adsurl = {https://ui.adsabs.harvard.edu/abs/2002JApA...23..185B}
}

@ARTICLE{Fidler+2015,
       author = {{Fidler}, Christian and {Rampf}, Cornelius and {Tram}, Thomas and {Crittenden}, Robert and {Koyama}, Kazuya and {Wands}, David},
        title = "{General relativistic corrections to N -body simulations and the Zel'dovich approximation}",
      journal = {\prd},
         year = 2015,
        month = dec,
       volume = {92},
       number = {12},
          eid = {123517},
        pages = {123517},
          doi = {10.1103/PhysRevD.92.123517},
archivePrefix = {arXiv},
       eprint = {1505.04756},
 primaryClass = {astro-ph.CO},
       adsurl = {https://ui.adsabs.harvard.edu/abs/2015PhRvD..92l3517F}
}

@ARTICLE{Challinor+2011,
       author = {{Challinor}, Anthony and {Lewis}, Antony},
        title = "{Linear power spectrum of observed source number counts}",
      journal = {\prd},
         year = 2011,
        month = aug,
       volume = {84},
       number = {4},
          eid = {043516},
        pages = {043516},
          doi = {10.1103/PhysRevD.84.043516},
archivePrefix = {arXiv},
       eprint = {1105.5292},
 primaryClass = {astro-ph.CO},
       adsurl = {https://ui.adsabs.harvard.edu/abs/2011PhRvD..84d3516C}
}

@ARTICLE{Yoo+2014,
       author = {{Yoo}, Jaiyul},
        title = "{Relativistic effect in galaxy clustering}",
      journal = {Classical and Quantum Gravity},
         year = 2014,
        month = dec,
       volume = {31},
       number = {23},
          eid = {234001},
        pages = {234001},
          doi = {10.1088/0264-9381/31/23/234001},
archivePrefix = {arXiv},
       eprint = {1409.3223},
 primaryClass = {astro-ph.CO},
       adsurl = {https://ui.adsabs.harvard.edu/abs/2014CQGra..31w4001Y}
}

@ARTICLE{Aragon_calvo+2024,
       author = {{Aragon-Calvo}, M.~A.},
        title = "{Hierarchical reconstruction of the cosmic web, the H-Spine method}",
      journal = {\mnras},
         year = 2024,
        month = mar,
       volume = {529},
       number = {1},
        pages = {74-88},
          doi = {10.1093/mnras/stae468},
archivePrefix = {arXiv},
       eprint = {2308.16186},
 primaryClass = {astro-ph.CO},
       adsurl = {https://ui.adsabs.harvard.edu/abs/2024MNRAS.529...74A}
}

@ARTICLE{Alam+2019,
       author = {{Alam}, Shadab and {Zu}, Ying and {Peacock}, John A. and {Mandelbaum}, Rachel},
        title = "{Cosmic web dependence of galaxy clustering and quenching in SDSS}",
      journal = {\mnras},
         year = 2019,
        month = mar,
       volume = {483},
       number = {4},
        pages = {4501-4517},
          doi = {10.1093/mnras/sty3477},
archivePrefix = {arXiv},
       eprint = {1801.04878},
 primaryClass = {astro-ph.CO},
       adsurl = {https://ui.adsabs.harvard.edu/abs/2019MNRAS.483.4501A}
}

@ARTICLE{Arnold+1982,
       author = {{Arnold}, V.~I. and {Shandarin}, S.~F. and {Zeldovich}, Ia. B.},
        title = "{The large scale structure of the universe I. General properties. One-and two-dimensional models}",
      journal = {Geophysical and Astrophysical Fluid Dynamics},
         year = 1982,
        month = jan,
       volume = {20},
       number = {1},
        pages = {111-130},
          doi = {10.1080/03091928208209001},
       adsurl = {https://ui.adsabs.harvard.edu/abs/1982GApFD..20..111A}
}
\clearemptydoublepage
%

\end{document}